\documentclass{mn2e}

\usepackage[small]{caption}
\usepackage{graphicx}
\usepackage{multirow}
\usepackage{amssymb}
\usepackage{float}
\usepackage{lscape}

\title[Suzaku observations of iron lines]{A {\sl Suzaku} survey of Fe\,K lines in Seyfert 1 AGN}

\author[A. Patrick et al.]{A.R. Patrick$^{1}$, J.N. Reeves$^{1,2}$, D. Porquet$^{3}$, A.G. Markowitz$^{4,5,6}$, V. Braito$^{7,8}$, A.P. Lobban$^{1,7}$\\
$^{1}$Astrophysics Group, School of Physical Sciences, Keele University, Keele, Staffordshire, ST5 5BG, UK\\
$^{2}$Department of Physics, University of Maryland Baltimore County, MD 21250, USA \\
$^{3}$Observatoire astronomique de Strasbourg, Universit�e de Strasbourg, CNRS, UMR 7550, 11 rue de l'Universite, F-67000 Strasbourg, France\\
$^{4}$Dr.\ Karl Remeis-Sternwarte and Erlangen Centre for Astroparticle Physics,
Frederic-Alexander Universit\"{a}t Erlangen-N\"{u}rnberg, \\ 
7 Sternwartstrasse, 96049 Bamberg, Germany\\
$^{5}$Alexander von Humboldt Fellow \\
$^{6}$Center for Astrophysics and Space Sciences, University of California, San Diego, M.C. 0424, La Jolla, CA, 92093-0424, USA\\
$^{7}$X-Ray Astronomy Observational Group, Department of Physics and Astronomy, Leicester University, Leicester LE1 7RH\\
$^{8}$INAF - Osservatorio Astronomico di Brera, via E. Bianchi 46, 23807, Merate (LC) \\
}

\pagerange{\pageref{firstpage}--\pageref{lastpage}} \pubyear{2012}

\begin{document}

\maketitle
\begin{abstract}
We construct full broadband models in an analysis of {\sl Suzaku} observations of nearby Seyfert 1 AGN ($z\leq0.2$) with exposures $>50$\,ks and with greater than 30000 counts in order to study their iron line profiles. This results in a sample of 46 objects and 84 observations. After a full modelling of the broadband {\sl Suzaku} and {\sl Swift}-BAT data (0.6-100\,keV) 
we find complex warm absorption is present in $59\%$ of the objects in this sample which has a significant bearing upon the derived Fe\,K region parameters. Meanwhile $35\%$ of the 46 objects require some degree of high column density partial coverer in order to fully model the hard X-ray spectrum. We also find that a large number of the objects in the sample ($22\%$) require high velocity, high ionization outflows in the Fe\,K region resulting from Fe\,{\rm XXV} and Fe\,{\rm XXVI}. A further four AGN feature highly ionized Fe\,K absorbers consistent with zero outflow velocity, making a total of 14/46 ($30\%$) AGN in this sample showing evidence for statistically significant absorption in the Fe\,K region. 

Narrow Fe\,K$\alpha$ emission from distant material at 6.4\,keV is found to be almost ubiquitous in these AGN. Examining the 6-7\,keV Fe\,K region we note that narrow emission lines originating from Fe\,{\rm XXV} at 6.63-6.70\,keV and from Fe\,{\rm XXVI} at 6.97\,keV are present in $52\%$ and $39\%$ of objects respectively. 

Our results suggest statistically significant relativistic Fe\,K$\alpha$ emission is detected in 23 of 46 objects ($50\%$) at $>99.5\%$ confidence, measuring an average emissivity index of $q=2.4\pm0.1$ and equivalent width $EW=96\pm10$\,eV using the \textsc{relline} model. When parameterised with a Gaussian we find an average line energy of $6.32\pm0.04$\,keV, $\sigma_{\rm width}=0.470\pm0.05¤$\,keV and $EW=97\pm19$\,eV. Where we can place constraints upon the black hole spin parameter $a$, we do not require a maximally spinning black hole in all cases.
\end{abstract}


\section{Introduction}
The analysis of the X-ray spectra of AGN can reveal information regarding the inner regions of the accretion disc, the AGN environment as a whole and subsequently the super massive black hole (SMBH) at its heart. It was suggested by Fabian et al. (1989) that emission occurring from the very inner regions of the accretion disc may be visible and subsequently broadened by Doppler motions and relativistic affects. The majority of AGN spectra show narrow line emission from neutral iron at 6.4\,keV (Fe\,K$\alpha$) likely originating from distant material e.g. the torus or the outer regions of the accretion disc (Krolik \& Kallman 1987; Nandra 2006), particularly strong due to the high abundance and fluorescent yield of iron. If Fe\,K$\alpha$ emission additionally arises from material close to the central SMBH it will become relativistically broadened (Fabian et al. 1989; Laor 1991), producing a both blue and red-wings to the traditionally narrow line profile. 

In some AGN spectra this relativistic Fe\,K$\alpha$ emission may be strong enough to be observed allowing its shape and strength measured using disk-line emission models such as \textsc{laor}, \textsc{kyrline}, \textsc{kerrdisk} and \textsc{relline} (Laor 1991; Dov\v{c}iak et al. 2004; Brenneman \& Reynolds 2006; Dauser et al. 2010). The application of these models allows properties such as the inclination and emissivity index of the disc to be measured in addition to the typical inner radius of emission and in some cases the spin of the central SMBH (Nandra et al. 2007; Patrick et al. 2011a). Gaining information regarding the distribution of SMBH spins is an essential tool in aiding our understanding of galaxy evolution and distinguishing between models such as prolonged or chaotic accretion and also the effect of mergers upon the SMBH spin (Hughes \& Blandford 2003; Volonteri et al. 2005; King \& Pringle 2007; Rezzolla et al. 2008). A spin distribution skewed towards higher values ($a\sim0.998$) would suggest prolonged accretion, whereas low SMBH spin ($a\sim0$) would indicate chaotic accretion models are more appropriate. In addition to this, the magnetic extraction of BH rotational energy through the Blandford-Znajek effect (Blandford \& Znajek 1977) could cause a reduction in the spin (i.e. towards zero) of the SMBH in some AGN (Berti \& Volonteri 2008). 


Recent publications have made steps towards making spin estimates of SMBHs in a variety of AGN, including those which feature low levels of intrinsic absorption thereby offering the simplest spectrum to analyse, avoiding complications regarding the degree of spectral curvature introduced with warm absorbing zones (Miniutti et al. 2009; Schmoll et al. 2009; Patrick et al. 2011a; Emmanoulopoulos et al. 2011). More complex AGN spectra have also been analysed and revealed further spin estimates e.g. MCG--06-30-15 (Miniutti et al. 2007; Patrick et al. 2011b), Mrk 79 (Gallo et al. 2011) and NGC 3783 (Brenneman et al. 2011; Patrick et al. 2011b). However, as discussed in Patrick et al. (2011a), the estimated SMBH spin is highly model dependent and strongly related to the treatment of features such as the soft excess or any intrinsic absorbing zones. Assuming a Comptonization origin of the soft excess results in a range of low to intermediate spins, whereas using a high degree of relativistic blurring to smooth the discrete soft emission lines into a continuum typically forces the spin to near maximal values and requires very high disc emissivities. 

This paper includes a sample of AGN from the public {\sl Suzaku} archive of all observations of Seyfert 1 AGN with total exposures $>50$\,ks and more than 30\,000 counts in order to increase the likelihood of detection and broadened emission from the inner regions if it is present. {\sl Suzaku} is the ideal instrument with which to do this work since it allows us to gather both soft and hard X-ray data simultaneously using the X-ray Imaging Spectrometer (Koyama et al. 2007) and Hard X-ray Detector (Takahashi et al. 2007) detectors which, when combined with further non-simultaneous hard X-ray data from {\sl Swift} BAT, gives a broad energy bandpass of 0.6-100.0\,keV. The crucial difference being that we can obtain data regarding the strength of the Compton reflection hump at $\sim30$\,keV (George \& Fabian 1991) which is beyond the capabilities of other current X-ray observatories. Only with hard X-ray data can the strength of the reflection component be appropriately constrained and hence its contribution to the Fe\,K region assessed prior to attempting to determine broadening in the Fe\,K region and eventually estimates upon SMBH spin. 

This is the final paper in a series of three in which a methodical and relatively uniform approach has been taken in an attempt to constrain accretion disc properties and SMBH spin from the Fe\,K regions from an analysis of the X-ray spectra of Seyfert 1 AGN. In Patrick et al. (2011a) a small sample of six 'bare' Seyfert 1 AGN (i.e. those featuring low intrinsic absorption: Ark 120, Fairall 9, MCG--02-14-009, Mrk 335, NGC 7469 and SWIFT J2127.4+5654) was analysed, finding that narrow ionized emission lines such as Fe\,{\rm XXVI} are relatively common (4/6 objects) while the emissivity index of the accretion disc indicated that no strongly centrally concentrated emission was required to model any relativistic broadening in the Fe\,K region with an average of $q\sim2.3\pm0.2$. 

Patrick et al. (2011b) undertook and analysis of high quality, long exposure ($>200$\,ks) observations of Seyfert 1 AGN with {\sl Suzaku} (Fairall 9, MCG--06-30-15, NGC 3516, NGC 3783 and NGC 4051) making use of the full 0.6-100.0\,keV bandpass in order to fully account for any warm absorber component to give the best possible opportunity to make estimates upon SMBH spin, also finding a low to moderate average emissivity index of $q\sim2.8\pm0.2$.  
Other authors have also made some progress towards making spin estimates on Seyfert 1 AGN e.g. Miniutti et al. (2009), Schmoll et al. (2009), Nardini et al. (2011), Gallo et al. (2011), Emmanoulopoulos et al. (2011) and Brenneman et al. (2011). 

The main aim of this paper is to assess the properties and total percentage of AGN which have been observed with {\sl Suzaku} that show evidence for broadened line emission in the Fe\,K region from the inner regions of the accretion disc resulting from an analysis of the broad-band X-ray spectrum. In this paper we expand our broad-band spectral analysis with {\sl Suzaku} to all the currently archived type 1-1.9 AGN, which have at least a 50\,ks total exposure and 30000 XIS band counts, sufficient for a broad band spectral analysis. This enables us to measure the iron line and reflection properties of a more substantial sample of 46 type 1 AGN, allowing the overall properties of the iron line and accretion disc to be investigated. We also aim to investigate {\sl Suzaku's} view of ionised emission and absorption lines in the Fe\,K region and the occurrence or warm absorbers, highly ionised absorbers and partially covering absorbers and the subsequent effects upon the AGN X-ray spectrum.


\begin{table*}
\caption{The {\sl Suzaku} Seyfert sample}
\begin{tabular}{l c c c c}
\hline
Object  & RA (J2000) & Dec (J2000) & Redshift & $N_{H}$ (Gal) ($10^{22}$cm$^{-2}$) \\ 
\hline
1H 0419--577 & 04 26 00.8 & --57 12 00.4 & 0.1040 & 0.0126 \\
3C 111 & 04 18 21.3 & +38 01 35.8 & 0.0485 & 0.2910 \\
3C 120 & 04 33 11.1 & +05 21 15.6 & 0.0330 & 0.1060 \\
3C 382 & 18 35 03.4 & +32 41 46.8 & 0.0579 & 0.0714 \\
3C 390.3 & 18 42 09.0 & +79 46 17.1 & 0.0561 & 0.0347 \\
3C 445 & 22 23 49.5 & --02 06 12.9 & 0.0559 & 0.0559 \\
4C 74.26 & 20 42 37.3 & +75 08 02.4 & 0.1040 & 0.1160 \\
Ark 120 & 05 16 11.4 & --00 08 59.4 & 0.0327 & 0.0978 \\
Ark 564 & 22 42 39.3 & +29 43 31.3 & 0.0247 & 0.0534 \\
Fairall 9 & 01 23 45.8 & --58 48 20.5 & 0.0470 & 0.0316 \\
IC 4329A & 13 49 19.3 & --30 18 34.0 & 0.0161 & 0.0461 \\
IRAS 13224--3809 & 13 25 19.4 & --38 24 52.7 & 0.0658 & 0.0534 \\
MCG--02-14-009 & 05 16 21.2 & --10 33 41.4 & 0.0285 & 0.0924 \\
MCG--02-58-22 & 23 04 43.65 & --08 41 08.6 & 0.0649 & 0.0291 \\
MCG--05-23-16 & 09 47 40.2 & --30 56 55.9 & 0.0085 & 0.0870 \\
MCG--06-30-15 & 13 35 53.8 & --34 17 44.1 & 0.0077 & 0.0392 \\
MCG+8-11-11 & 05 54 53.6 & +46 26 21.6 & 0.0205 & 0.1840 \\
MR 2251--178 & 22 54 05.8 & --17 34 55.0 & 0.0640 & 0.0640 \\
Mrk 79 & 07 42 32.8 & +49 48 34.8 & 0.0222 & 0.0527 \\
Mrk 110 & 09 25 12.9 & +52 17 10.5 & 0.0353 & 0.0130 \\
Mrk 205 & 12 21 44.0 & +75 18 38.5 & 0.0708 & 0.0280 \\
Mrk 279 & 13 53 03.5 & +69 18 29.6 & 0.0305 & 0.0152 \\
Mrk 335 & 00 06 19.5 & +20 12 10.5 & 0.0258 & 0.0356 \\
Mrk 359 & 01 27 32.5 & +19 10 43.8 & 0.0174 & 0.0426 \\
Mrk 509 & 20 44 09.7 & --10 43 24.5 & 0.0344 & 0.0344 \\
Mrk 766 & 12 18 26.5 & +29 48 46.3 & 0.0129 & 0.0178 \\
Mrk 841 & 15 04 01.2 & +10 26 16.2 & 0.0364 & 0.0222 \\ 
NGC 1365 & 03 33 36.4 & --36 08 25.5 & 0.0055 & 0.0134 \\
NGC 2992 & 09 45 42.1 & --14 19 35.0 & 0.0077 & 0.0487 \\
NGC 3147 & 10 16 53.7 & +73 24 02.7 & 0.0093 & 0.0285 \\
NGC 3227 & 10 23 30.6 & +19 51 54.2 & 0.0039 & 0.0199 \\
NGC 3516 & 11 06 47.5 & +72 34 06.9 & 0.0088 & 0.0345 \\
NGC 3783 & 11 39 01.7 & -37 44 18.9 & 0.0097 & 0.0991 \\
NGC 4051 & 12 03 09.6 & +44 31 52.8 & 0.0023 & 0.0115 \\
NGC 4151 & 12 10 32.6 & +39 24 20.6 & 0.0033 & 0.0230 \\
NGC 4593 & 12 39 39.4 & --05 20 39.3 & 0.0090 & 0.0189 \\
NGC 5506 & 14 13 14.9 & --03 12 27.3 & 0.0062 & 0.0408 \\
NGC 5548 & 14 17 59.5 & +25 08 12.4 & 0.0172 & 0.0155 \\
NGC 7213 & 22 09 16.3 & --47 09 59.8 & 0.0058 & 0.0106 \\
NGC 7314 & 22 35 46.2 & --26 03 01.7 & 0.0048 & 0.0150 \\
NGC 7469 & 23 03 15.6 & +08 52 26.4 & 0.0163 & 0.0445 \\
PDS 456 & 17 28 19.8 & --14 15 55.9 & 0.1840 & 0.1960 \\
PG 1211+143 & 12 14 17.7 & +14 03 12.6 & 0.0809 & 0.0274 \\
RBS 1124 & 12 31 36.4 & +70 44 14.1 & 0.2080 & 0.0152 \\
SWIFT J2127.4+5654 & 21 27 45.0 & +56 56 39.7 & 0.0144 & 0.7650 \\
TON S180 & 00 57 19.9 & --22 22 59.1 & 0.0620 & 0.0136 \\
\hline
\end{tabular}
\label{tab:sample}
\end{table*} 

\section{Observations \& Data Reduction}
\subsection{Observations and sample selection}
The objects included within this sample are listed in Table \ref{tab:sample} and are all the Seyfert 1-1.9 AGN with exposures $>50$\,ks and greater than 30\,000 0.6-10.0\,keV counts which have been observed with {\sl Suzaku} with data publically available in the {\sl Suzaku} data archive\,\footnote{http://heasarc.gsfc.nasa.gov/} as of 09/11. We also include data from some type 1 radio-loud (BLRGs -- non-Blazar) AGN provided they fit the above exposure and count criteria. High energy X-ray data from {\sl Swift}/BAT from the 58 month BAT catalogue is also used in addition to that obtained from the HXD detector onboard {\sl Suzaku} (but allowing the relative cross-normalisation to vary), therefore the total energy range covered is 0.6-100.0\,keV. Details of the observations included are listed in Table \ref{tab:observations}. 

In some instances, objects may have been observed on numerous occasions, provided that there is little variation between the data sets they are combined and a single analysis is performed. If the observations do indeed vary a separate analysis is performed on each data set, although with similar model components and inferred geometries where possible, e.g. the inclination angle of the accretion disc would be linked between observations. 


\subsection{Data reduction}
The {\sl Suzaku} data in this paper were reduced using version 6.10 of the HEASOFT data reduction and analysis package. The XIS source spectra were extracted using 3.0\arcmin circular regions centred  on the source. Background spectra were also extracted using 3.0\arcmin circular regions, this time centred on a region of the CCD not featuring any of the source or Fe\,55 calibration regions. Only data from the front-illuminated XIS\,0, 2 and 3 detectors were used due to their greater sensitivity at Fe\,K energies, however data from the back-illuminated XIS\,1 remains consistent with the front-illuminated detectors. Observations since November 2006, however, do not include data from XIS\,2 as it is no longer operational.  

Objects which have been observed on multiple occasions but where the X-ray spectra (or spectral shape) does not appear have varied by a significant amount (e.g. other than simple changes in power law normalisation) are combined into a single data file and then a suitable model is fit to the data. This is the case for IC 4329A, MCG--06-30-15, Mrk 509, Mrk 841, NGC 2992 and NGC 5548. However, if there are discernible differences between multiple observations of the same object (e.g. differing spectra shapes or changes in absorption), these data sets are analysed separately while retaining as many similar model components as possible. Such an analysis is performed for Fairall 9, NGC 1365, NGC 3227, NGC 3516, NGC 3783 and NGC 4051. Both of the 2006 observations of NGC 5506 are combined into a single spectrum while the 2007 observation is kept separate. Similarly, the later three observations of 3C 120 are combined whereas the first observation (OBSID: 700001010, see Table \ref{tab:observations}) is analysed separately. MCG--06-30-15 was observed on three occasions in January 2006, this paper makes use of all three observations and the time averaged spectrum is used in the main analysis. Analysis of the individual observations yields Fe\,K line profiles which were consistent within errors and as such all three of the January 2006 observations have been coadded. Systematic errors are not included in datasets other than NGC 1365 in which systematic errors are set at 2\%.

\begin{figure}
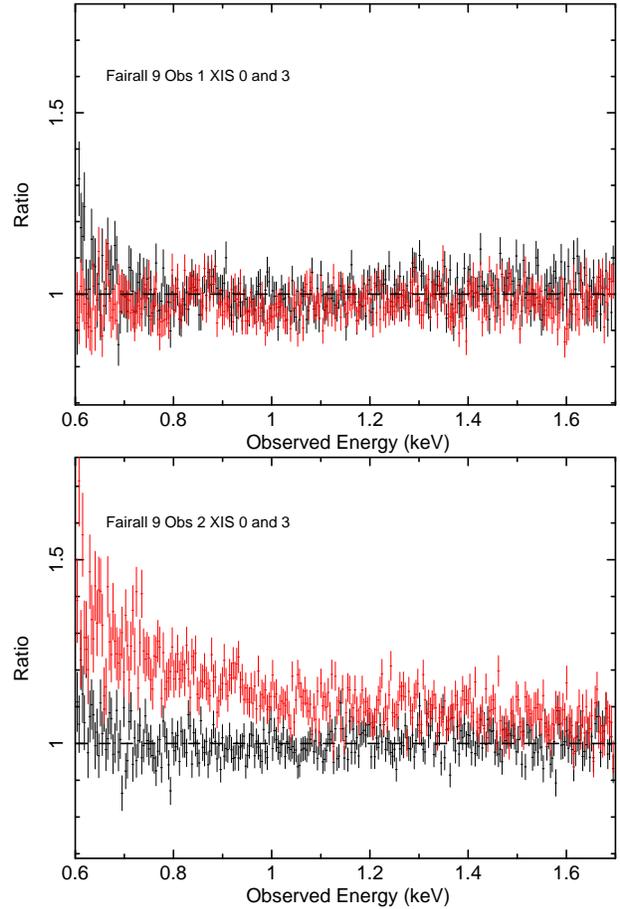

\rotatebox{-90}{\includegraphics[width=6cm]{xis03_nodiverge.ps}}
\rotatebox{-90}{\includegraphics[width=6cm]{xis03_diverge.ps}}
\caption{Ratio plots of 0.6-1.0\,keV of XIS 0 (red) and XIS 3 (black) for the 2007 (upper panel) and 2010 (lower panel) observations of Fairall 9. These show the contamination which has recently affected XIS 0 at soft X-ray energies causing a divergence with XIS 3, note that this is not apparent in the first observation. The soft X-ray spectrum was modelled with a basic (\textsc{powerlaw+compTT})*\textsc{wabs} model.}
\label{fig:XIS03_diverge}
\end{figure}

\begin{figure}
\rotatebox{-90}{\includegraphics[width=6cm]{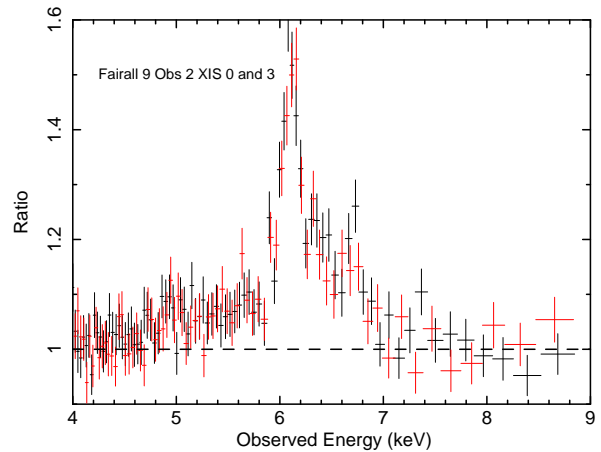}}
\caption{Ratio plot of the 4-9\,keV Fe\,K regions of XIS 0 (red) and XIS 3 (black) for the 2010 observation of Fairall 9. Indicating that while contamination has a significant effect at soft X-ray energies, the Fe\,K region remains entirely consistent between XIS 0 and XIS 3.}
\label{fig:XIS03_FeK}
\end{figure}

The HEASOFT tool \textsc{xisrmfgen} was used to generate the XIS redistribution matrix files (rmf) and the ancillary response files were generated using \textsc{xissimarfgen}. The data from each of the front-illuminated XIS detectors were then co-added using \textsc{mathpha}, \textsc{addrmf} and \textsc{addarf} in order to increase signal to noise. We ignore all XIS data below 0.6\,keV, above 10.0\,keV and between 1.7-2.0\,keV due to calibration uncertainties of the detectors around the Si\,K edge. 
Some of the most recent observations with {\sl Suzaku} (2011 onwards) show evidence for contamination on the CCD of XIS 0 which primarily affects the soft X-ray energies resulting in a divergence of XIS 0 with XIS 3 below 1\,keV. This is notable in the two observations of Fairall 9; both XIS 0 and 3 are consistent in the 2007 observation, however, the 2010 observation displays a notable difference between XIS 0 and 3 (see Figure \ref{fig:XIS03_diverge}). In this analysis, we do not co-add XIS 0 and 3 for the contaminated Fairall 9 observation, instead preferring to include them as separate data sets although limiting XIS 0 to $>1.0$\,keV, thus retaining some of the sensitivity at Fe\,K energies. Both detectors, however, remain consistent at Fe\,K energies, see Figure \ref{fig:XIS03_FeK}.

Spectra from the HXD were extracted from the cleaned HXD/PIN events files and subsequently corrected for instrument dead time using the tool \textsc{hxddtcor}. The tuned HXD/PIN non X-ray background (NXB) events files were used for background subtraction (Fukazawa et al. 2009) and generated with $10\times$ the actual background rate in order to reduce photon noise, also with identical good time intervals (GTIs) as used in the source events. A simulated cosmic X-ray background was also produced for each observation using XSPEC v 12.6.0q with a spectral form identical to that used in Gruber et al. (1999), this was then added to the corrected NXB to form a single background file for each observation. The appropriate response and flat field files were used from the {\sl Suzaku} CALDB suitable for the respective epochs of each observation. HXD data is used over the 15.0-60.0\,keV range along with the appropriate PIN/XIS-FI cross-normalisation according to the epoch and nominal pointing position\,\footnote{http://heasarc.gsfc.nasa.gov/docs/suzaku/analysis/watchout.html}. 

While the BAT data from {\sl Swift} are not necessarily simultaneous with the {\sl Suzaku} observations and is of relatively low statistical weight, we find throughout the paper that the BAT spectral indices are consistent with those obtained from the XIS and HXD/PIN data. That is, due to variability of the reflection fraction over time scales of years the spectral shape may vary and hence may not be equivalent to the spectral shape of the hard X-rays obtained with {\sl Suzaku}. However, removing the BAT data and refitting still yields the original photon index $\Gamma$ as well as the other spectral parameters and there appears to be little evidence that the spectral shape $>15$\,keV varies over the timescales considered here. It is therefore appropriate to consider that there is a simple change in the normalisation of the hard X-ray component (as accounted for by the cross-normalisation between XIS and BAT being allowed to vary within the models) and the inclusion of BAT data from {\sl Swift} serves to improve our view of the X-ray spectral shape above 10\,keV. The BAT data therefore improves the accuracy to which we can estimate the powerlaw continuum and the reflection component.

\section{Analysis \& Results}
Spectral analysis and model fitting is performed from within XSPEC v 12.6.0q (Arnaud 1996), all models are modified by Galactic absorption which is accounted for by the \textsc{wabs} multiplicative model (Morrison \& McCammon 1983). The respective Galactic column densities were obtained using the \textsc{nh} {\sl ftool} for each source giving the weighted average $N_{\rm H}$ value of the LAB Survey of Galactic H\,{\rm I} (Kalberla et al. 2005), using abundances from Anders \& Grevesse (1989). Data are fit over the full 0.6-100.0\,keV range available, excluding those regions affected by the uncertainties in the XIS calibration mentioned above. The $\chi^{2}$ minimisation technique is used throughout, all errors are quoted at the 90\% confidence level ($\triangle\chi^{2}=2.71$ for one interesting parameter) and include statistical and not instrumental systematic errors. Where the significance of components is quoted in terms of $\triangle\chi^{2}$, the component in question has been removed from the model and the data refit to ensure that the order in which components are added to the model does not affect the quoted statistical significance. Throughout this paper, a positive $\triangle\chi^{2}$ corresponds to a poorer fit, whereas a negative $\triangle\chi^{2}$ corresponds to an improvement in the fit. 

\subsection{Spectral models}
The data used in this paper are selected such that there are a sufficient number of counts for detailed spectral analysis to be conducted. The wide 0.6-100.0\,keV energy bandpass as used in this sample means that a full treatment must be given to all the main components of the X-ray spectrum of these Seyfert 1 AGN, namely the soft excess, Compton reflection, Galactic absorption, warm absorbers and any highly ionized outflows or relativistic broadening which may be present. In order to model the X-ray spectra in this sample as consistently and uniformly as possible we construct models following the criteria in the following sections. 

Model components are added according the residuals in the data and the statistical requirement of such a component. For example, a model to account for the soft excess is considered to be required, provided that its inclusion exceeds the 90\% confidence level according to $\triangle\chi^{2}$ given the appropriate number of free parameters. For example, an \textsc{xstar} grid typically has two: column density and ionisation parameter although some objects may require a third due to an outflow velocity.

\subsubsection{Modelling of the soft excess}
A number of AGN spectra show an excess over a simple \textsc{powerlaw} at lower X-ray energies, typically $<2.0$\,keV, this relatively common feature has been termed the 'soft excess'. The origin of the soft excess is as yet unknown, however, there are a number of different interpretations and methods with which to model it. Perhaps the most basic method of modelling the soft excess is as a black body to replicate direct thermal emission from successive annuli of the accretion disc (Malkan \& Sargent 1982). The inferred constant temperature of the soft excess component (which scales with $M_{\rm BH}^{-1/4}$) is in disagreement with the typical accretion disc properties of a SMBH and is arguably more suitable for an intermediate mass BH (Gierli\'{n}ski \& Done 2004). 

A variation upon this concept is that the soft excess is produced by the Compton up-scattering of EUV seed photons from the disc in a hot plasma or corona lying above the disc. For example, the \textsc{compTT} model (Titarchuk 1994) allows for a large variation in photon seed temperature while still producing a relatively constant output photon temperature as is required by many of the observed soft excesses in AGN. This method has been successful in modelling the soft excess of a large number of AGN, including a sample of PG quasars (Porquet et al. 2004) and in Mrk 509 by Mehdipour et al. (2011). This interpretation of the soft excess is also used throughout this paper, consistent with the method in Patrick et al. (2011a,b); similar results can also be obtained with a second steep \textsc{power law} component. 

The above methods assume a smooth shape to account for the soft excess, however alternatives have been suggested e.g. an atomic origin (Brenneman \& Reynolds 2006). Soft X-ray emission lines emitted from regions close to the central SMBH are relativistically blurred and broadened, much in the same way as the often observed broad 6.4\,keV Fe\,K$\alpha$ emission line and its red-wing. A series of discrete emission line features are then relativistically blurred to such an extent that they merge to form a smooth continuum (Ross, Fabian \& Ballantyne 2002). A number of AGN have been modelled in this fashion e.g. Ark 120, Fairall 9, Mrk 335 \& RBS 1124 (Schmoll et al. 2009; Miniutti et al. 2010; Nardini et al. 2011; Patrick et al. 2011a), although a significant amount of relativistic blurring is required to smooth the characteristically narrow features into a broad continuum forces the accretion disc emissivity index to high values ($q\gtrapprox4.5$) and SMBH spin towards near maximal ($a\sim0.998$). Indeed Schmoll et al. (2009) note that ignoring the XIS data below 2\,keV (to avoid fitting the soft excess) relaxes the spin constraint to low values. These parameters are often at odds with those similarly derived from the asymmetric line profile from the Fe\,K region, for example Miniutti et al. (2010) find $q=4.1^{+5.3}_{-0.9}$ and $a\geq0.6$  in RBS 1124 while the authors note that there is no evidence for a strong broad line in the Fe\,K region. However, given the assumption that the material responsible for both the soft emission lines and Fe\,K$\alpha$ emission is located in the same region, it is logical to assume that estimated accretion disc and SMBH parameters should be consistent (Patrick et al. 2011a). 

Note that relativistic blurring of the reflection component at soft X-ray energies can still contribute towards the soft excess, however, the soft X-ray flux from relativistically blurred emission of the inner regions is unlikely to prove particularly strong in comparison to the flux from the Comptonization component due to any residuals in the Fe\,K region driving the fit with the majority of the soft X-ray flux already accounted for with the \textsc{compTT} model. Therefore, if the Fe\,K line is to be used as a diagnostic for determining accretion disc parameters and constraining SMBH spin, the treatment of the soft excess has a significant role to play in ensuring that it is not driving the fit if blurred reflection models are used in the fitting process. As our aim is to parameterise the properties of the disc and reflection based on the Fe\,K line profile, independent of the nature of the soft excess, we retain this approach of the Comptonization origin of the soft excess here. 



\subsubsection{Compton reflection and emission lines}

As the narrow iron K$\alpha$ line is ubiquitous in virtually all these AGN, an {\sl unblurred} \textsc{reflionx} (Ross \& Fabian 2005) component, measured down to an ionisation parameter of $\xi=1$ and additional narrow fluorescent emission lines representative of reflection off distant material (e.g. the torus or outer regions of the accretion disc) are included in all AGN spectra showing evidence for a hard excess. Previous studies of the iron line regions in Seyfert AGN by Bianchi et al. (2004), Nandra et al. (2007) and Patrick et al. (2011a,b) also suggest that ionized species of iron are relatively common at energies of 6.7\,keV and 6.97\,keV for Fe\,{\rm XXV} and Fe\,{\rm XXVI} respectively. Neutral narrow Fe\,K$\alpha$ and the accompanying Fe\,K$\beta$ emission at 6.4\,keV and 7.056\,keV respectively are ubiquitous in AGN spectra (Nandra et al. 2007).

As stated above, in this paper we use the \textsc{reflionx} model to account for the distant near neutral reflection continuum with the input photon index $\Gamma$ tied to that of the intrinsic \textsc{powerlaw}. Although soft emission lines are included in \textsc{reflionx}, these are also added on an ad-hoc basis when required and as such are modelled using narrow Gaussians of fixed width $\sigma=0.01$\,keV e.g. O\,{\rm VIII} or Ne\,{\rm IX} emission from distant photo-ionized gas e.g. the BLR or NLR. Fe\,K$\beta$ emission at 7.056\,keV with flux $F_{\rm K\beta}=0.13\times\,F_{\rm K\alpha}$ is not included self-consistently in \textsc{reflionx} which is therefore modelled using a narrow Gaussian of fixed width $\sigma=0.01$\,keV and with flux fixed at the value obtained during an initial parametrisation of the Fe\,K region and the narrow Fe\,K$\alpha$ flux when modelled with a Gaussian. Ionized lines in the Fe\,K region (such as Fe\,{\rm XXV} at $\sim6.63-6.7$\,keV and Fe\,{\rm XXVI} at 6.97\,keV) are accounted for with narrow Gaussians ($\sigma=0.01$\,keV) if and when required, although in some circumstances these can be indistinguishable from the blue-wing of a relativistically broadened Fe\,K$\alpha$ line profile. 

While the iron abundance $Z_{\rm Fe}$ is left free to vary, in some AGN the strength of the narrow Fe\,K$\alpha$ core may force $Z_{\rm Fe}$ to unfeasibly high values (e.g. $Z_{\rm Fe}\gtrapprox3$) while improperly modelling the hard X-ray reflection spectrum due to the greater number of counts in the 5-6\,keV region. To avoid this scenario, we fix $Z_{\rm Fe}$ at Solar abundance and add an additional narrow Gaussian of fixed width ($\sigma=0.01$\,keV, to prevent interference with any broad component from the disc) at 6.4\,keV to model the Fe\,K$\alpha$ core while maintaining a good fit to the HXD and BAT data. Rather than simply being an ad-hoc solution, this could be representative of Fe\,K$\alpha$ emission additionally arising from Compton-thin matter such as the BLR or NLR as well as, for example scattering off a distant Compton-thick torus. 

\subsubsection{Warm absorption}
The X-ray spectra of many AGN feature one or more zones of warm absorbing 
gas, while primarily affecting the spectrum at soft X-ray energies, with 
higher column densities (e.g. $>10^{22}$\,cm$^{-2}$) it can add subtle 
curvature above 2.5\,keV. Previous studies (e.g. Miniutti et al. 2007; 
Nandra et al. 2007) restrict their analysis to 2.5-10.0\,keV to avoid 
complications with the warm absorber, instead choosing to focus in 
upon the Fe\,K region. However, as found in Reeves et al. (2004), 
Turner et al. (2005), Miller et al. (2009) and Patrick et al. (2011b) 
effects of the warm absorber extend even to the Fe\,K region, 
contributing significantly to the strength of the observed red-wing below 
6.4\,keV. For example, in MCG--06-30-15 Miniutti et al. (2007) model the 
time averaged January 2006 spectra without absorption, resulting in an 
apparent strong broad Fe\,K$\alpha$ line and near maximal SMBH spin; 
while a full treatment of the warm absorber appears to reduce the 
strength of the broad component and subsequently BH spin to more 
intermediate values (Zycki et al. 2010; Patrick et al. 2011b).

To model the soft X-ray warm absorber components in this paper, we use 
an \textsc{xstar} (Kallman et al. 2004) generated grid illuminated by an  
X-ray photon index of $\Gamma=2.0$, in agreement with the mean values 
found in radio-quiet type I AGN (Scott et al. 2011). 
The abundances are fixed at Solar values (except Ni, 
which was set to zero), where \textsc{xstar} uses the values of 
Grevesse \& Sauval (1998). The turbulent velocity is set to
$100\,{\rm km}\,{\rm s}^{-1}$. 
The grid is well suited to 
accounting for a variety of absorption zones due to its wide range in 
column density
($5\times10^{18}{\rm cm}^{-2}<N_{\rm H}<5\times10^{24}{\rm cm}^{-2}$, 
in steps of $\Delta N_{\rm H} = 10^{19}$\,cm$^{-2}$) 
and ionization parameter ($0<{\rm log}\,\xi<5$, with 
grid steps $\Delta(\log \xi) = 0.5$). An electron density of 
$n_{\rm e}=10^{10}$\,cm$^{-3}$ is assumed, although the data
are largely insensitive to the density at CCD resolution. 
The computed \textsc{xstar} spectra were generated over the energy range 
from $0.05-120$\,keV, so are suited to broad-band spectra, especially 
at hard X-ray energies.

During the fitting 
process absorption zones are added as required, in some objects more 
than one zone may be statistically required and in exceptional 
circumstances a grid with a higher turbulent velocity may be 
required to model highly ionized absorption in the Fe\,K region (see below), 
as at low turbulences the iron K absorption can become easily saturated. 
Absorption zones are added until a good overall fit is found 
to the data and there are no clear residuals remaining in the soft X-ray 
spectrum.

\subsubsection{Highly ionized absorption}
Further to typical warm absorption, highly ionized absorbing zones of gas (where present) also play an important role in the Fe\,K region and in the determination of the measured line parameters and strength of the observed red-wing (Reeves et al. 2004; Turner et al. 2005). Absorption lines indicative of such zones are fairly common in X-ray spectra of AGN e.g. 1s-2p resonance lines from Fe\,{\rm XXV} and/or Fe\,{\rm XXVI} at their rest-frame energies, i.e. 6.7\,keV and 6.97\,keV respectively, although there is evidence for {\sl blue-shifted} absorption lines in many AGN (Braito et al. 2007; Tombesi et al. 2010a; Lobban et al. 2011). 

In the event that absorption features in the Fe\,K region are found, which 
are not adequately modelled by the above low turbulence grid, then 
these are accounted for using a model representative of a 
highly ionized absorption zone. We use an \textsc{xstar} generated grid 
with a turbulent velocity of $1000\,{\rm km}\,{\rm s}^{-1}$, with an 
input X-ray photon index of $\Gamma=2.0$ and using the input SED from 
Tombesi et al. (2011). This grid covers a range in column density and 
ionization parameter of 
$1\times10^{20}{\rm cm}^{-2}<N_{\rm H}<1\times10^{24}{\rm cm}^{-2}$ 
(in steps $\Delta N_{\rm H} = 6\times10^{19}$\,cm$^{-2}$)
and $0<{\rm log}\,\xi<6$ (with $\Delta (\log \xi) = 0.32$) respectively. 
The electron density assumed is also $n_{\rm e}=10^{10}$\,cm$^{-3}$.

\begin{table*}
\caption{Model applicability to each AGN. Comparison of components and features use to model each AGN; \textsc{compTT} is used to model the soft excess, an unblurred \textsc{reflionx} is used to model the reflection component if statistically required. $^{1}$ Only an increase in the neutral Galactic column is required. $^{2}$ No \textsc{xstar} grid is required in MCG--05-23-16, however additional neutral absorption is used to absorb only the intrinsic \textsc{powerlaw}. Residuals in the Fe\,K region are considered to be present if there is an improvement of $\triangle\chi^{2}>6.3$ with the addition of a broad Gaussian (see Table \ref{tab:broad}). $^{3}$ Only required in the 2006 observation of Mrk 766.}
\begin{tabular}{l c c c c c c }
\hline
Object  & Soft excess & Reflection & Partial covering & Warm absorber & High $\xi$ absorption & Residuals at Fe\,K \\ 
\hline
1H 0419--577 & \checkmark & \checkmark & \checkmark & \checkmark & X & X \\
3C 111 & X & X & X & X$^{1}$ & \checkmark & \checkmark \\
3C 120 & \checkmark & \checkmark & X & X$^{1}$ & X & \checkmark \\
3C 382 & \checkmark & \checkmark & X & \checkmark & X & \checkmark \\
3C 390.3 & \checkmark & \checkmark & X & X & X & \checkmark \\
3C 445 & X & \checkmark & \checkmark & \checkmark & \checkmark & \checkmark \\
4C 74.26 & X & \checkmark & X & \checkmark & X & \checkmark \\
Ark 120 & \checkmark & \checkmark & X & X & X & \checkmark \\
Ark 564 & \checkmark & \checkmark & \checkmark & \checkmark & X & X \\
Fairall 9 & \checkmark & \checkmark & X & X & X & \checkmark \\
IC 4329A & X & \checkmark & X & \checkmark & X & \checkmark\\
IRAS 13224--3809 & \checkmark & X & \checkmark & \checkmark & X & \checkmark \\
MCG--02-14-009 & X & \checkmark & X & X & X & X \\
MCG--02-58-22 & \checkmark & \checkmark & X & \checkmark & X & X \\
MCG--05-23-16 & \checkmark & \checkmark & X & X$^{2}$ & X & \checkmark \\
MCG--06-30-15 & \checkmark & \checkmark & \checkmark & \checkmark & \checkmark & \checkmark \\
MCG+8-11-11 & X & \checkmark & X & \checkmark & X & \checkmark \\
MR 2251--178 & \checkmark & X & X & \checkmark & \checkmark & \checkmark \\
Mrk 79 & X & X & X & \checkmark & X & \checkmark \\
Mrk 110 & \checkmark & \checkmark & X & X & X & X \\
Mrk 205 & \checkmark & \checkmark & \checkmark & X & X & X \\
Mrk 279 & X & \checkmark & \checkmark & \checkmark & X & X \\
Mrk 335 & \checkmark & \checkmark & X & X & X & \checkmark \\
Mrk 359 & \checkmark & \checkmark & X & X & X & \checkmark \\
Mrk 509 & \checkmark & \checkmark & X & \checkmark & X & \checkmark \\
Mrk 766 & \checkmark & \checkmark & \checkmark & \checkmark & \checkmark & \checkmark$^{3}$ \\
Mrk 841 & \checkmark & \checkmark & X & \checkmark & X & \checkmark\\ 
NGC 1365 & X & \checkmark & \checkmark & \checkmark & \checkmark & \checkmark \\
NGC 2992 & \checkmark & \checkmark & X & X$^{1}$ & X & \checkmark  \\
NGC 3147 & X & \checkmark & X & X & X & \checkmark \\
NGC 3227 & \checkmark & \checkmark & \checkmark & \checkmark & \checkmark & \checkmark \\
NGC 3516 & X & \checkmark & \checkmark & \checkmark & \checkmark & \checkmark \\
NGC 3783 & \checkmark & \checkmark & X & \checkmark & \checkmark & \checkmark \\
NGC 4051 & \checkmark & \checkmark & \checkmark & \checkmark & \checkmark & \checkmark \\
NGC 4151 & \checkmark & \checkmark & \checkmark & \checkmark & \checkmark & X \\
NGC 4593 & X & \checkmark & X & \checkmark & X & \checkmark \\
NGC 5506 & X & \checkmark & \checkmark & \checkmark & X & \checkmark \\
NGC 5548 & X & \checkmark & X & \checkmark & \checkmark & X \\
NGC 7213 & \checkmark & \checkmark & X & X & X & X \\
NGC 7314 & \checkmark & X & X & X$^{1}$ & X & \checkmark \\
NGC 7469 & \checkmark & \checkmark & X & X & X & \checkmark \\
PDS 456 & \checkmark & X & \checkmark & X & \checkmark & X \\
PG 1211+143 & \checkmark & X & X & \checkmark & \checkmark & X \\
RBS 1124 & \checkmark & \checkmark & X & X & X & X \\
SWIFT J2127.4+5654 & X & \checkmark & X & X$^{1}$ & X & \checkmark \\
TON S180 & \checkmark & \checkmark & \checkmark & \checkmark & X & X \\
\hline
Fraction & 31/46 & 39/46 & 16/46 & 27/46 & 14/46 & 32/46 \\ 
\hline
\end{tabular}
\label{tab:comparison}
\end{table*} 

\subsubsection{Partial covering}
Some models used in the analysis of these AGN use partial covering geometries whereby a fraction of the observed X-rays are absorbed by a surrounding gas in the line-of-sight (in addition to typical fully-covering absorbers) as described in Section 3.1.3, while some fraction of the continuum `leaks' or scatters through and is unattenuated by the partially covering material (e.g. Miller et al. 2008). The partial coverer used here takes the form of (\textsc{powerlaw + xstar*powerlaw}) with the photon index of both the powerlaw components tied and normalisations of both powerlaws free to vary. The parameters of the warm absorber (ionization and column density) are also allowed to vary. The column density of the partially covering medium can have a significant affect upon the spectrum. For example, high column partial coverers ($N_{\rm H}\sim10^{24}\,{\rm cm}^{-2}$) predominantly affect the hard X-ray energies and can contribute towards a hard X-ray excess in cases where a hard excess remains (i.e. in addition to \textsc{reflionx}). Lower column density partial coverers may greatly affect spectral curvature at lower X-ray energies and in some cases can entirely remove any `broad' residuals in the Fe\,K region (Miller et al. 2009). While using a partial coverer in this way provides an alternate explanation for observed curvature or a red-wing in the Fe\,K region, we restrict the analysis to a maximum of one partially covering zone in order to not arbitrarily model away spectral features present in the spectra. The absorption grid used is the same 
as used to model the warm absorber, with a turbulence of 100\,km\,s$^{-1}$.

\subsubsection{Relativistic line emission}
The majority of X-ray lines originate from material sufficiently distant to the SMBH (e.g. the ubiquitous narrow 6.4\,keV line) such that effects such as relativistic Doppler motions and gravitational redshift have a negligible effect upon the observed spectrum. If, however, emission comes from the very inner regions of the accretion disc it will of course be subject to these effects. As discussed above, this is one interpretation for the commonly observed red-wing in the Fe\,K regions of some AGN which may remain after a full modelling of the broad-band spectrum, taking into account warm absorbers which may be present. If an excess remains at $\sim5-6.4$\,keV, we initially model this using the relativistic line emission model \textsc{relline} (Dauser et al. 2010). This allows properties such as the emissivity index and inclination of the disc to be measured in addition to placing estimates upon the spin of the central SMBH (or the inner radius of emission). A more comprehensive modelling of the inner regions of the accretion disc would include the accompanying blurred reflection spectrum. For example a convolution of the \textsc{relline} kernel (i.e. \textsc{relconv}, Dauser et al. 2010) with \textsc{reflionx} allows relativistic effects to be applied to both the hard and soft X-ray reflection spectrum as well as to the Fe\,K$\alpha$ emission line at 6.4\,keV, we investigate such an approach in Section 3.4. 

\subsection{Baseline model}
The baseline model is intended to model the entire 0.6-100.0\,keV spectrum, accounting for features such as the underlying continuum, soft excess, {\sl distant} reflection and both warm and neutral absorption. With this model we aim to assess the remaining residuals in the Fe\,K region which may be attributed to relativistic line emission from the inner regions of the accretion disc, i.e. we use a combination of the models and methods outlined in Section 3.1 to form a baseline model which does not include broadened line emission. The baseline model acts as the null hypothesis whereby the X-ray spectrum of these AGN can be described and adequately fit by emission or reflection from purely distant material. Figure \ref{fig:FeK_none} therefore shows the Fe K regions of these AGN prior to any modelling of broad emission, distant emission lines or highly ionised absorption.

\begin{figure*}
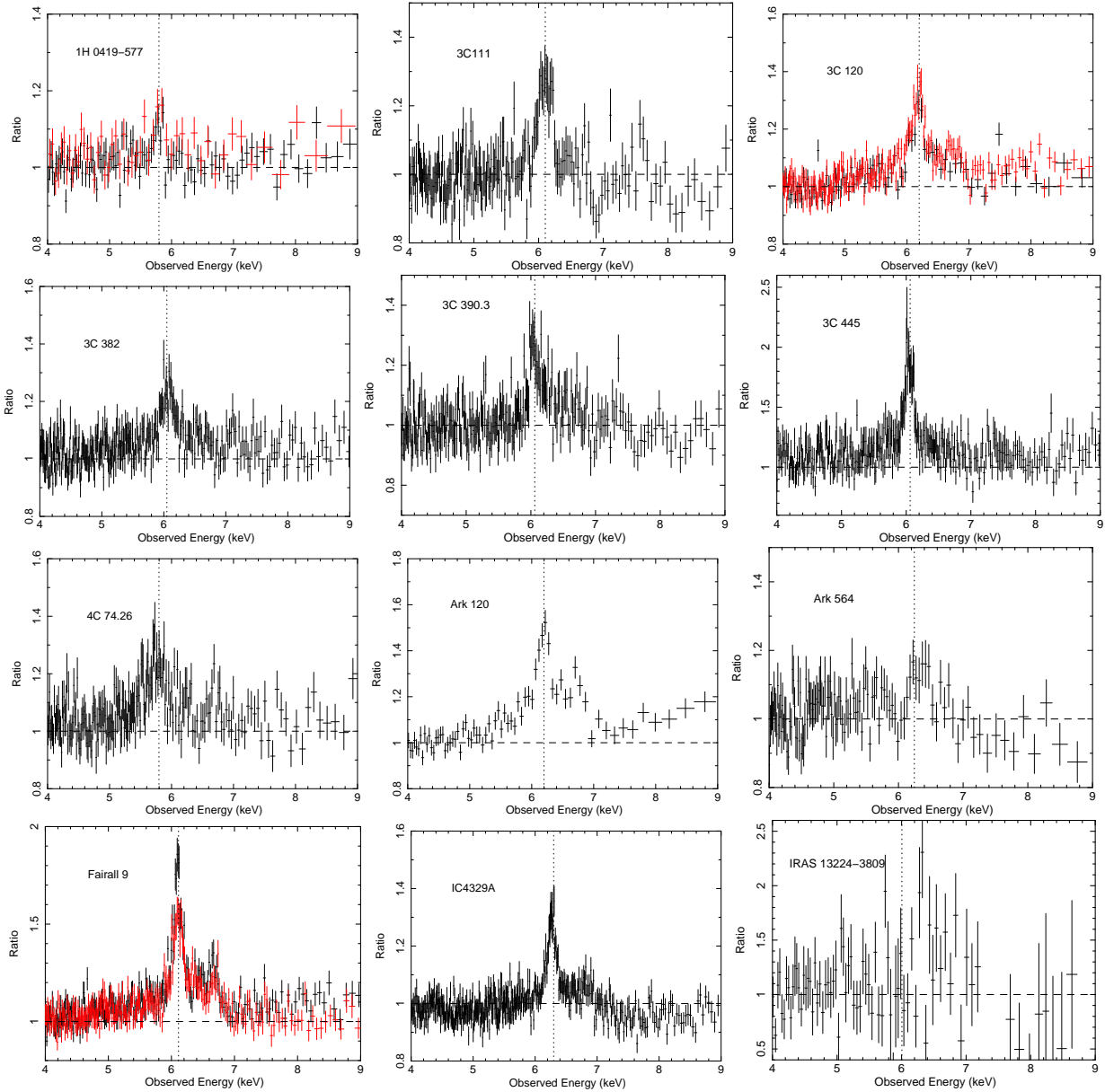

\rotatebox{-90}{\includegraphics[width=4cm]{1H0419_FeK_none.ps}}
\rotatebox{-90}{\includegraphics[width=4cm]{3C111_FeK_none.ps}}
\rotatebox{-90}{\includegraphics[width=4cm]{3C120_FeK_none.ps}}
\rotatebox{-90}{\includegraphics[width=4cm]{3C382_FeK_none.ps}}
\rotatebox{-90}{\includegraphics[width=4cm]{3C390_FeK_none.ps}}
\rotatebox{-90}{\includegraphics[width=4cm]{3C445_FeK_none.ps}}
\rotatebox{-90}{\includegraphics[width=4cm]{4C74_FeK_none.ps}}
\rotatebox{-90}{\includegraphics[width=4cm]{Ark120_FeK_none.ps}}
\rotatebox{-90}{\includegraphics[width=4cm]{Ark564_FeK_none.ps}}
\rotatebox{-90}{\includegraphics[width=4cm]{Fairall9_FeK_none.ps}}
\rotatebox{-90}{\includegraphics[width=4cm]{IC4329_FeK_none.ps}}
\rotatebox{-90}{\includegraphics[width=4cm]{IRAS13224_FeK_none.ps}}
\caption{Ratio plots of the 4-9\,keV residuals modelled with the baseline model minus \textsc{reflionx}, narrow lines and Fe\,K absorbers, i.e. the Fe\,K region is left totally unmodelled with only warm absorption at soft X-ray energies being taken into account in addition to the intrinsic powerlaw and \textsc{compTT} (where required). The data has been refit after the removal of components. The vertical dashed lines represent 6.4\,keV in the observed frames. Red data points indicate those from further observations. }
\label{fig:FeK_none}
\end{figure*}

\begin{figure*}
\rotatebox{-90}{\includegraphics[width=4cm]{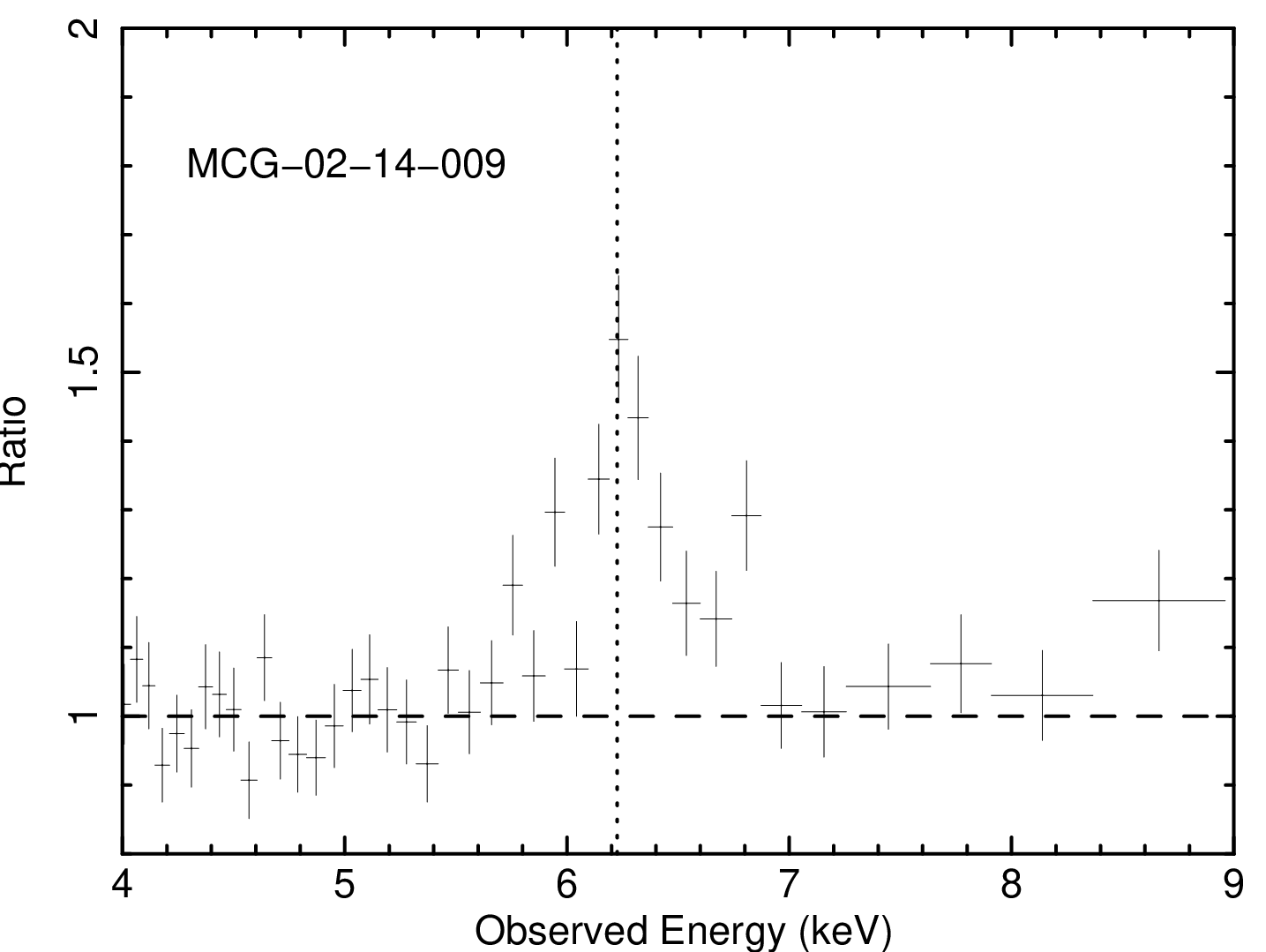}}
\rotatebox{-90}{\includegraphics[width=4cm]{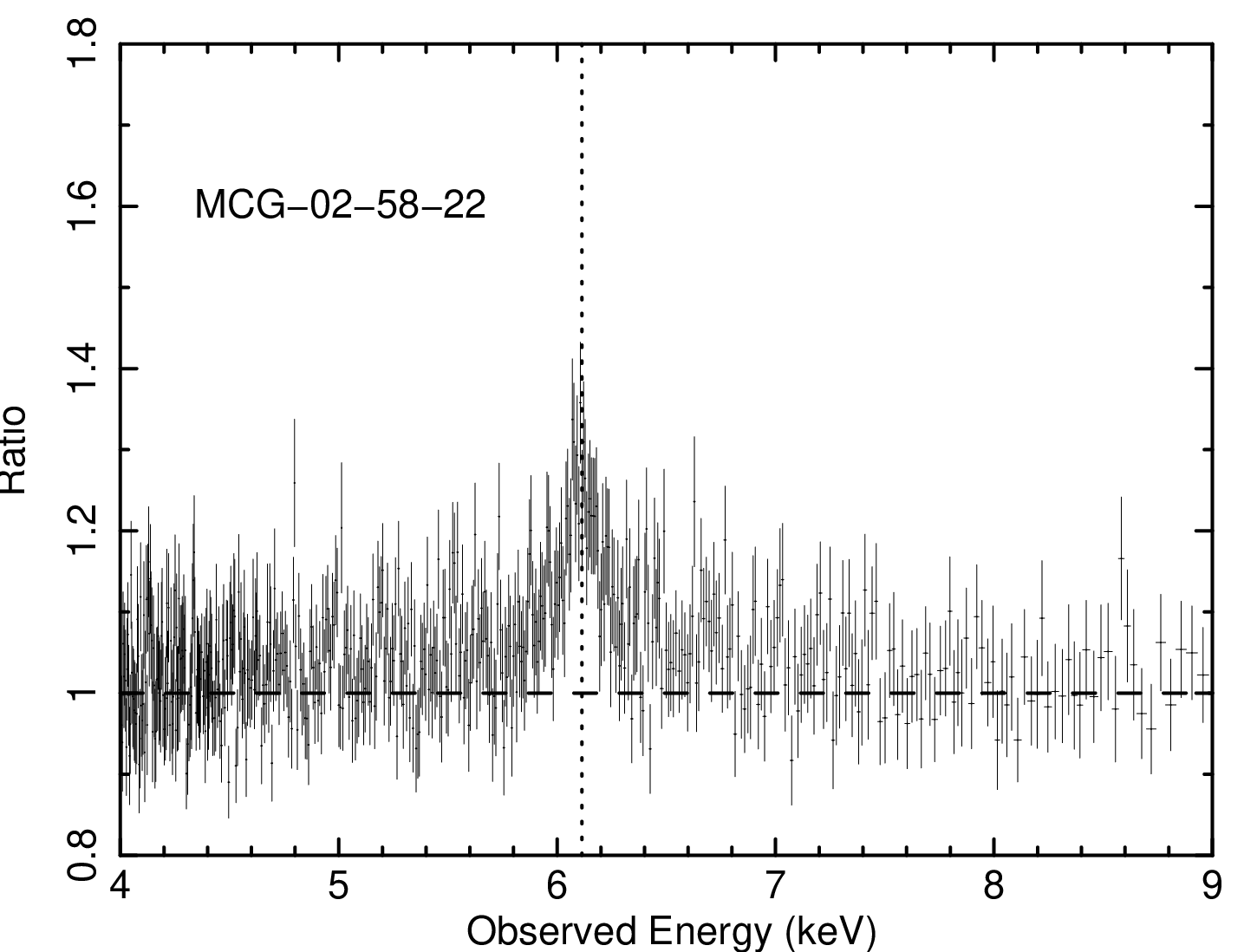}}
\rotatebox{-90}{\includegraphics[width=4cm]{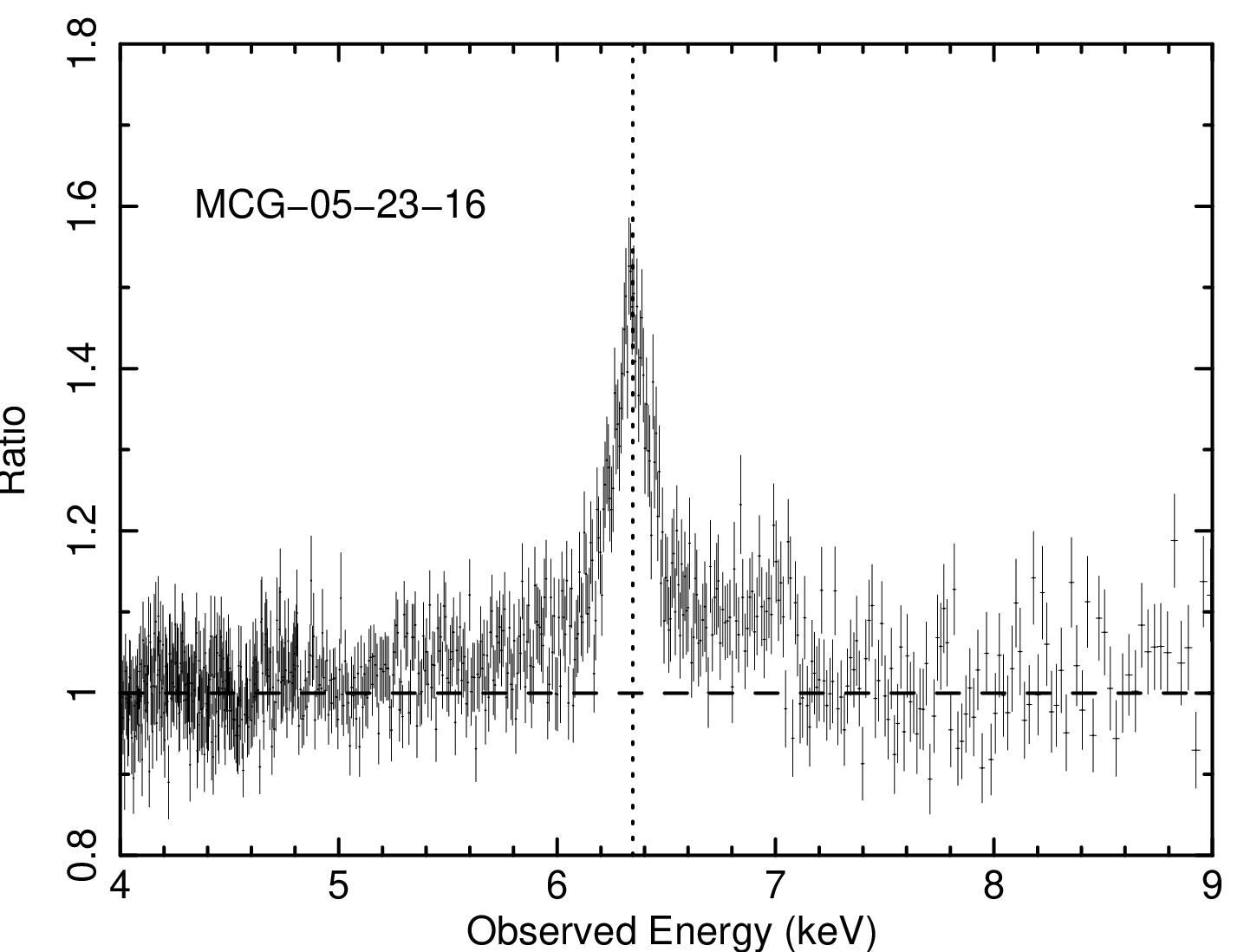}}
\rotatebox{-90}{\includegraphics[width=4cm]{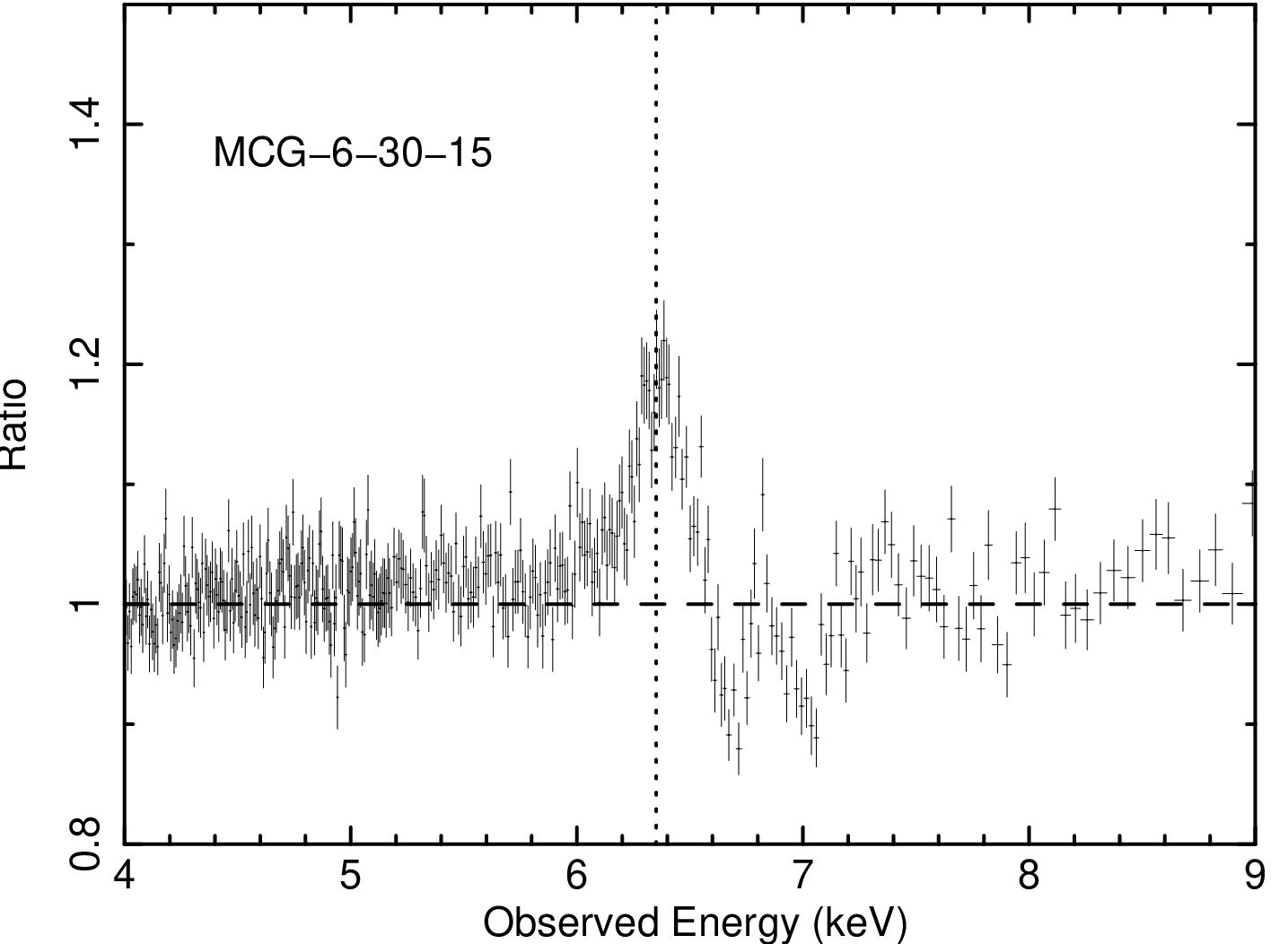}}
\rotatebox{-90}{\includegraphics[width=4cm]{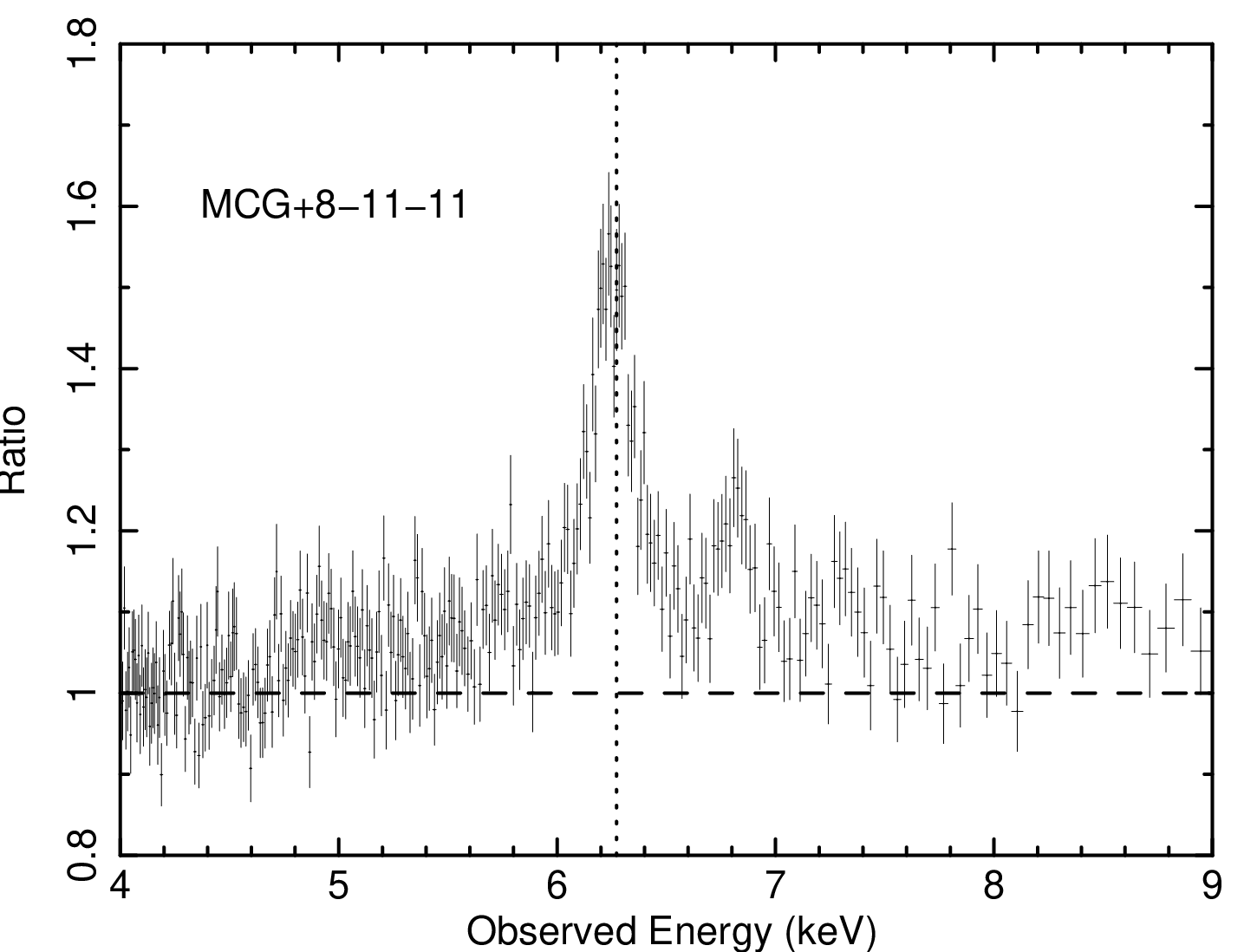}}
\rotatebox{-90}{\includegraphics[width=4cm]{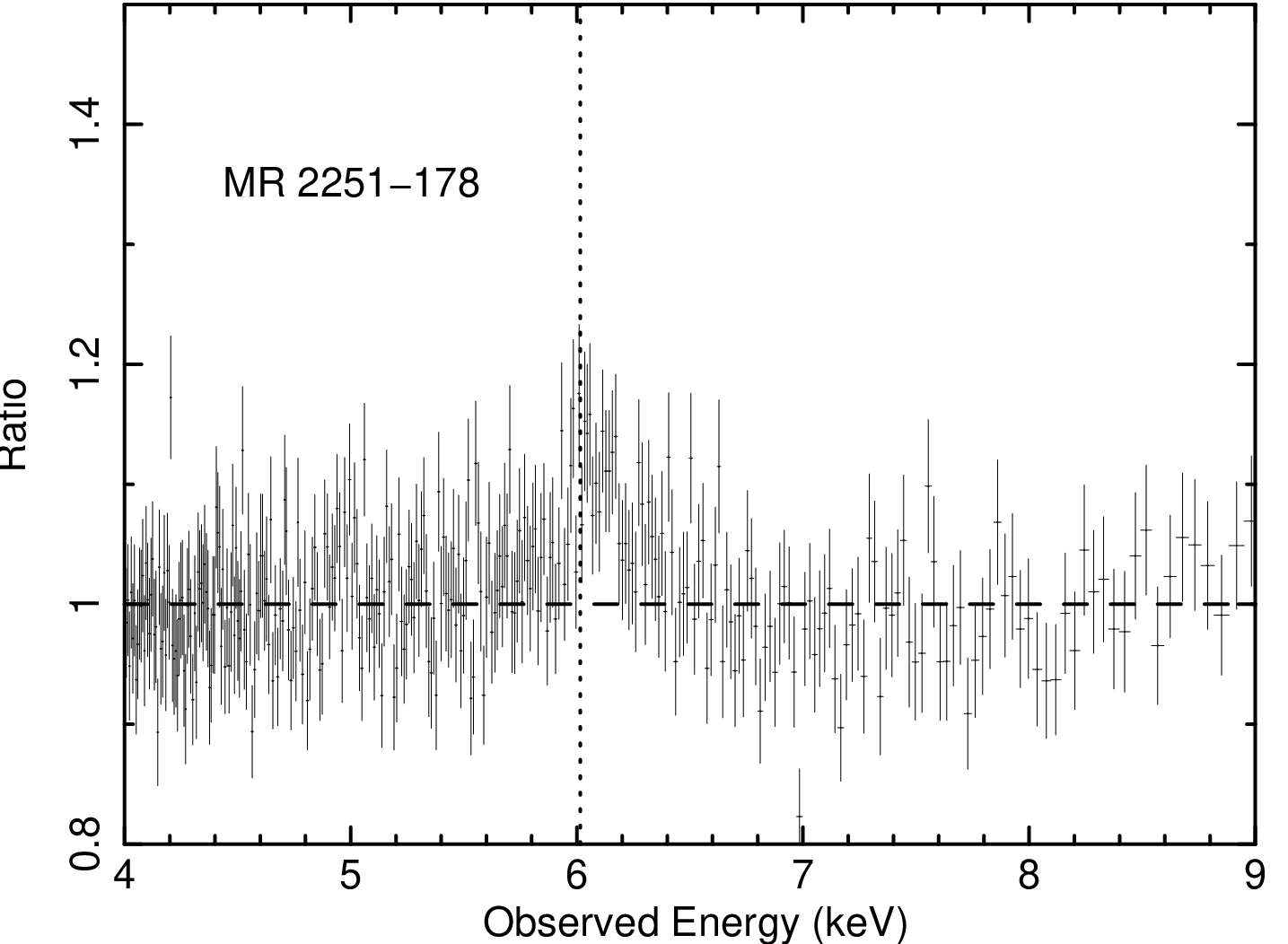}}
\rotatebox{-90}{\includegraphics[width=4cm]{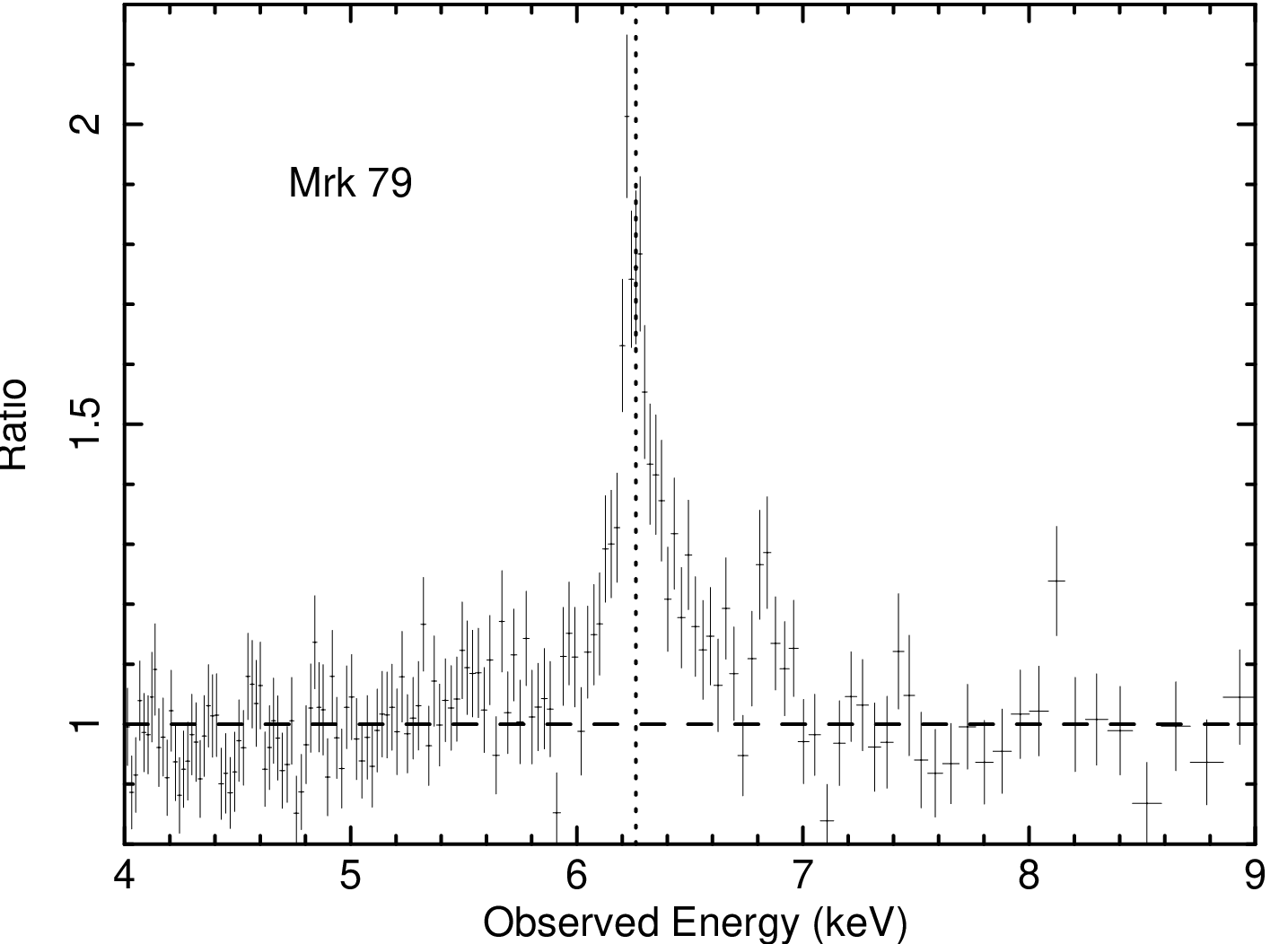}}
\rotatebox{-90}{\includegraphics[width=4cm]{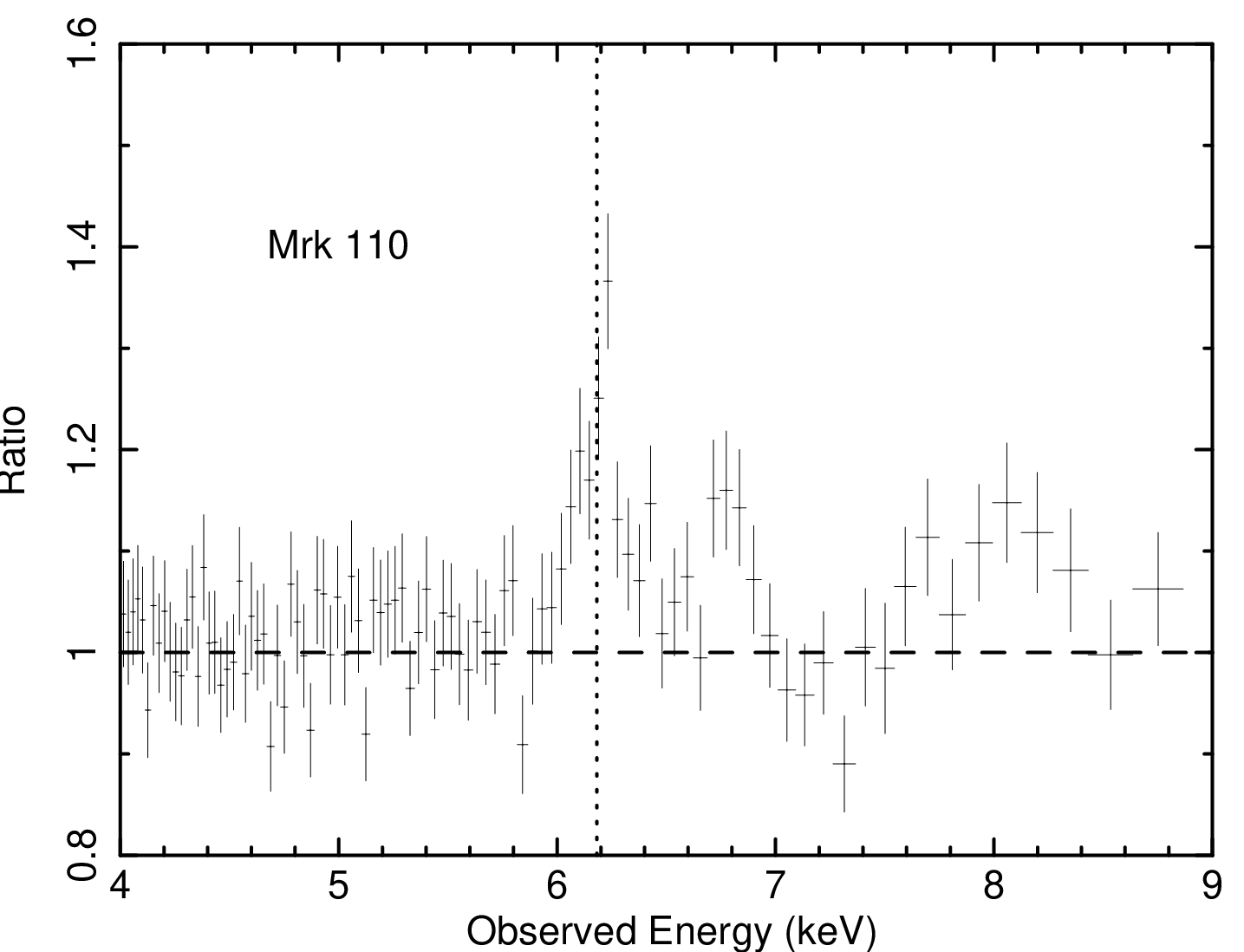}}
\rotatebox{-90}{\includegraphics[width=4cm]{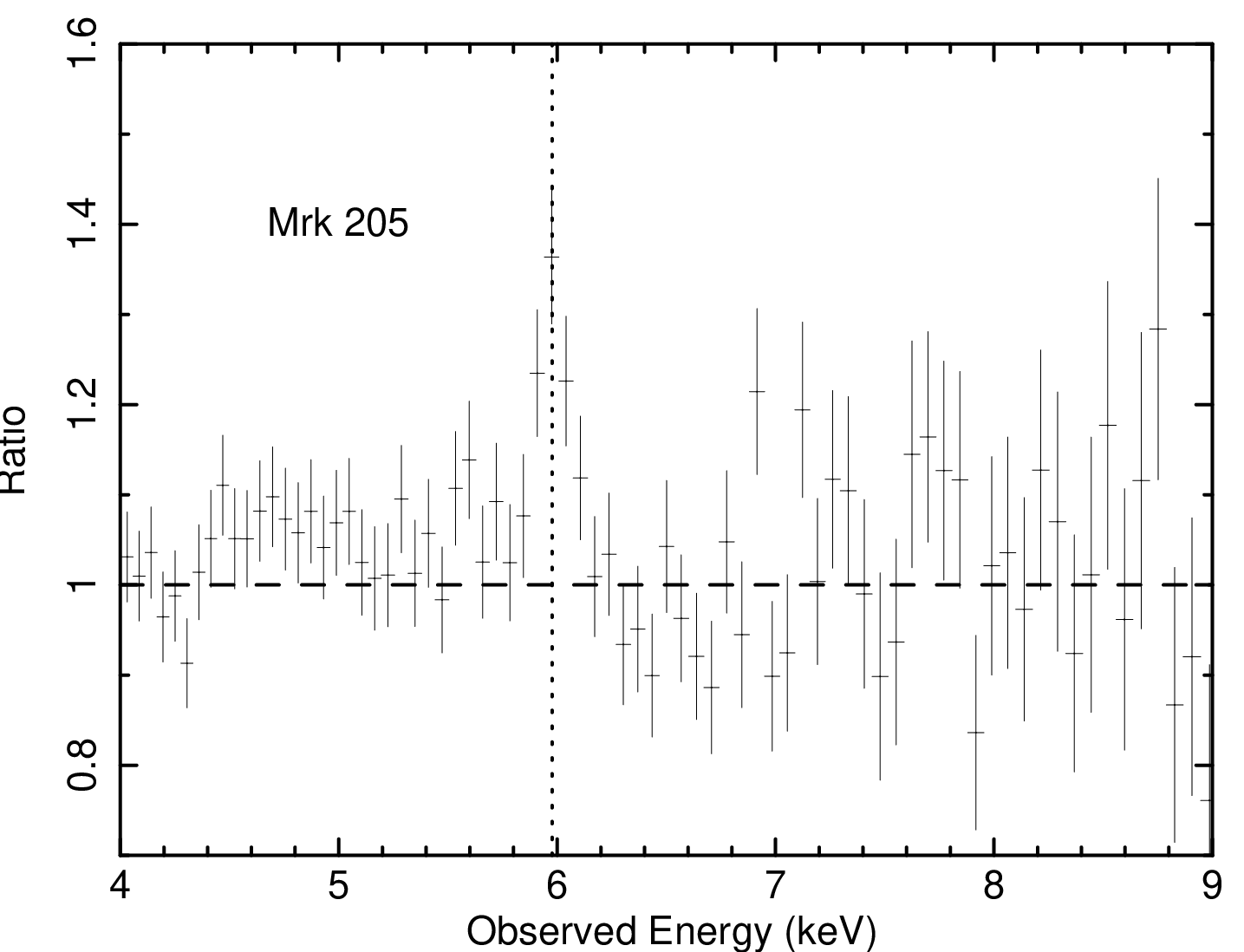}}
\rotatebox{-90}{\includegraphics[width=4cm]{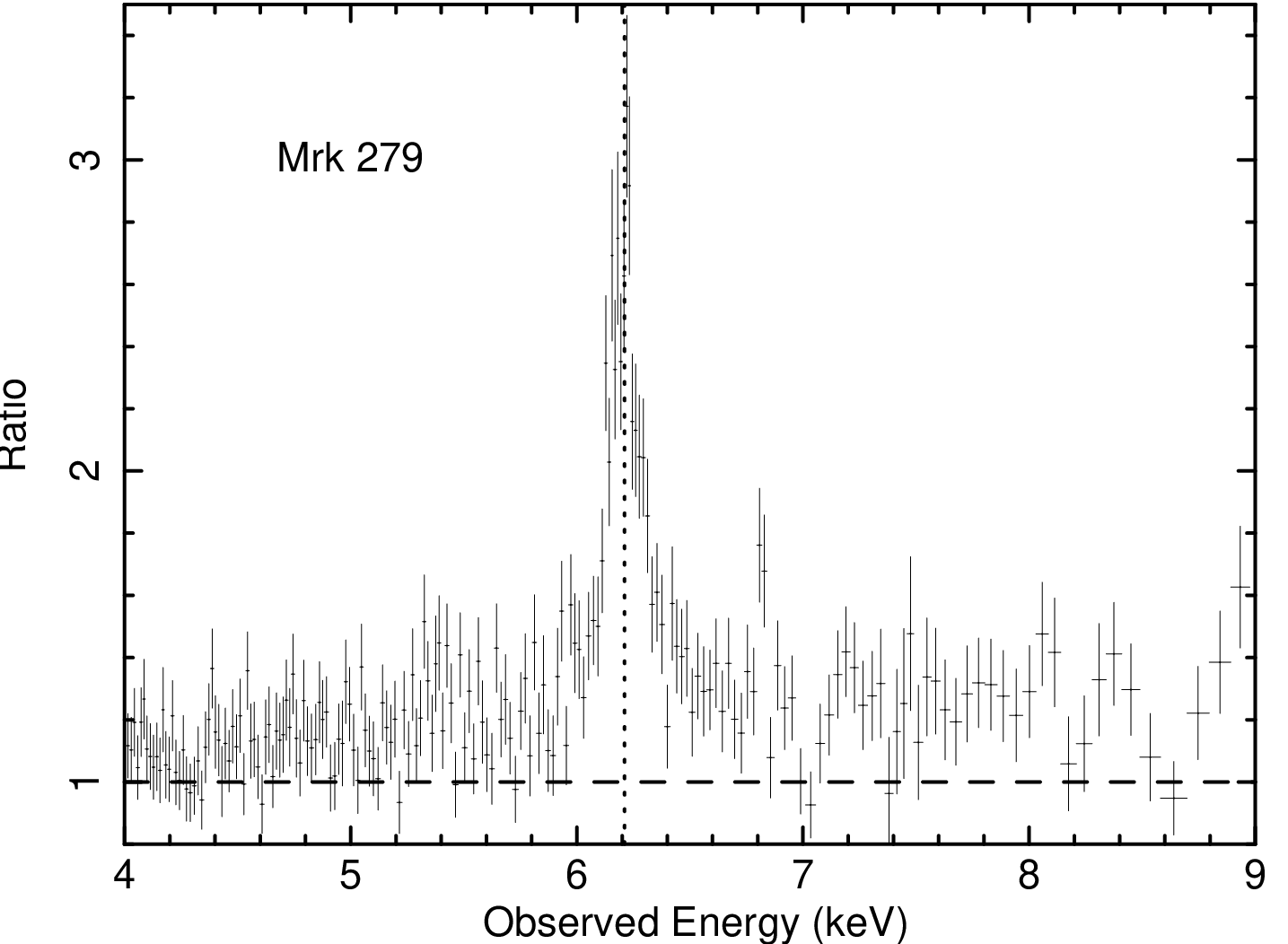}}
\rotatebox{-90}{\includegraphics[width=4cm]{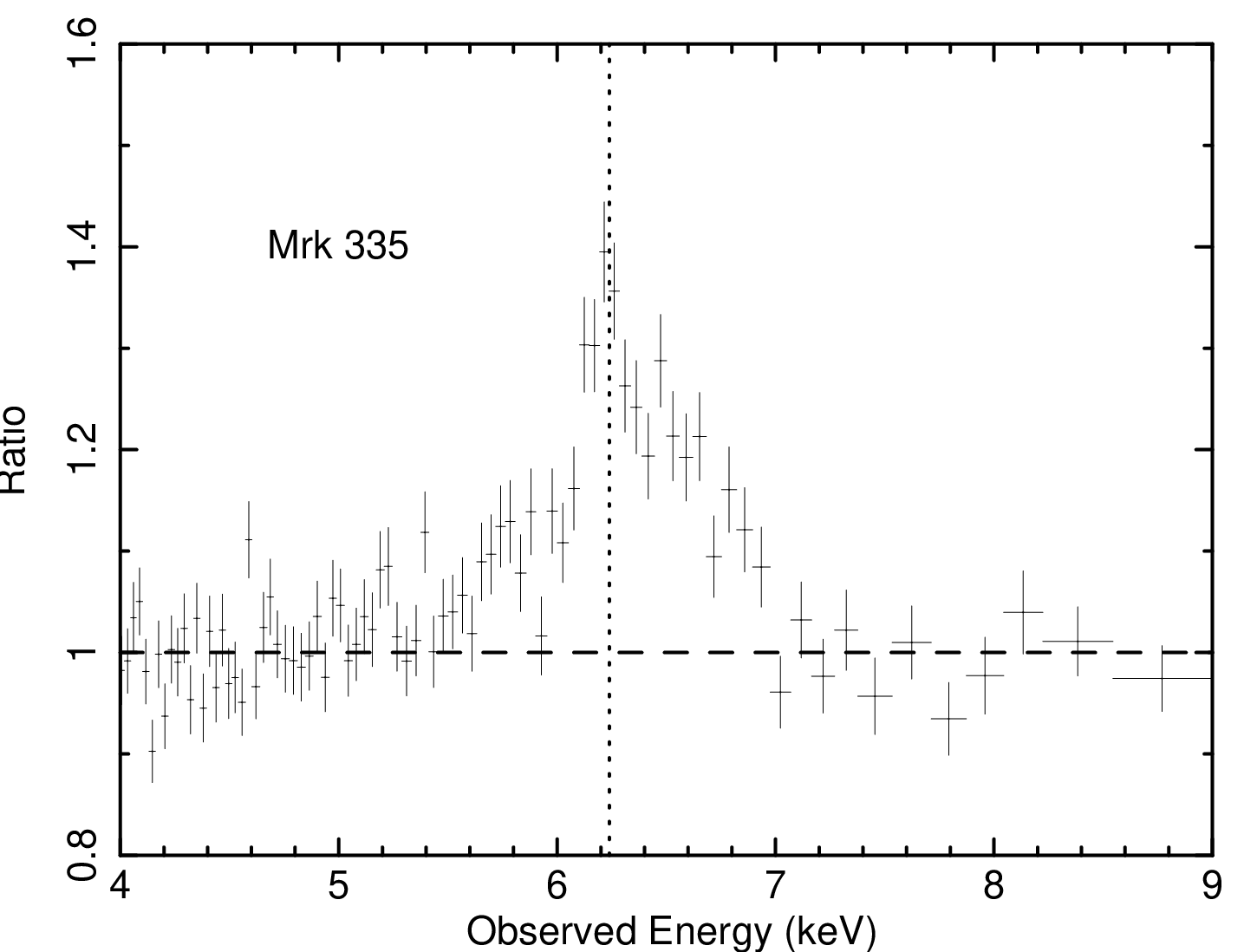}}
\rotatebox{-90}{\includegraphics[width=4cm]{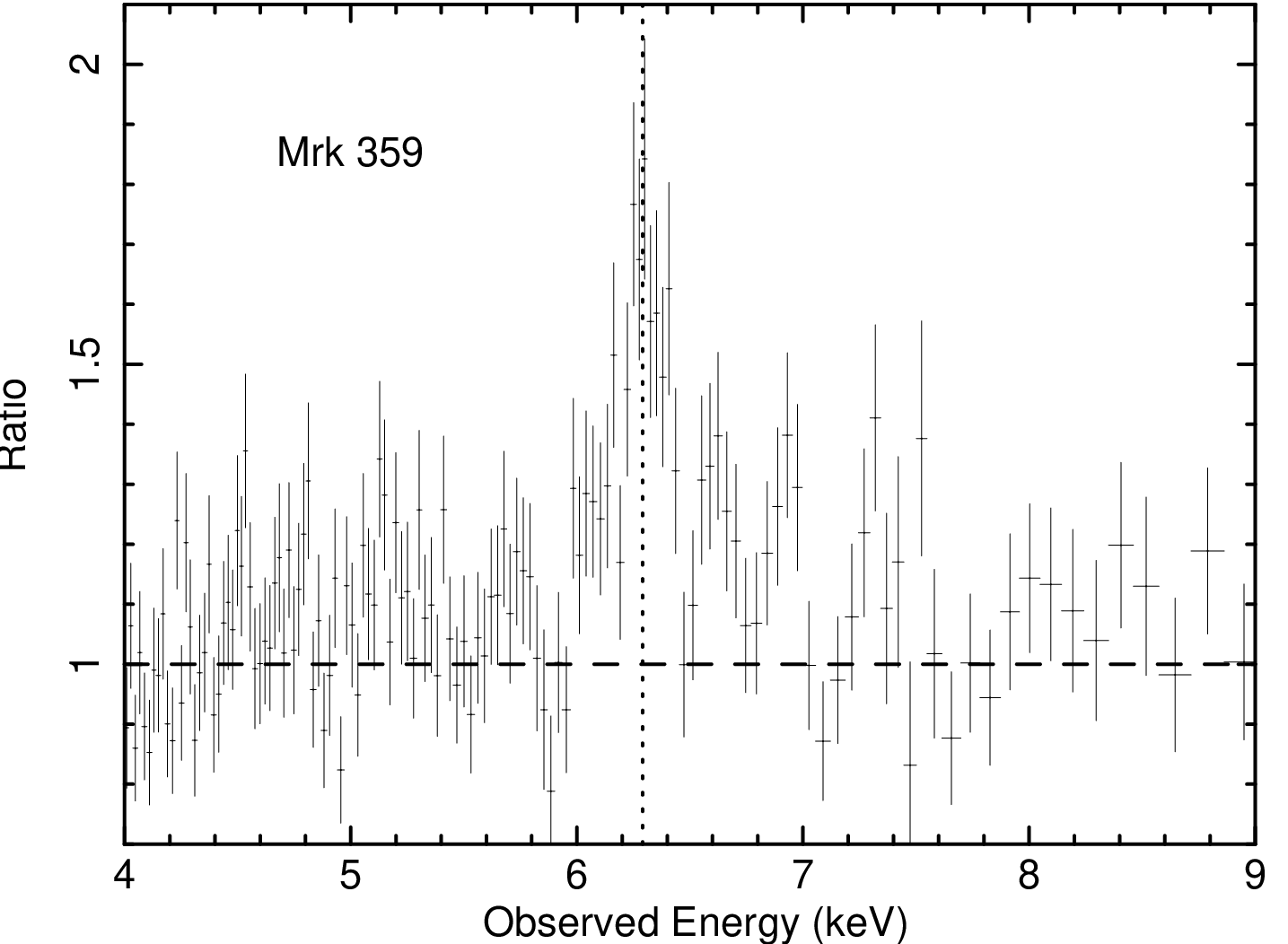}}

\contcaption{Ratio plots of the 4-9\,keV residuals without any modelling of the reflection component i.e. the Fe\,K region is left totally unmodelled with only warm absorption at soft X-ray energies being taken into account in addition to the intrinsic powerlaw and \textsc{compTT} (where required). The data has been refit after the removal of components. The vertical dashed lines represent 6.4\,keV in the observed frames. Red data points indicate those from further observations.}
\end{figure*}

\begin{figure*}
\rotatebox{-90}{\includegraphics[width=4cm]{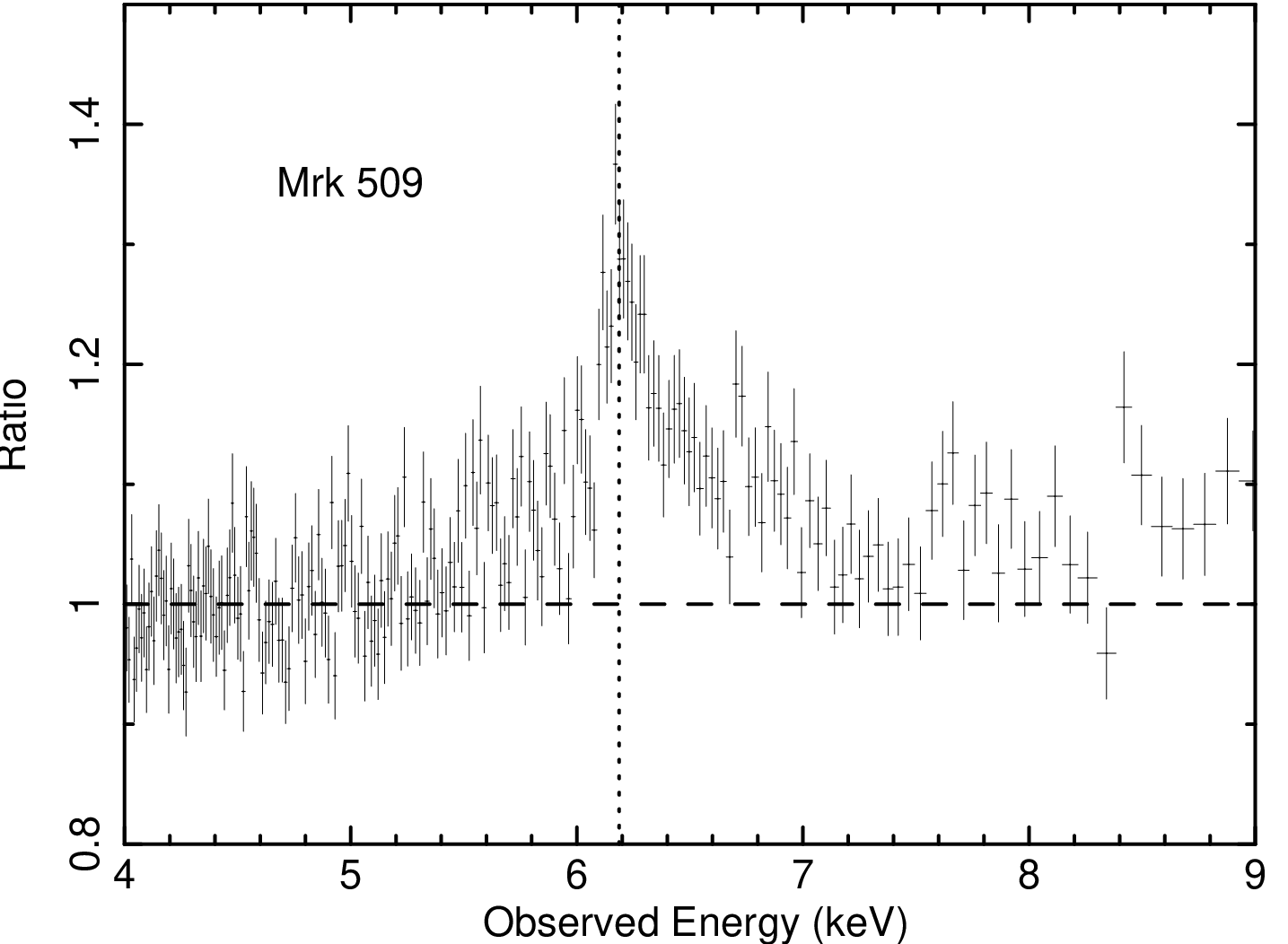}}
\rotatebox{-90}{\includegraphics[width=4cm]{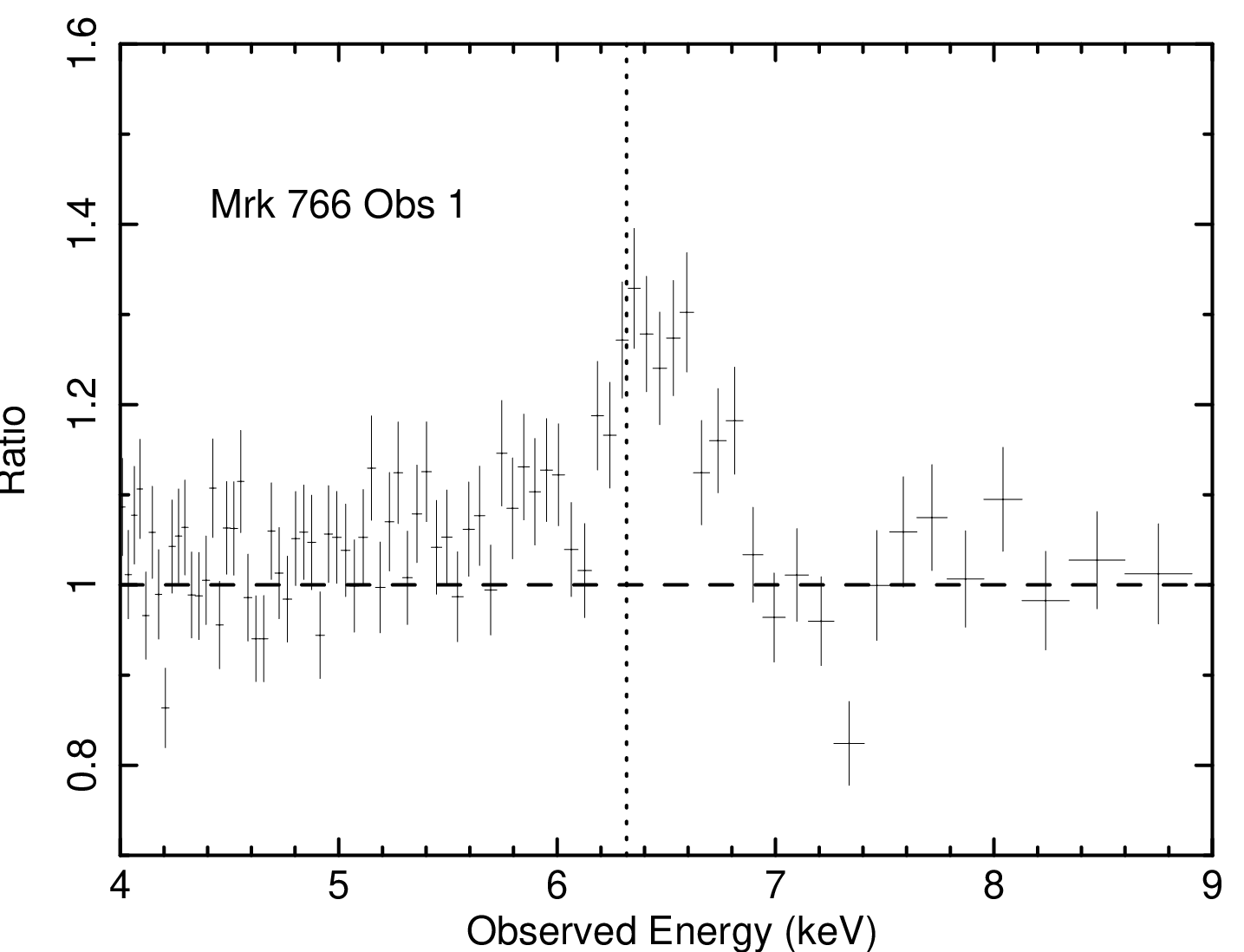}}
\rotatebox{-90}{\includegraphics[width=4cm]{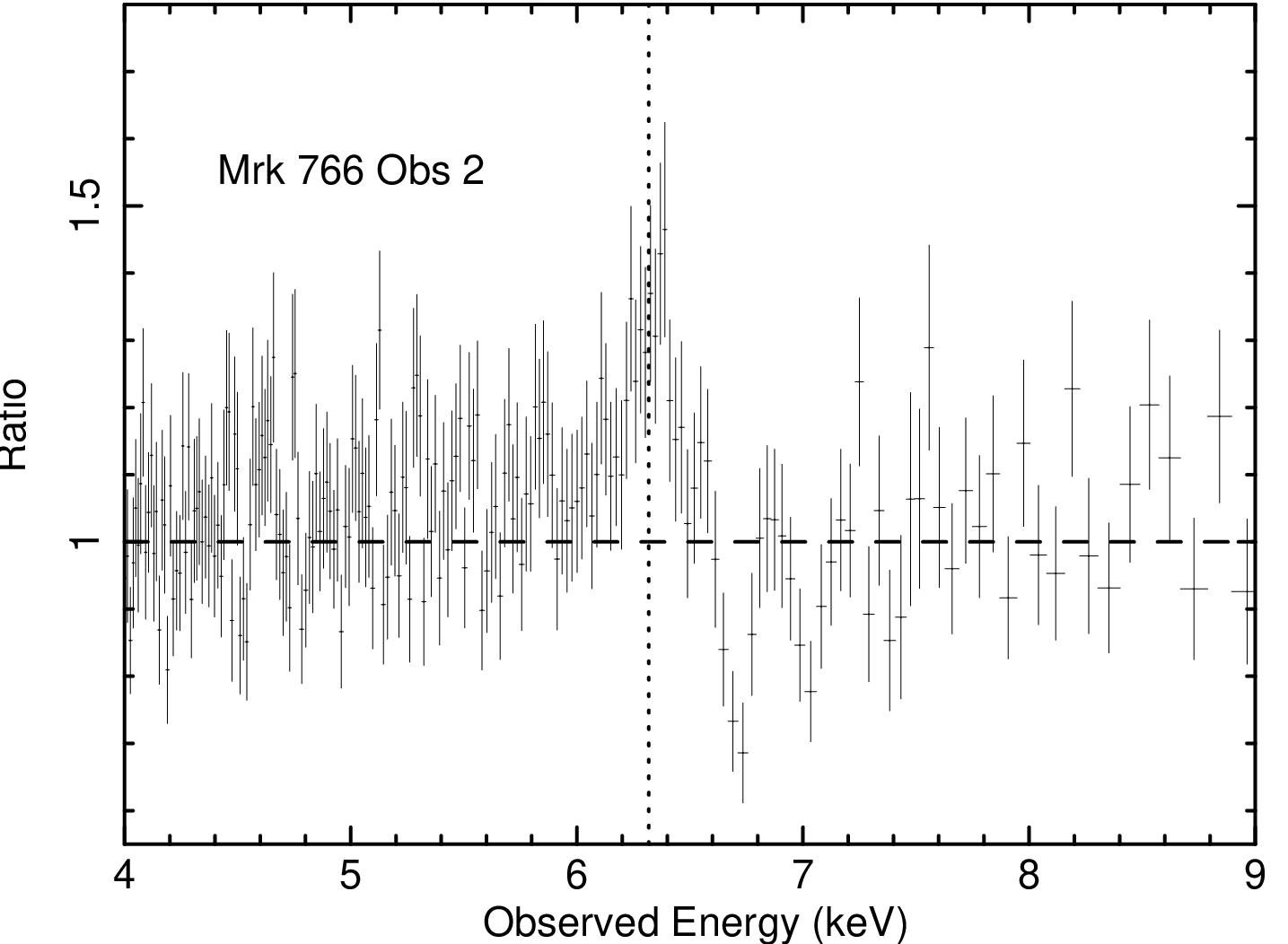}}
\rotatebox{-90}{\includegraphics[width=4cm]{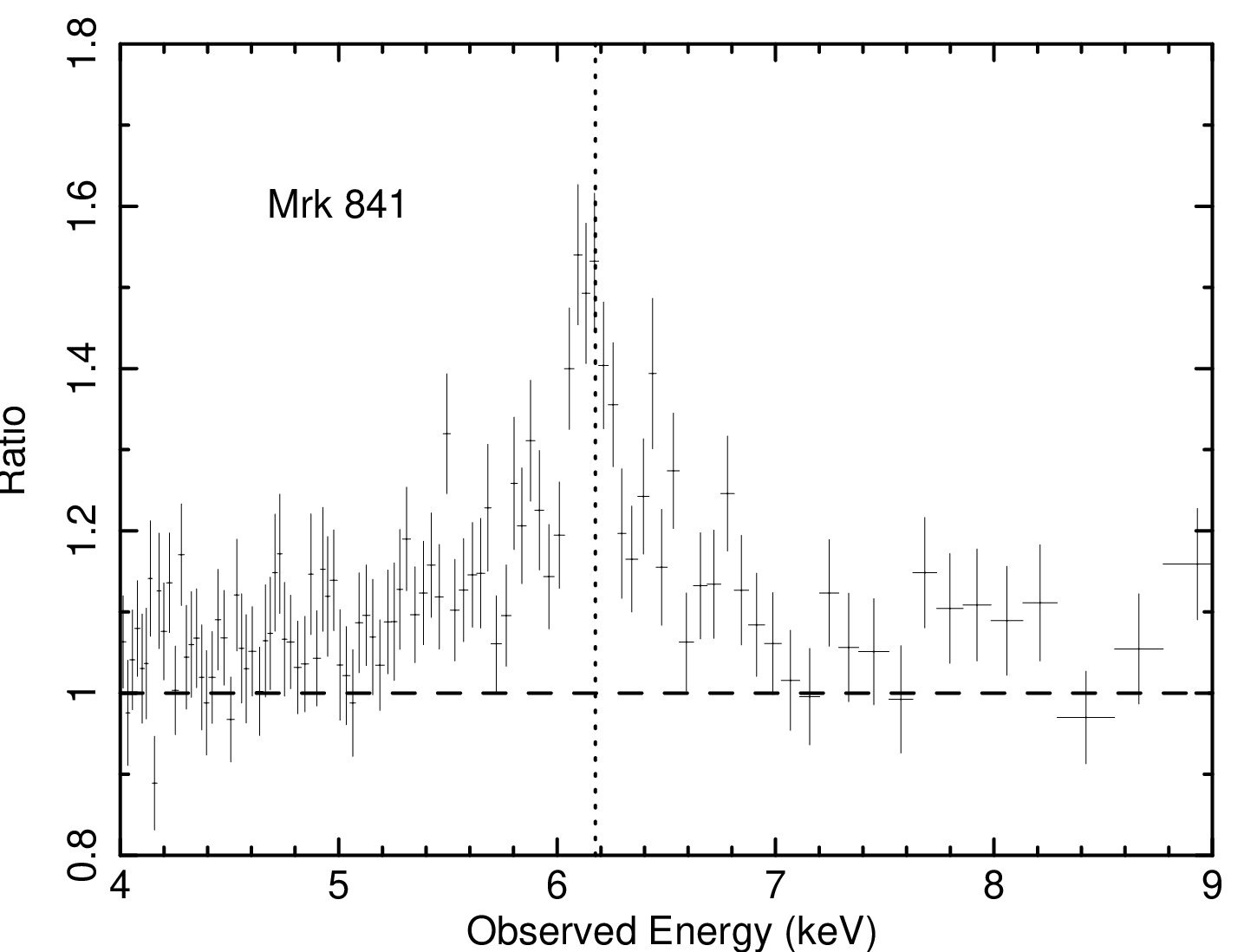}}
\rotatebox{-90}{\includegraphics[width=4cm]{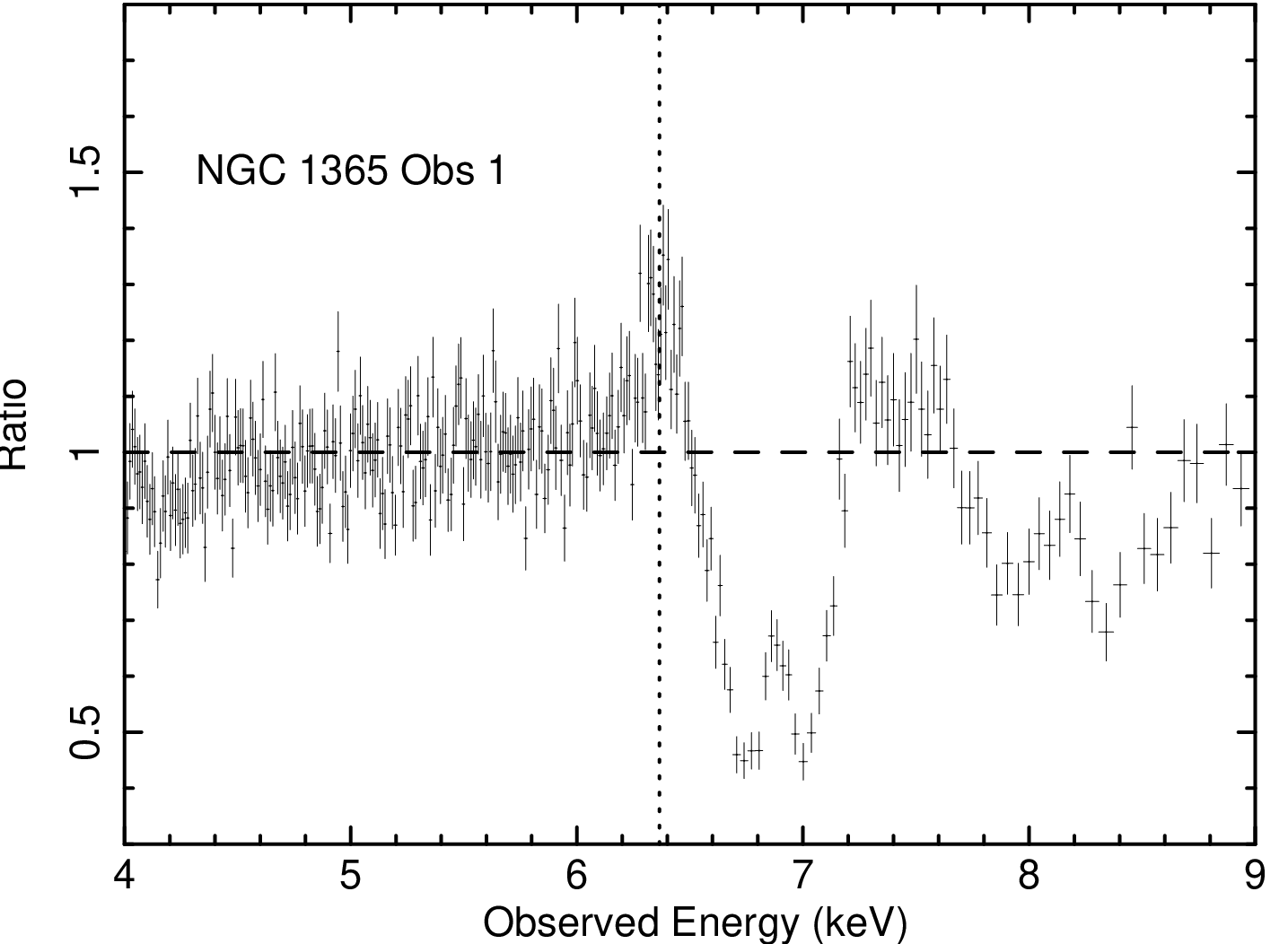}}
\rotatebox{-90}{\includegraphics[width=4cm]{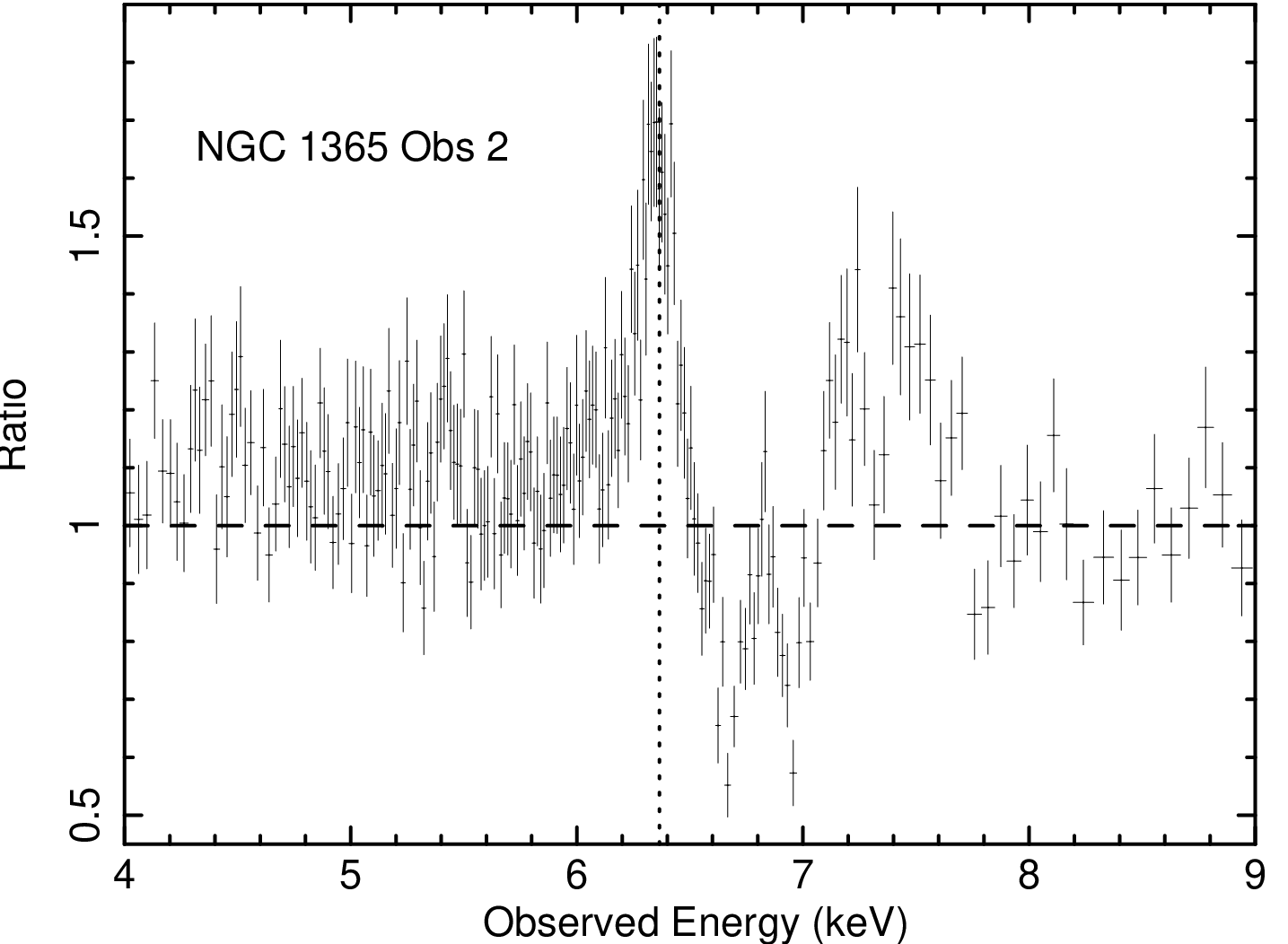}}
\rotatebox{-90}{\includegraphics[width=4cm]{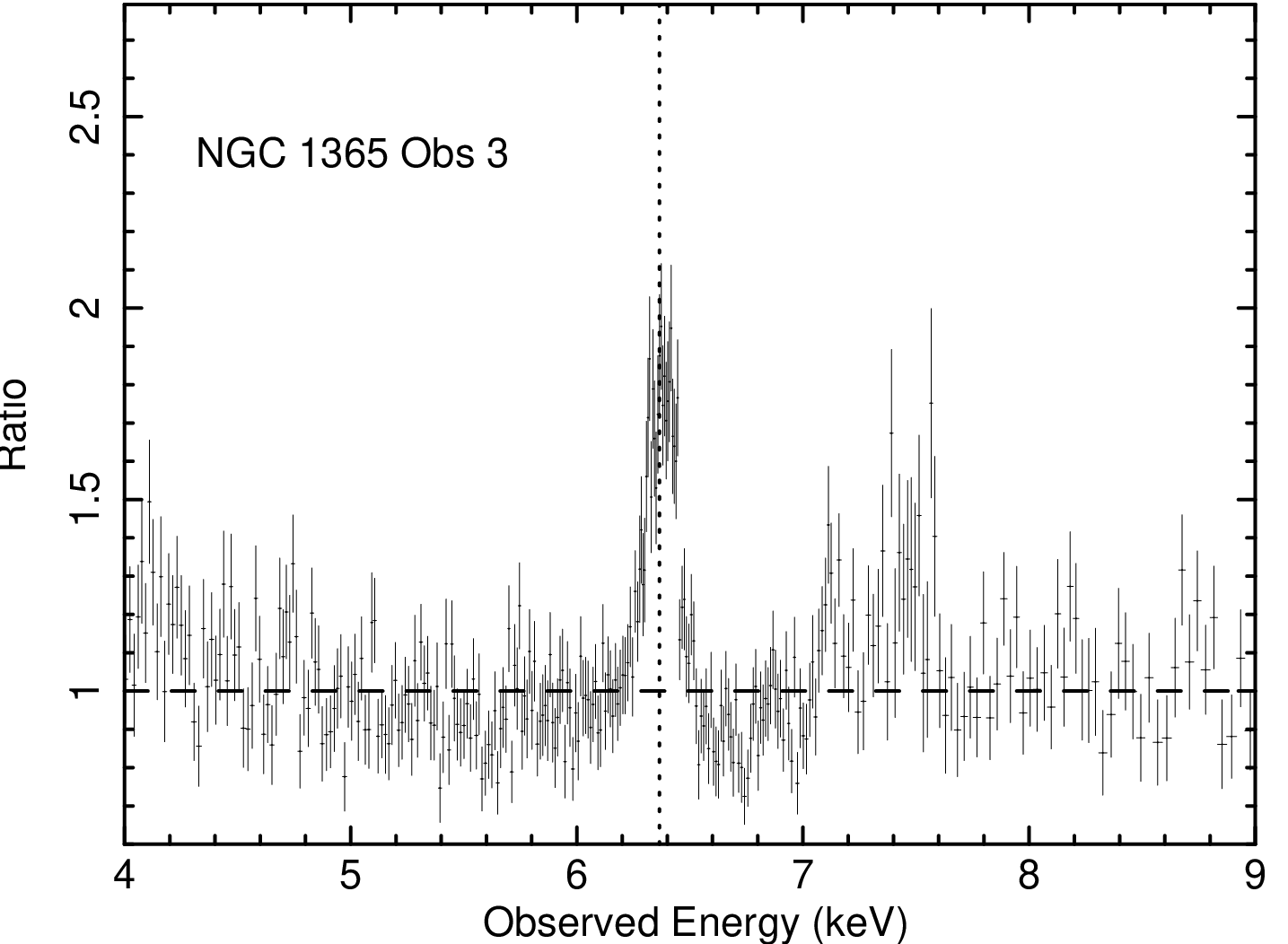}}
\rotatebox{-90}{\includegraphics[width=4cm]{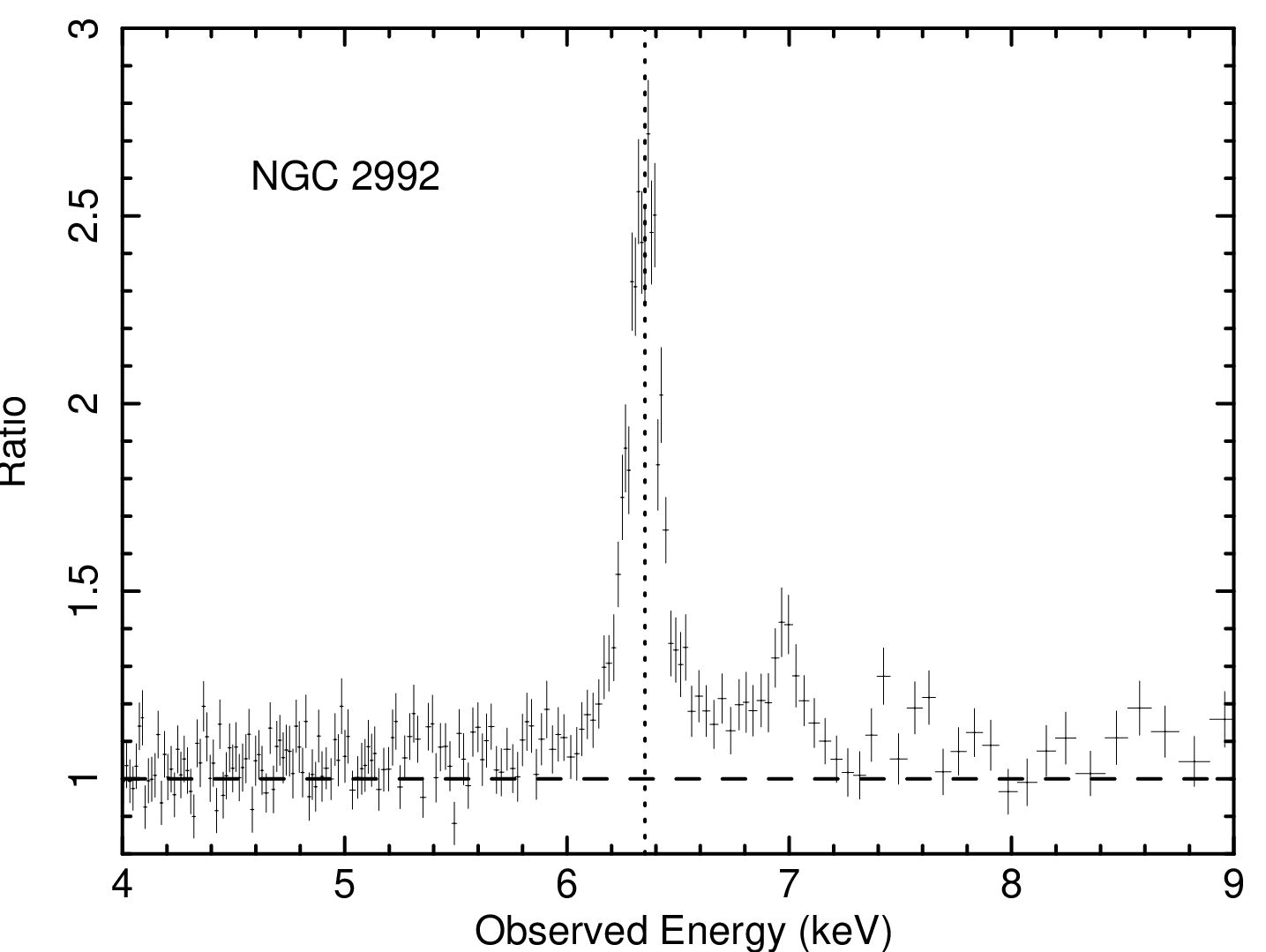}}
\rotatebox{-90}{\includegraphics[width=4cm]{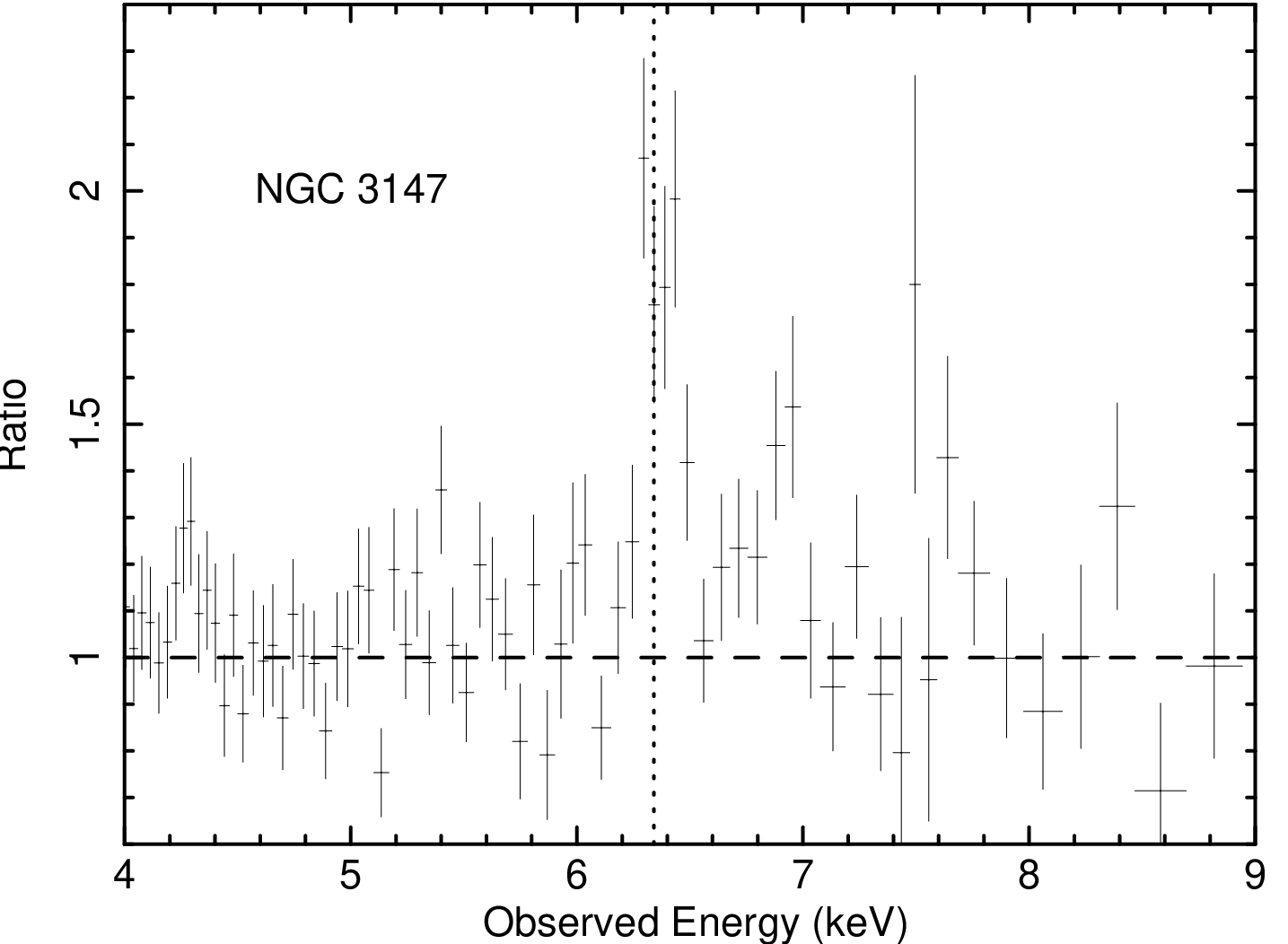}}
\rotatebox{-90}{\includegraphics[width=4cm]{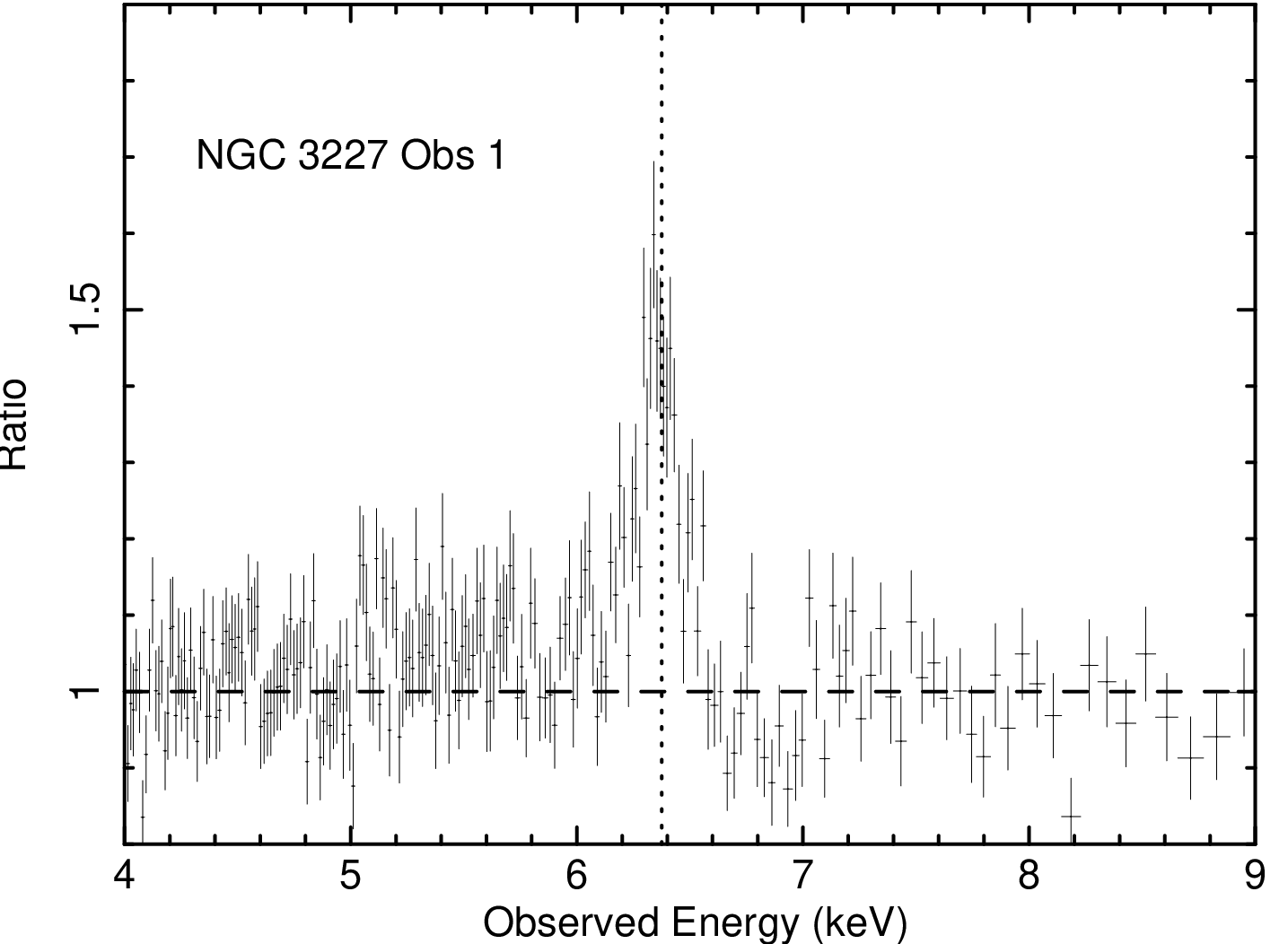}}
\rotatebox{-90}{\includegraphics[width=4cm]{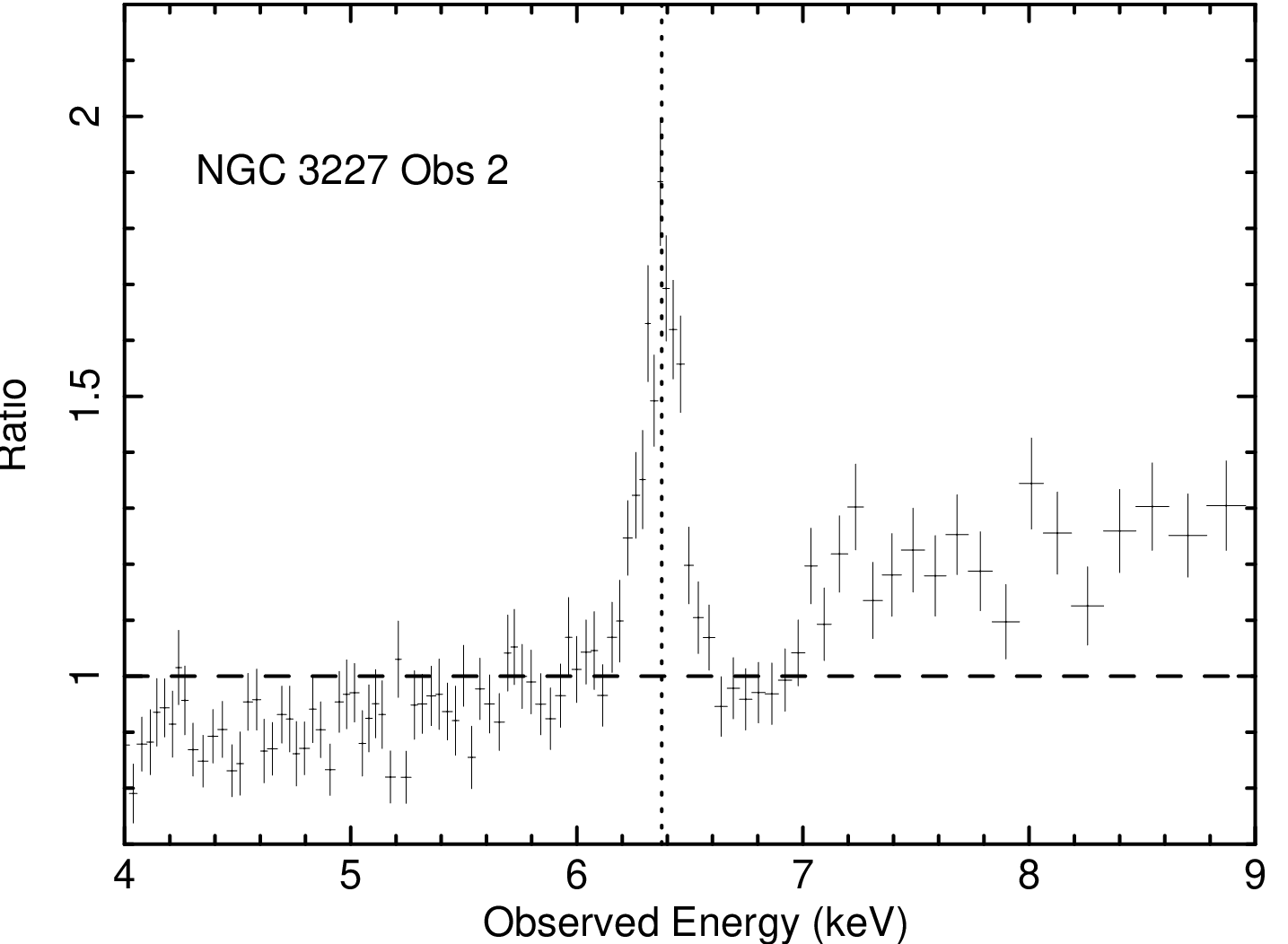}}
\rotatebox{-90}{\includegraphics[width=4cm]{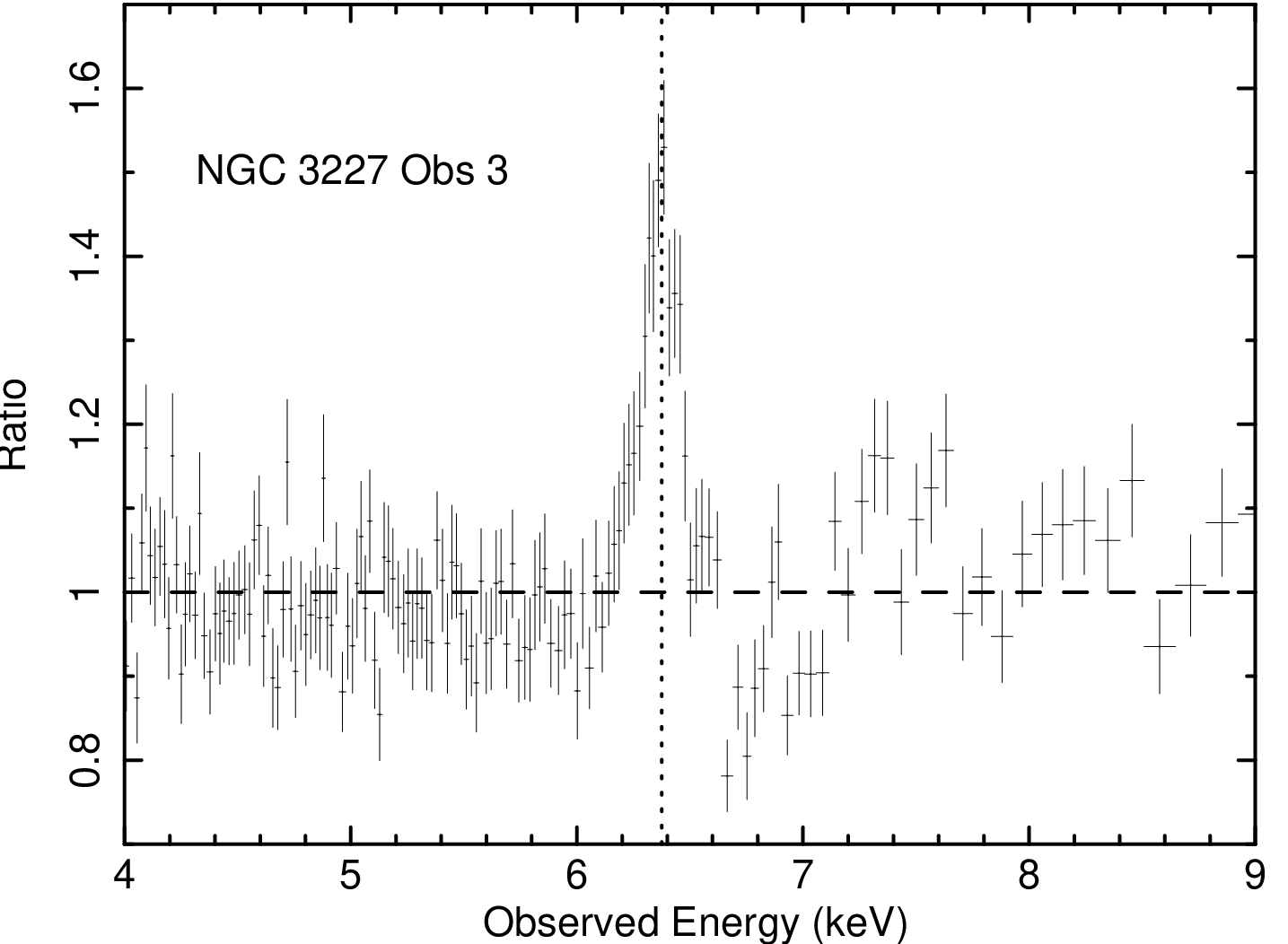}}

\contcaption{Ratio plots of the 4-9\,keV residuals without any modelling of the reflection component i.e. the Fe\,K region is left totally unmodelled with only warm absorption at soft X-ray energies being taken into account in addition to the intrinsic powerlaw and \textsc{compTT} (where required). The data has been refit after the removal of components. The vertical dashed lines represent 6.4\,keV in the observed frames. Red data points indicate those from further observations.}
\end{figure*}

\begin{figure*}
\rotatebox{-90}{\includegraphics[width=4cm]{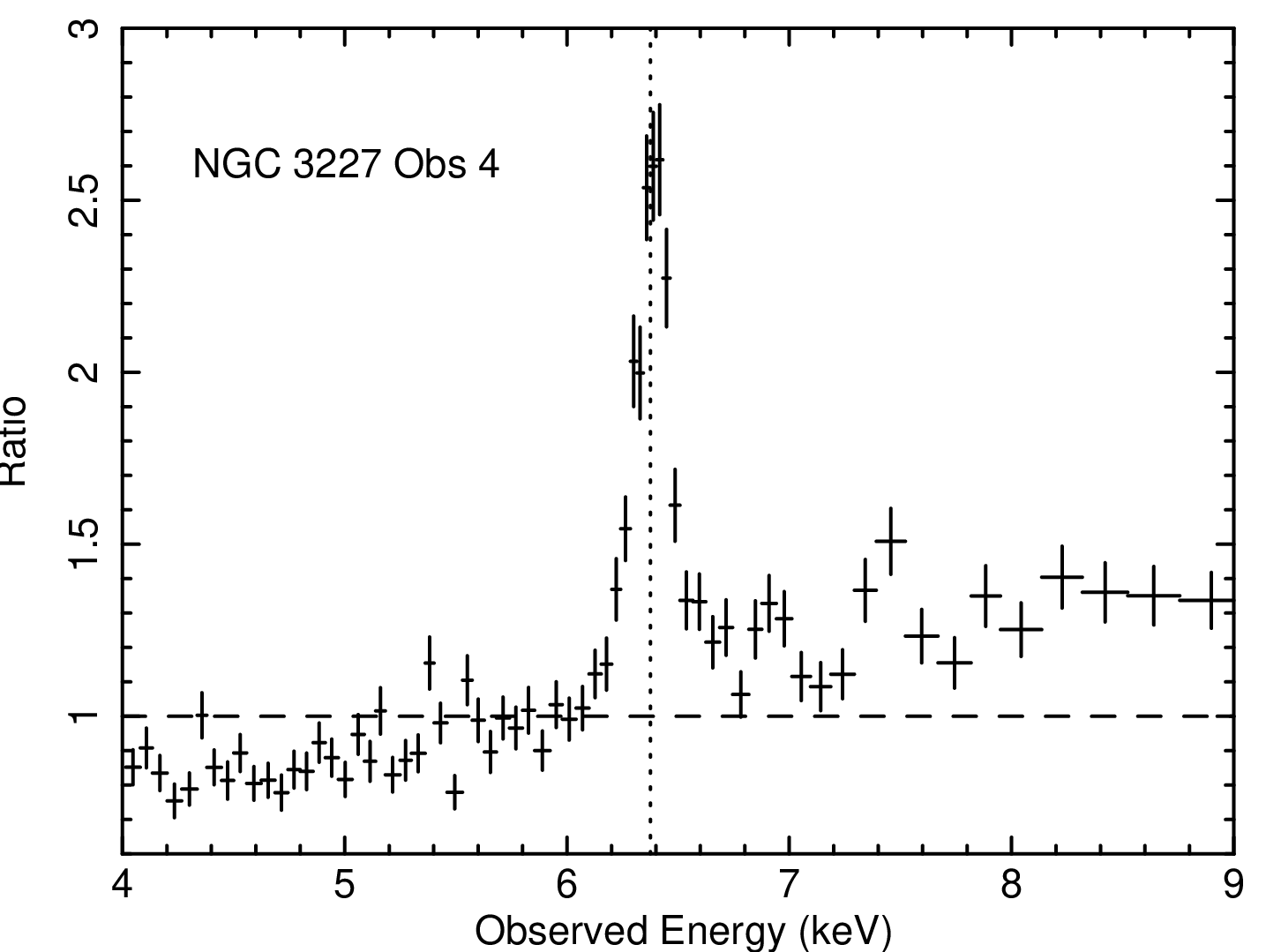}}
\rotatebox{-90}{\includegraphics[width=4cm]{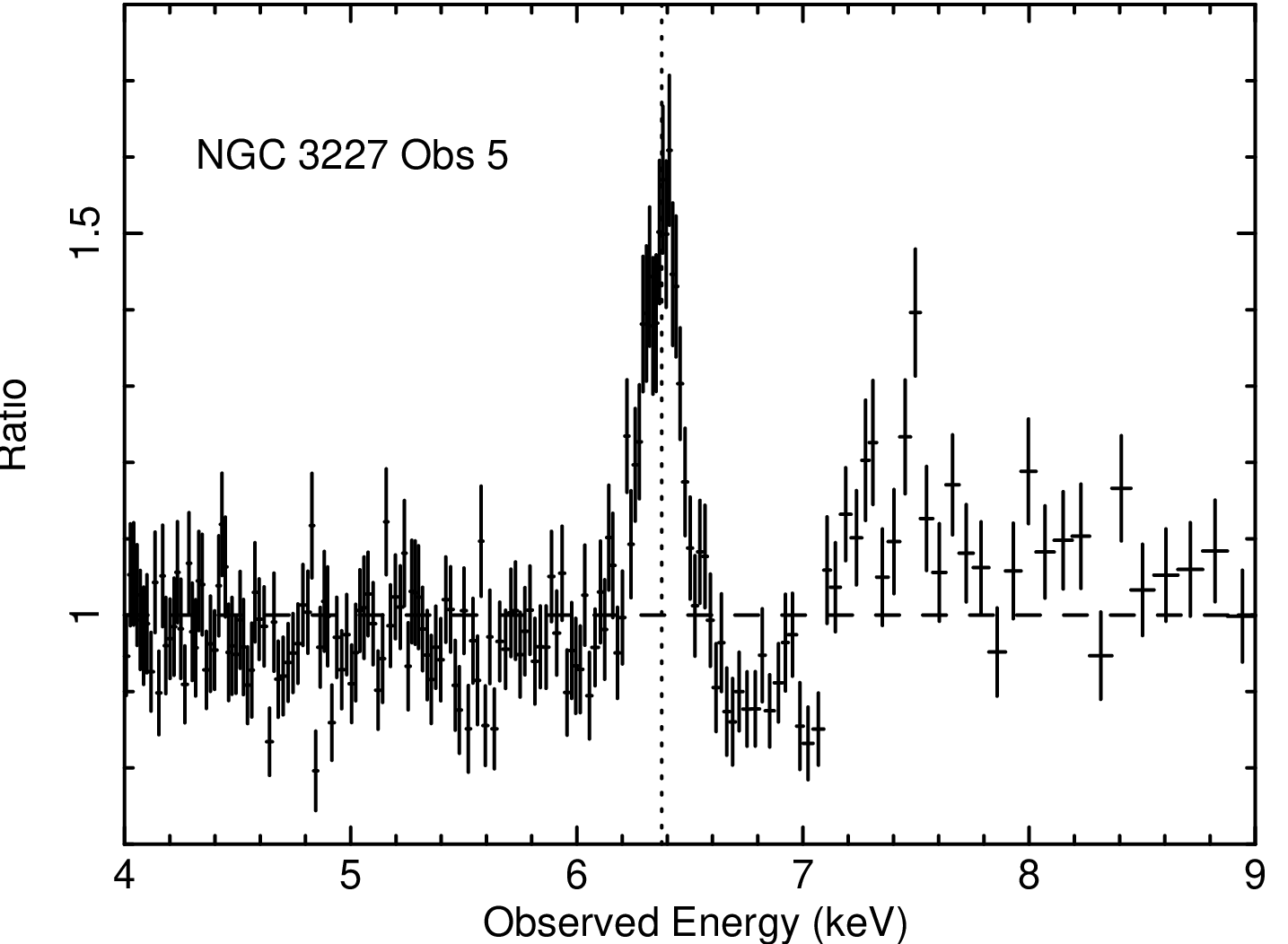}}
\rotatebox{-90}{\includegraphics[width=4cm]{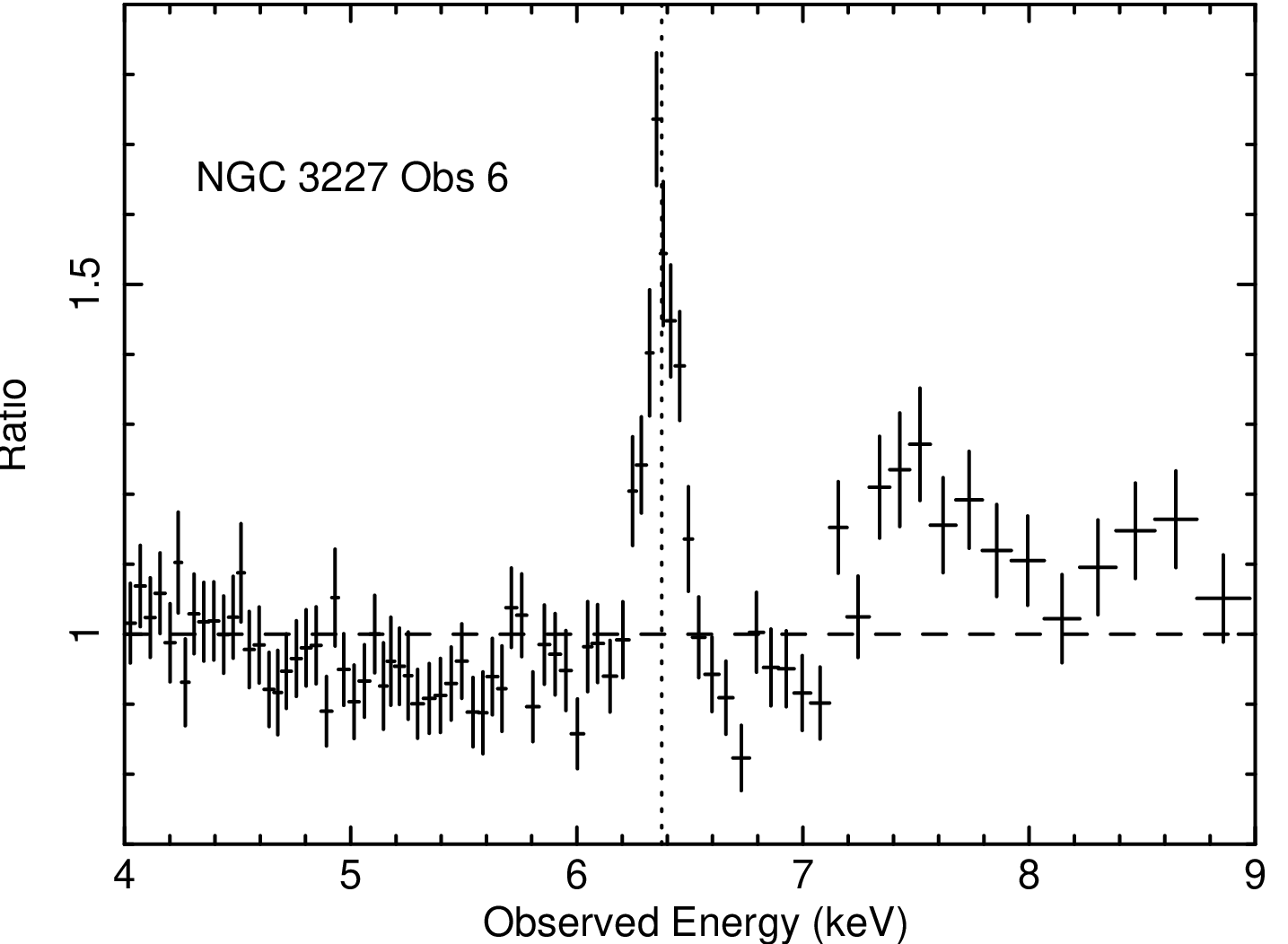}}
\rotatebox{-90}{\includegraphics[width=4cm]{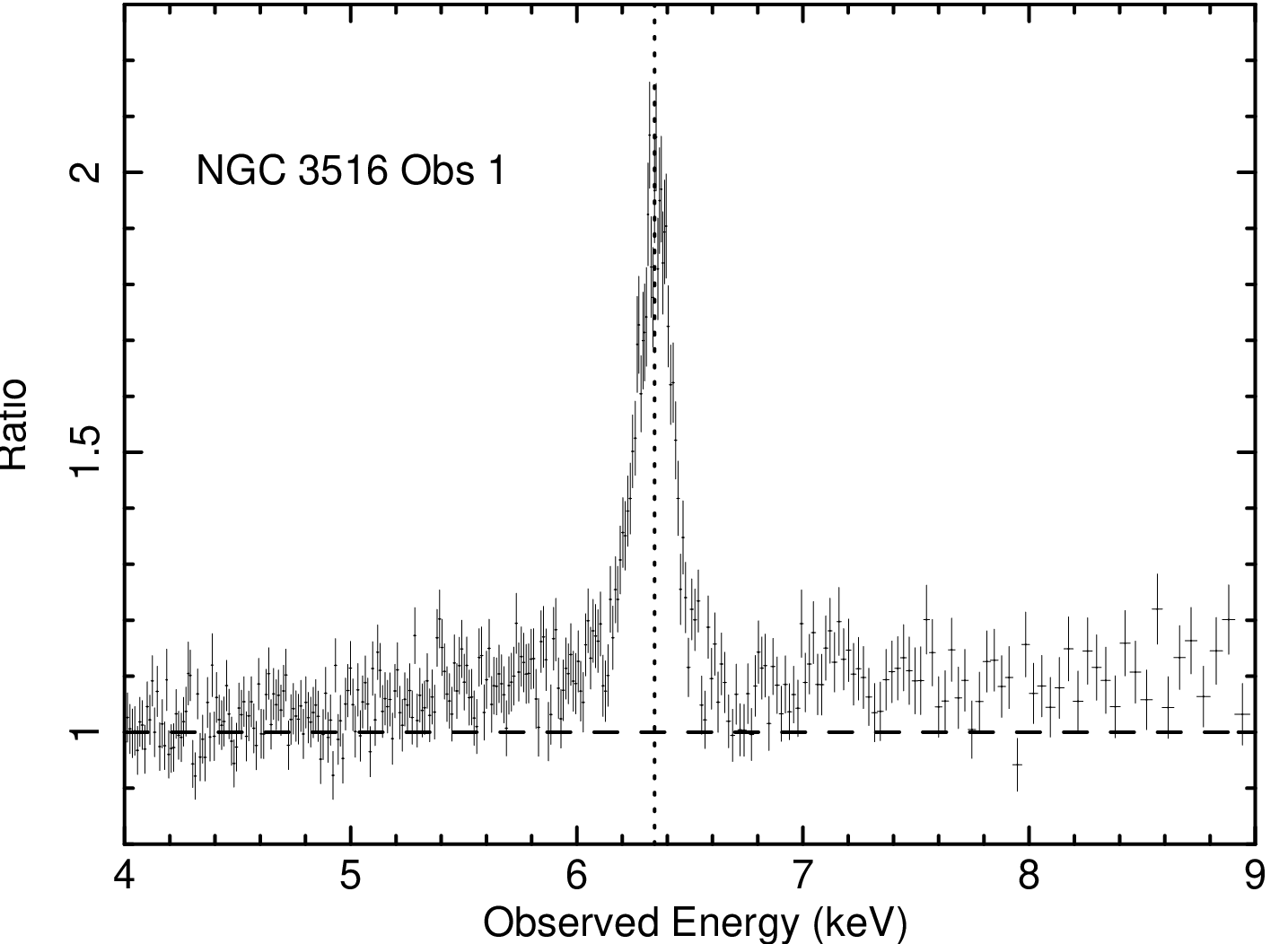}}
\rotatebox{-90}{\includegraphics[width=4cm]{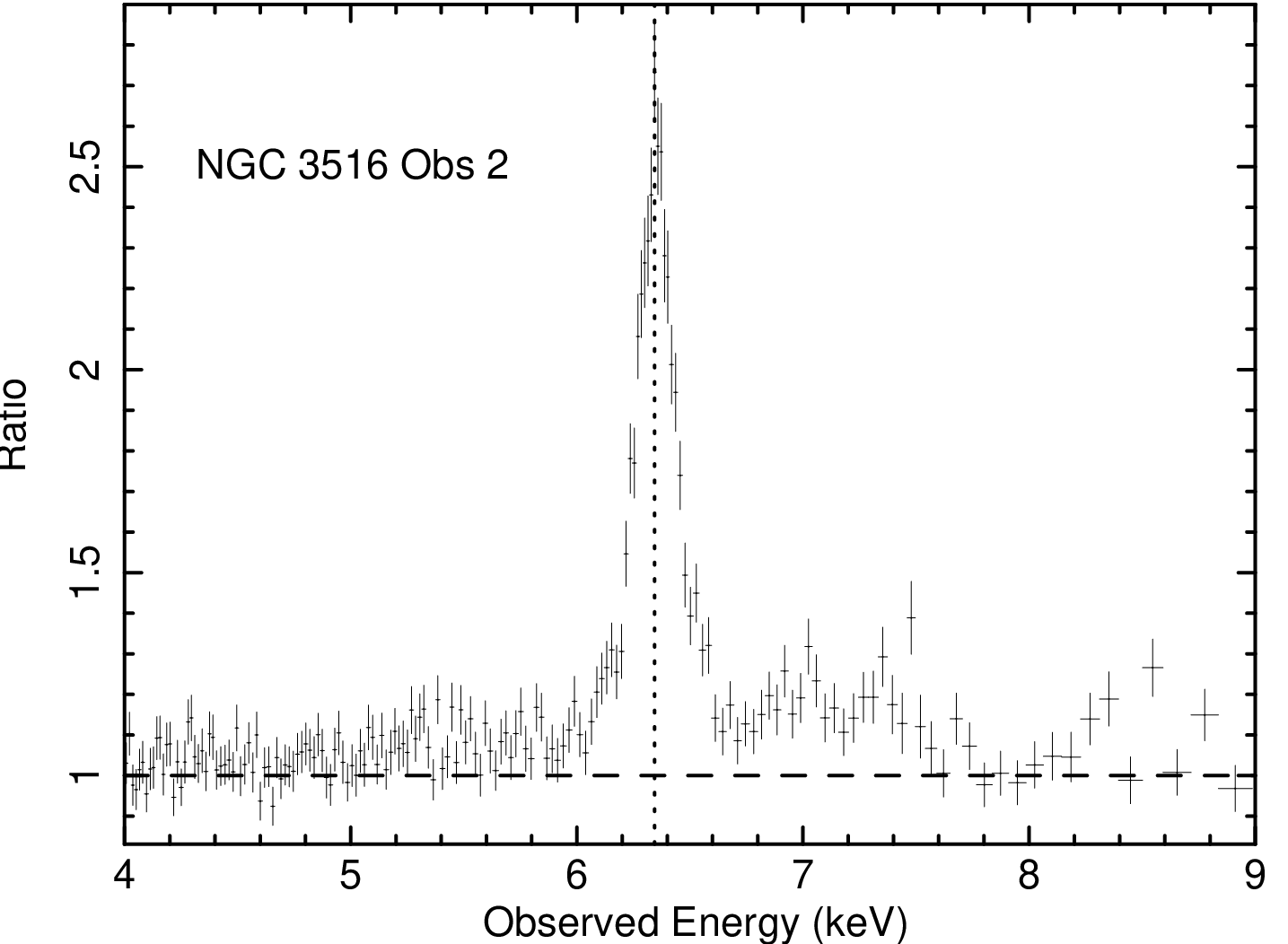}}
\rotatebox{-90}{\includegraphics[width=4cm]{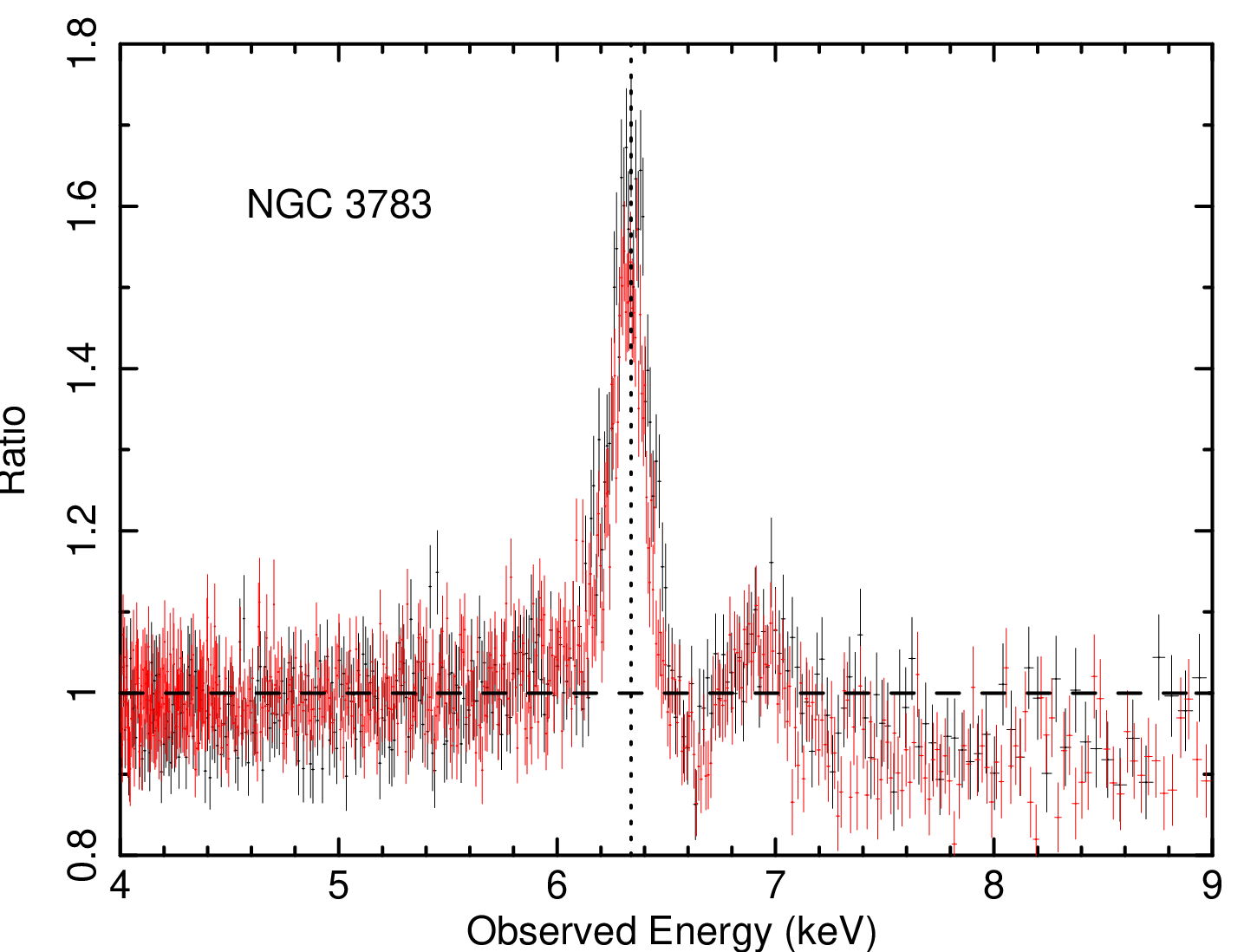}}
\rotatebox{-90}{\includegraphics[width=4cm]{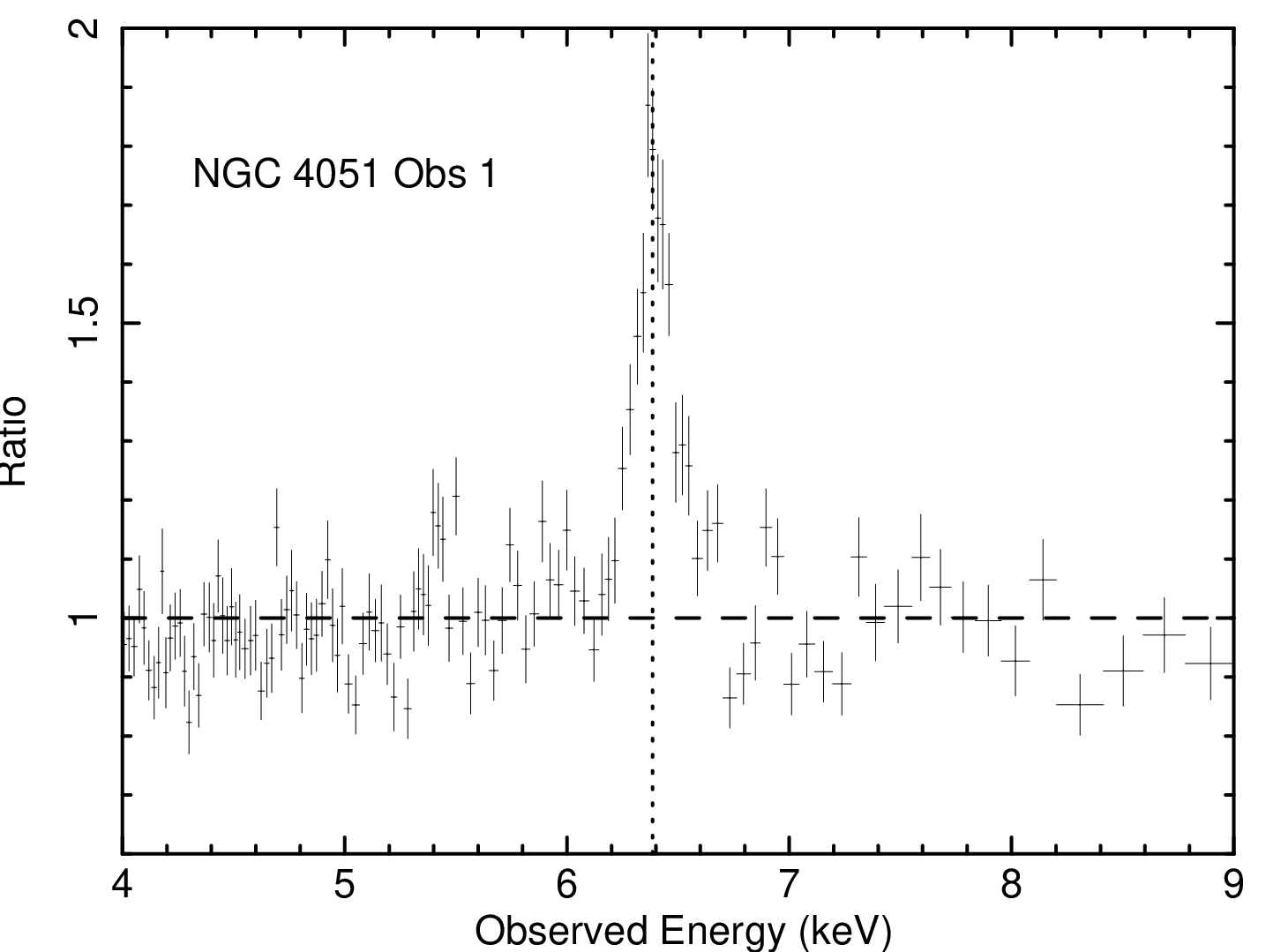}}
\rotatebox{-90}{\includegraphics[width=4cm]{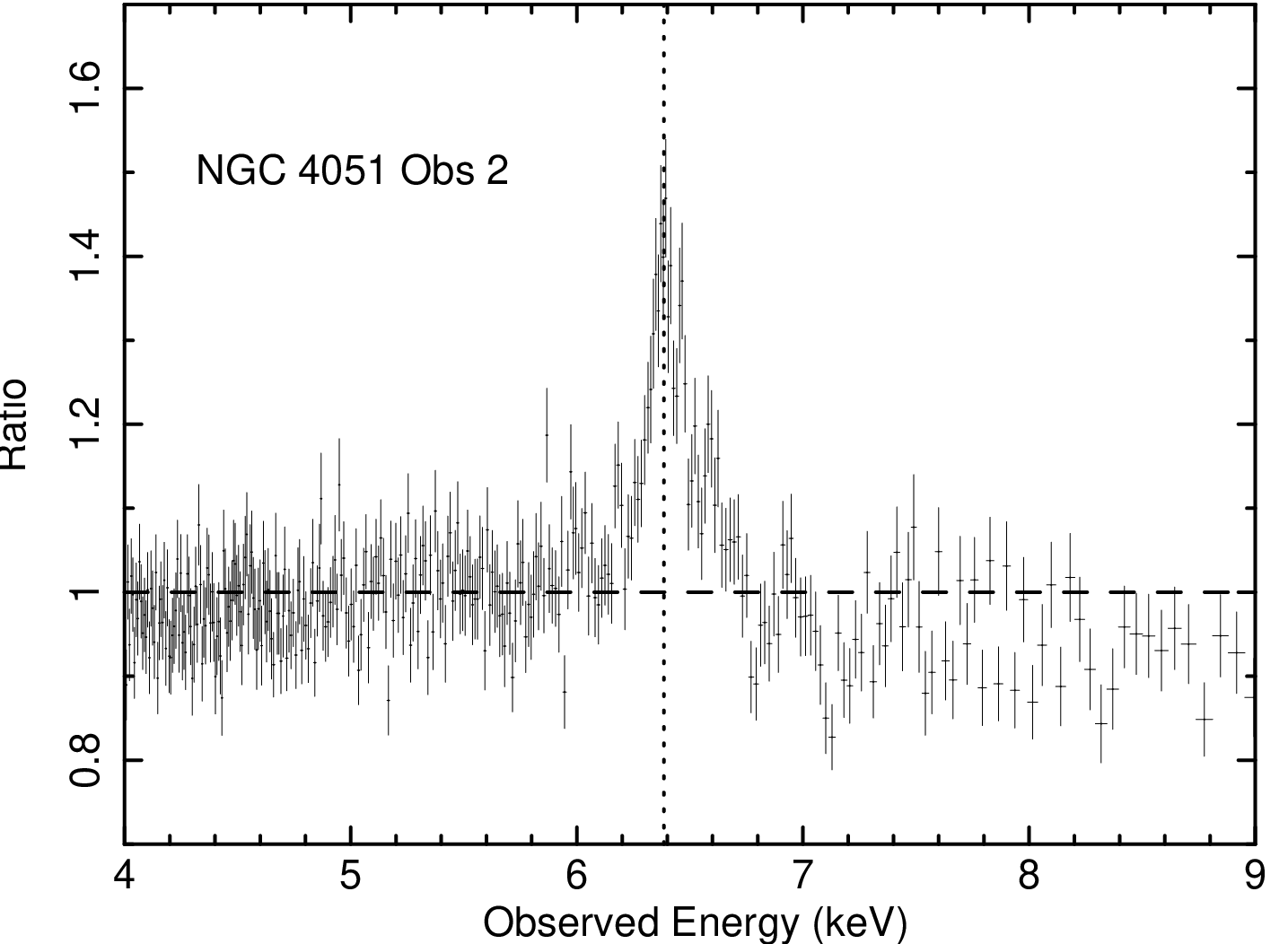}}
\rotatebox{-90}{\includegraphics[width=4cm]{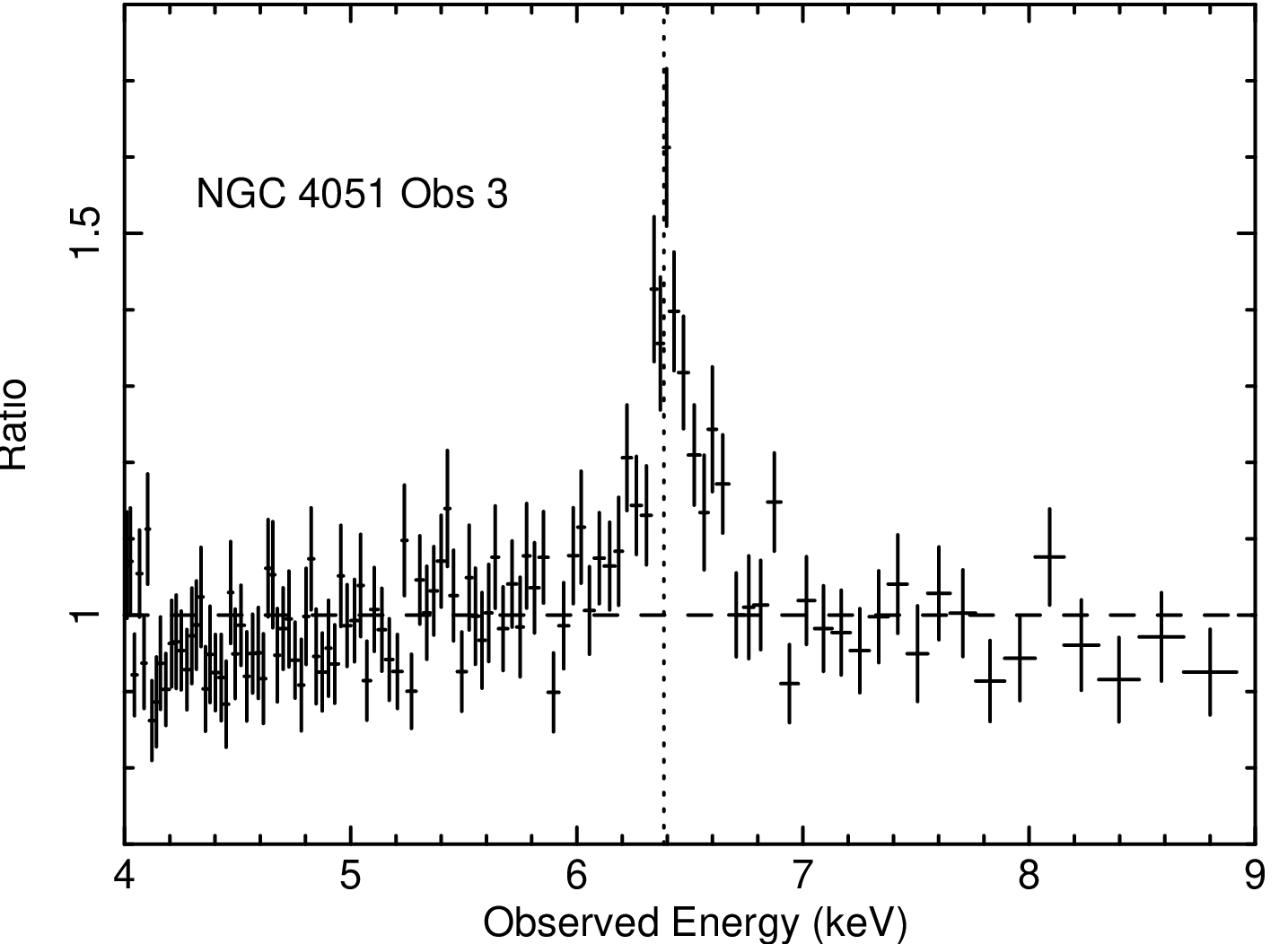}}
\rotatebox{-90}{\includegraphics[width=4cm]{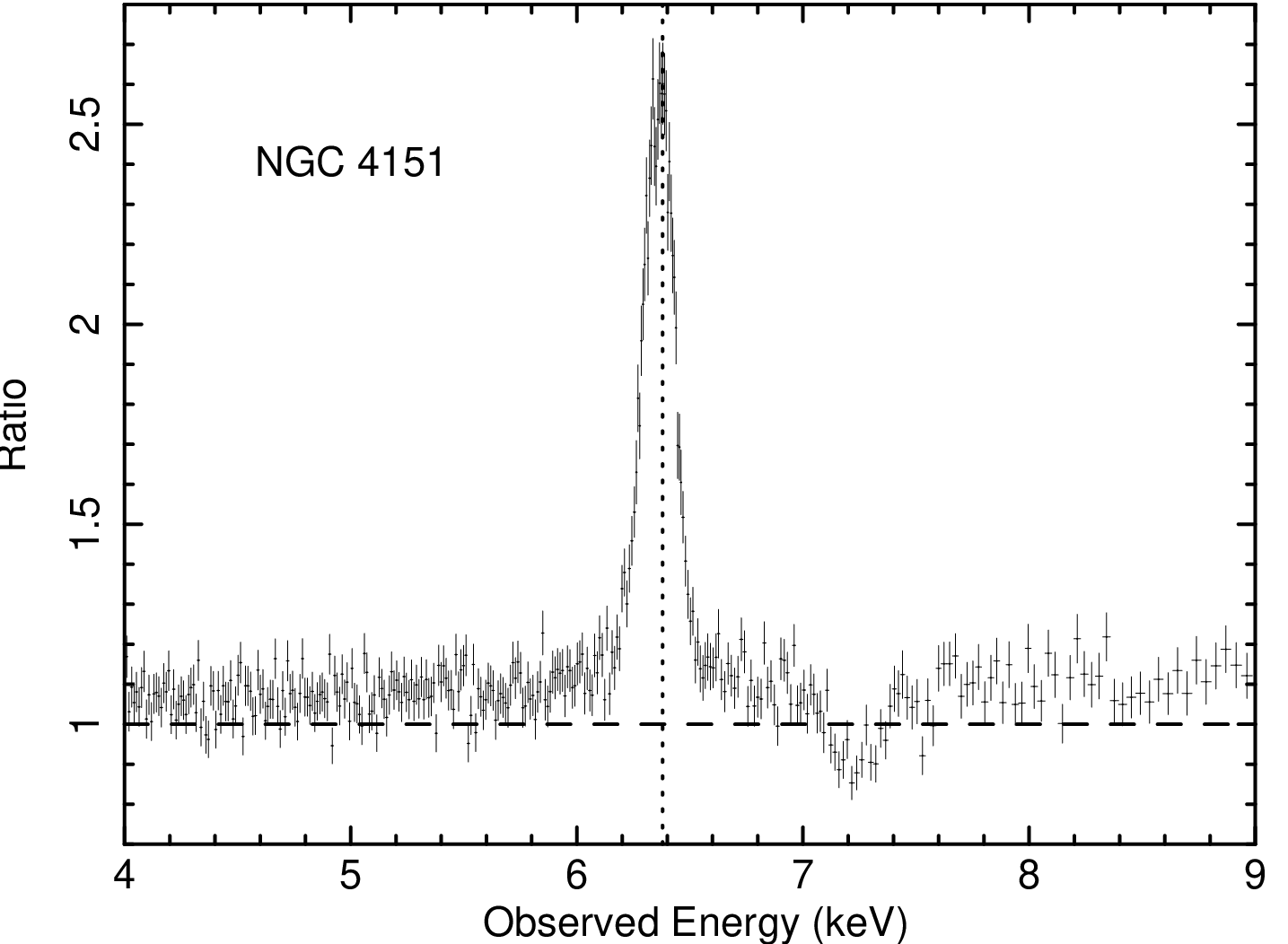}}
\rotatebox{-90}{\includegraphics[width=4cm]{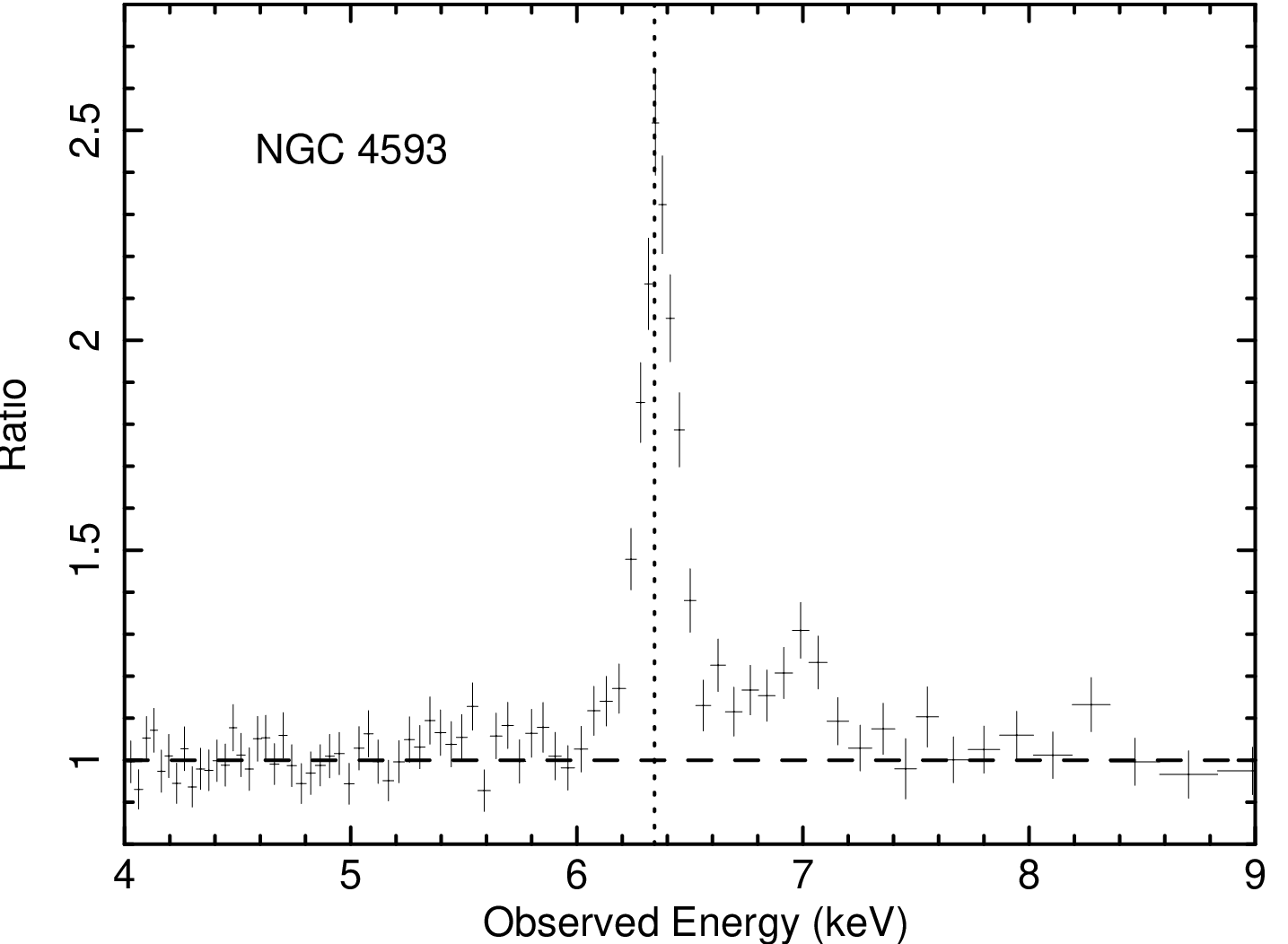}}
\rotatebox{-90}{\includegraphics[width=4cm]{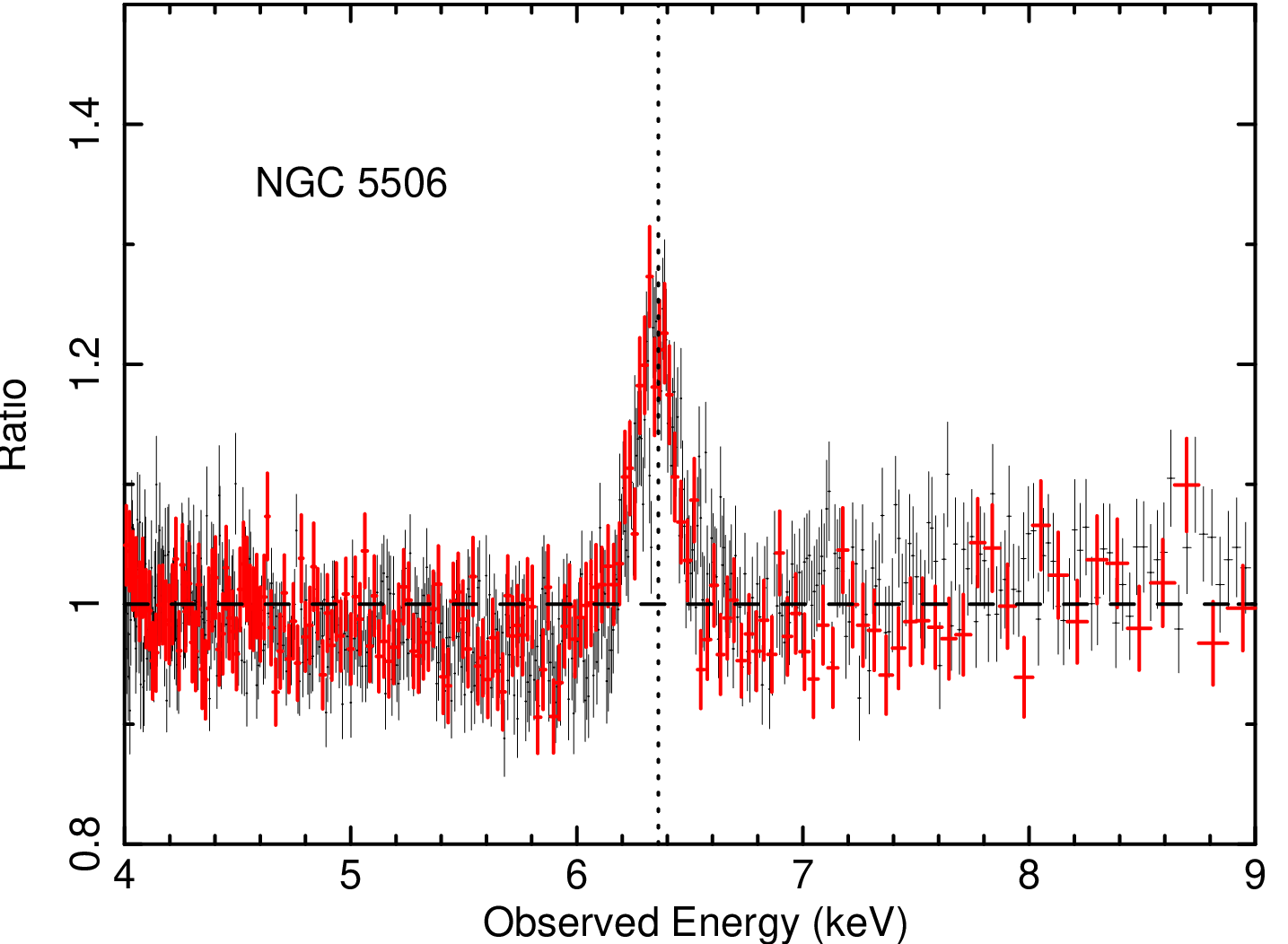}}

\contcaption{Ratio plots of the 4-9\,keV residuals without any modelling of the reflection component i.e. the Fe\,K region is left totally unmodelled with only warm absorption at soft X-ray energies being taken into account in addition to the intrinsic powerlaw and \textsc{compTT} (where required). The data has been refit after the removal of components. The vertical dashed lines represent 6.4\,keV in the observed frames. Red data points indicate those from further observations.}
\end{figure*}

\begin{figure*}
\rotatebox{-90}{\includegraphics[width=4cm]{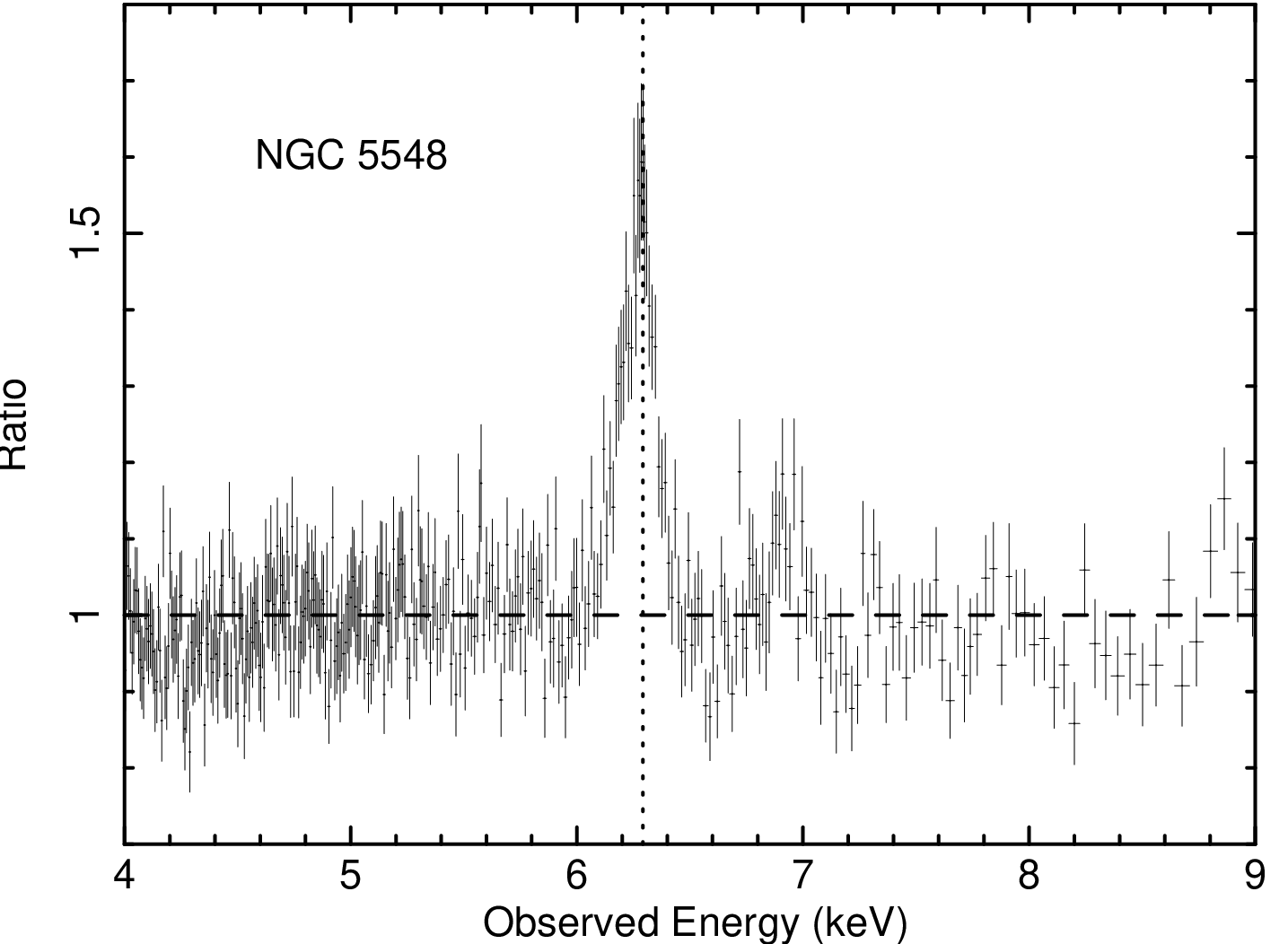}}
\rotatebox{-90}{\includegraphics[width=4cm]{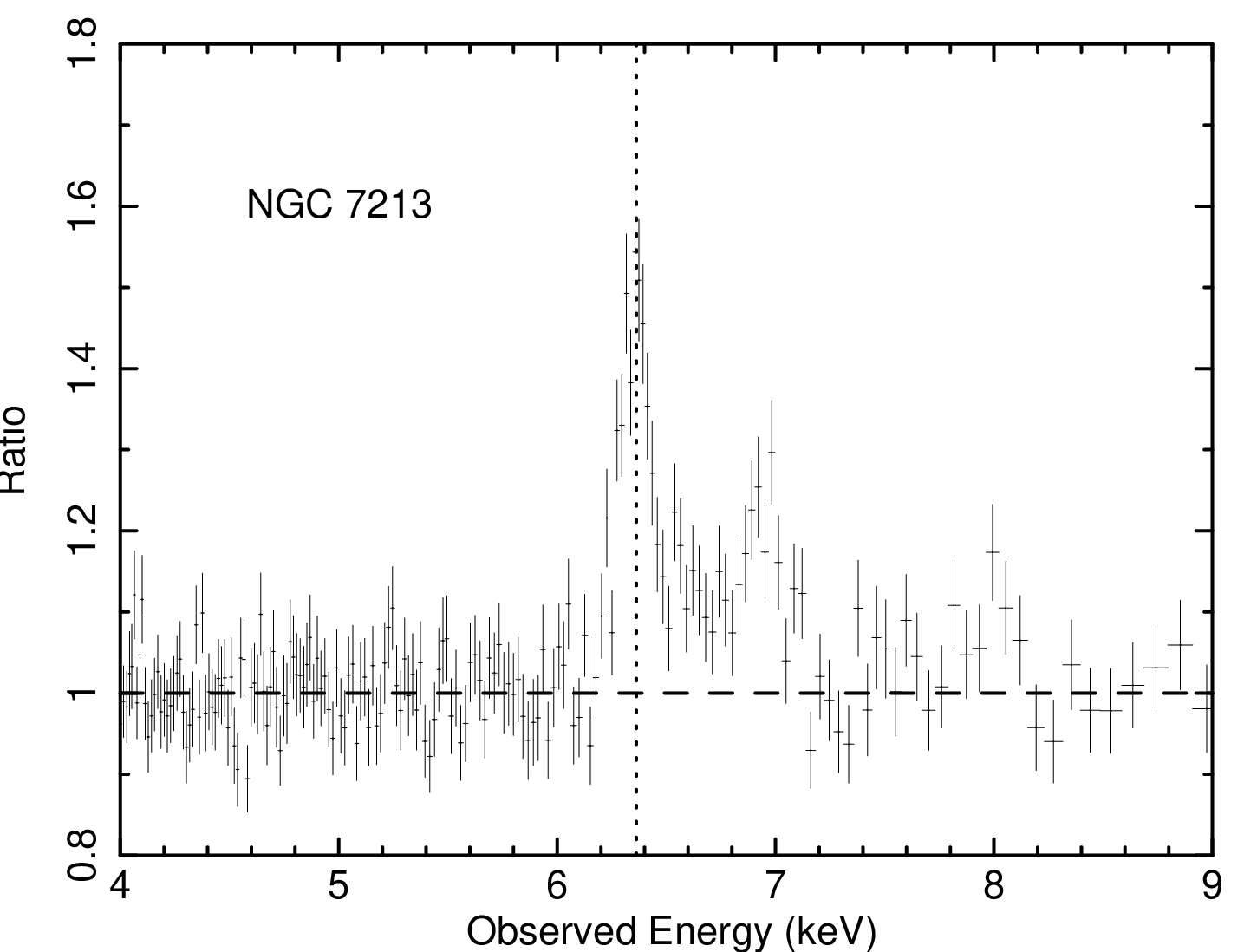}}
\rotatebox{-90}{\includegraphics[width=4cm]{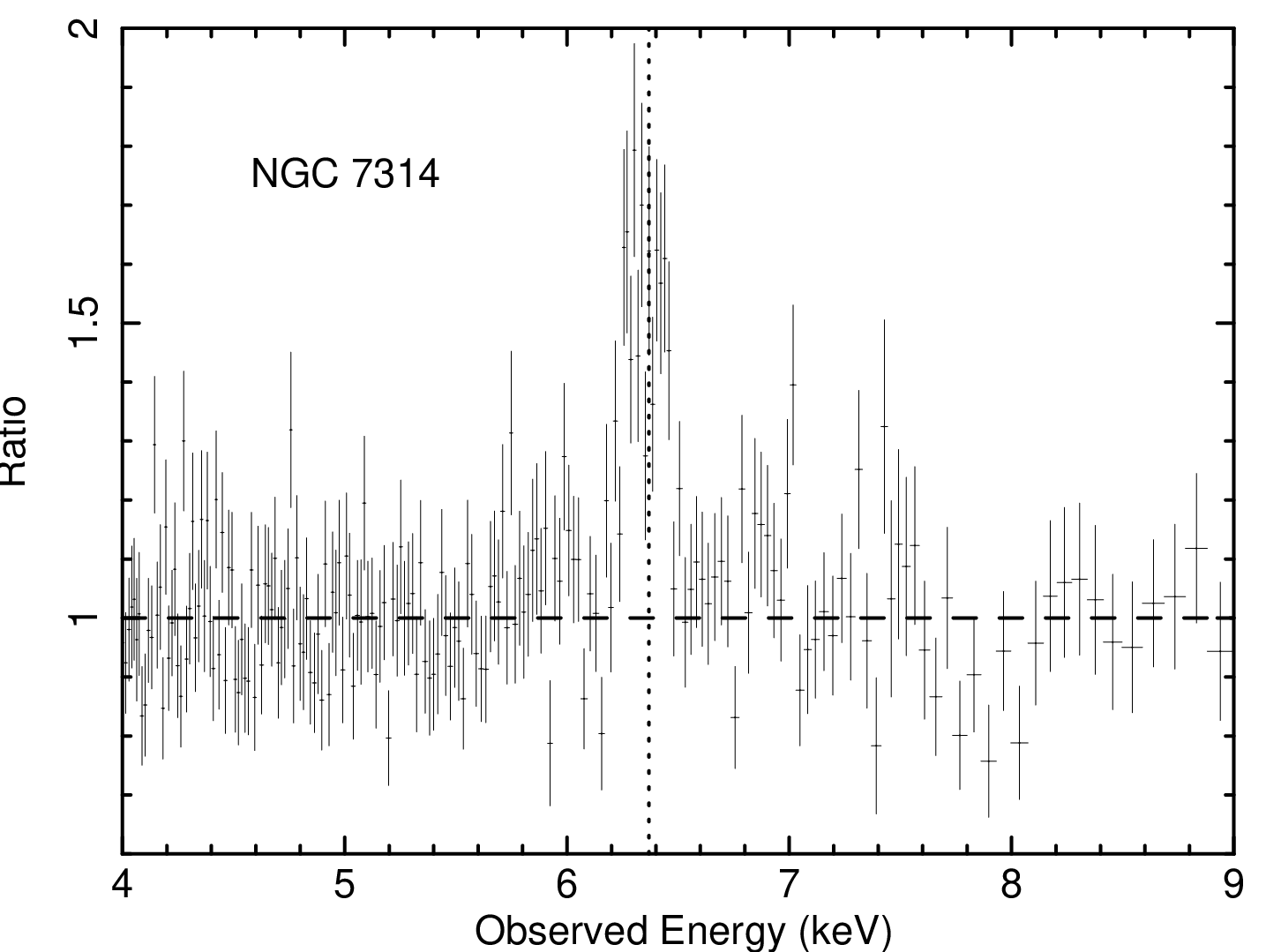}}
\rotatebox{-90}{\includegraphics[width=4cm]{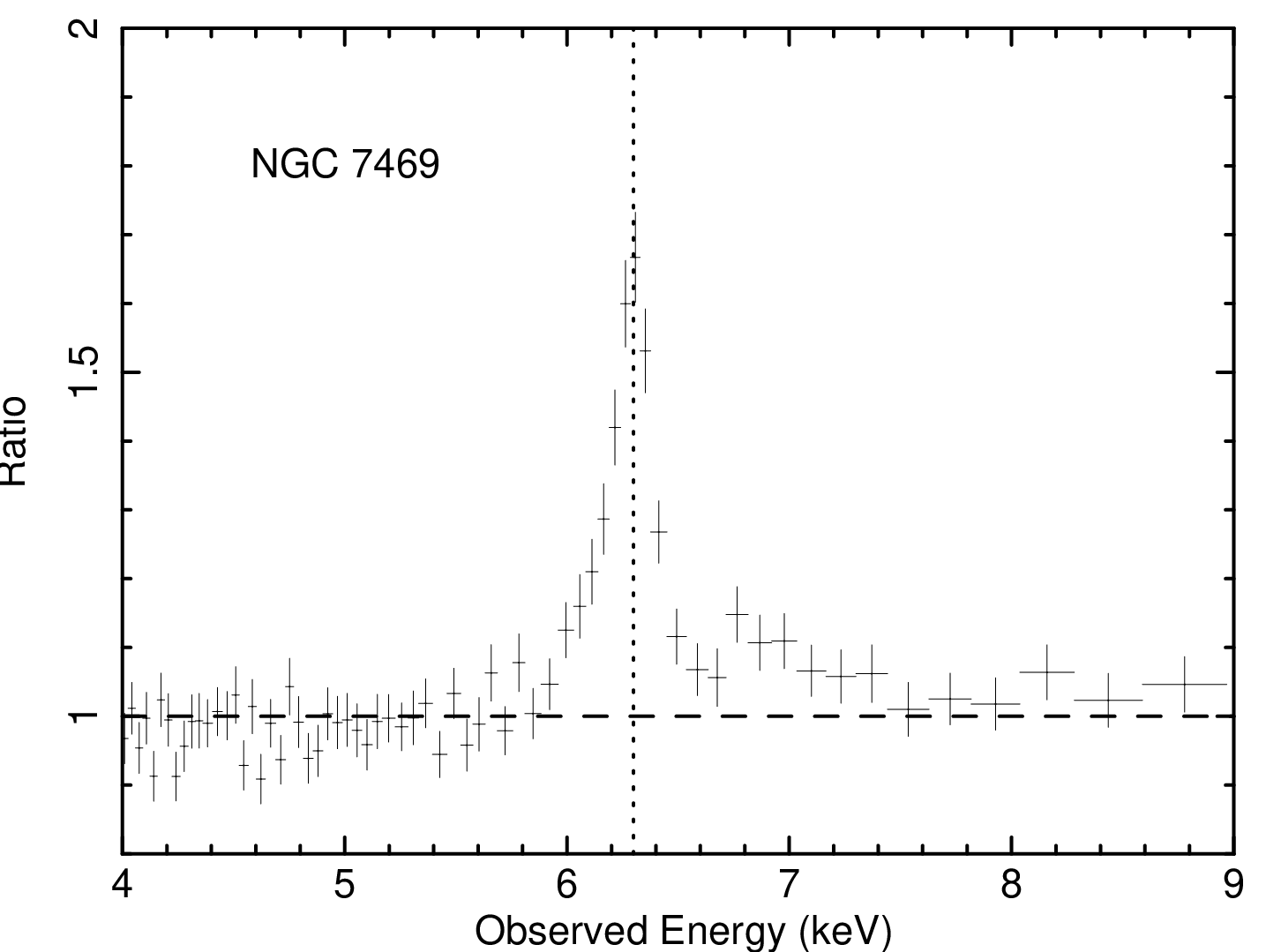}}
\rotatebox{-90}{\includegraphics[width=4cm]{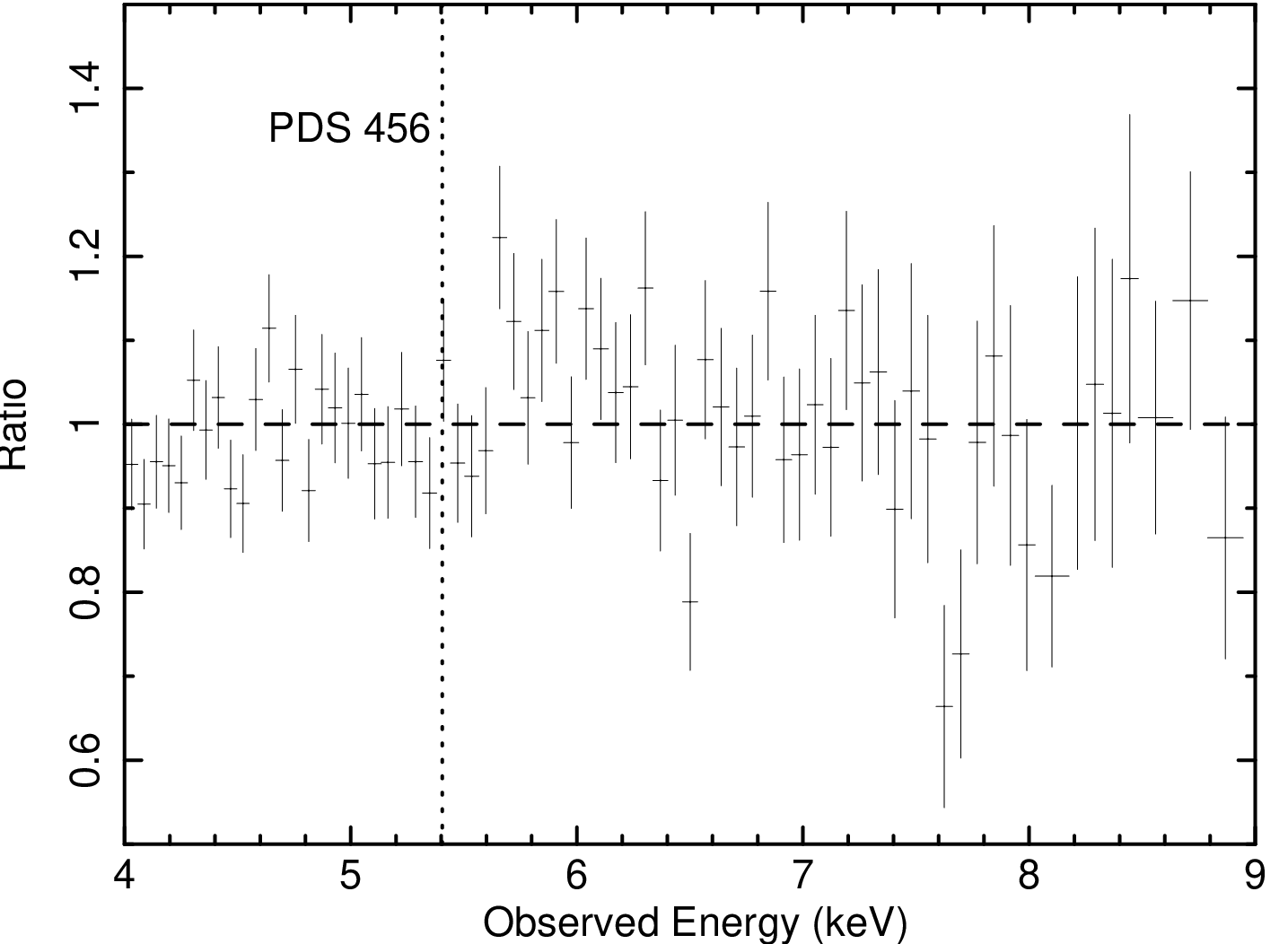}}
\rotatebox{-90}{\includegraphics[width=4cm]{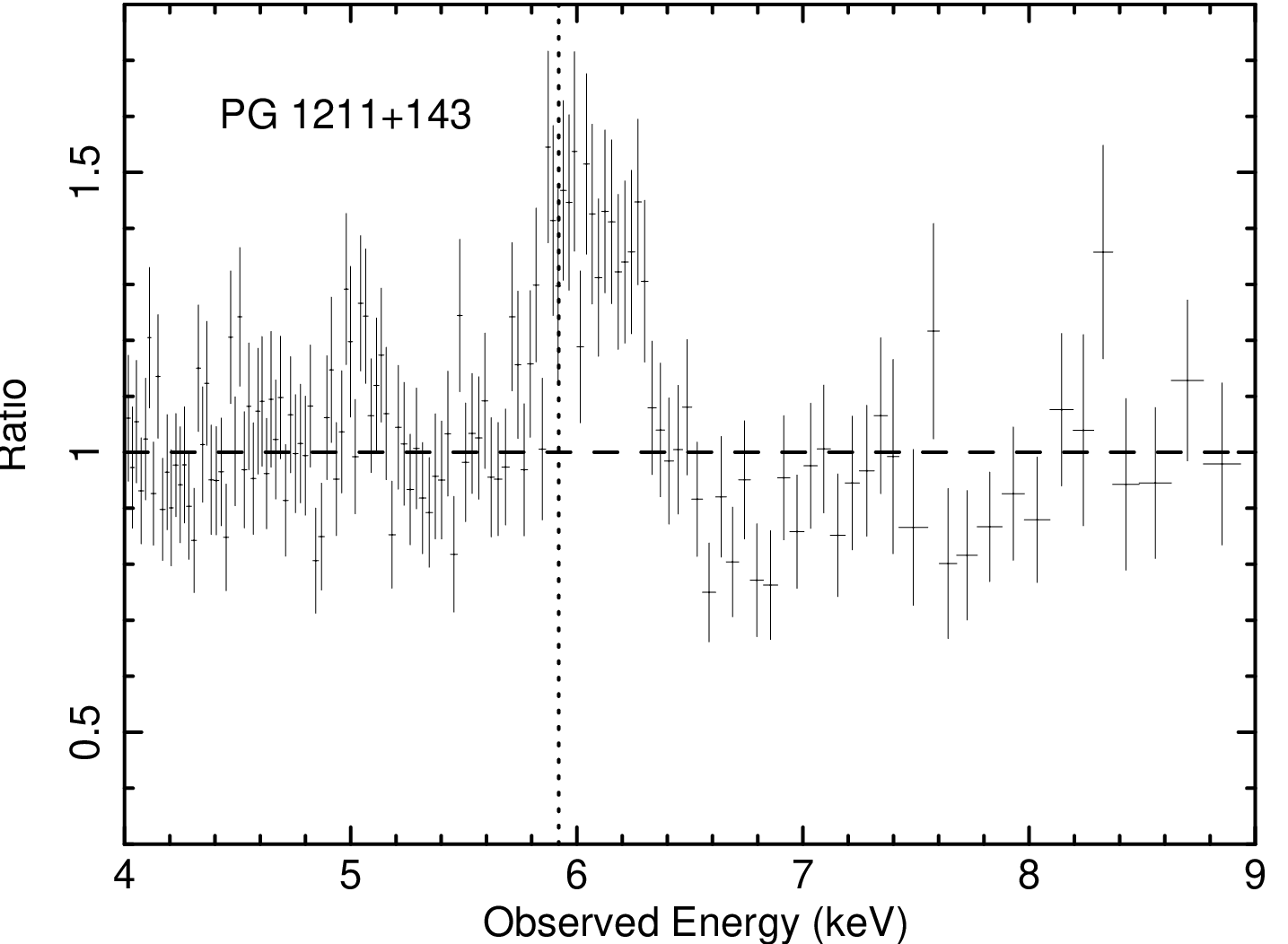}}
\rotatebox{-90}{\includegraphics[width=4cm]{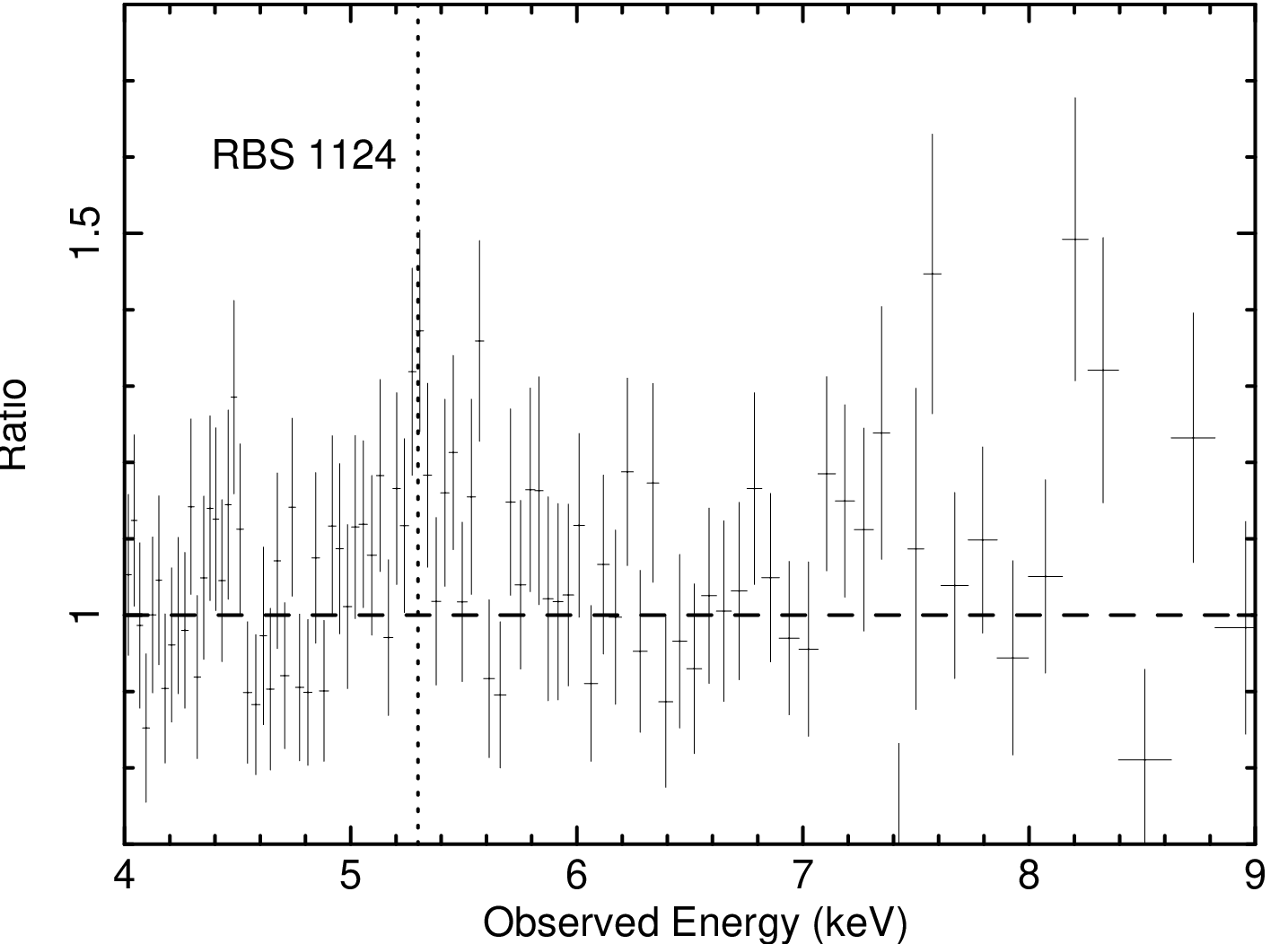}}
\rotatebox{-90}{\includegraphics[width=4cm]{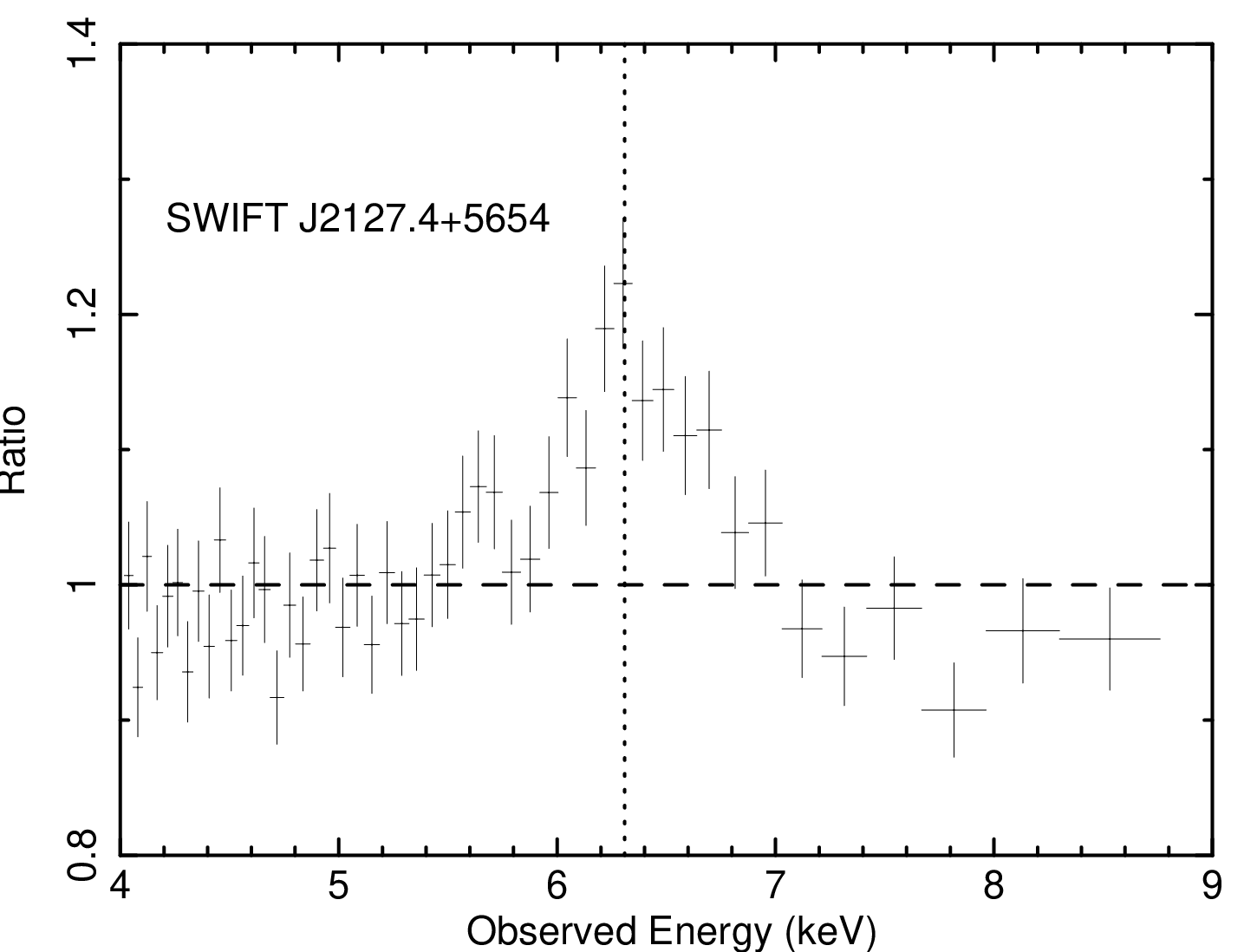}}
\rotatebox{-90}{\includegraphics[width=4cm]{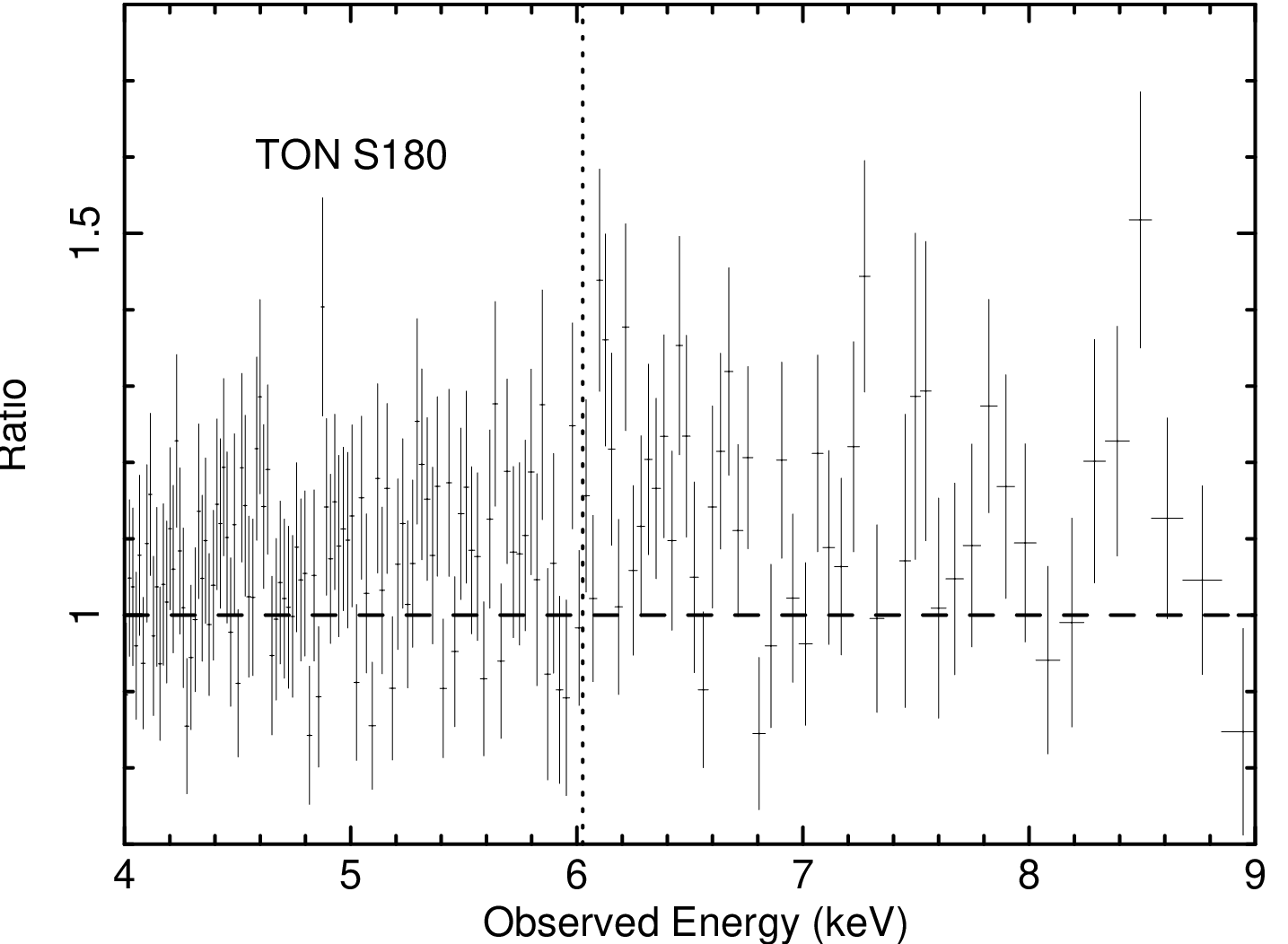}}

\contcaption{Ratio plots of the 4-9\,keV residuals without any modelling of the reflection component i.e. the Fe\,K region is left totally unmodelled with only warm absorption at soft X-ray energies being taken into account in addition to the intrinsic powerlaw and \textsc{compTT} (where required). The data has been refit after the removal of components. The vertical dashed lines represent 6.4\,keV in the observed frames. Red data points indicate those from further observations.}
\end{figure*}

\subsubsection{Analysis of `bare' Seyfert spectra}
As noted in Patrick et al. (2011b) the warm absorber below 2.5\,keV has a significant affect upon the X-ray spectrum at higher energies, particularly important when attempting to test for the strength or indeed presence of a relativistically broadened red-wing in the Fe\,K region. Subsequently the broadband 0.6-100.0\,keV analysis conducted here suggests that the majority of Seyfert 1 AGN in this sample show evidence for at least one warm absorber zone or additional neutral absorption over the simple Galactic absorbing column. 

Only 11/46 objects in this sample can be considered 'bare' in that no absorption (either neutral, warm or partially covering) whatsoever is required: 3C 390.3, Ark 120, Fairall 9, MCG--02-14-009, Mrk 110, Mrk 335, Mrk 359, NGC 3147, NGC 7213, NGC 7469 and RBS 1124. These AGN are therefore straight forward to model, without any complications due to curvature from warm absorbing components, representing the most fiducial of AGN X-ray spectra simply consisting of \textsc{(powerlaw+compTT+reflionx)*wabs}. Slightly more complex are those objects which are free from warm absorption, however, requiring an additional neutral absorbing column over and above that from the standard Galactic absorption calculated by the \textsc{nh} {\sl ftool}. This is accounted for by multiplying the typical 'bare' AGN model by a single \textsc{zphabs} fixed at the redshift of the object with column density free to vary, as used in 6/46 of AGN in the sample: 3C 111, 3C 120, MCG--05-23-16, NGC 2992, NGC 7314 and SWIFT J2127.4+5654. Note that the higher neutral absorption in 3C 111 is likely due to the presence of a giant molecular cloud in the line of sight, see Bania et al. (1991) and Rivers, Markowitz \& Rothschild (2011b). MCG--05-23-16 does not require an additional totally covering zone of neutral absorption, instead requiring a neutral \textsc{zphabs} geometry whereby only the \textsc{powerlaw} component is absorbed. Two further AGN (Mrk 205 and PDS 456) require a partial coverer, but not a warm absorber. Hence a total of 19/46 objects in this sample do not indicate the presence of a {\sl warm} absorbing component. 


\begin{table*}
\caption{Broad Gaussian parametrisation of Fe\,K region in addition to the baseline model. $^{a}$ Flux given in units $(10^{-5}\,{\rm ph\,{\rm cm}^{-2}\,s^{-1}})$. $^{b}$ Parameters other than normalisation tied between multiple observations, all are tied in NGC 5506. -- Indicates that no broad Gaussian can be fit i.e. no discernible reduction in $\chi^{2}$ whatsoever.}
\begin{tabular}{l c c c c c c c}
\hline
Object & \multicolumn{5}{c}{Broad line} & Fe\,{\rm XXV} & Fe\,{\rm XXVI} \\
& LineE (keV) & $\sigma_{\rm Broad}$ (keV) & $EW_{\rm Broad}$ (eV) & Flux$^{a}$ & $\triangle\chi^{2}$ & emission & emission \\
\hline
1H 0419--577 & -- & -- & -- & -- & -- & X & X  \\
3C 111 & $6.28^{+0.09}_{-0.05}$ & $<0.136$ & $19^{+23}_{-9}$ & $0.55^{+0.67}_{-0.27}$ & -11 & \checkmark & \checkmark \\
3C 120 & $6.38^{+0.01}_{-0.01}$ & $0.124^{+0.025}_{-0.022}$ & $76^{+6}_{-9}$  & $4.04^{+0.35}_{-0.49}$ & -75 & \checkmark & X \\
& & & $84^{+7}_{-10}$ & \\
3C 382 & $6.45^{+0.09}_{-0.12}$ & $0.198^{+0.210}_{-0.090}$ & $42^{+18}_{-17}$ & $2.45^{+1.07}_{-1.01}$ & -20 & X & X \\
3C 390.3 & $6.49^{+0.09}_{-0.10}$ & $0.320^{+0.256}_{-0.106}$ & $86^{+31}_{-20}$ & $3.29^{+1.17}_{-0.78}$ & -40 & X & X \\
3C 445 & $6.04^{+0.12}_{-0.11}$ & $<0.436$ & $23^{+30}_{-13}$ & $1.07^{+0.58}_{-0.46}$ & -7 & \checkmark & X \\
4C 74.26 & $6.12^{+0.06}_{-0.07}$ & $<0.186$ & $22^{+12}_{-9}$ & $1.07^{+0.58}_{-0.46}$ & -20 & \checkmark & X \\
Ark 120 & $6.36^{+0.08}_{-0.09}$ & $0.320^{+0.110}_{-0.090}$ & $105^{+26}_{-24}$ & $0.70^{+0.32}_{-0.32}$ & -12 & X & \checkmark \\
Ark 564 & $6.42^{+0.09}_{-0.08}$ & $<0.161$ & $20^{+14}_{-13}$ & $0.34^{+0.25}_{-0.22}$ & -3 & \checkmark & X \\
Fairall 9 & $5.91^{+0.24}_{-0.21}$ & $0.505^{+0.187}_{-0.173}$ & $49^{+20}_{-18}$ & $1.45^{+0.58}_{-0.55}$ & -36 & \checkmark & \checkmark \\
IC 4329A & $6.34^{+0.15}_{-0.15}$ & $0.551^{+0.119}_{-0.100}$ & $61^{+15}_{-15}$ & $7.94^{+1.96}_{-1.91}$ & -53 & X & \checkmark \\
IRAS 13224--3809 & $6.19^{+0.58}_{-0.65}$ & $1.014^{+0.613}_{-0.346}$ & $632^{+303}_{-260}$ & $0.37^{+0.18}_{-0.15}$ & -9 & \checkmark & X \\
MCG--02-14-009 & $6.33^{+0.21}_{-0.24}$ & $0.280^{+0.280}_{-0.130}$ & $92^{+59}_{-56}$ & $0.49^{+0.31}_{-0.30}$ & -3 & X & \checkmark \\
MCG--02-58-22 & -- & -- & -- & -- & -- & X & X \\
MCG--05-23-16 & $6.34^{+0.08}_{-0.10}$ & $0.512^{+0.085}_{-0.078}$ & $108^{+16}_{-17}$ & $12.25^{+1.80}_{-1.87}$ & -15 & \checkmark & X \\
MCG--06-30-15 & $5.93^{+0.07}_{-0.14}$ & $0.840^{+0.06}_{-0.06}$ & $149^{+21}_{-9}$ & $1.31^{+0.68}_{-0.66}$ & -302 &X & \checkmark \\
MCG+8-11-11 & $6.35^{+0.03}_{-0.04}$ & $0.170^{+0.056}_{-0.042}$ & $67^{+13}_{-12}$ & $5.30^{+1.02}_{-0.97}$ & -73 & X & \checkmark \\
MR 2251--178 & $6.46^{+0.10}_{-0.17}$ & $0.307^{+0.199}_{-0.100}$ & $49^{+14}_{-14}$ & $2.65^{+0.76}_{-0.75}$ & -26 & X & X \\
Mrk 79 & $6.17^{+0.22}_{-0.27}$ & $0.529^{+0.208}_{-0.159}$ & $136^{+44}_{-45}$ & $2.83^{+0.91}_{-0.93}$ & -51 & \checkmark & \checkmark \\
Mrk 110 & -- & -- & -- & -- & -- & \checkmark & X \\
Mrk 205 & -- & -- & -- & -- & -- & X & X \\
Mrk 279 & $6.59^{+0.26}_{-0.19}$ & $<0.333$ & $20^{+25}_{-19}$ & $0.15^{+0.19}_{-0.14}$ & -6 & X & X \\
Mrk 355 & $6.27^{+0.13}_{-0.17}$ & $0.500^{+0.130}_{-0.110}$ & $134^{+42}_{-38}$ & $2.28^{+0.72}_{-0.65}$ & -53 & \checkmark & \checkmark \\
Mrk 359 & $6.40^{+0.06}_{-0.06}$ & $<0.177$ & $88^{+39}_{-41}$ & $0.52^{+0.23}_{-0.25}$ & -10 & \checkmark & X \\
Mrk 509 & $6.60^{+1.28}_{-0.14}$ & $0.690^{+1.363}_{-0.151}$ & $120^{+28}_{-27}$ & $6.32^{+1.50}_{-1.42}$ & -57 & \checkmark & X \\
Mrk 766 & $6.66^{+0.11}_{-0.07}$ & $0.144^{+0.109}_{-0.115}$ & $73^{+27}_{-25}$ & $0.81^{+0.30}_{-0.28}$ & -14 & \checkmark & X \\
Mrk 841 & $5.89^{+0.22}_{-0.24}$ & $0.402^{+0.288}_{-0.173}$ & $80^{+50}_{-37}$ & $1.47^{+0.92}_{-0.68}$ & -20 & \checkmark & X \\
NGC 1365 & $6.48^{+0.02}_{-0.01}$ & $<0.028$ & $<13$ & $<0.29$ & -63 & \checkmark & \checkmark \\
NGC 2992 & $6.51^{+0.14}_{-0.11}$ & $0.323^{+0.153}_{-0.100}$ & $73^{+28}_{-25}$ & $1.39^{+0.53}_{-0.47}$ & -17 & X & X \\
NGC 3147 & $6.45^{+0.04}_{-0.03}$ & $<0.066$ & $110^{+51}_{-41}$ & $0.21^{+0.10}_{-0.08}$ & -15 & X & \checkmark \\
NGC 3227$^{b}$ & $6.34^{+0.09}_{-0.09}$ & $0.707^{+0.100}_{-0.087}$ & $80^{+24}_{-23}$ & $4.50^{+1.33}_{-1.28}$ & -124 & X & \checkmark \\
& & & $176^{+32}_{-31}$ & $5.38^{+0.97}_{-0.94}$ & \\
& & & $26^{+25}_{-25}$ & $1.17^{+1.13}_{-1.13}$ & \\
& & & $281^{+39}_{-38}$ & $5.71^{+0.79}_{-0.77}$ & \\
& & & $55^{+23}_{-22}$ & $2.22^{+0.92}_{-0.90}$ & \\
& & & $43^{+32}_{-31}$ & $1.35^{+1.00}_{-0.98}$ & \\
NGC 3516$^{b}$ & $6.32^{+0.12}_{-0.12}$ & $0.874^{+0.118}_{-0.101}$ & $151^{+22}_{-21}$ & $6.95^{+1.01}_{-0.97}$ & -84 & \checkmark & X \\
& & & $81^{+29}_{-29}$ & $1.89^{+0.69}_{-0.68}$ & \\
NGC 3783$^{b}$ & $6.07^{+0.20}_{-0.18}$ & $0.761^{+0.221}_{-0.104}$ & $97^{+38}_{-32}$ & $6.38^{+2.51}_{-2.10}$ & -68 & X & \checkmark \\
& & & $57^{+15}_{-15}$ & $4.89^{+1.25}_{-1.25}$ & \\
NGC 4051$^{b}$ & $6.23^{+0.17}_{-0.19}$ & $0.742^{+0.151}_{-0.124}$ & $139^{+39}_{-36}$ & $1.80^{+0.50}_{-0.47}$ & -70 & \checkmark & X \\
& & & $74^{+24}_{-23}$ & $2.12^{+0.68}_{-0.66}$ &  \\ 
& & & $112^{+37}_{-37}$ & $2.48^{+0.83}_{-0.81}$ & \\
NGC 4151 & -- & -- & -- & -- & -- & X & X \\
NGC 4593 & $6.65^{+0.18}_{-0.14}$ & $0.368^{+0.289}_{-0.136}$ & $87^{+40}_{-35}$ & $1.26^{+0.58}_{-0.51}$ & -12 & \checkmark & X \\
NGC 5506$^{b}$ & $6.50^{+0.07}_{-0.09}$ & $0.317^{+0.074}_{-0.060}$ & $19^{+5}_{-4}$ & $7.83^{+1.91}_{-1.51}$ & -32 & \checkmark & \checkmark \\
& & & & $20^{+5}_{-4}$ & \\
NGC 5548 & $6.36^{+0.03}_{-0.10}$ & $<0.098$ & $26^{+31}_{-22}$ & $0.60^{+0.72}_{-0.52}$ & -5 & X & X \\ 
NGC 7213 & $6.74^{+0.16}_{-0.16}$ & $<1.294$ & $12^{+30}_{-9}$ & $0.30^{+0.74}_{-0.22}$ & -4 & \checkmark & \checkmark \\
NGC 7314 & $6.36^{+0.35}_{-0.19}$ & $<0.057$ & $58^{+37}_{-34}$ & $0.66^{+0.43}_{-0.39}$ & -7 & X & \checkmark \\
NGC 7469 & $6.32^{+0.06}_{-0.11}$ & $0.150^{+0.07}_{-0.03}$ & $62^{+928}_{-44}$ & $1.55^{+0.70}_{-1.10}$ & -24 & X & X \\
PDS 456 & -- & -- & -- & -- & -- & \checkmark & \checkmark \\
PG 1211+143 & $6.49^{+0.07}_{-0.06}$ & $0.142^{+0.068}_{-0.061}$ & $146^{+28}_{-44}$ & $0.87^{+0.17}_{-0.26}$ & -5 & \checkmark & X \\
RBS 1124 & -- & -- & -- & -- & -- & X & X \\
SWIFT J2127.4+5654 & $6.32^{+0.24}_{-0.25}$ & $0.390^{+0.450}_{-0.260}$ & $71^{+53}_{-48}$ & $3.43^{+2.56}_{-2.32}$ & -12 & \checkmark & \checkmark \\
Ton S180 & -- & -- & -- & -- & -- & X & X \\
\hline
\end{tabular}
\label{tab:broad}
\end{table*}

\subsubsection{Absorbed Seyfert spectra -- fully covering}
The remaining objects in the sample (27/46) feature some degree of complex warm absorption, which is modelled with successive zones of a multiplicative \textsc{xstar} grid (see Table \ref{tab:comparison}). The majority of these AGN are well modelled using one or more fully covering \textsc{xstar} grids with turbulent velocity $v_{\rm turb}=100\,{\rm km}\,{\rm s}^{-1}$ fixed at the redshift of the object (i.e. from Table \ref{tab:sample}) as described in Section 3.1.3. 

\begin{figure}
\rotatebox{-90}{\includegraphics[width=6cm]{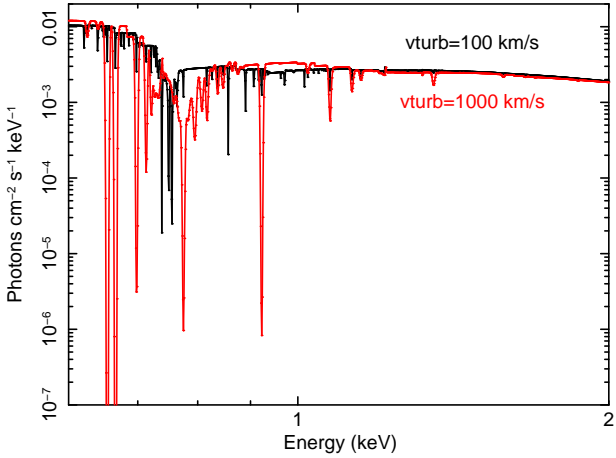}}
\caption{Comparison of the differences between the soft X-ray modelling of the spectrum when using an \textsc{xstar} grid with turbulent velocities of $v_{\rm turb}=100\,{\rm km}\,{\rm s}^{-1}$ and $v_{\rm turb}=1000\,{\rm km}\,{\rm s}^{-1}$. These represent a powerlaw with $\Gamma=2$ absorbed by a fully covering absorption zone with column density $N_{\rm H}=1\times10^{22}\,{\rm cm}^{-2}$ and ionisation ${\rm log}\,\xi=1$.}
\label{fig:vturb}
\end{figure}

Rather using the typical $v_{\rm turb}=100\,{\rm km}\,{\rm s}^{-1}$ \textsc{xstar} grid, some AGN in this sample require the use of an \textsc{xstar} generated grid with $v_{\rm turb}=1000\,{\rm km}\,{\rm s}^{-1}$ i.e. the turbulent velocity exceeds the local thermal velocity of the absorbing ion, proving to be the dominant factor for absorption line broadening, see Figure \ref{fig:vturb}. This scenario is required when modelling the soft X-ray absorber in only 3/28 of objects featuring complex absorption: 3C 445, Ark 564 and IC 4329A. 
IC 4329A requires two low $v_{\rm turb}$ \textsc{xstar} grids in addition to the single high $v_{\rm turb}$ \textsc{xstar} grid. The warm absorber properties are parameterised in Table \ref{tab:baseline}.

\subsubsection{Absorbed Seyfert spectra -- partially covering}
\begin{figure}
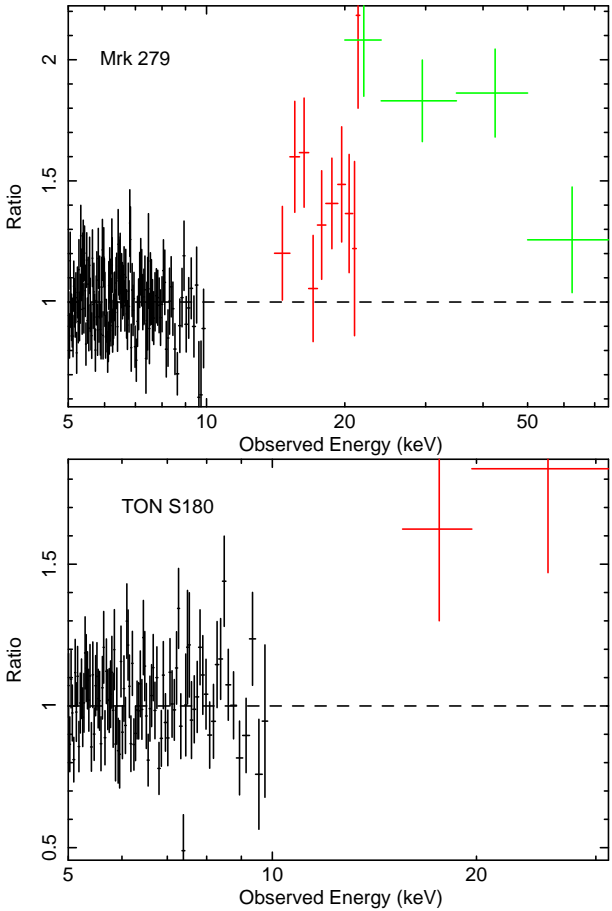

\rotatebox{-90}{\includegraphics[width=6cm]{Mrk279_hard.ps}}
\rotatebox{-90}{\includegraphics[width=6cm]{TONS180_hard.ps}}
\caption{Examples of the hard excesses which sometimes remain after modelling the full broad-band spectrum with a reflection component. In the baseline model we have accounted for these using high column density partial covering geometries. The black data points are XIS, the red are HXD data and the green are BAT data from {\sl Swift}.}
\label{fig:hard_residuals}
\end{figure}

A reasonable fraction of the AGN in this sample (16/46) require the application of a partial covering geometry whereby some fraction of the underlying powerlaw is subject to an extra zone of absorbing gas (parameters tabulated in Table \ref{tab:baseline}). The majority of the partial covering scenarios here involve high column density warm absorbing zones which have their largest effect at harder X-ray energies. For example, some AGN still exhibit a hard X-ray excess at high energies $>10$\,keV after the use of a single \textsc{reflionx} unblurred distant reflection component and require a partial coverer with $N_{\rm H}\gtrsim10^{24}\,{\rm cm}^{-2}$ to fully model the HXD and BAT data, see Figure \ref{fig:hard_residuals}. In some cases extra modelling of the hard X-ray spectrum may be due to a second blurred reflector e.g. Mrk 205, see Section 3.4. 
Due to the nature of a partial coverer with such a high column, there is not expected to be much interference with residuals in the Fe\,K region except at the Fe\,K edge 
(see discussion in Patrick et al. 2011b, Section 3.3.1). In some objects, residuals remain above 10\,keV even after applying a reflection component such as an unblurred \textsc{reflionx} can be significant, see Figure \ref{fig:hard_residuals}, also see Turner et al. (2009) as example of the presence of a strong Compton-thick partial coverer.

Five AGN (NGC 1365, NGC 3227, NGC 3516 and NGC 4151) are best fit with a lower column density partial coverer with $N_{\rm H}\sim10^{23}\,{\rm cm}^{-2}$ which adds curvature (i.e. a small, shallow bump) at $>2$\,keV. Of course, in this region of the X-ray spectrum, there is likely to be some degree of interplay between the partial covering parameters (e.g. covering fraction, column density and ionization) and the strength of 'broad' residuals in the Fe\,K region. Note that the partial coverer properties in NGC 1365 vary between each of the three {\sl Suzaku} observations, in particular the second observation (OBSID: 705031010) requires an additional neutral absorber i.e. \textsc{powerlaw + xstar*zphabs*powerlaw}. Only IRAS 13224--3809 features a partial coverer with a lower column density of $N_{\rm H}\sim10^{22}\,{\rm cm}^{-2}$ which models small residuals remaining in the 2-4\,keV spectrum. The 2-10\,keV {\sl Suzaku} spectrum of IRAS 13224--3809 only has a small number of counts and subsequently no great deal of information can be gathered regarding any partial covering scenario. It should be noted that the majority of the partially covering zones in this sample are of low to neutral ionisation. 

\subsubsection{Highly ionized absorption and outflows}
A total of 14/46 (30\%) of the AGN in this sample feature highly ionised absorption at a minimum of the 90\% confidence level. A substantial proportion (10/46) of AGN in this sample are found to possess highly ionized zones which are somewhat blue-shifted with respect to absorption lines from Fe\,{\rm XXV} and Fe\,{\rm XXVI} at rest-frame energies of 6.7\,keV and 6.97\,keV (3C 111, 3C 445, MCG--06-30-15, MR 2251--178, Mrk 766, NGC 1365, NGC 3227, NGC 3516, NGC 3783, NGC 4051, NGC 4151, NGC 5548, PDS 456 and PG 1211+143). All of these zones (bar those in NGC 3783 and NGC 5548) are outflowing i.e. the absorption lines are blue-shifted to higher energies with typical velocities of a few thousand ${\rm km}\,{\rm s}^{-1}$ and column densities $1\times10^{21}\,{\rm cm}^{-2}\lesssim\,N_{\rm H}\lesssim1\times10^{24}\,{\rm cm}^{-2}$, see Table \ref{tab:FeK}. We note that the absorption zones in NGC 3227, NGC 3516, NGC 3783 and NGC 5548 have outflow velocities which are consistent with zero, i.e. 10/46 (22\%) of the AGN in this sample require highly ionised outflows at greater than the 90\% confidence level.

\begin{table*}
\caption{Summary of relline fits to objects in which some statistical improvement is made. Line energy fixed at 6.4\,keV. * Denotes a frozen parameter. $^{a}$ Flux given in units $(10^{-5}\,{\rm ph\,{\rm cm}^{-2}\,s^{-1}})$. $^{b}$ Inner radius of emission is quoted as an alternative to the spin parameter $a$, i.e. spin is frozen at maximal ($a=0.998$) and the inner radius is allowed to vary away from $R_{\rm ISCO}$. $^{c}$ Relline component is only employed in the 2006 of Mrk 766 observation. $^{d}$ Emissivity frozen at best-fit value from Patrick et al. (2011b).}
\begin{tabular}{p{2cm} c c c c c c c c c c}
\hline
Object & $EW$\,(eV) & $q$ & $a$ & $i^{\circ}$ & Flux${^a}$ & $R_{\rm in}$\,$^{b}$ & Fe\,{\rm XXV} & Fe\,{\rm XXVI} & $\triangle\chi^{2}$ & $\chi^{2}_{\nu}$ \\
\hline
3C 111 & $37^{+30}_{-17}$ & $<2.7$ & -- & $22^{+12}_{-5}$ & $0.98^{+0.79}_{-0.44}$ & $<59$ & \checkmark & \checkmark & -10 & 1091.2/1094 \\
3C 120 & $78^{+14}_{-16}$ & $1.5^{+0.2}_{-0.3}$ & -- & $17^{+1}_{-1}$ & $4.17^{+0.76}_{-0.85}$ & $<17$ & \checkmark & X & -45 & 3586.4/3449 \\
 & $93^{+17}_{-19}$ & & & & 
 \\
3C 382 & $57^{+20}_{-18}$ & $2.7^{+1.2}_{-0.8}$ & -- & $30^{+2}_{-4}$ & $2.65^{+0.93}_{-0.85}$ & $15^{+14}_{-6}$ & X & X & -27 & 959.9/934 \\
3C 390.3 & $142^{+24}_{-30}$ & $2.3^{+0.6}_{-0.3}$ & -- & $49^{+3}_{-3}$ & $4.96^{+0.84}_{-1.06}$ & $<15$ & X & X & -42 & 1457.2/1482 \\
3C 445 & $71^{+48}_{-33}$ & $<1.6$ & -- & $>45$ & $1.05^{+0.71}_{-0.49}$ & $<332$ & X & X & -6 & 452.2/421 \\
4C 74.26 & $48^{+25}_{-26}$ & $<1.7$ & -- & $72^{+17}_{-35}$ & $1.82^{+0.94}_{-0.99}$ & $<89$ & X & X & -5 & 1339.7/1301 \\
Ark 120 & $95^{+32}_{-26}$ & $2.3^{+0.4}_{-0.4}$ & $<0.94$ & $33^{+2}_{-17}$ & $3.62^{+1.22}_{-0.99}$ & $25^{+19}_{-7}$ & X & \checkmark & -22 & 724.4/648 \\
Fairall 9 & $54^{+26}_{-25}$ & $2.9^{+0.5}_{-0.4}$ & $0.60^{+0.19}_{-0.62}$ & $31^{+4}_{-5}$ & $1.49^{+0.72}_{-0.70}$ & $4^{+2}_{-1}$ & \checkmark & \checkmark & -46 & 3558.7/3271 \\
 & $69^{+18}_{-17}$ & & & & 
 \\
IC 4329A & $69^{+13}_{-14}$ & $2.3^{+0.3}_{-0.4}$ & -- & $51^{+4}_{-3}$ & $8.25^{+1.54}_{-1.72}$ & $37^{+8}_{-9}$ & X & \checkmark & -46 & 2374.8/2198 \\
MCG-02-14-009 & $142^{+47}_{-46}$ & $2.0^{+0.4}_{-0.4}$ & $<0.88$ & $24^{+10}_{-9}$ & $6.64^{+2.20}_{-2.15}$ & $>13$ & X & \checkmark & -12 & 611.6/538 \\
MCG--05-23-16 & $94^{+11}_{-15}$ & $2.5^{+0.4}_{-0.5}$ & $<0.50$ & $24^{+3}_{-3}$ & $10.16^{+1.19}_{-1.65}$ & $9^{+5}_{-2}$ & X & X & -112 & 1471.2/1462 \\
MCG--06-30-15 & $161^{+46}_{-44}$ & $2.7^{+0.2}_{-0.1}$ & $0.49^{+0.20}_{-0.12}$ & $44^{+6}_{-2}$ & $5.73^{+1.63}_{-1.57}$ & $5^{+1}_{-1}$ & X & \checkmark & -59 & 1967.8/1819 \\
MCG+8-11-11 & $72^{+14}_{-14}$ & $2.1^{+0.3}_{-0.4}$ & -- & $18^{+2}_{-2}$ & $5.50^{+1.07}_{-1.07}$ & $<18$ & X & \checkmark & -79 & 970.7/932 \\
MR 2251--178 & $55^{+22}_{-12}$ & $2.4^{+1.4}_{-1.0}$ & -- & $36^{+10}_{-4}$ & $2.74^{+1.08}_{-0.61}$ & $<35$ & X & X & -30 & 947.3/898 \\
Mrk 79 & $199^{+40}_{-37}$ & $2.7^{+0.8}_{-0.6}$ & $<0.80$ & $34^{+3}_{-3}$ & $3.08^{+0.62}_{-0.57}$ & $19^{+7}_{-9}$ & X & \checkmark & -51 & 551.1/539 \\
Mrk 335 & $146^{+39}_{-39}$ & $2.6^{+0.5}_{-0.3}$ & $0.70^{+0.12}_{-0.01}$ & $38^{+2}_{-2}$ & $2.52^{+0.67}_{-0.67}$ & $32^{+12}_{-16}$ & \checkmark & \checkmark & -40 & 803.0/719 \\
Mrk 359 & $76^{+62}_{-60}$ & 3.0* & -- & $26^{+12}_{-7}$ & $0.46^{+0.38}_{-0.37}$ & $>56$ & \checkmark & X & -4 & 606.2/559 \\
Mrk 509 & $76^{+11}_{-11}$ & $2.1^{+0.3}_{-0.3}$ & -- & $41^{+4}_{-3}$ & $4.56^{+0.66}_{-0.65}$ & $45^{+31}_{-24}$ & X & X & -31 & 1945.3/1868 \\
Mrk 766$^{c}$ & $162^{+97}_{-62}$ & $>3.0$ & -- & $39^{+3}_{-4}$ & $1.89^{+1.13}_{0.72}$ & $<28$ & \checkmark & X & -17 & 1040.1/995 \\
Mrk 841 & $161^{+43}_{-70}$ & 3.0* & -- & $>32$ & $2.85^{+0.77}_{-1.23}$ & $>8$ & \checkmark & X & -20 & 917.9/854 \\
NGC 1365 & $94^{+36}_{-36}$ & 3.0* & -- & $52^{+11}_{-2}$ & $1.59^{+0.61}_{-0.61}$ & $5^{+18}_{-3}$ & \checkmark & \checkmark & -100 & 2162.8/1975 \\
 & $<50$ & & & & $<0.41$ & \\
 & $<1$ & & & & $<0.19$ & \\
NGC 2992 & $71^{+28}_{-31}$ & $2.1^{+0.8}_{-0.7}$ & -- & $>26$ & $1.33^{+0.52}_{-0.58}$ & $<169$ & X & X & -12 & 1076.6/1075 \\
NGC 3147 & $<17$ & 3.0* & -- & -- & $<0.11$ & -- & X & \checkmark & -2 & 278.6/263 \\
NGC 3227 & $47^{+23}_{-22}$ & $2.7^{+0.5}_{-0.4}$ & -- & $33^{+2}_{-2}$ & $2.05^{+0.98}_{-0.98}$ & $11^{+3}_{-6}$ & X & \checkmark & -120 & 4465.6/4189 \\
 & $125^{+31}_{-31}$ & & & & $2.93^{+0.73}_{-0.72}$ & \\
 & $<50$ & & & & $<1.74$ & \\
 & $26^{+4}_{-4}$ & & & & $3.85^{+0.62}_{-0.61}$ & \\
 & $48^{+23}_{-23}$ & & & & $1.49^{+0.71}_{-0.70}$ & \\
 & $<54$ & & & & $<1.32$ & \\
NGC 3516 & $58^{+9}_{-9}$ & $3.1^{+0.4}_{-0.2}$ & $<0.30$ & $<41$ & $2.79^{+0.42}_{-0.43}$  & $8^{+1}_{-1}$ & \checkmark & X & -55 & 1208.3/1121 \\
 & $14^{+2}_{-2}$ & & & & $0.39^{+0.14}_{-0.13}$ \\
NGC 3783 & $57^{+12}_{-16}$ & 3.2*$^{d}$ & $<0.24$ & $<23$ & $3.72^{+0.80}_{-1.06}$ & $8^{+1}_{-2}$ & X & \checkmark & -64 & 2490.5/2302 \\
 & $46^{+10}_{-13}$ & \\
NGC 4051 & $81^{+32}_{-25}$ & 3.0* & -- & $<20$ & $1.09^{+0.42}_{-0.34}$ & $9^{+1}_{-1}$ & \checkmark & X & -18 & 3187.2/2939 \\
 & $40^{+15}_{-12}$ & & & & 
 \\
 & $41^{+16}_{-13}$ & & & & 
 \\
NGC 5506 & $30^{+10}_{-11}$ & $1.9^{+0.4}_{-0.5}$ & -- & $20^{+5}_{-3}$ & $4.82^{+1.67}_{-1.73}$ & $<21$ & \checkmark & \checkmark & -50 & 3139.3/2867 \\
 & $53^{+12}_{-12}$ & & & & $7.92^{+1.85}_{-1.78}$ & \\
NGC 7314 & $56^{+44}_{-41}$ & $<4.6$ & -- & $<89$ & $0.58^{+0.46}_{-0.43}$ & $<164$ & X & \checkmark & -4 & 588.3/541 \\
NGC 7469 & $91^{+9}_{-8}$ & $1.7^{+0.4}_{-0.6}$ & $0.69^{+0.09}_{-0.09}$ & $23^{+15}_{-7}$ & $2.23^{+0.22}_{-0.20}$ & $81^{+82}_{-37}$ & X & X & -25 & 815.5/808 \\
SWIFT J2127.4+5654 & $178^{+82}_{-69}$ & $2.6^{+1.0}_{-0.4}$ & $0.70^{+0.10}_{-0.14}$ & $43^{+5}_{-10}$ & $6.35^{+2.91}_{-2.46}$ & $<33$ & X & X & -37 & 852.3/867 \\
\hline
\end{tabular}
\label{tab:relline}
\end{table*}

\subsubsection{Soft excess and reflector properties}
The majority of the AGN in this sample show both soft and hard X-ray excesses above the basic intrinsic powerlaw. A substantial 31/46 AGN indicate an observed soft excess below $\sim2$\,keV (modelled as described in Section 3.1.1), however this is lower than the fraction obtained in Porquet et al. (2004) in a selection of PG quasars who found that 19/21 objects exhibited a significant soft excess. It should be noted, however, that nearly all of the objects without a soft excess feature complex warm or neutral absorption. This can lead to any weak soft excess which may be present being `hidden' or simply absorbed, thereby leave little trace or indication of its presence. Only MCG--02-14-009 and NGC 3147 are without additional absorption (either neutral or ionized) and still show no indications of a soft excess, whereas SWIFT J2127.4+5654 has sufficiently strong neutral absorption (no warm absorption) to mask the presence of a soft excess. 

A large number of AGN in this sample also exhibit strong excesses at hard X-ray energies, indicative of reflection off distant material e.g. the putative torus or outer accretion disc. 39/46 of objects in the sample are modelled using the \textsc{reflionx} reflection model, with properties such as ionization and iron abundance $Z_{\rm Fe}$ left free to vary. The vast majority of AGN modelled using \textsc{reflionx} here are well fit with a neutral or close to neutral reflector ($\xi<60$\,erg\,{\rm cm}\,s$^{-1}$) with only Ton S180 featuring a moderately ionized reflector ($\xi\sim270$\,erg\,{\rm cm}\,s$^{-1}$). The remainder of the objects without a strong reflection component still feature neutral Fe\,K$\alpha$ emission, however, IRAS 13224--3809 features neither a reflection component nor 6.4\,keV Fe\,K$\alpha$ emission.

In the majority of these AGN, the reflection component is well modelled with Solar iron abundance and in 23/39 of objects with evidence for strong reflection it is fixed at $Z_{\rm Fe}=1.0$. This fraction includes the scenario in which some proportion of the observed narrow Fe\,K$\alpha$ flux additionally arises from material which may be in the broad line region (BLR) i.e. the Fe\,K$\alpha$ core also originates from both reflection off distant matter and the Compton-thin BLR. We should note that if we assume Fe\,K emission from the BLR, this does not necessitate the addition of a further reflection component such as \textsc{reflionx}. This geometry is most evident when the strength of the narrow core is disproportionately stronger than the strength of other reflector features such as the Compton hump, often indicated by what may be an unfeasibly high iron abundance e.g. $Z_{\rm Fe}>3.0$ being forced into the fits. We account for this by freezing $Z_{\rm Fe}=1.0$ and including an additional narrow Gaussian $\sim6.4$\,keV (such as in MCG--05-23-16, Mrk 110, Mrk 509, NGC 2992, NGC 3147, NGC 3516, NGC 4151, NGC 4593 and NGC 7213). We note, however, that if this additional Fe\,K$\alpha$ component originates from the BLR, it might be expected that the $\sigma_{\rm width}$ be higher e.g. up to $\sigma\sim0.10$\,keV. While the exact origin of the line may be unknown, we keep $\sigma=0.01$\,keV for consistency with the other additional narrow lines used in the models, noting that some moderate broadening consistent with BLR widths cannot be ruled out. 

Only 4/39 of the remaining objects still require a slightly super-Solar iron abundance; namely 3C 120, Fairall 9, Mrk 335 and Ton S180. Slightly sub-Solar iron abundance reflectors are found in 12/46 of objects and a good fit is found to both the Compton hump in the HXD and the flux of the narrow Fe\,K$\alpha$ core with few residuals remaining. NGC 3227 and NGC 4051 (Lobban et al. 2011) are best fit with a sub-Solar iron abundance in conjunction with a narrow 6.4\,keV Gaussian. This is in a similar fashion to above whereby the narrow Fe\,K$\alpha$ flux exceeds that expected from a simple Solar abundance reflector, however, both of these objects appear to still have weaker relative reflection components i.e. a sub-Solar reflector plus narrow Gaussian is preferred. This may represent the scenario whereby the Fe\,K$\alpha$ line either additionally or predominantly arises in Compton-thin material such as the BLR or NLR rather than the outer regions of the accretion disc or torus. The baseline model parameters for each object and observation are summarised in Table \ref{tab:baseline} and Table \ref{tab:FeK}. Of the objects which do not require a reflection component, only IRAS 13224--3809 does not feature a narrow neutral Fe\,K line of any kind, while the narrow iron line in both 3C 111 and MR 2251--178 is near neutral at $\sim6.5$\,keV. 


\subsubsection{Additional distant ionized emission}
We examine the Fe\,K regions of these AGN to determine whether emission lines due to ionized H-like and He-like species of iron are present, making up the final piece in our baseline models for each object. Lines at 6.63-6.70\,keV and 6.97\,keV are (likely due to Fe\,{\rm XXV} and Fe\,{\rm XXVI} respectively) emitted from highly ionized gas existing at large distances from the central SMBH. The presence of such lines has been well documented in many Seyfert 1 spectra and it should come as no surprise that they appear relatively common in this sample (Bianchi et al. 2004; Nandra et al. 2007; Bianchi et al. 2009; Patrick et al. 2011a). Approximately half (24/46) of the objects in this sample show evidence for Fe\,{\rm XXV} emission at $\sim6.7$\,keV, while fewer (18/46) show evidence for Fe\,{\rm XXVI} at 6.97\,keV (Table \ref{tab:FeK}). However, the lower fraction of H-like Fe detections may be due to its proximity to Fe\,K$\beta$ at 7.056\,keV which is included self consistently in all objects in addition to the narrow 6.4\,keV Fe\,K$\alpha$ core.



\subsection{Broad residuals in the Fe\,K region}
We first attempt to parameterise the strength of any broad residuals which may remain in the Fe\,K regions of these objects by including a broad Gaussian with line energy, $\sigma_{\rm width}$ and normalisation left free to vary. If this broad Gaussian component proves to be statistically significant at the 90\% level (three additional parameters i.e. an improvement to the fit of $\triangle\chi^{2}>6.3$), we then proceed to add more physical models of the line emission form the inner regions of the accretion disc. Figure \ref{fig:FeK_none} shows ratio plots of the 4-9\,keV region prior to any model of reflection or disc emission components i.e. the Fe\,K band is left unmodelled. 

Indeed, not all of the objects in the sample require any further modelling and an excellent fit is found to the data. For example, no reasonable statistical improvement is made when adding a broad Gaussian component to the following AGN in this sample: 1H 0419--577, Ark 564, MCG--02-14-009, MCG--02-58-22, Mrk 110, Mrk 205, Mrk 279, NGC 4151, NGC 5548, NGC 7213, PDS 456, PG 1211+143, RBS 1124 and Ton S180 (see Table \ref{tab:broad}). The remaining AGN in the sample show at least some indications of a broad red-wing in the Fe\,K region and therefore perhaps emission from the inner regions of the accretion disc (32/46 objects), however, it should be noted that this is only a simple parametrisation and the true fraction of AGN may differ from this when a more physical model is used. From Table \ref{tab:broad} we can estimate the mean parameters of the simple broad Gaussian at LineE\,$=6.32\pm0.04$\,keV, $\sigma_{\rm width}=0.470\pm0.051$\,keV and equivalent width $EW=97\pm19$\,eV. Of course, attempting to model an asymmetric line emission profile with a simple Gaussian can lead to a mis-modelling of the Fe\,K region, for example, broader Gaussians (such as those found here) are heavily influenced by the way in which the narrow Fe\,K$\alpha$ core and Compton shoulder are modelled. Since the use of a broad Gaussian is essentially simply a means of adding curvature to the 5-7\,keV region, in a sample of complex AGN featuring warm absorption/ highly ionized outflows there is likely to be a reasonable amount of interplay between the width and strength of the broad line with the absorber properties. 

We next replace the broad Gaussian with the \textsc{relline} relativistic line emission model (Dauser et al. 2010). This produces a more physical asymmetric line profile taking into account effects such as relativistic Doppler broadening producing both red and blue-wings. The shape of this line profile allows properties such as the inclination and emissivity index $q$ of the accretion disc to be measured. We assume the emission line has a rest frame energy of 6.4\,keV and that the inner radius of the accretion disc extends down to $R_{\rm ISCO}$ with a single uniform emissivity index throughout the disc i.e. the inner and outer $q$ values are tied. The emissivity index scales as $R^{-q}$ where $q\sim3$ would be expected from the inner regions of the accretion disc and $q>5$ would suggest that emission is very centrally concentrated, invoking a significant amount of light bending e.g. Miniutti \& Fabian (2004). In some AGN, the broad line may not be sufficiently strong enough to constrain the emissivity index and in such cases we fix $q=3.0$. The \textsc{relline} model also allows estimates to be placed upon the black hole spin parameter $a$ (assuming $R_{\rm in}=R_{\rm ISCO}$), which varies between $a=-0.998$ for a maximally rotating {\sl retrograde} SMBH and $a=0.998$ for a maximally spinning {\sl prograde} BH. 
Objects with multiple observations are modelled allowing for the normalisation of the \textsc{relline} component to vary, but all other parameters tied e.g. in NGC 1365, NGC 3227, NGC3516 and NGC 5506. The \textsc{relline} component in 3C 120, Fairall 9, NGC 3783 and NGC 4051 is consistent with a constant flux held between observations (i.e. the line does not vary) and the normalisation is tied as such, see Table \ref{tab:relline}.

After applying the \textsc{relline} model a good fit is found in all the remaining objects, some of which are significantly improved with the introduction of a relativistic line emission model. Emission from the inner regions of the disc is not formally statistically required at the 90\% confidence level in 20/46 objects (see Table \ref{tab:relline} for a summary of \textsc{relline} fits for objects which required a broad Gaussian in Table \ref{tab:broad}, note that not all objects listed in the table formally require the \textsc{relline} model). Of the remaining 26/46 objects which do require emission from the inner regions of the accretion disc (23/46 of the objects in this sample require the \textsc{relline} model at $>99.5\%$ confidence), we find a moderate average strength of $EW=96\pm10$\,eV and a low accretion disc emissivity index of $q=2.4\pm0.1$ at an inclination of $i=33\pm2^{\circ}$ (see Figure \ref{fig:relline_eeuf}). Alternatively fixing $a=0.998$ and allowing $R_{\rm in}$ to vary yields an average $R_{\rm in}=21\pm6\,R_{\rm g}$. 

With the advent of high quality broadband X-ray data, tentative steps can be made towards placing estimates upon black hole spin. In particular, objects with multiple or deep observations may possess spectra with a sufficient number of counts and temporal information with which to form models allowing for consistent spin estimates and the variation in spectral shapes between observations, e.g. see Figure \ref{fig:eeuf_var}. Here we make a total of 11 tentative constraints upon the spin parameter $a$ (upper and lower bounds in 5 AGN), see Section 4.3.3.

\subsection{Dual reflector fits}
If strong emission arises from reflection off the inner regions of the accretion disc producing a relativistically broadened Fe\,K$\alpha$ line profile, the same degree of broadening and relativistic affects can also be applied to the entire reflection continuum. In this scenario we form dual reflector fits (for those objects which do formally require a \textsc{relline} component, see Table \ref{tab:relline}) consisting of a distant unblurred \textsc{reflionx} and a second inner \textsc{reflionx} convolved with the \textsc{relconv} kernel (Dauser et al. 2010). However, we still include a \textsc{compTT} component to model the soft excess continuum. For example, the analysis of a small sample of `bare' Seyferts in Patrick et al. (2011a) suggests that in order to produce the significant amount of blurring required to smooth the discrete soft emission lines into a continuum, both the spin parameter $a$ and emissivity of the disc are forced to high and extreme values ($a=0.998$ and $q>4$). Retaining the use of the \textsc{compTT} to model the soft excess ensures that the main feature driving the fit of the blurred reflector are the broad residuals in the Fe\,K region regardless of the interpretation of the soft excess. 

Due to the additional hard X-ray flux produced by a dual reflector model, we test this as an alternative to high column partial covering scenarios, i.e. partial covering geometries are avoided during fitting unless no good fit can be found without a partially covering model. The iron abundance is tied between the inner and distant reflectors for consistency. The ionisation of the two reflectors are initially allowed to vary but tied if both are approximately equal, this is found to be the case in all objects.

Simply replacing the \textsc{relline} component in each model with the \textsc{relconv*reflionx} convolution produces a good fit to the data in objects without partial covering, with estimated accretion disc parameters consistent with those obtained in the previous model (Table \ref{tab:dual}). This should come as no surprise since the Fe\,K region remains the main driver behind the fit. The dual reflector approach as an alternative to partial covering provides a reasonable fit to Mrk 766 ($\triangle\chi^{2}=1109.6/1002$), however, this is notably worse than when the use of a partial covering geometry is retained ($\chi^{2}_{\nu}=1055.1/995$). A poor fit is obtained to all three observations of NGC 4051 without invoking partial covering ($\chi^{2}_{\nu}=3321.9/2943$ versus $\chi^{2}_{\nu}=3177.1/2939$) while keeping the iron abundance of both reflectors tied ($Z_{\rm Fe}=1.0$), however, disc parameters are forced to extreme values (i.e. $q>7$ and $a=0.998$). In addition to this, the 2-100\,keV blurred reflector flux of the 2005 observation is approximately 2-3 times that of the 2008 observations despite being in a far more absorbed state (see Figure \ref{fig:eeuf_var}), we therefore retain the use of a partial coverer due to both physical and statistical arguments (see relevant section in the appendix for more details on NGC 4051).

A poor and statistically inferior fit to the broadband data is found in some objects when a high column partial coverer is replaced with a dual reflector as an alternate mechanism for accounting for additional hard X-ray flux above 10\,keV. For those objects which feature a broad iron line, a partial covering geometry is still statistically required in Mrk 766, NGC 1365, NGC 3227, NGC 3516 and NGC 4051 (Table \ref{tab:dual}). Treating these AGN as above and replacing \textsc{relline} with \textsc{relconv*reflionx} while still including a partial coverer again produces a good fit to the data with consistent accretion disc parameters to those in Table \ref{tab:relline} (see Table \ref{tab:dual}). Removing the partial coverer and accounting for the hard excess with the inner reflector yields an improved fit to MCG--06-30-15 and NGC 5506 (i.e. 2/16), however, the parameters obtained are consistent with those estimated with a partial covering geometry.

\begin{figure*}
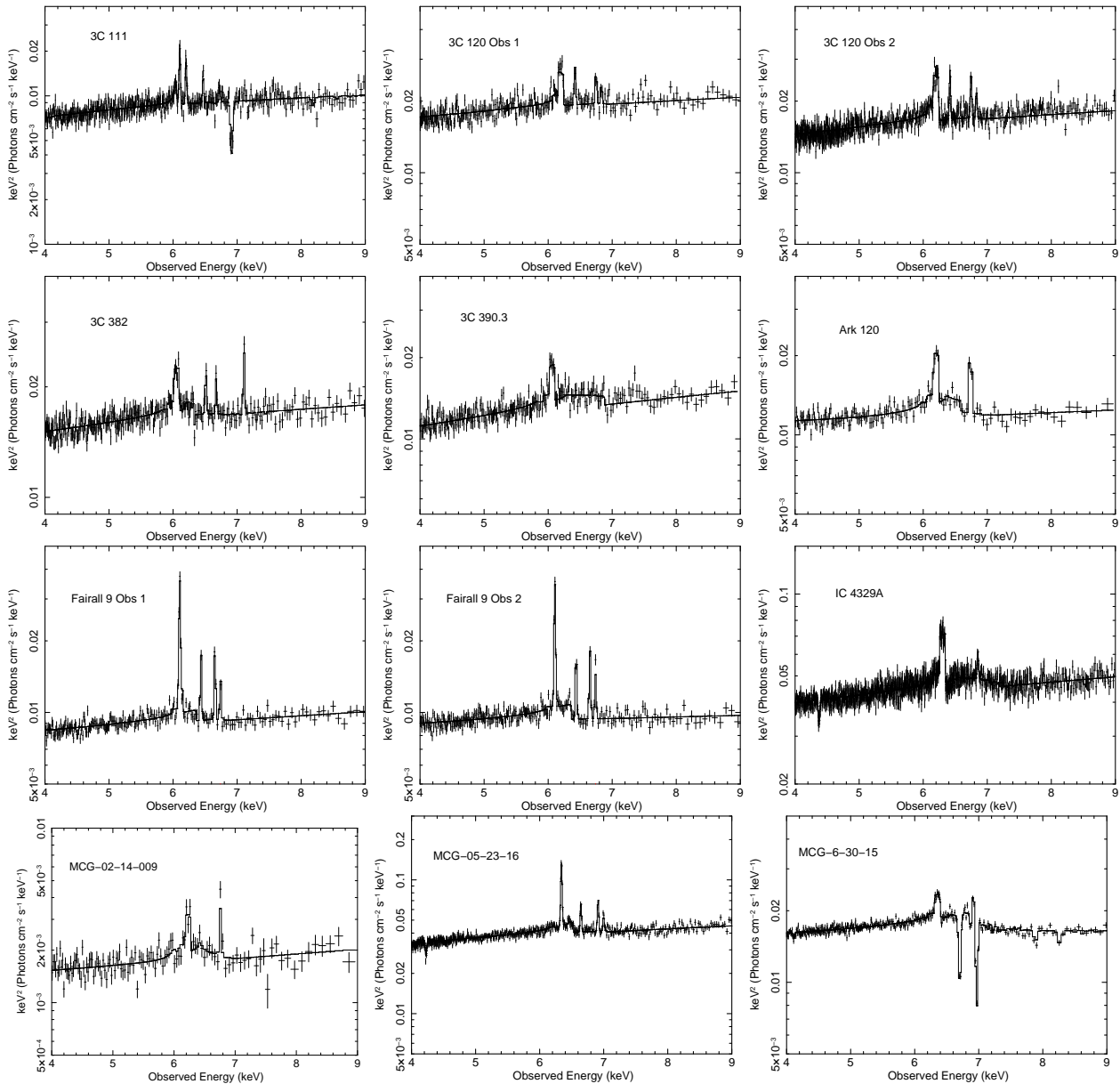


\rotatebox{-90}{\includegraphics[width=4cm]{3C111_relline_eeuf.ps}}
\rotatebox{-90}{\includegraphics[width=4cm]{3C120_relline_eeuf_obs1.ps}}
\rotatebox{-90}{\includegraphics[width=4cm]{3C120_relline_eeuf_obs2.ps}}
\rotatebox{-90}{\includegraphics[width=4cm]{3C382_relline_eeuf.ps}}
\rotatebox{-90}{\includegraphics[width=4cm]{3C390_relline_eeuf.ps}}
\rotatebox{-90}{\includegraphics[width=4cm]{Ark120_relline_eeuf.ps}}
\rotatebox{-90}{\includegraphics[width=4cm]{Fairall9_relline_eeuf_obs1.ps}}
\rotatebox{-90}{\includegraphics[width=4cm]{Fairall9_relline_eeuf_obs2.ps}}
\rotatebox{-90}{\includegraphics[width=4cm]{IC4329_relline_eeuf.ps}}
\rotatebox{-90}{\includegraphics[width=4cm]{MCG-02-14_relline_eeuf.ps}}
\rotatebox{-90}{\includegraphics[width=4cm]{MCG-05-23_relline_eeuf.ps}}
\rotatebox{-90}{\includegraphics[width=4cm]{MCG-06-30_relline_eeuf.ps}}
\caption{$\nu\,F\nu$ plots of the baseline+\textsc{relline} model in the 4-9\,keV region. Note the strong absorption in the Fe\,K region in many objects and relatively weak red-wing. The solid black line shows the total model.}
\label{fig:relline_eeuf}
\end{figure*}

\begin{figure*}
\rotatebox{-90}{\includegraphics[width=4cm]{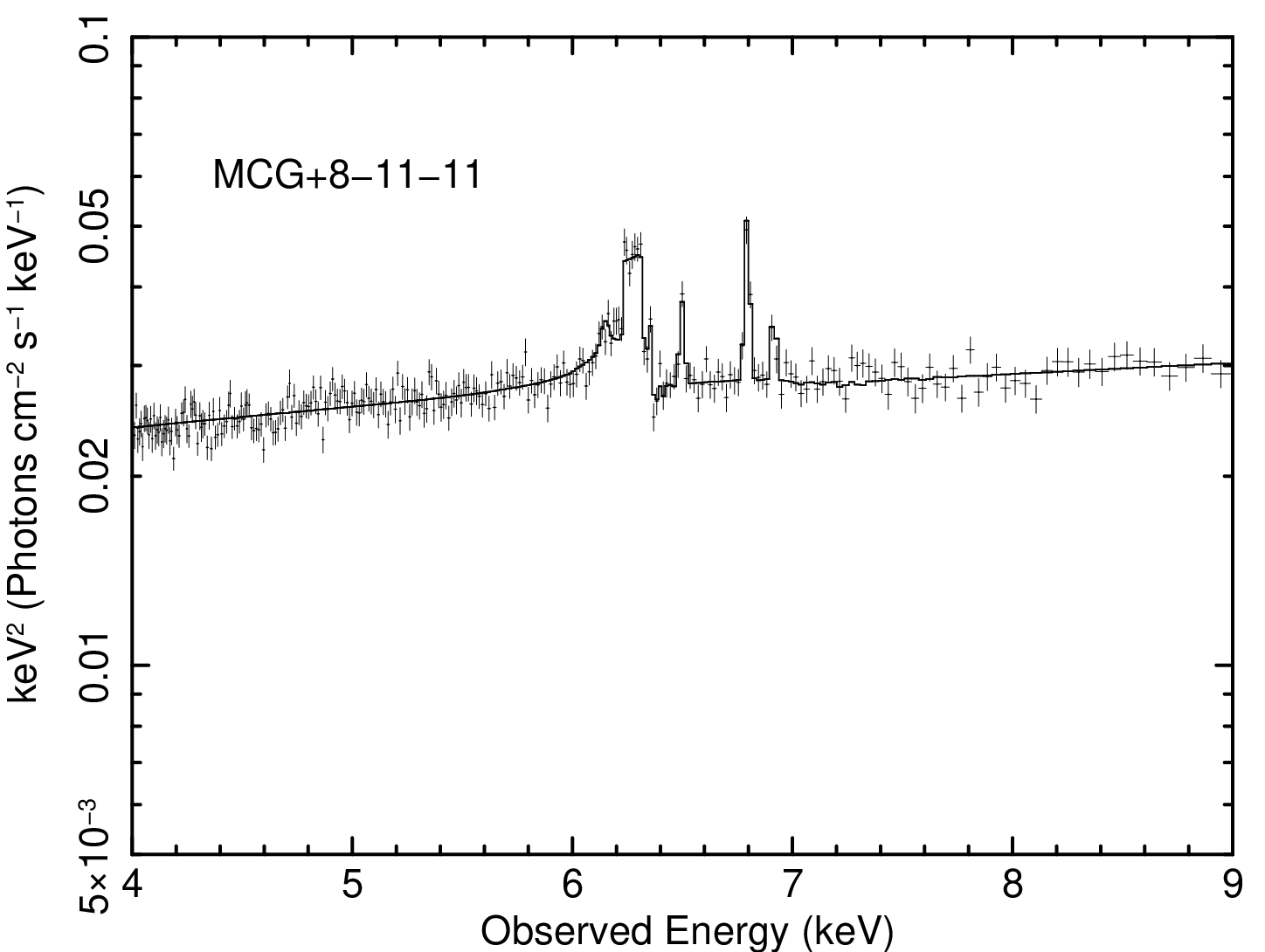}}
\rotatebox{-90}{\includegraphics[width=4cm]{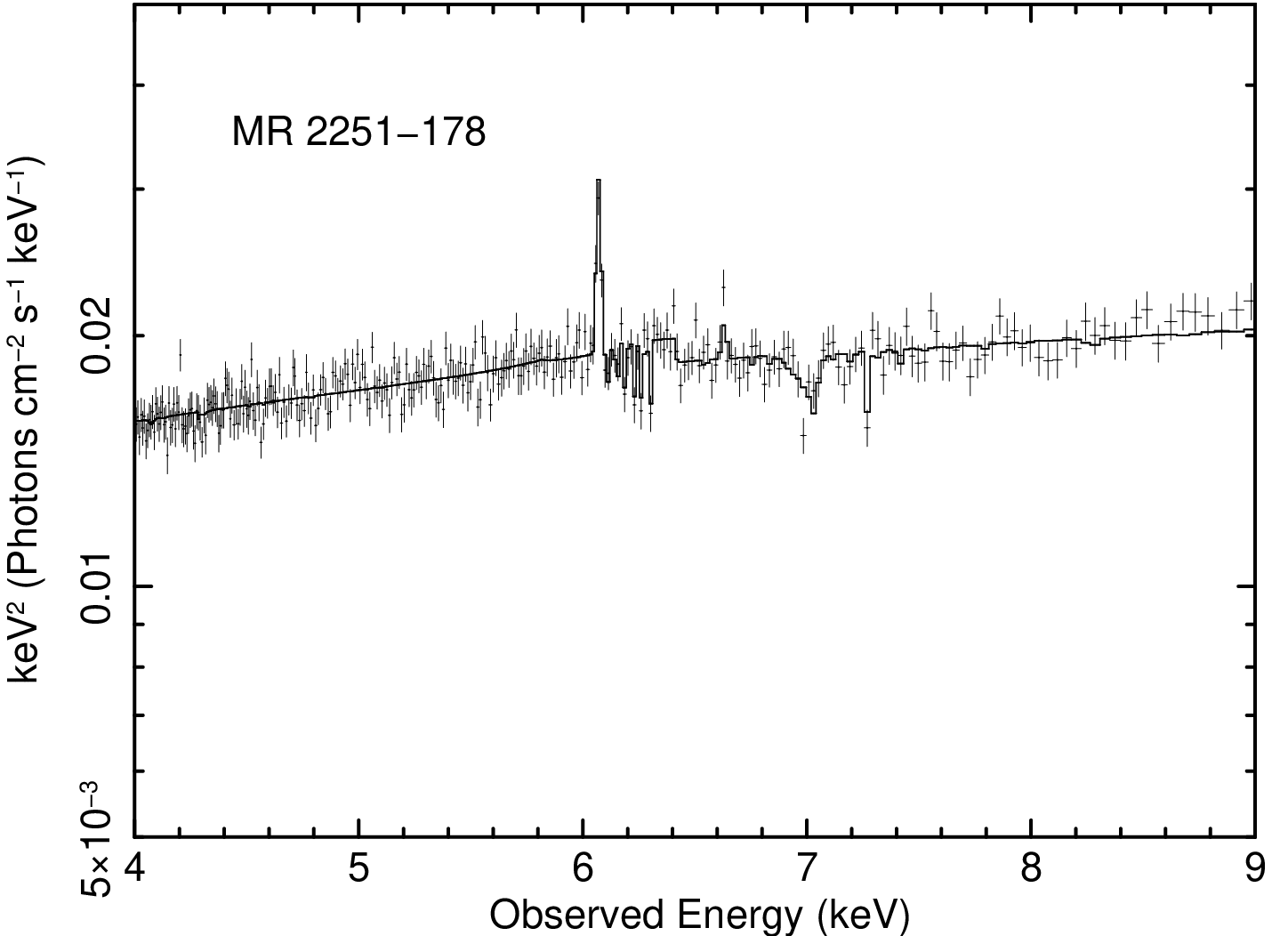}}
\rotatebox{-90}{\includegraphics[width=4cm]{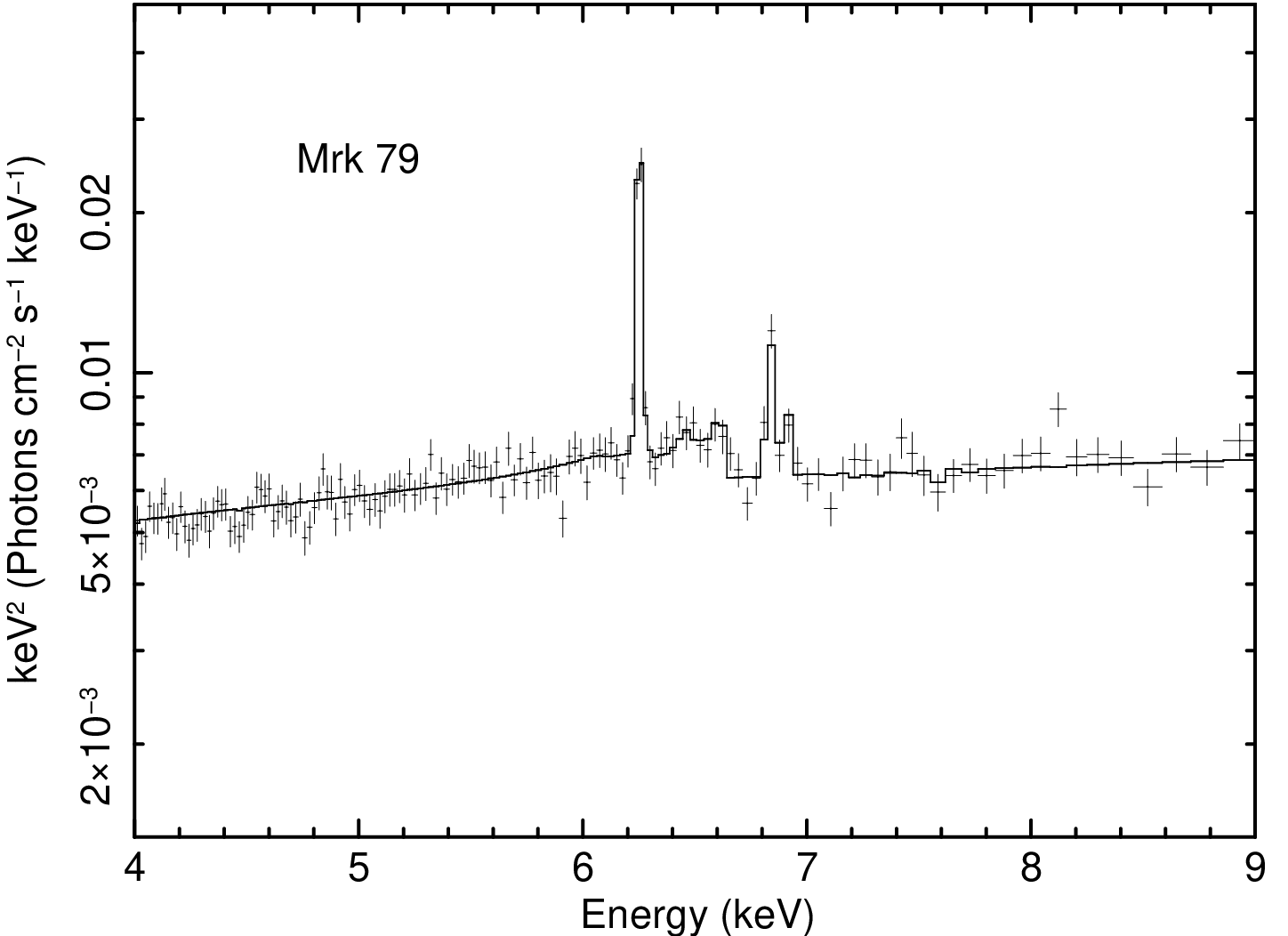}}
\rotatebox{-90}{\includegraphics[width=4cm]{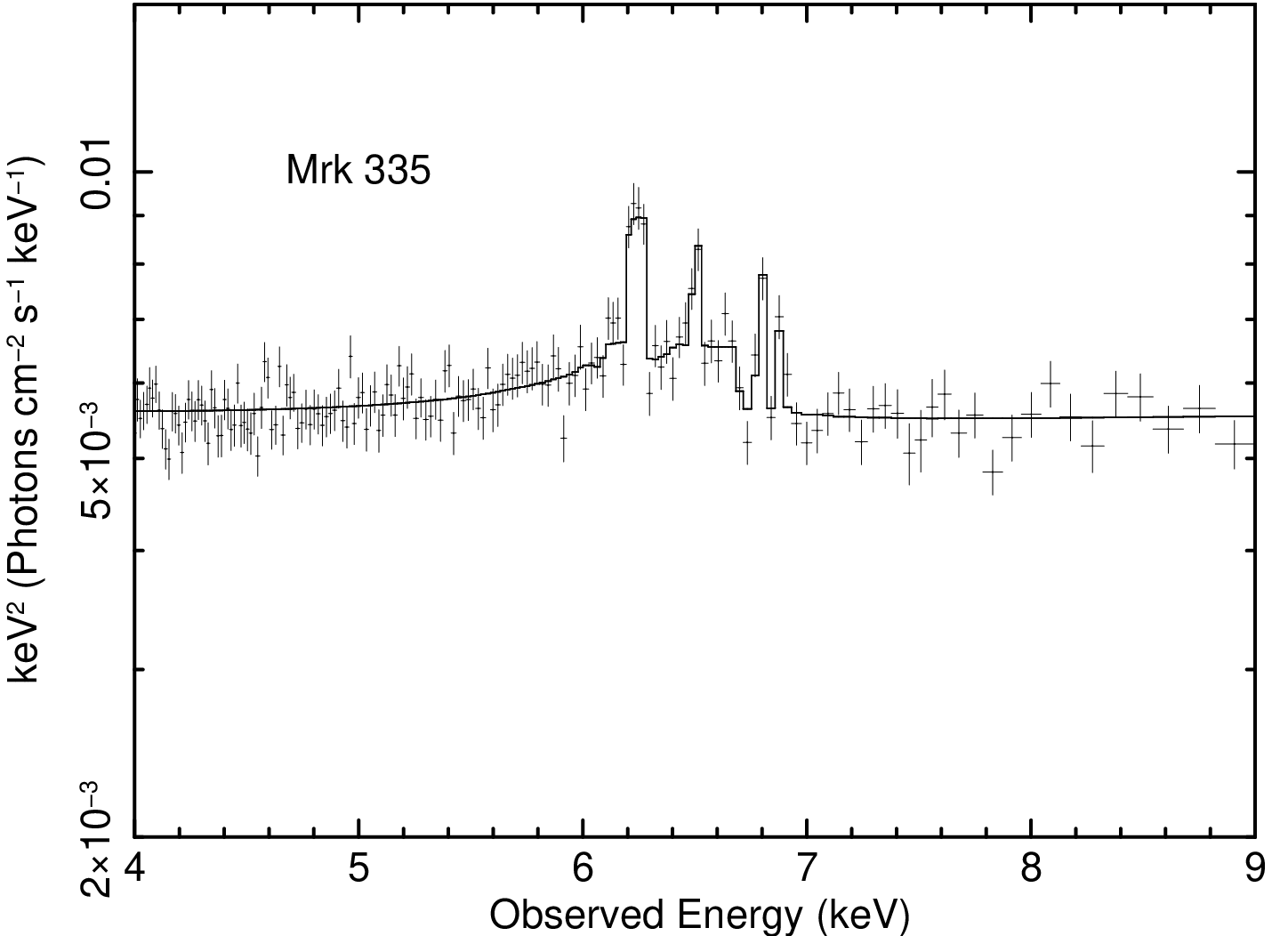}}
\rotatebox{-90}{\includegraphics[width=4cm]{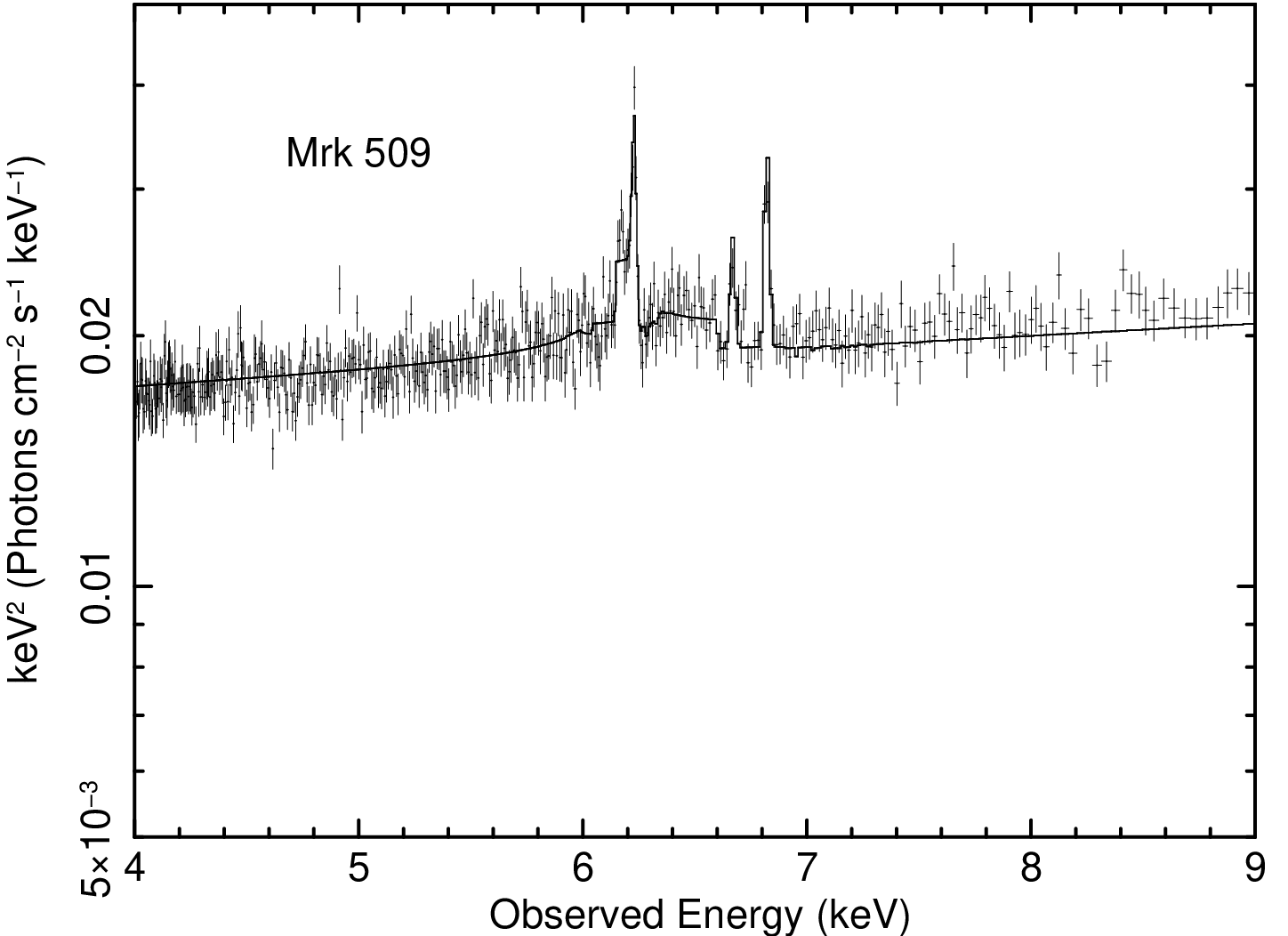}}
\rotatebox{-90}{\includegraphics[width=4cm]{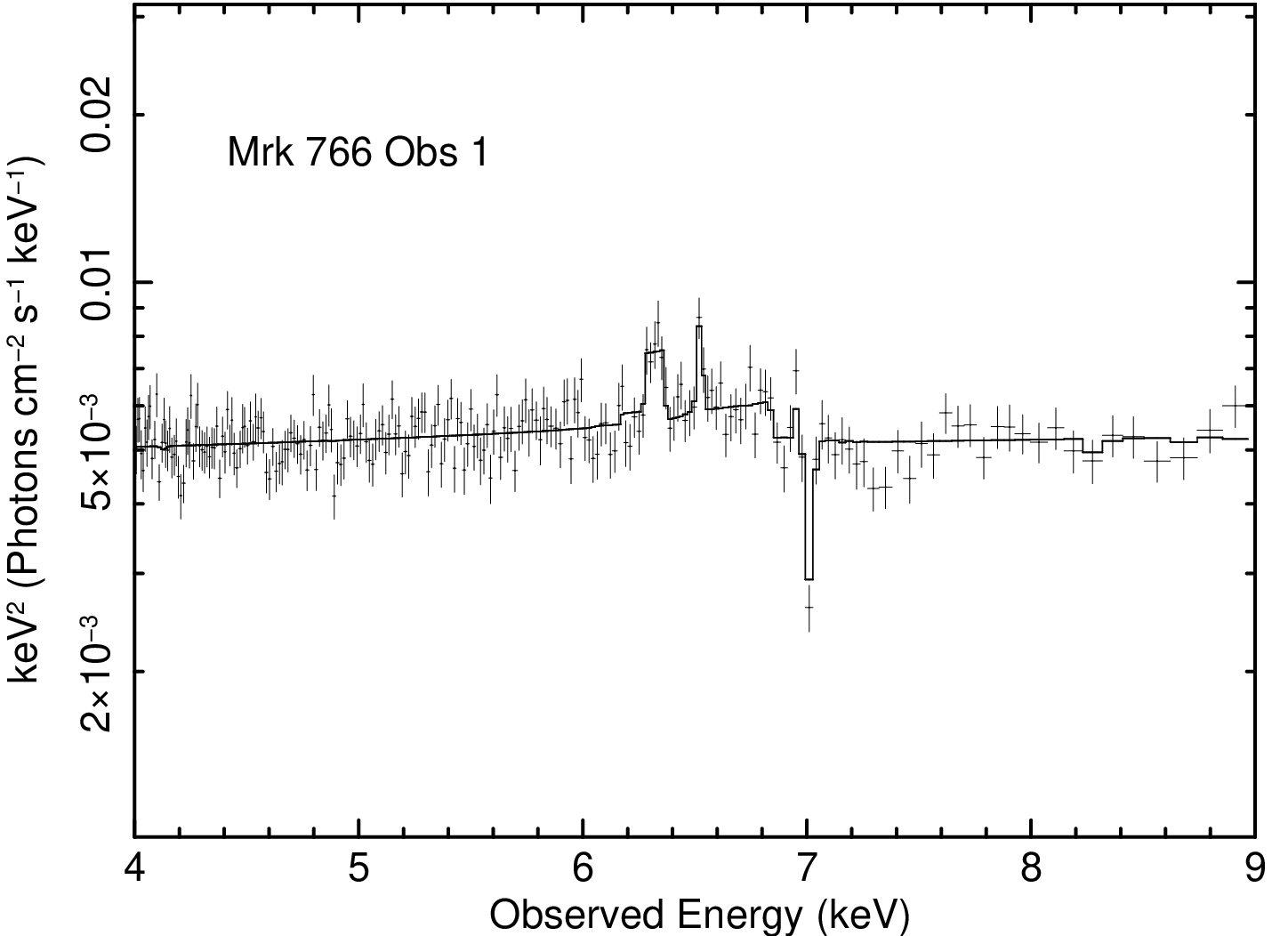}}
\rotatebox{-90}{\includegraphics[width=4cm]{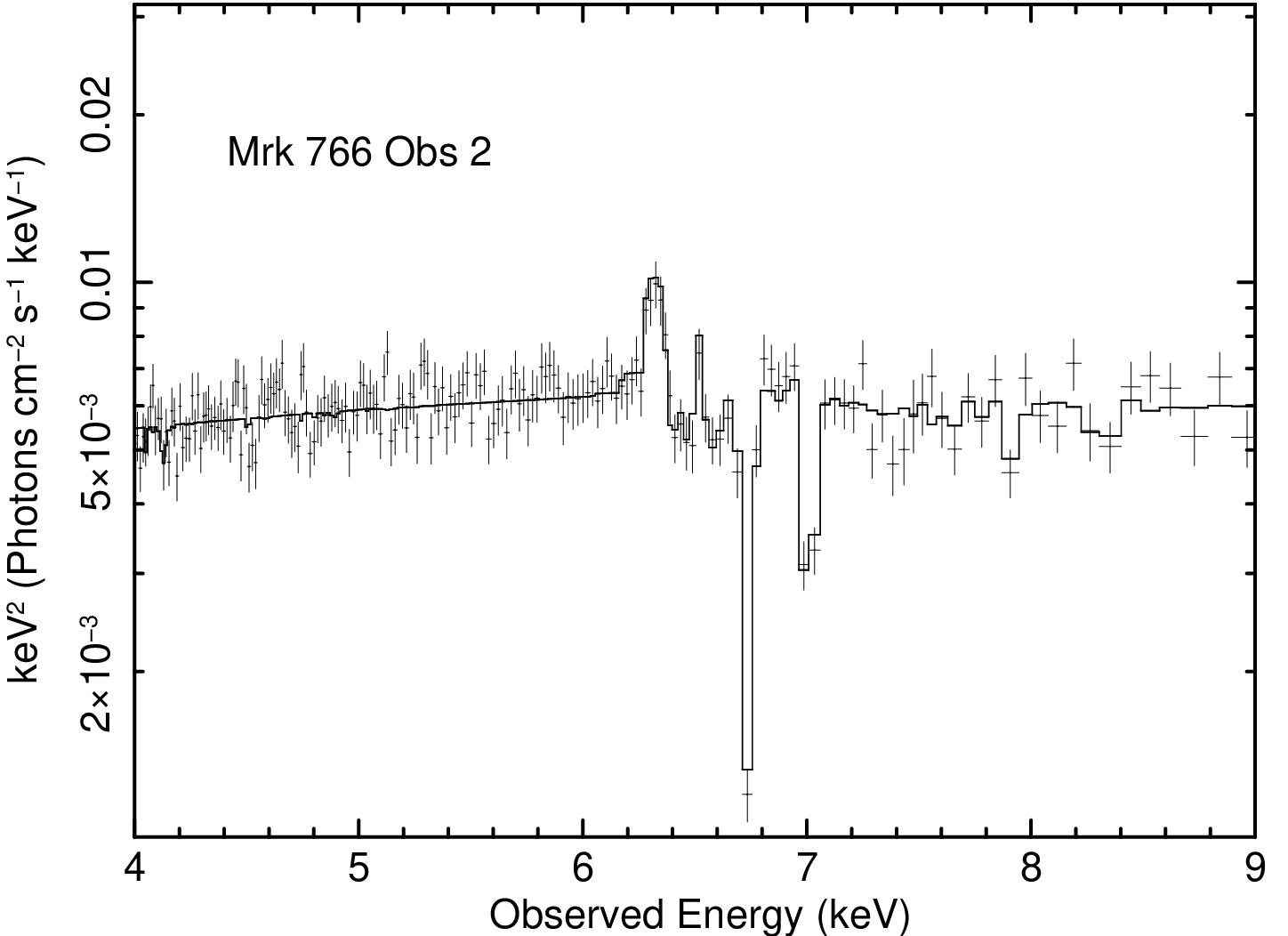}}
\rotatebox{-90}{\includegraphics[width=4cm]{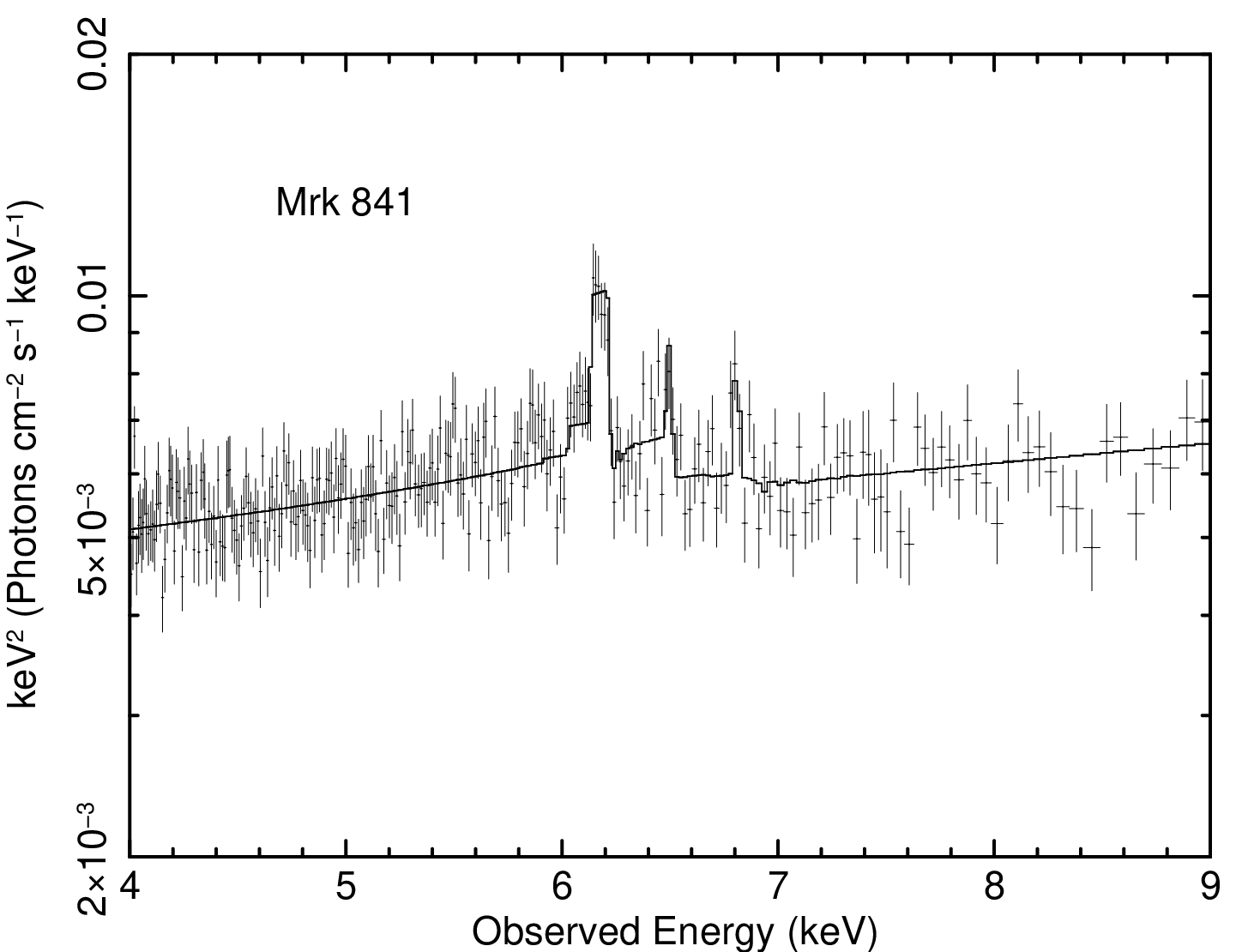}}
\rotatebox{-90}{\includegraphics[width=4cm]{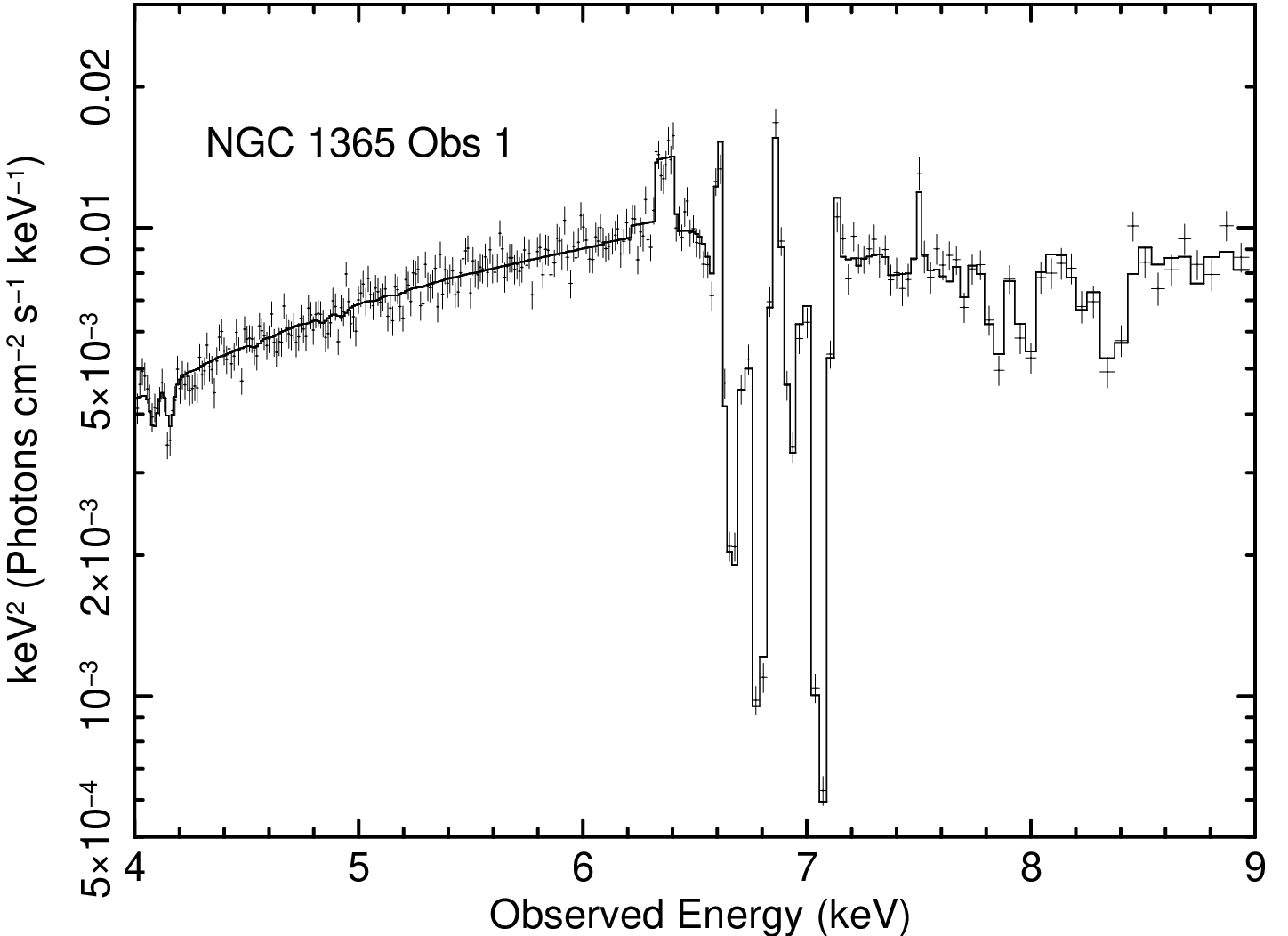}}
\rotatebox{-90}{\includegraphics[width=4cm]{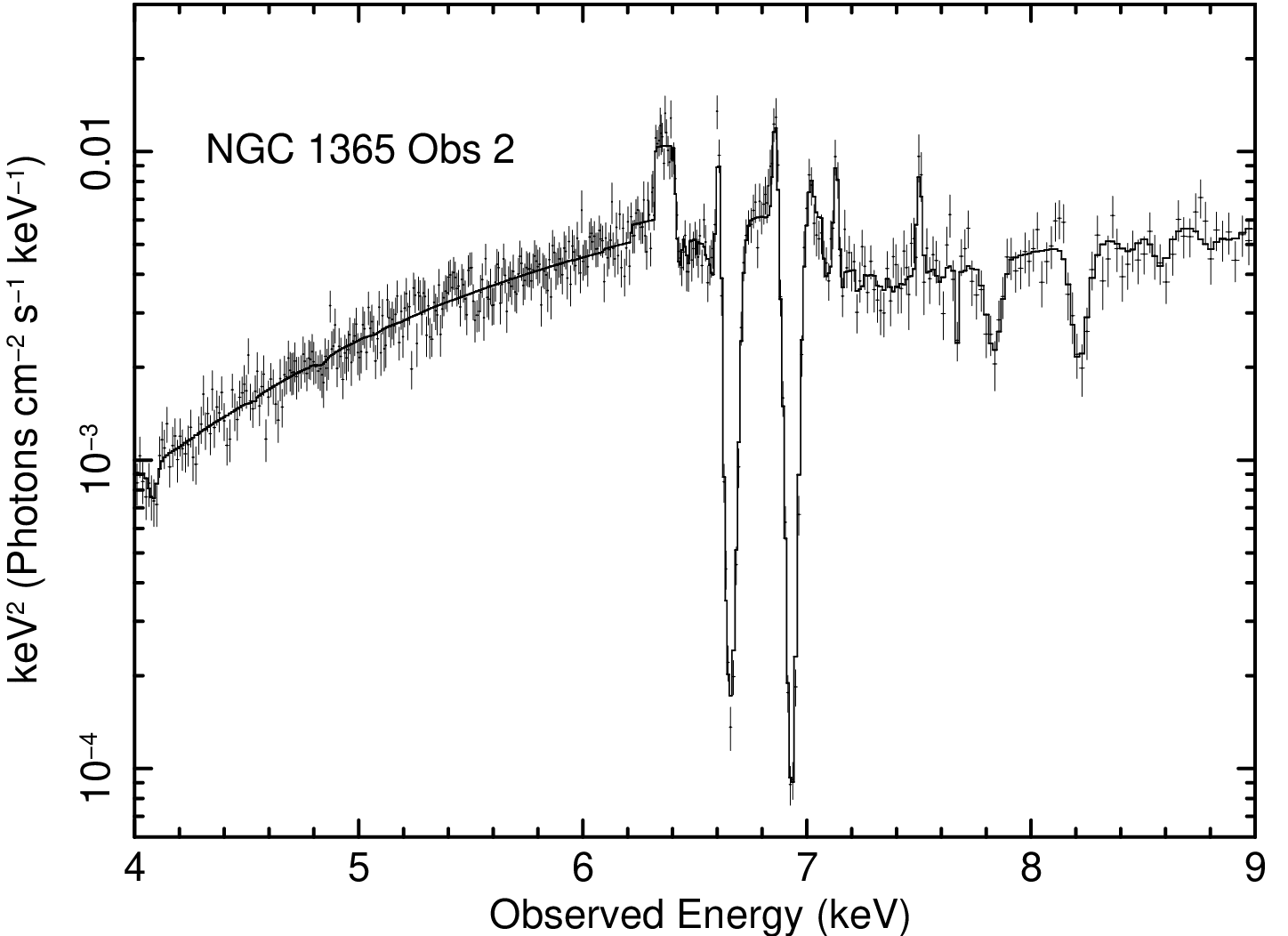}}
\rotatebox{-90}{\includegraphics[width=4cm]{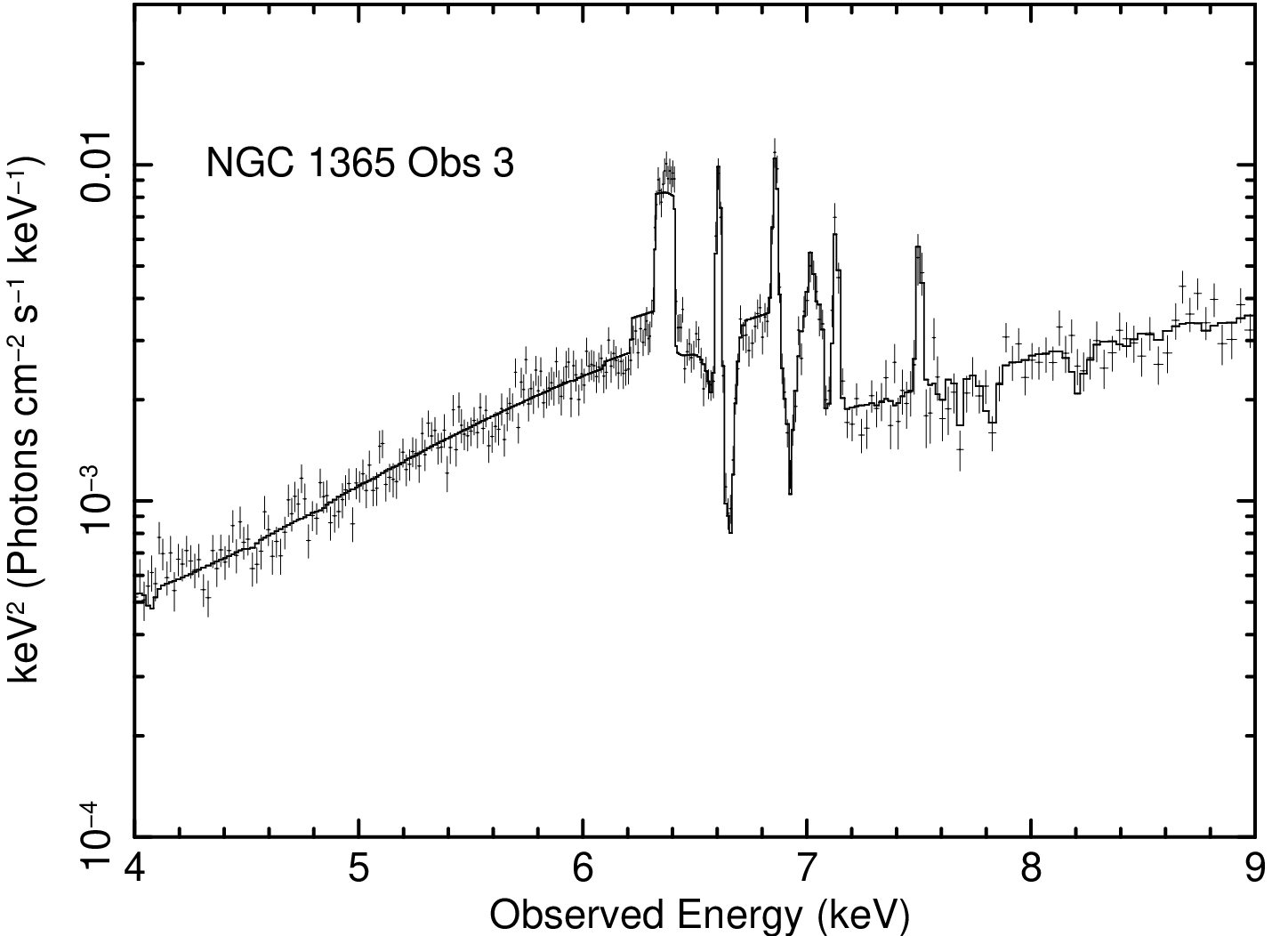}}
\rotatebox{-90}{\includegraphics[width=4cm]{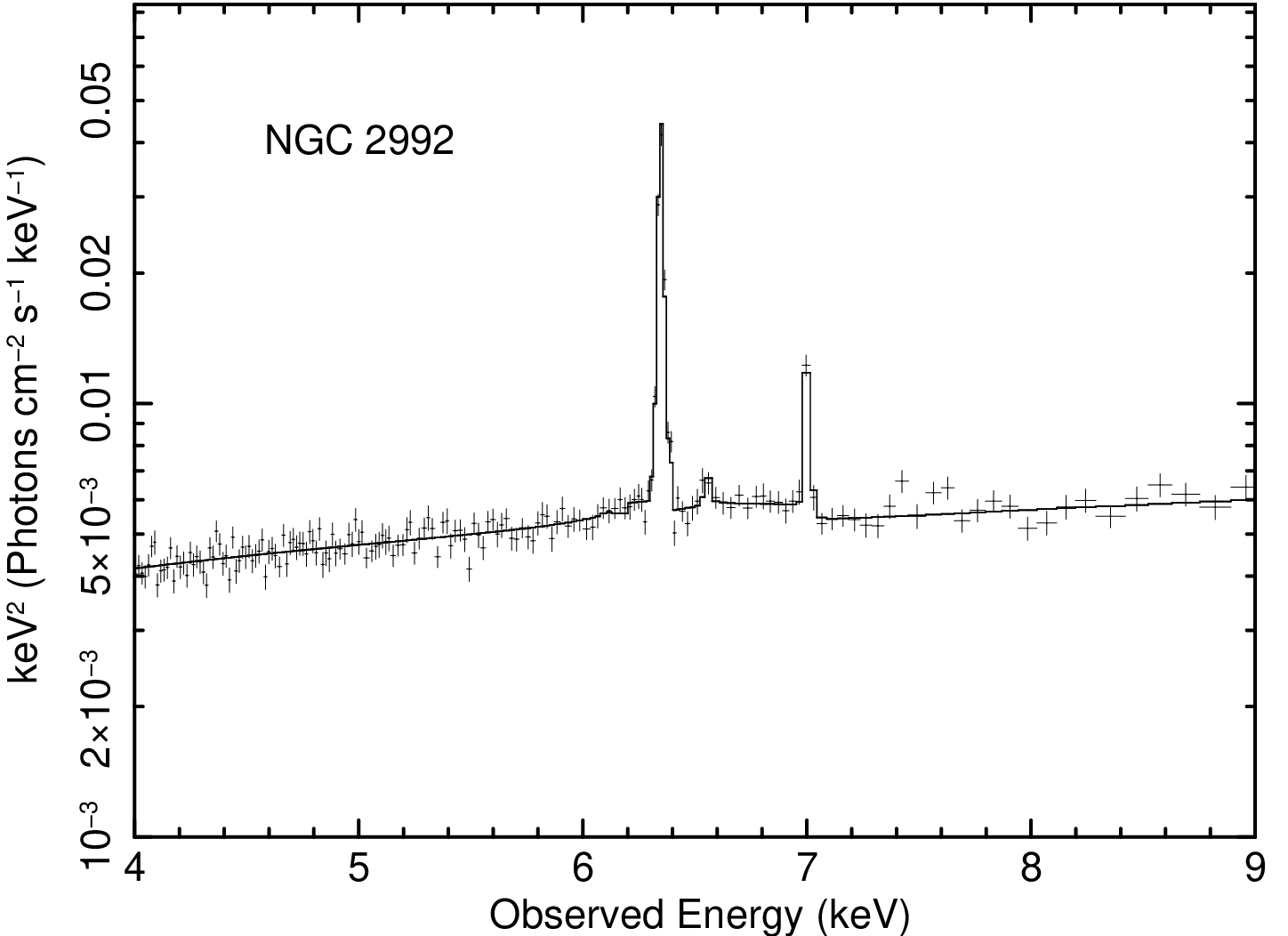}}

\contcaption{$\nu\,F\nu$ plots of the baseline+\textsc{relline} model in the 4-9\,keV region. Note the strong absorption in the Fe\,K region in many objects and relatively weak red-wing. The solid black line shows the total model.}
\end{figure*}

\begin{figure*}
\rotatebox{-90}{\includegraphics[width=4cm]{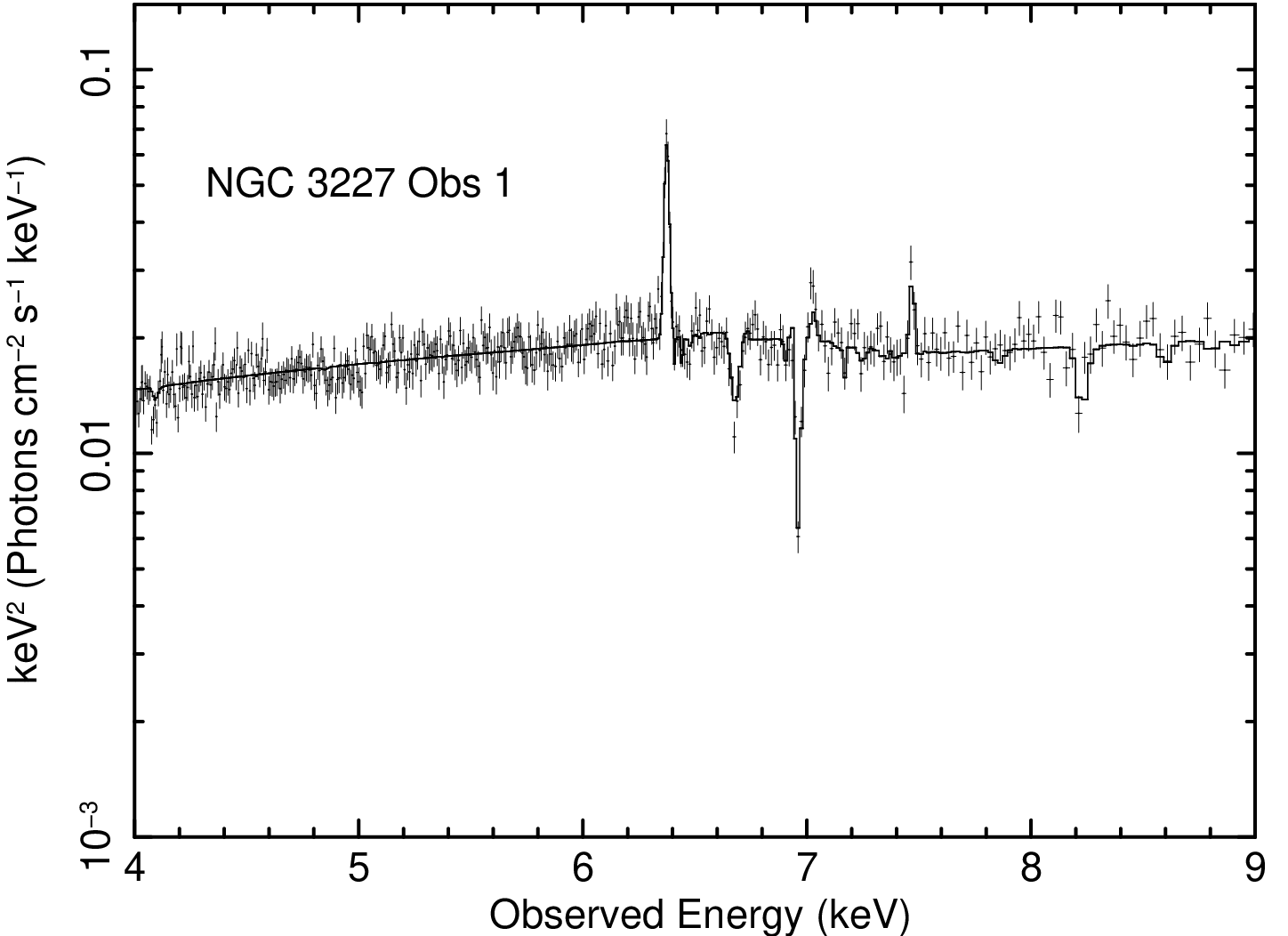}}
\rotatebox{-90}{\includegraphics[width=4cm]{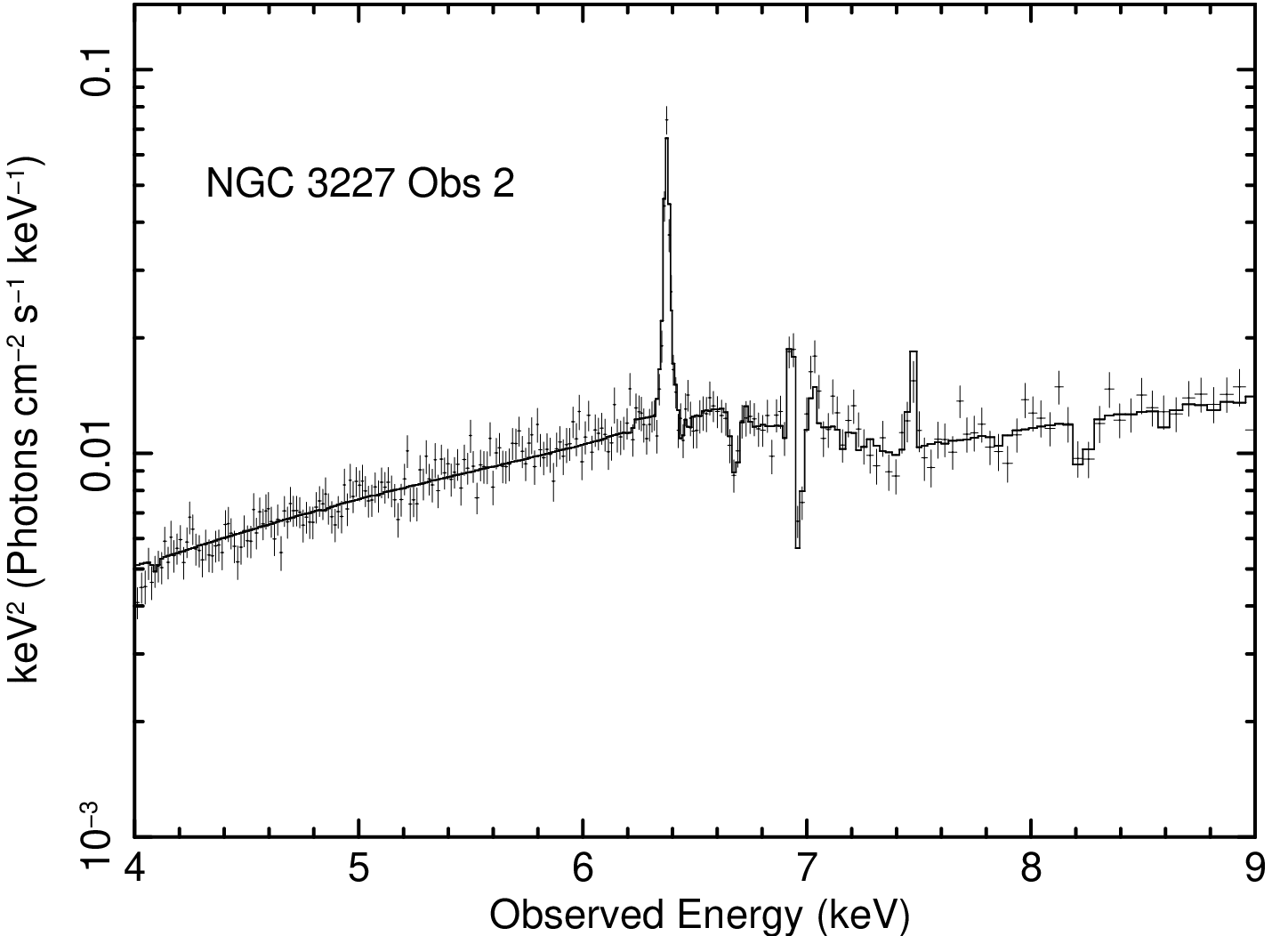}}
\rotatebox{-90}{\includegraphics[width=4cm]{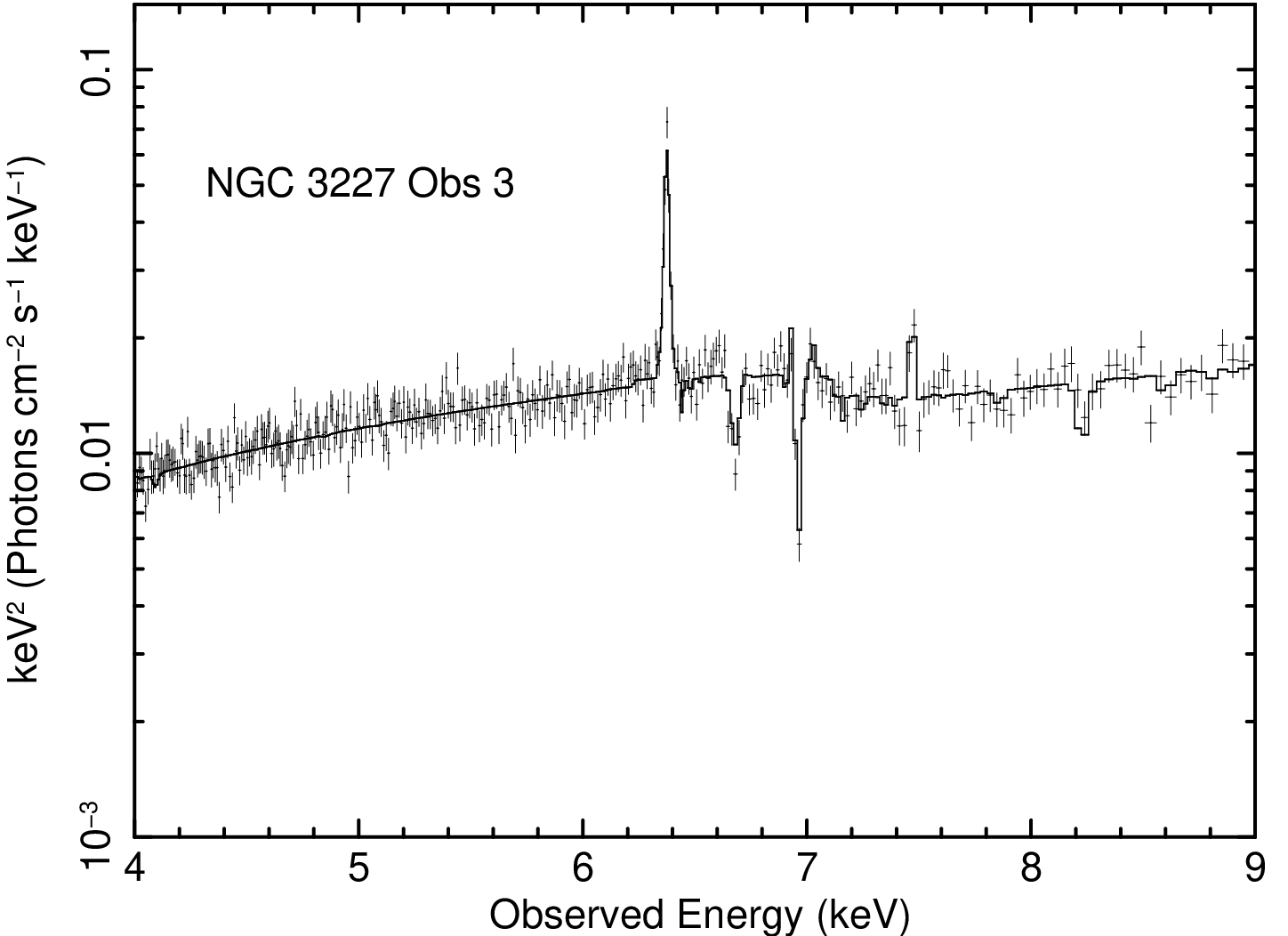}}
\rotatebox{-90}{\includegraphics[width=4cm]{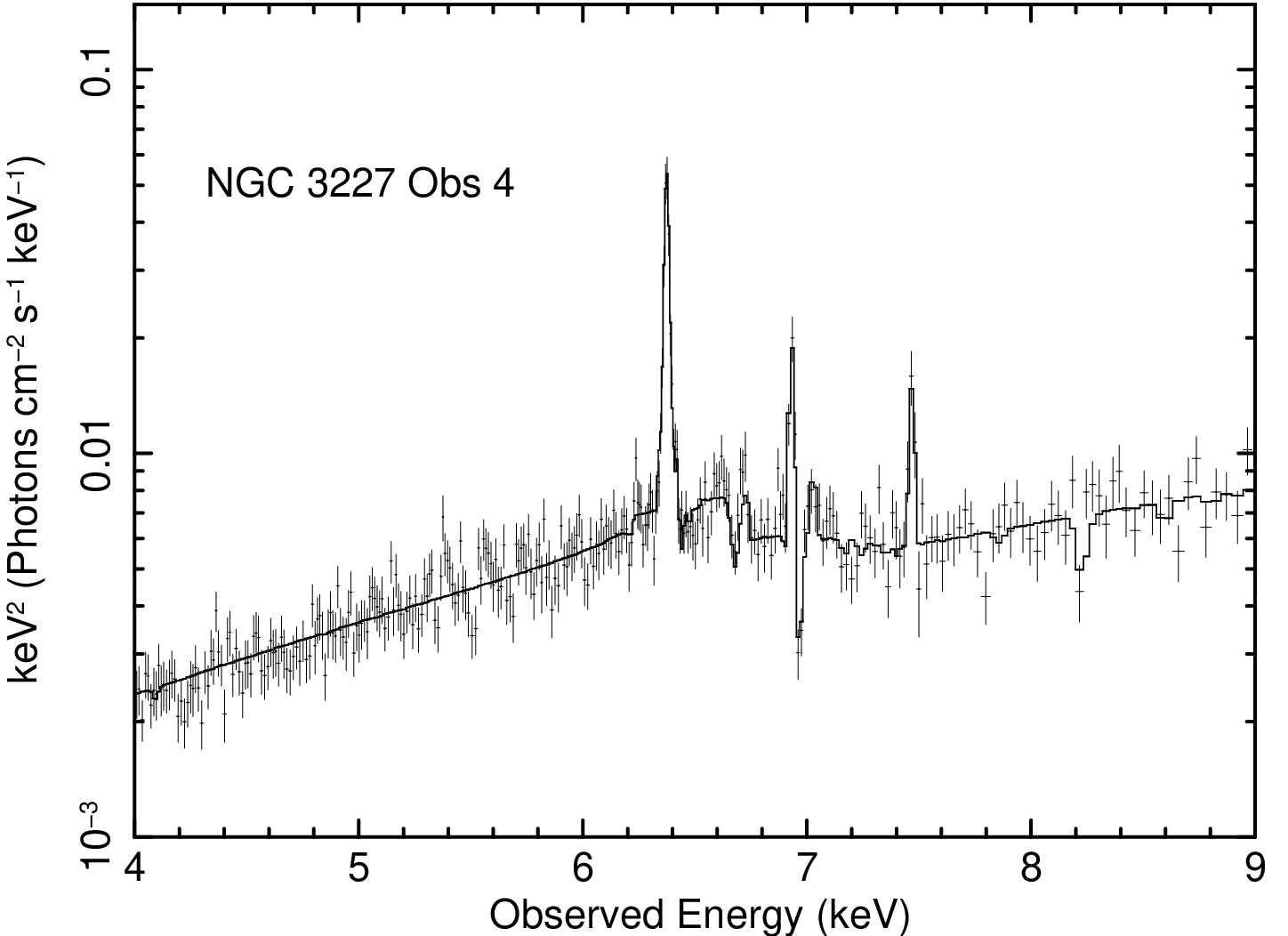}}
\rotatebox{-90}{\includegraphics[width=4cm]{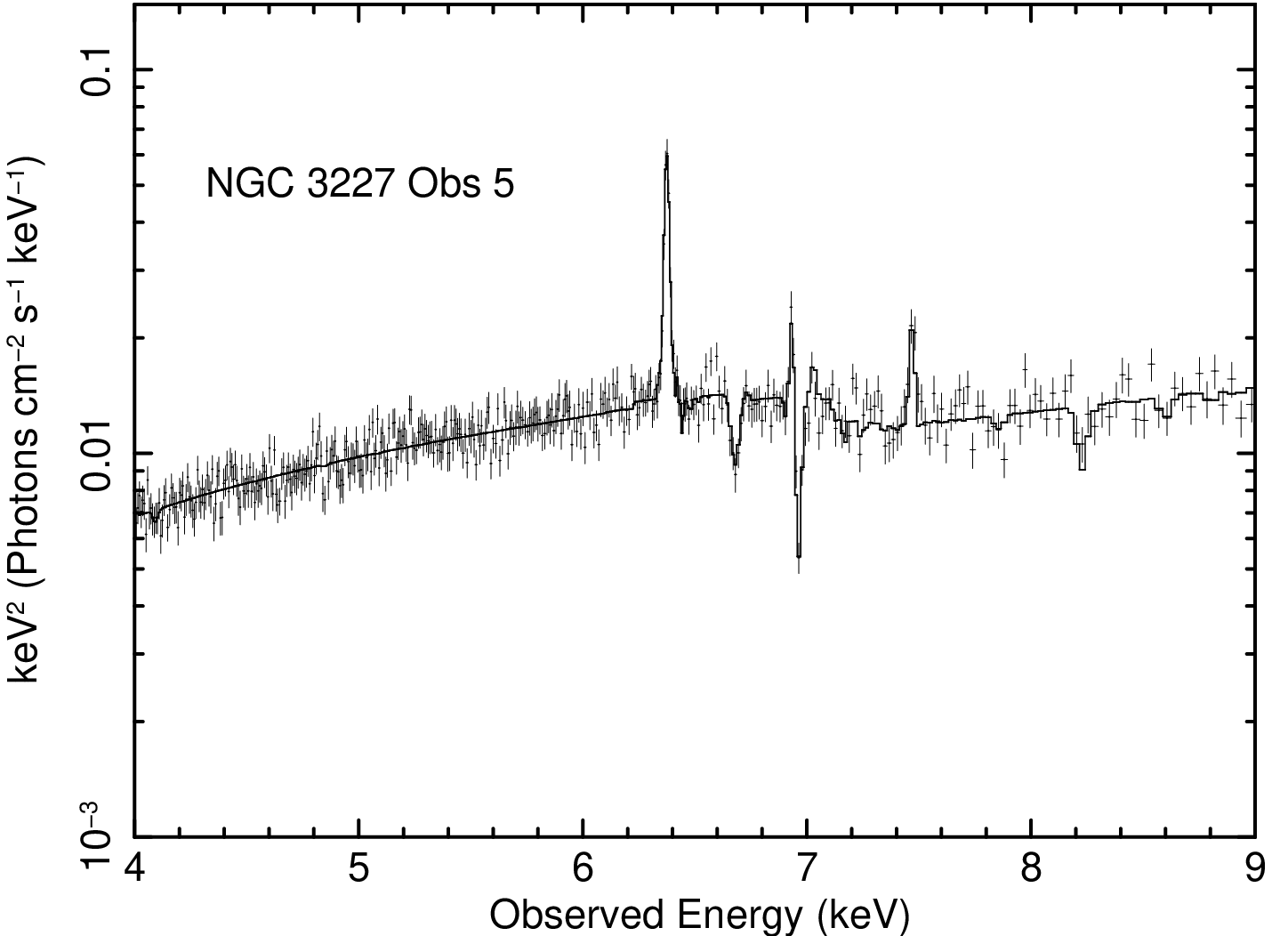}}
\rotatebox{-90}{\includegraphics[width=4cm]{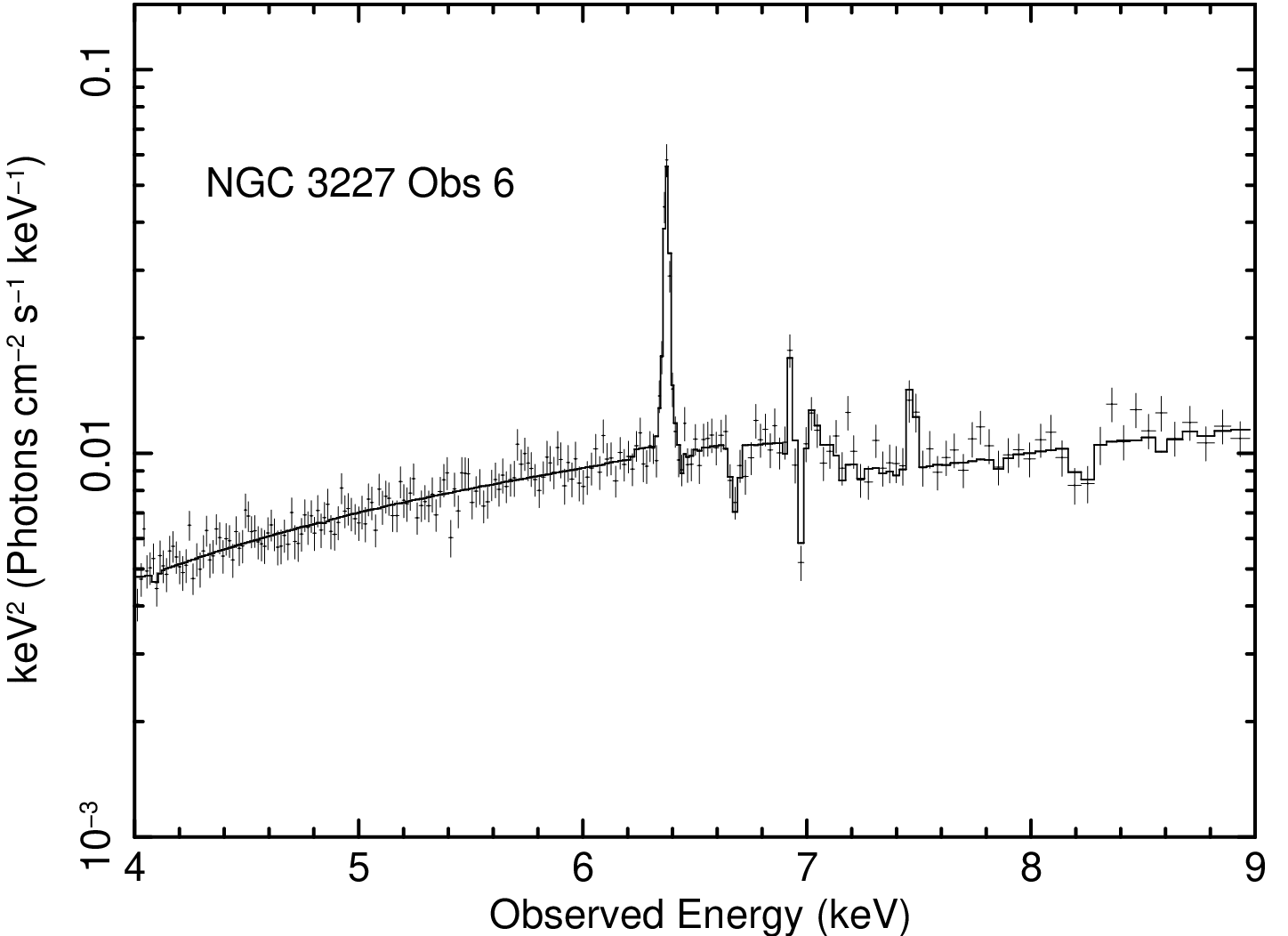}}
\rotatebox{-90}{\includegraphics[width=4cm]{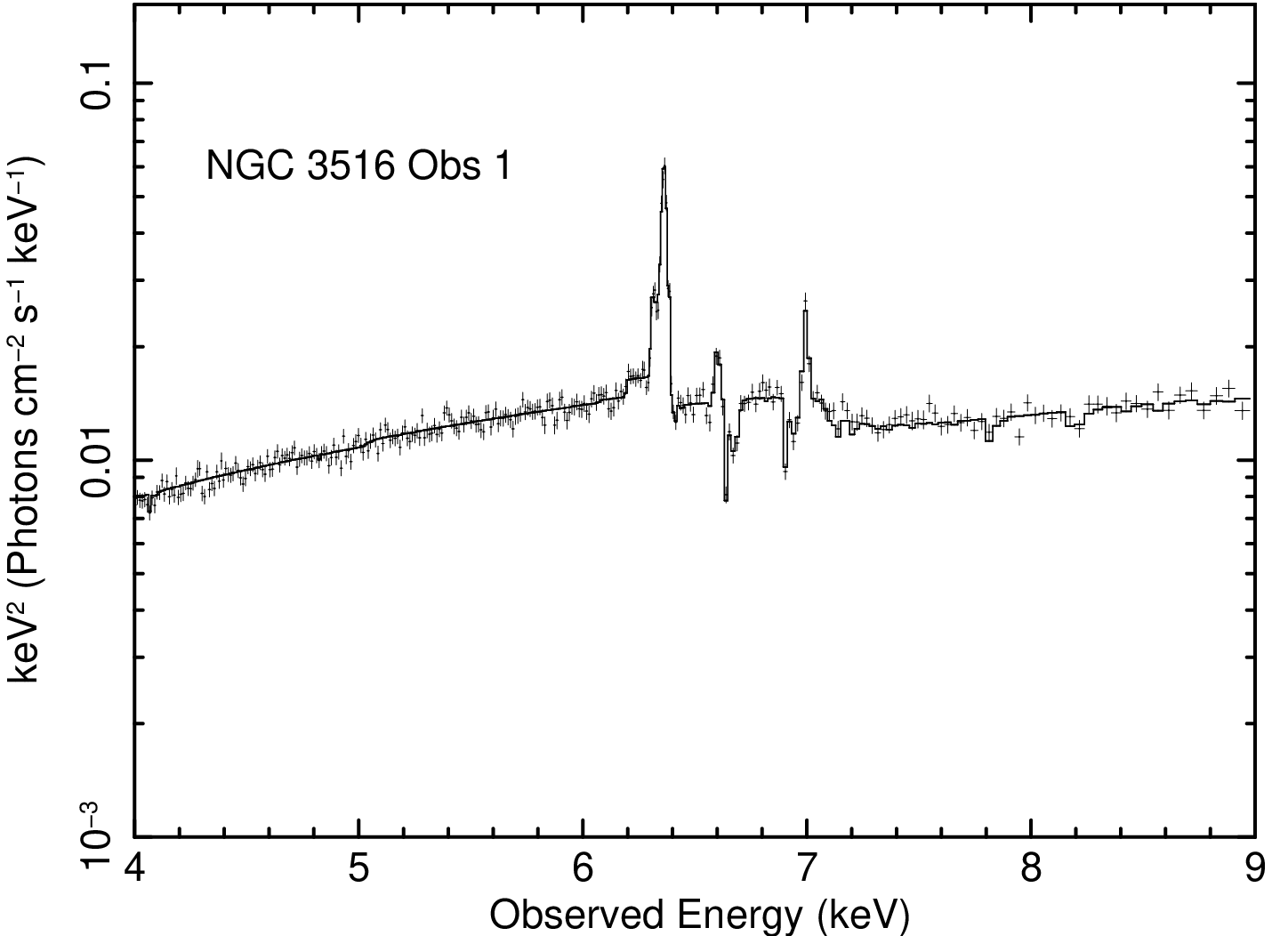}}
\rotatebox{-90}{\includegraphics[width=4cm]{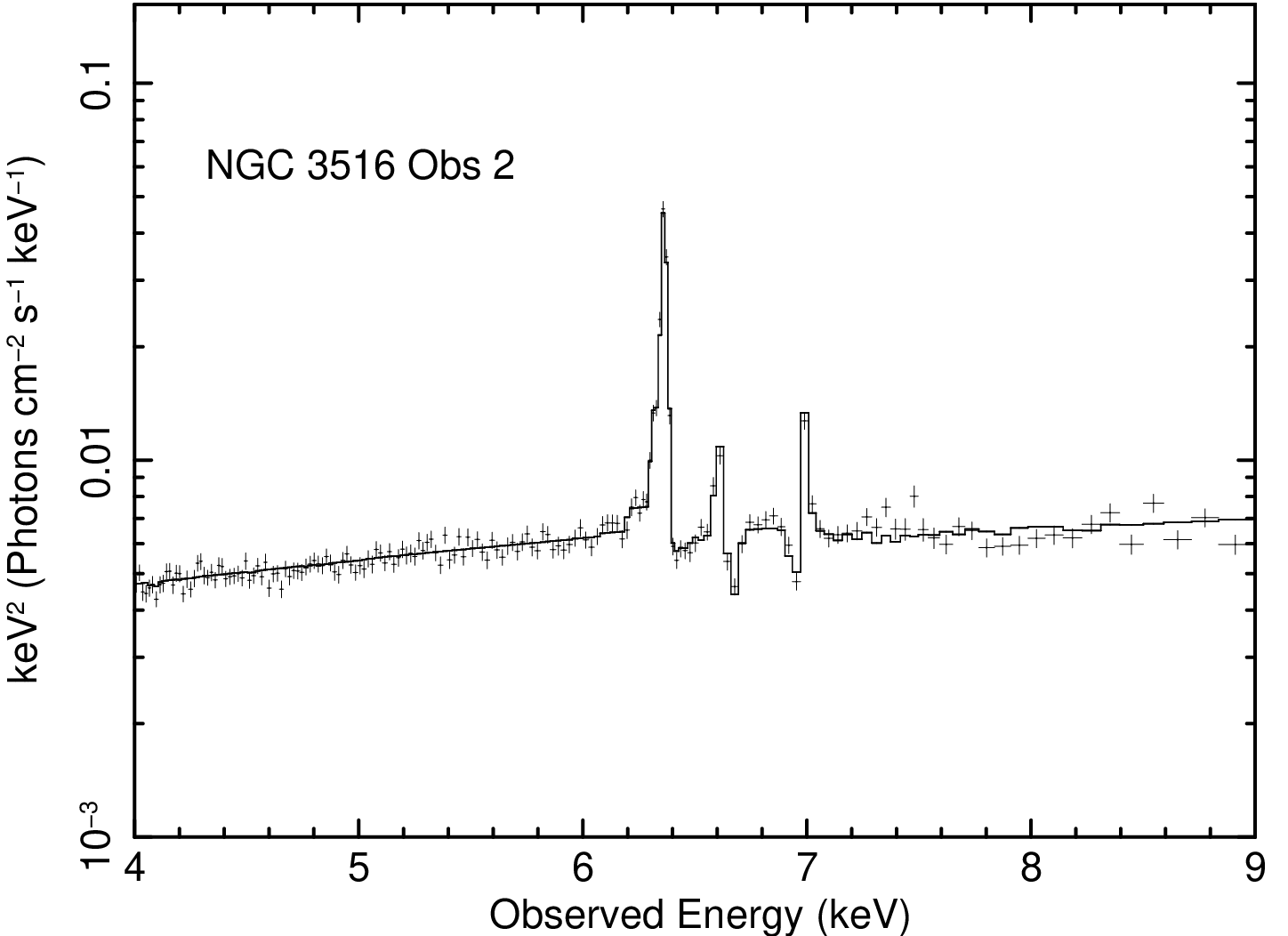}}
\rotatebox{-90}{\includegraphics[width=4cm]{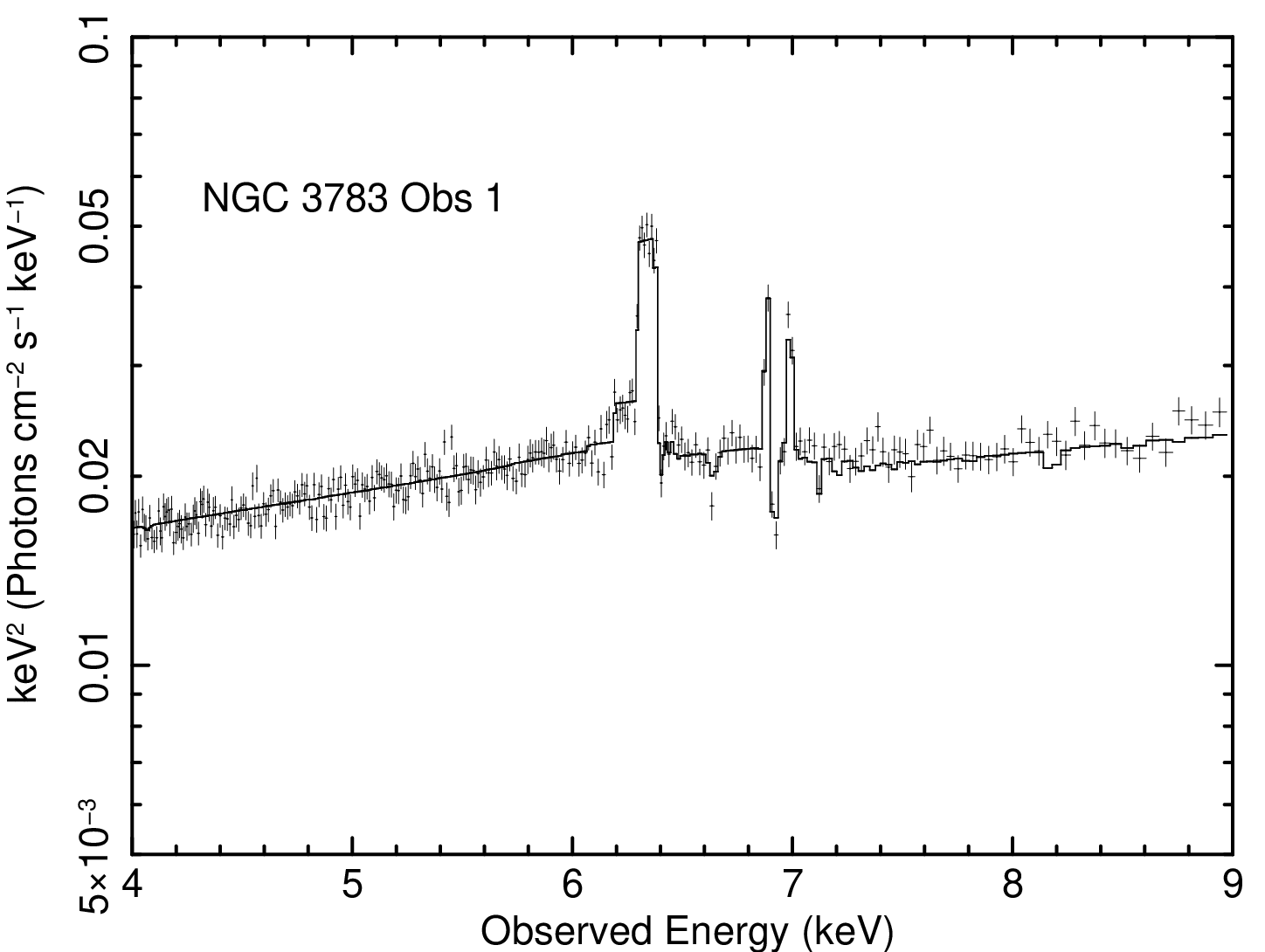}}
\rotatebox{-90}{\includegraphics[width=4cm]{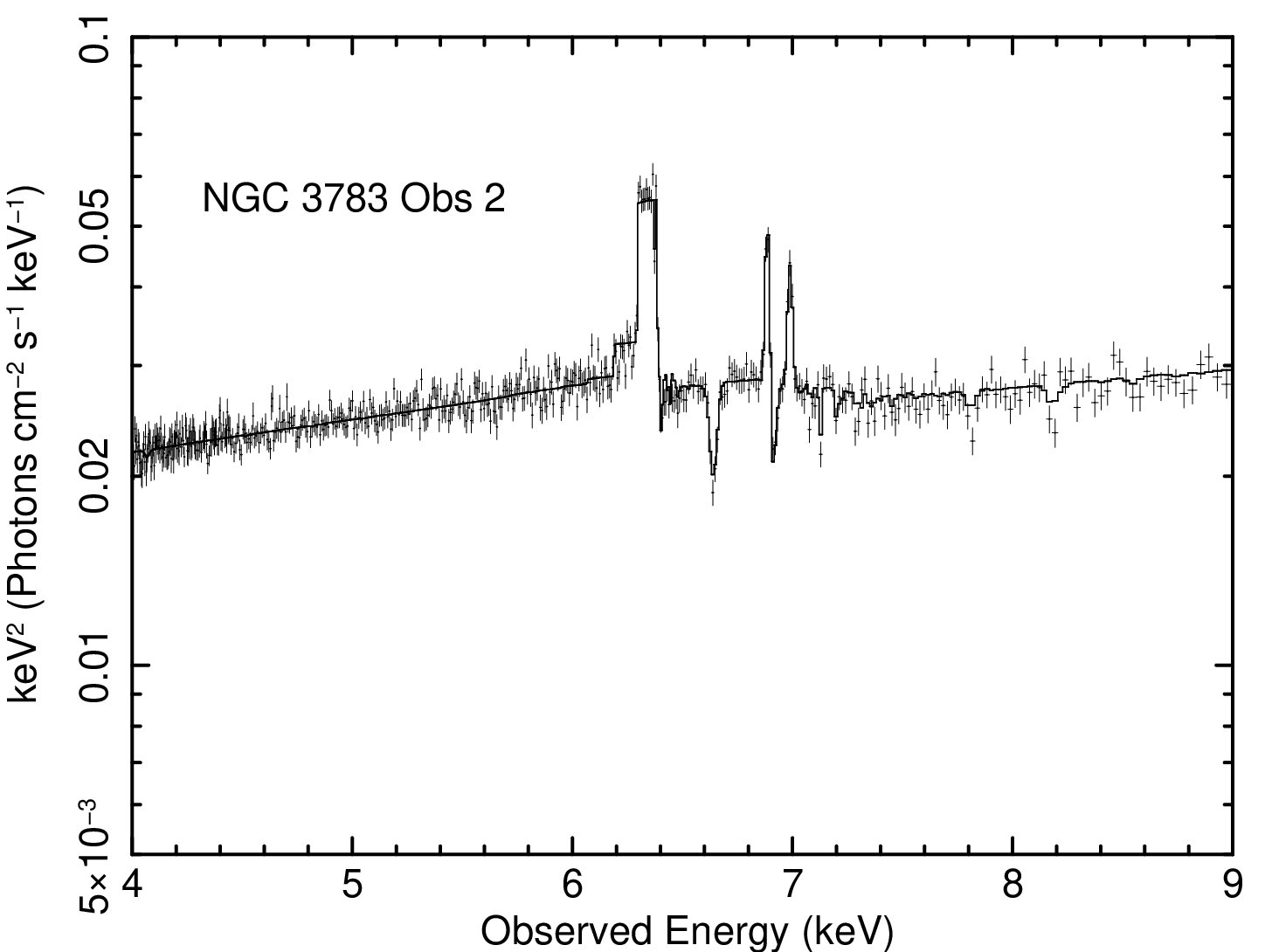}}
\rotatebox{-90}{\includegraphics[width=4cm]{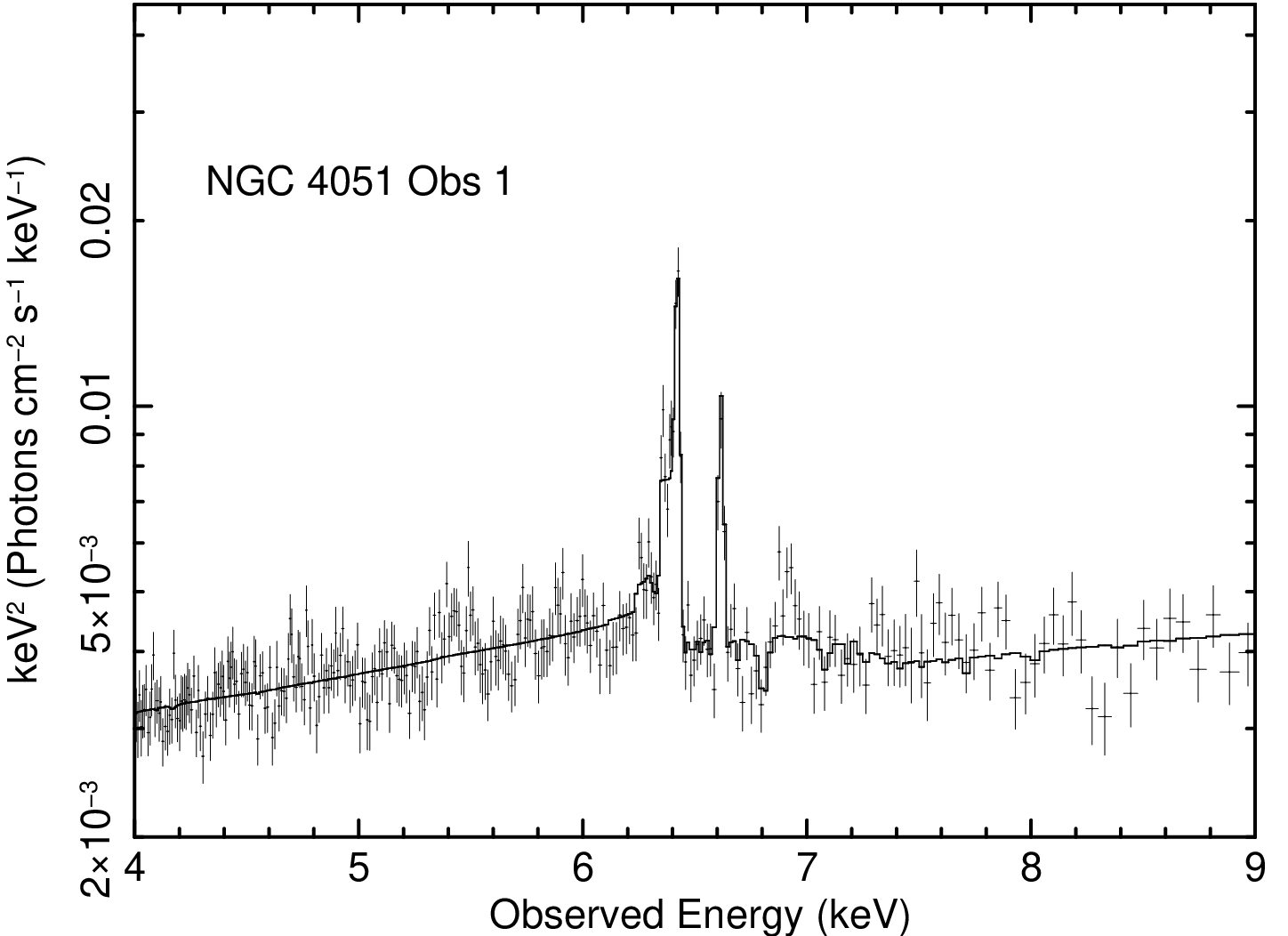}}
\rotatebox{-90}{\includegraphics[width=4cm]{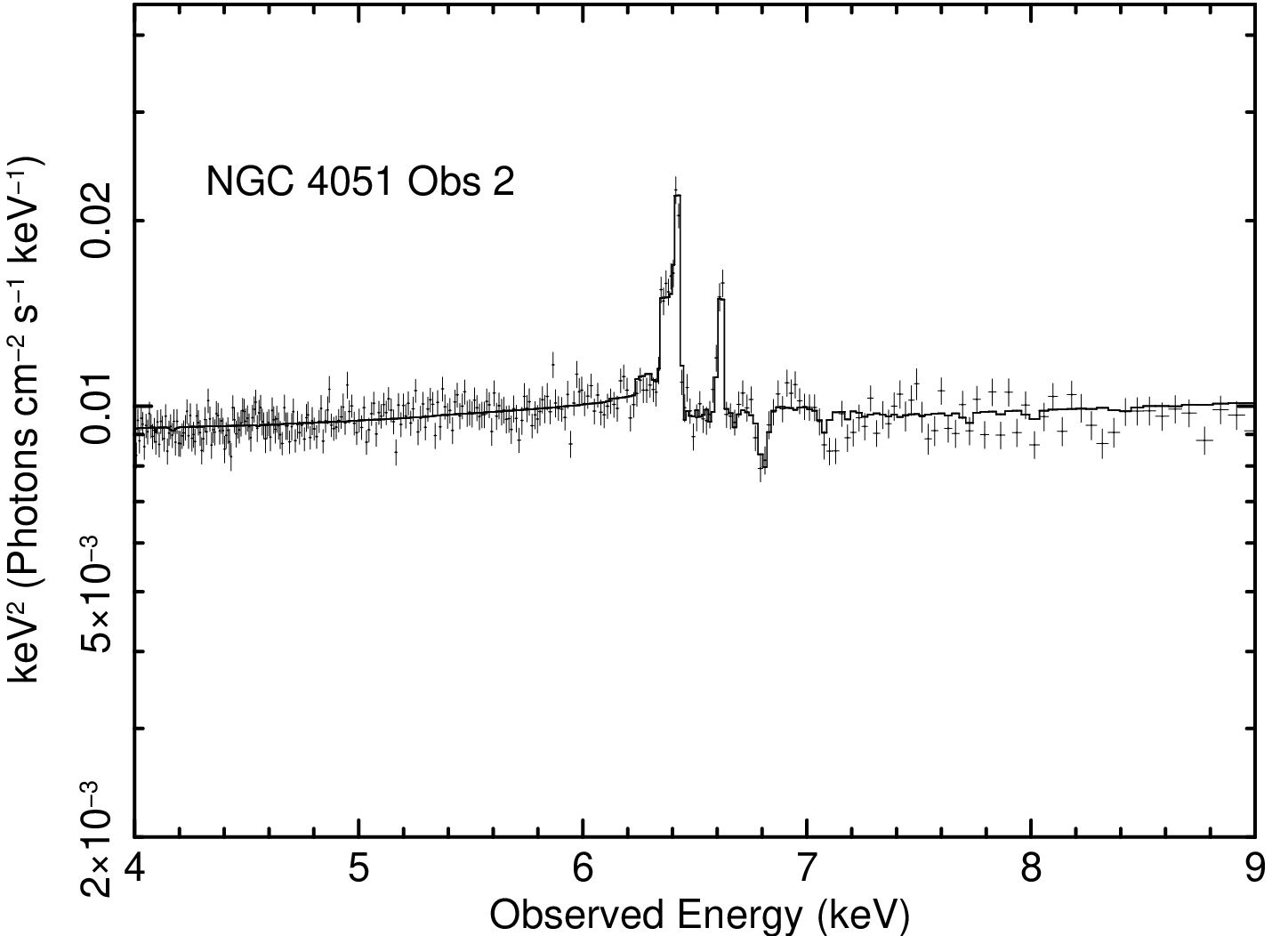}}

\contcaption{$\nu\,F\nu$ plots of the baseline+\textsc{relline} model in the 4-9\,keV region. Note the strong absorption in the Fe\,K region in many objects and relatively weak red-wing. The solid black line shows the total model.}
\end{figure*}

\begin{figure*}

\rotatebox{-90}{\includegraphics[width=4cm]{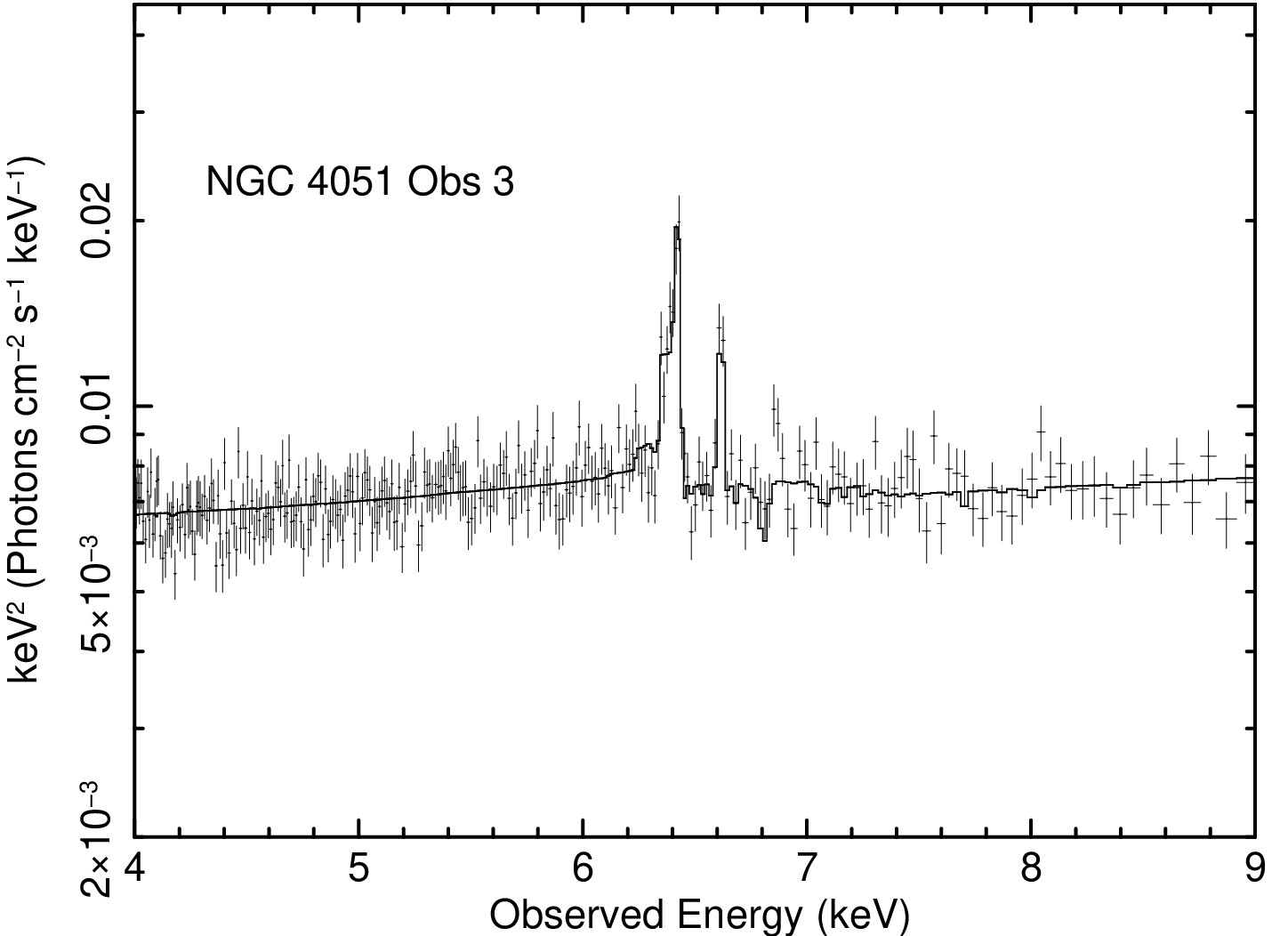}}
\rotatebox{-90}{\includegraphics[width=4cm]{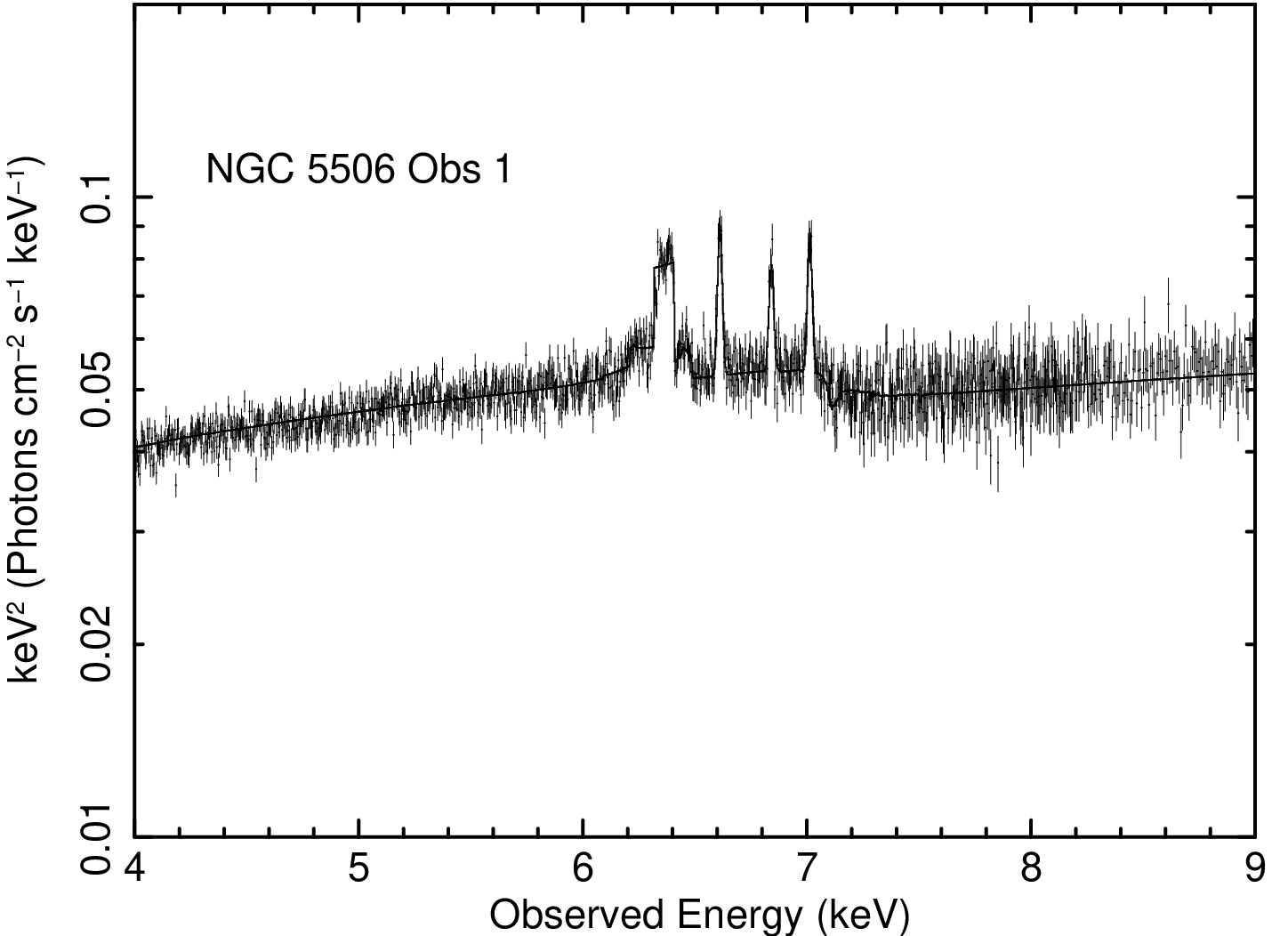}}
\rotatebox{-90}{\includegraphics[width=4cm]{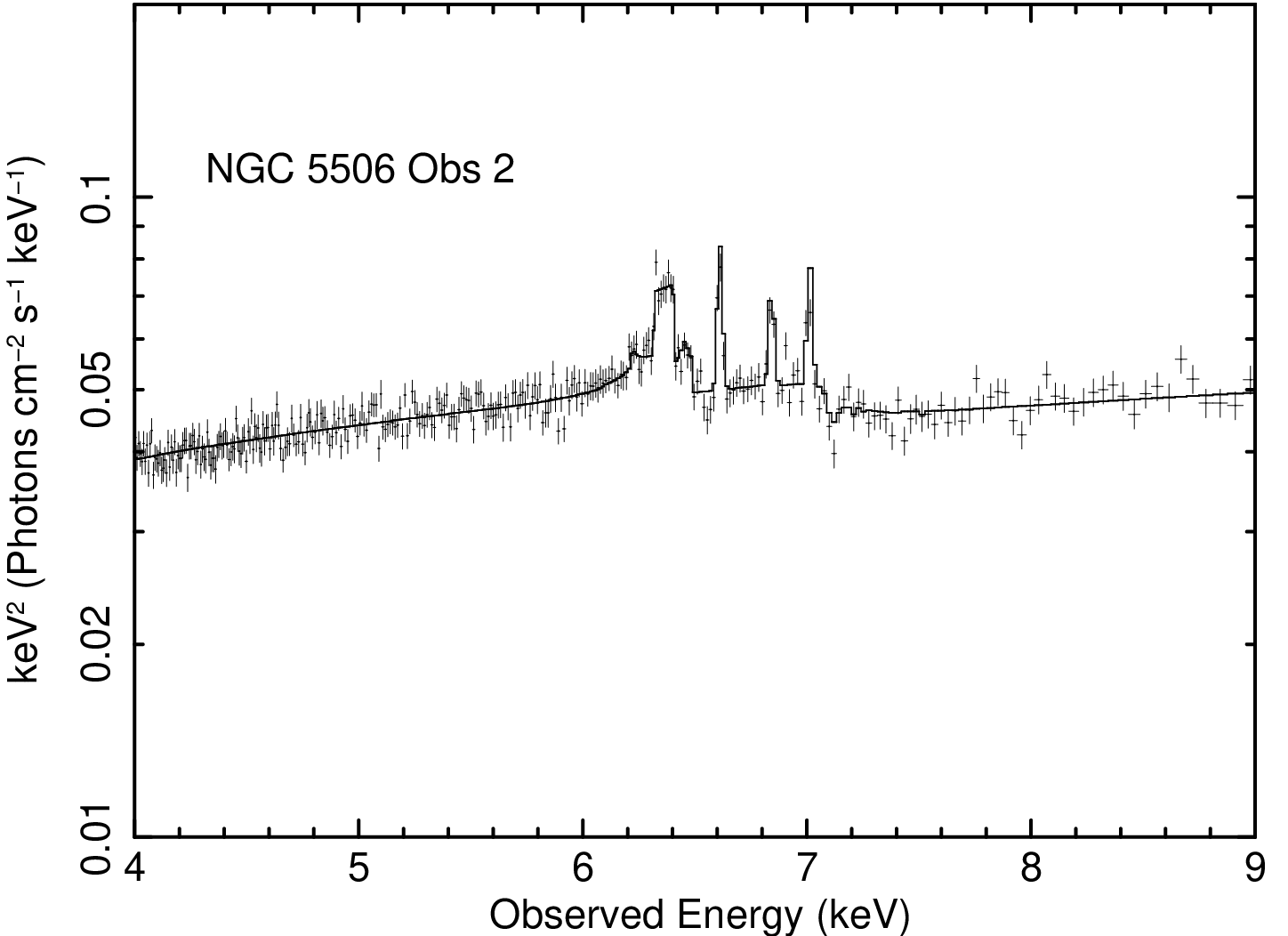}}
\rotatebox{-90}{\includegraphics[width=4cm]{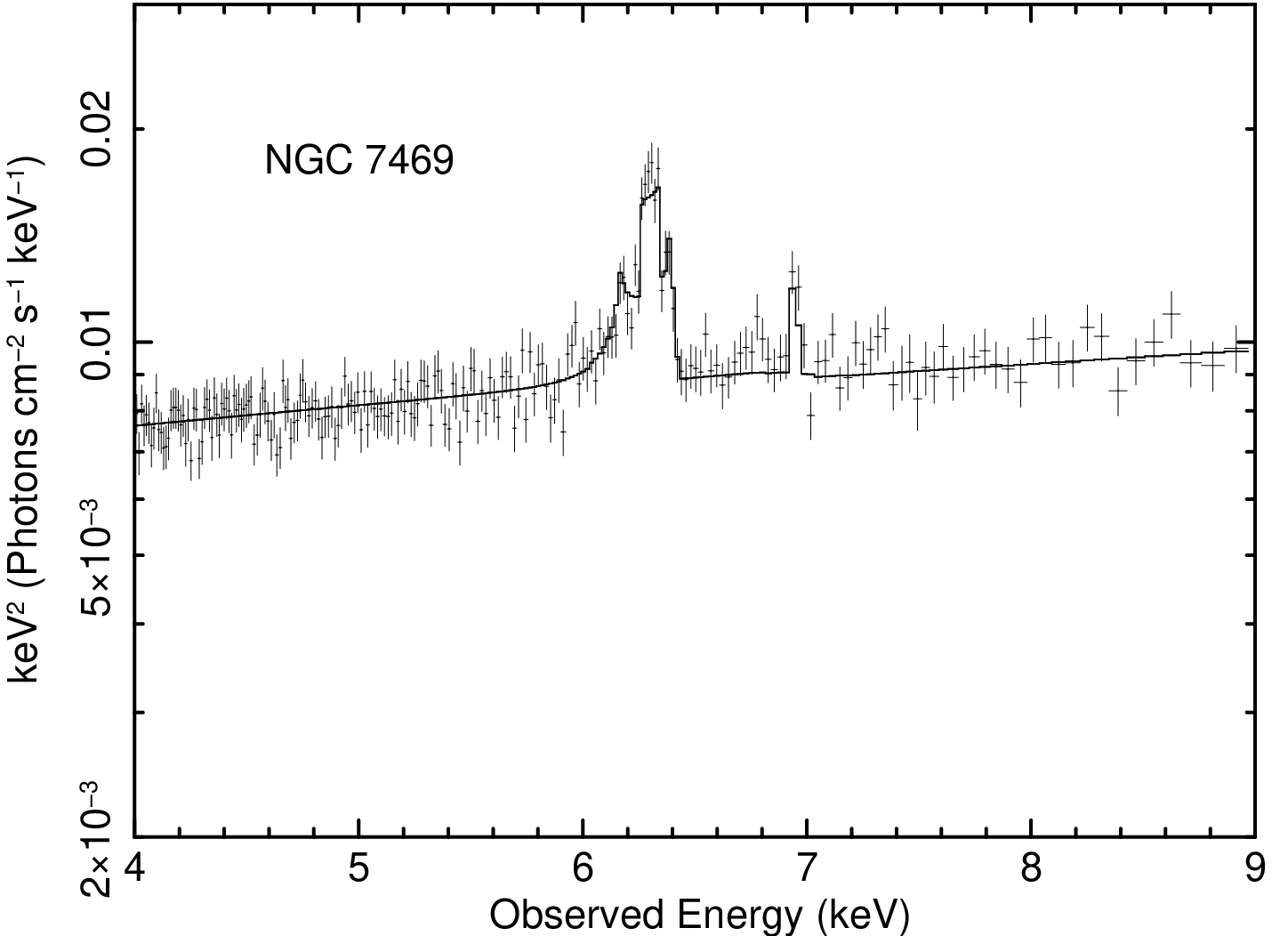}}
\rotatebox{-90}{\includegraphics[width=4cm]{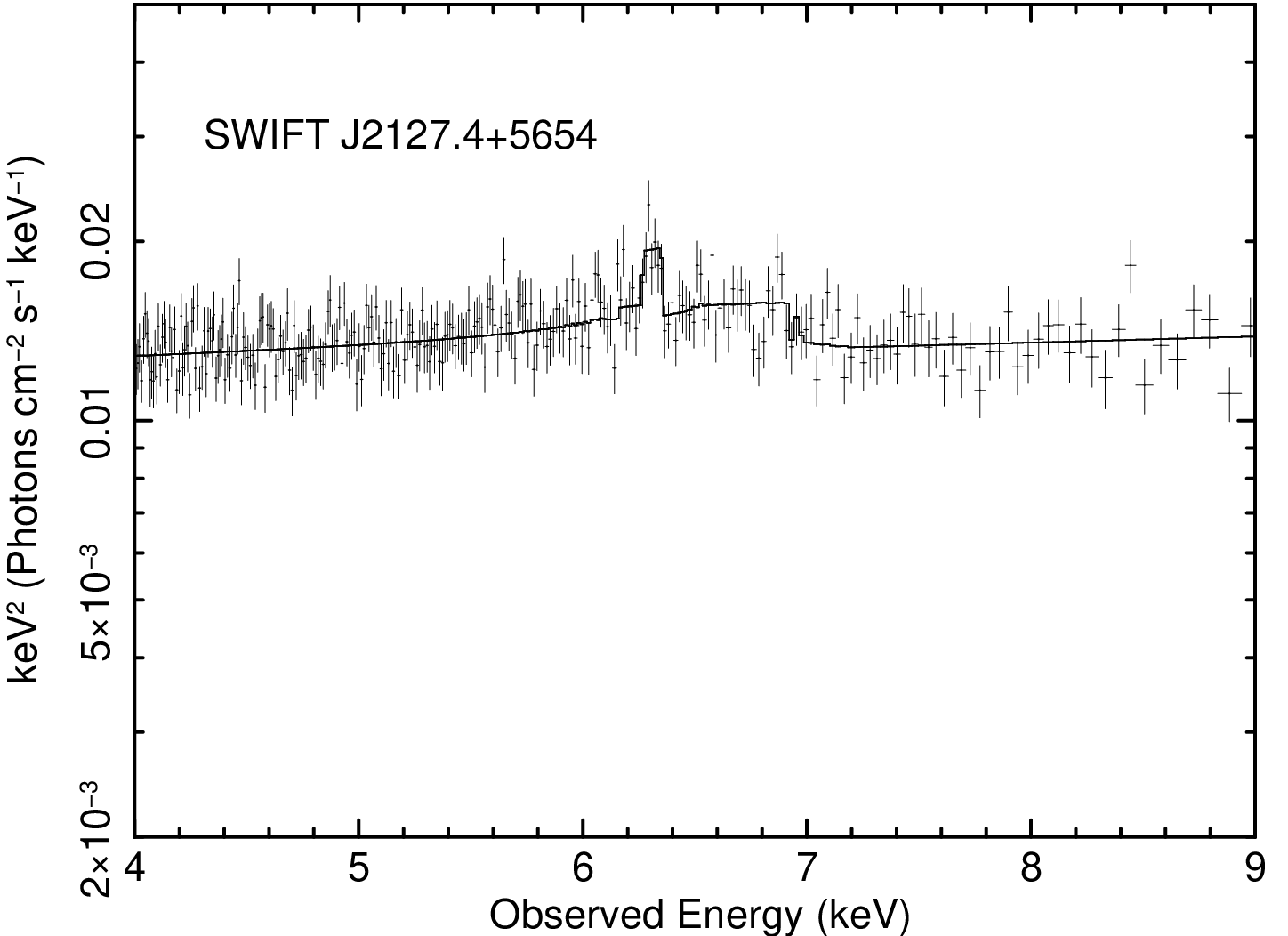}}

\contcaption{$\nu\,F\nu$ plots of the baseline+\textsc{relline} model in the 4-9\,keV region. Note the strong absorption in the Fe\,K region in many objects and relatively weak red-wing. The solid black line shows the total model.}
\end{figure*}

\section{Discussion}
This sample of 46 objects includes all the Seyfert 1 AGN matching our selection criteria outlined in Section 2.1 with observations which are publically available in the {\sl Suzaku} archive. The aim of this paper is to form detailed broad-band models for each object, fully modelling and accounting for any absorption where required which then allows us to parameterise the Fe\,K regions and ultimately see in how many objects we can estimate or constrain SMBH spin. The main motivation for conducting this analysis is to produce an in-depth study of the Fe\,K regions of these AGN and to assess the strength and prevalence of broad emission from the inner regions of the accretion disc. In order to accurately examine both emission and absorption properties in the Fe\,K region, it is essential that high energy data is used to measure the strength of the Compton hump and hard X-ray excess as part of the formation of a broad-band model; the {\sl Suzaku} X-ray observatory is unique its ability to provide such data simultaneously with soft X-ray data. The data included in this sample have been selected with sufficient S/N such that, if broadened line emission from the inner regions of the accretion disc exists (down to tens of eV in equivalent width), we should be able to detect it. 

It should be noted that the broad-band X-ray spectra of these AGN are complex with many of them featuring multiple absorption zones which have a significant role in adding spectral curvature and hence have a significant affect upon and `broad' residuals in the Fe\,K region. Nandra et al. (2007) also note that the continua are complex in a study of the 2.5-10.0\,keV energy band with data obtained with {\sl XMM-Newton}, however, here we take this study a step further by also considering a detailed modelling of the 0.6-10.0\,keV soft X-ray data, allowing us to form more complete models of the entire X-ray spectrum up to 100\,keV. Low energy X-ray data below 2.5\,keV is essential to perform a detailed analysis of the Fe\,K region of any object since first forming an appropriate baseline model, with soft X-ray warm absorption included, is an important prerequisite. This is due to the warm absorption zones which are required to model the X-ray data below 2.5\,keV can add significant spectral curvature above 2.5\,keV and therefore influence the Fe\,K region and residuals which may otherwise be interpreted as a broad red-wing from the very inner regions of the accretion disc, for example see MCG--06-30-15 (Patrick et al. 2011b).

\begin{figure*}
\rotatebox{-90}{\includegraphics[width=5cm]{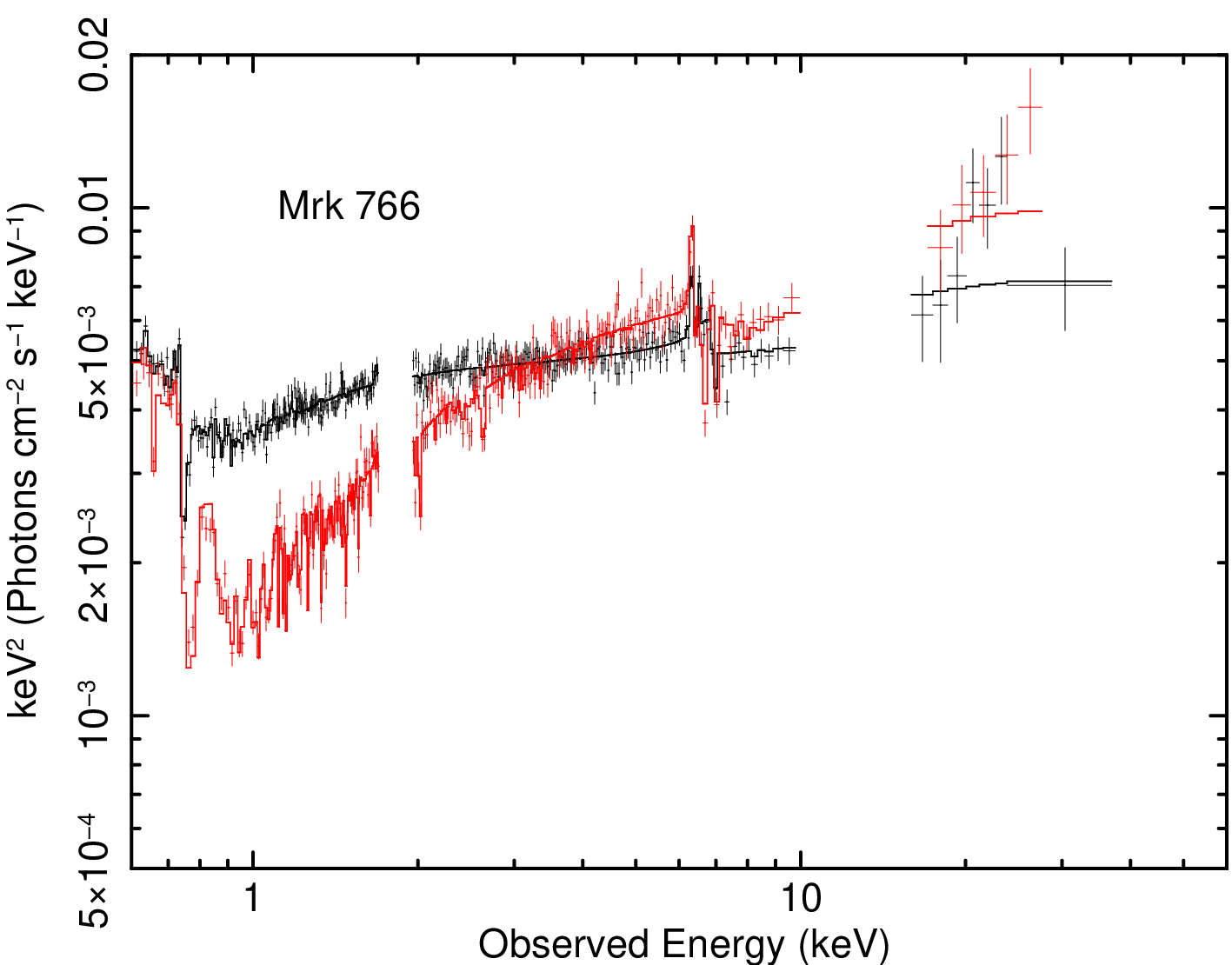}}
\rotatebox{-90}{\includegraphics[width=5cm]{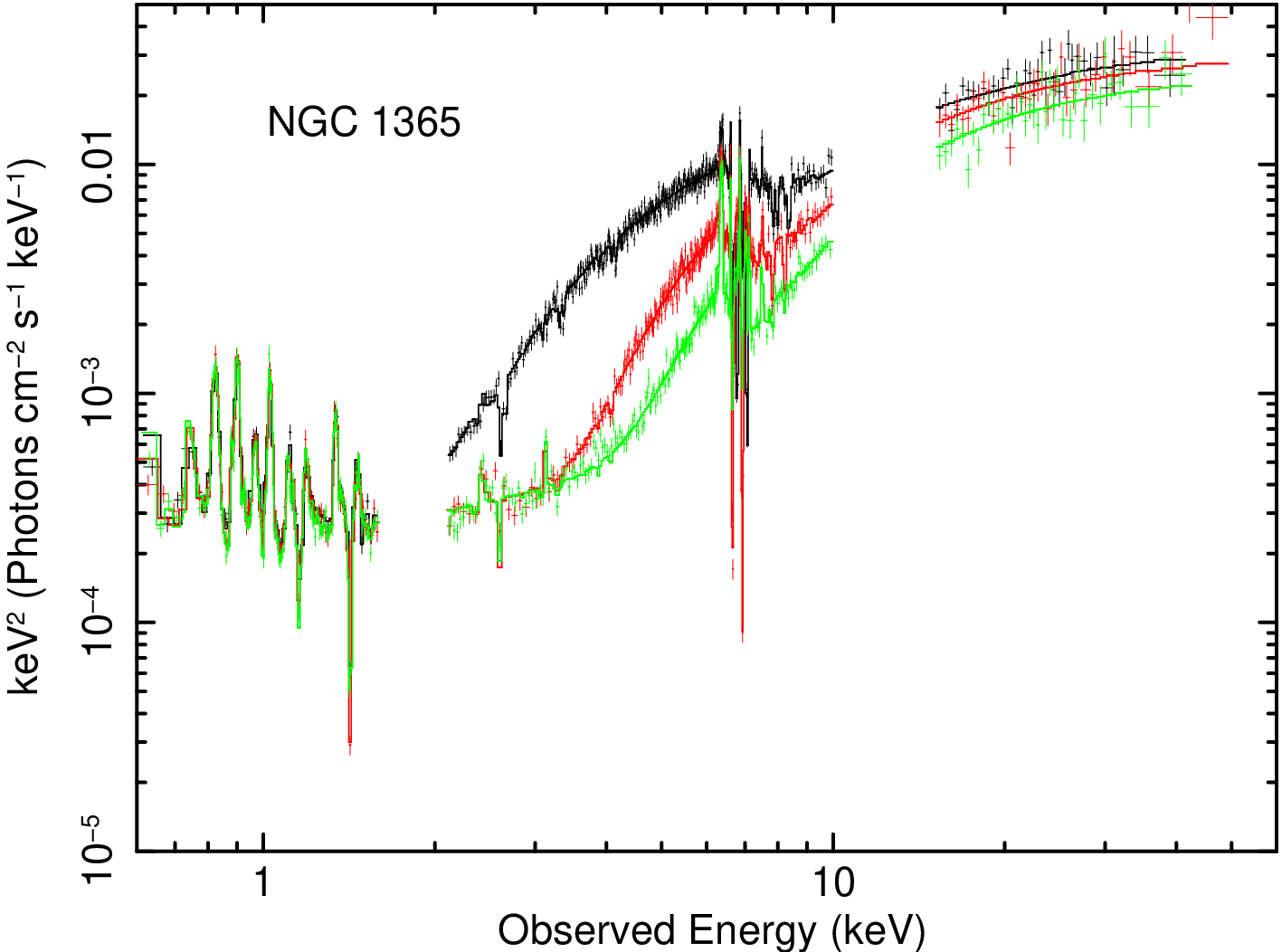}}
\rotatebox{-90}{\includegraphics[width=5cm]{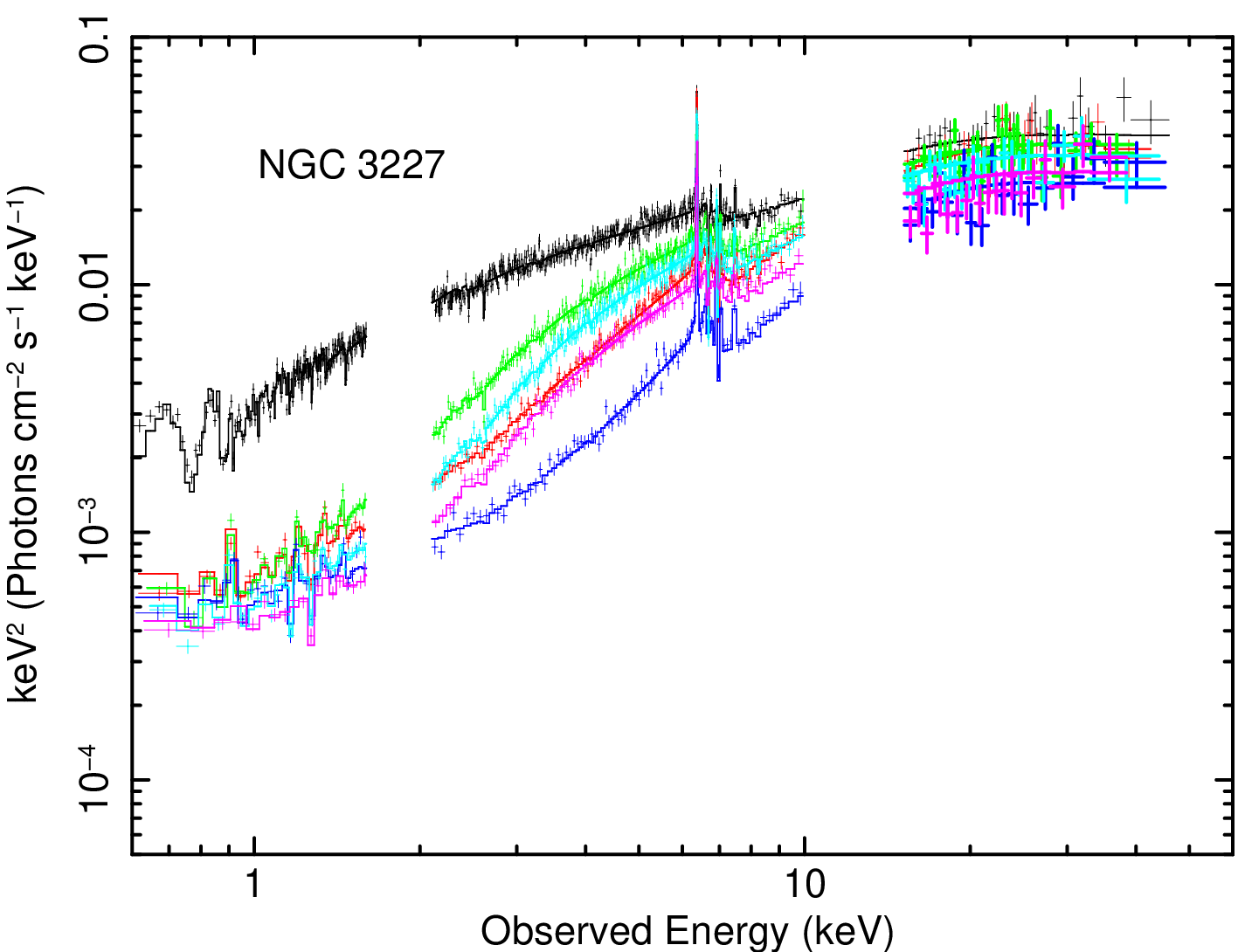}}
\rotatebox{-90}{\includegraphics[width=5cm]{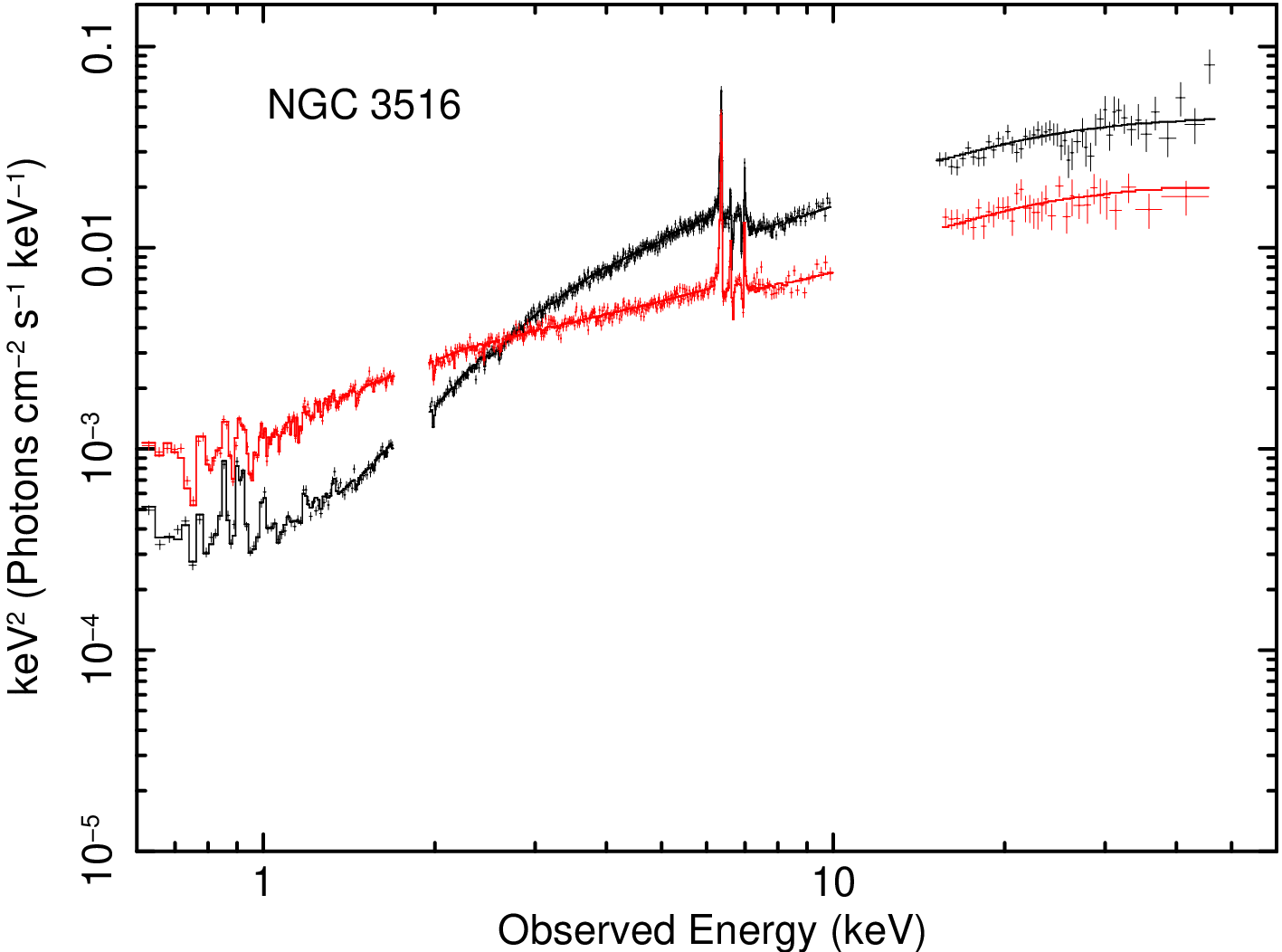}}
\rotatebox{-90}{\includegraphics[width=5cm]{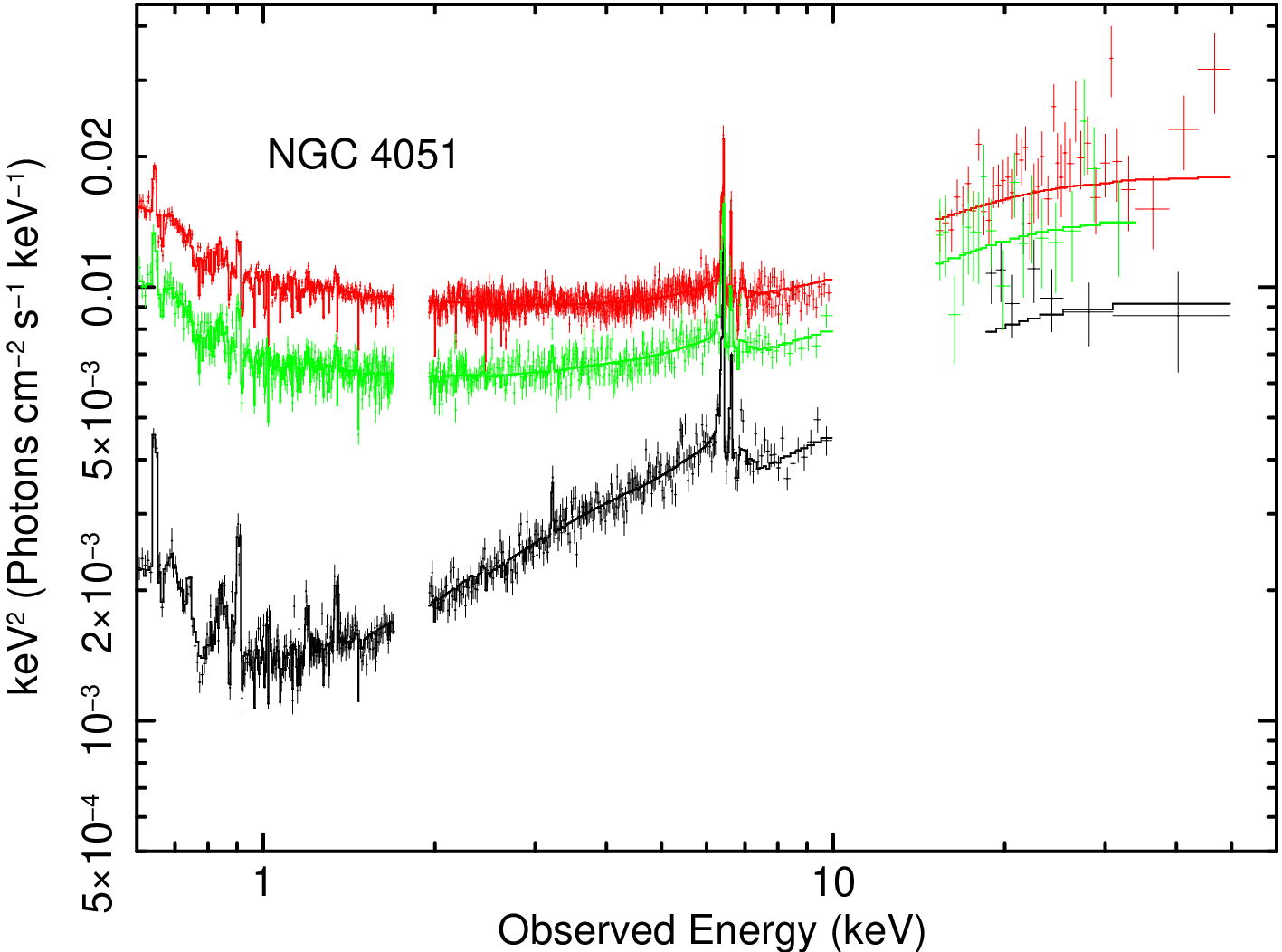}}
\caption{$\nu\,F\nu$ plots of the full broad-band models of some objects displaying significant variability. Note that the 2005 (black) observation of NGC 3516 is higher flux, yet appears to be more absorbed whereas in the majority of these examples it intuitively appears that simply the covering fraction has varied between observations.}
\label{fig:eeuf_var}
\end{figure*}

\subsection{Ionized emission in the Fe\,K region}
Narrow ionized emission from distant material is found to be relatively common (in the baseline model) amongst these Seyfert 1 AGN with 24/46 (52\%) of objects featuring Fe\,{\rm XXV} emission at 6.63-6.7\,keV and 18/46 (39\%) featuring Fe\,{\rm XXVI} emission at $\sim6.97$\,keV. Meanwhile 10/46 objects feature both Fe\,{\rm XXV} and Fe\,{\rm XXVI} emission (Table \ref{tab:FeK}). These fractions are much higher than those obtained by Nandra et al. (2007) who only find significant emission at 6.7\,keV in NGC 5506 and significant 6.97\,keV emission in both NGC 3783 and NGC 4593. Comparing to Fukazawa et al. (2011), we also find a higher fraction here, however, this may be due to our more stringent selection criteria (e.g. $>30000$ counts and exposure $>50$\,ks) and hence an increased ability to detect a weaker ionised line in the data if it is indeed present. The findings here, however, concur with those found by Bianchi et al. (2009) and Patrick et al. (2011a) who similarly find that narrow ionized emission lines are a common feature of a number of Seyfert 1 spectra.

After the inclusion of a component such as \textsc{relline} to account for any broad iron line residuals, however, these fractions can be somewhat reduced due to interplay with the blue-wing of a relativistically broadened line profile falling in the 6.6-7.0\,keV region in some AGN. For example, a feature at $\sim6.7$\,keV can often be modelled as either narrow ionized emission or as part of a broad line profile with fairly typical emissivity and inclination parameters. It is likely, therefore, that the true fraction of Seyfert 1 spectra with narrow ionized emission (Fe\,{\rm XXV} in particular) is lower after the inclusion of a broad line component. In 4 objects the feature at $\sim6.7$\,keV is preferentially described by a relativistic line profile for the above reasons, leaving the final number of AGN in this sample with distant Fe\,{\rm XXV} emission as 18/46 objects. Calorimeter resolution spectra from {\sl Astro-H} will help to resolve these issues. 

\begin{figure}
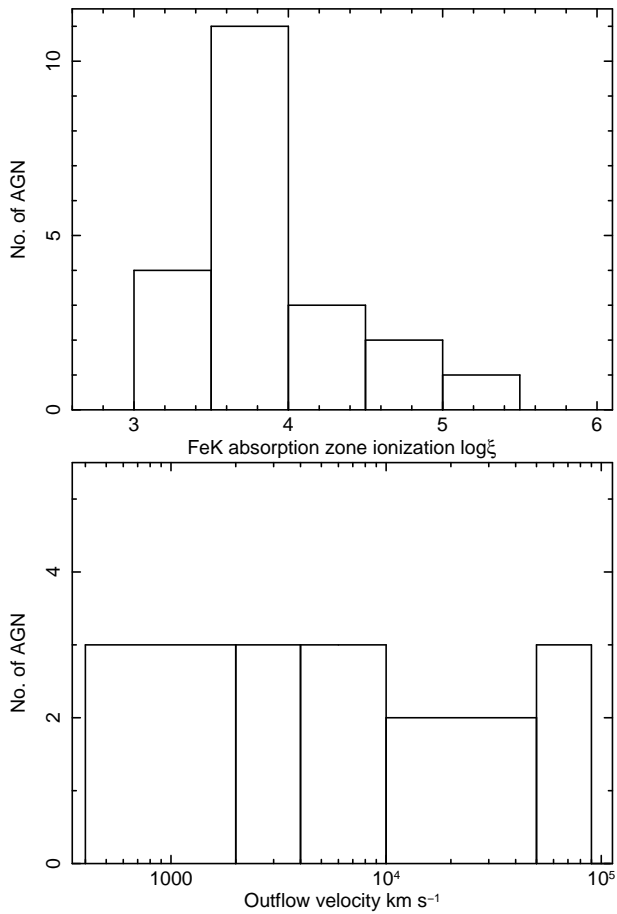

\rotatebox{-90}{\includegraphics[width=6cm]{FeKabs_xi_histo.ps}}
\rotatebox{-90}{\includegraphics[width=6cm]{vout_histo_rebin.ps}}
\caption{The distribution of ionization and outflow velocity of absorption zones at Fe\,K energies. Note that the outflow velocity covers a wide range velocities with no clear peak within the parameter space.}
\label{fig:FeK_abs_histos}
\end{figure}

\subsection{Absorption in the Fe\,K region}
In addition to absorption at soft X-ray energies, absorption in the Fe\,K region due to 1s-2p resonance lines from Fe\,{\rm XXV} and Fe\,{\rm XXVI} are found to be relatively common amongst the AGN in this sample. Evidence for highly ionized absorption is found 14/46 (30\%) of objects: 3C 111, 3C 445, MCG--6-30-15, MR 2251--178, Mrk 766, NGC 1365, NGC 3227, NGC 3516, NGC 3783, NGC 4051, NGC 4151, NGC 5548, PDS 456 and PG 1211+143; consistent with the 36\% detected by Tombesi et al. (2010b) in a larger sample of 101 {\sl XMM-Newton} observations searching for ultra-fast outflows. The majority (10/14) of these lines are blue-shifted, suggesting that they originate from an outflowing zone of absorbing gas. When modelled with a high turbulent velocity \textsc{xstar} grid ($v_{\rm turb}=1000\,{\rm km\,s}^{-1}$) we find a mean column density $N_{\rm H}=(1.74\pm0.85)\times10^{23}\,{\rm cm}^{-2}$, mean ionization ${\rm log}(\xi)=3.97\pm0.13$ and a wide range in outflow velocity ranging between $400\,{\rm km\,s}^{-1}<v_{\rm out}<84600\,{\rm km\,s}^{-1}$ with no clear peak in the distribution of $v_{\rm out}$ values (as shown in Figure \ref{fig:FeK_abs_histos}). More details regarding highly ionised and outflowing zones will appear in a forthcoming paper (Gofford et al. 2012, in prep), which performs a more exhaustive statistical search for blue shifted iron K absorption lines.

We must note that these highly ionised zones can influence the strength of any broad residuals in the Fe\,K region, for example, Reeves et al. (2004) found that no broad residuals remained in an {\sl XMM-Newton} observation of NGC 3783 after the application of the required \textsc{xstar} grid to model the highly ionized absorption. The measured strength of any `broad' residual is therefore reduced given an appropriate high $\xi$ absorption zone and hence the detection and modelling of such zones is essential if robust statistics regarding the Fe\,K region are to be formed. Based upon the work by Tombesi et al. (2010a \& 2010b) the actual fraction of AGN with statistically significant outflows may indeed be higher than the 30\% found here due to the presence of UFOs (Ultra Fast Outflows) with absorption lines blue-shifted to $>7$\,keV since we have not systematically searched for these from 7-10\,keV (see Gofford et al. 2012, in prep). The fraction of AGN with highly ionized zones presented here may also in fact be larger. It is possible that the presence of narrow emission lines from ionized species of iron may reduce our ability to detect absorption lines in CCD resolution spectra from the same species of iron with current instruments. 

\begin{figure*}
\rotatebox{-90}{\includegraphics[width=5cm]{Relline_q_histo.ps}}
\rotatebox{-90}{\includegraphics[width=5cm]{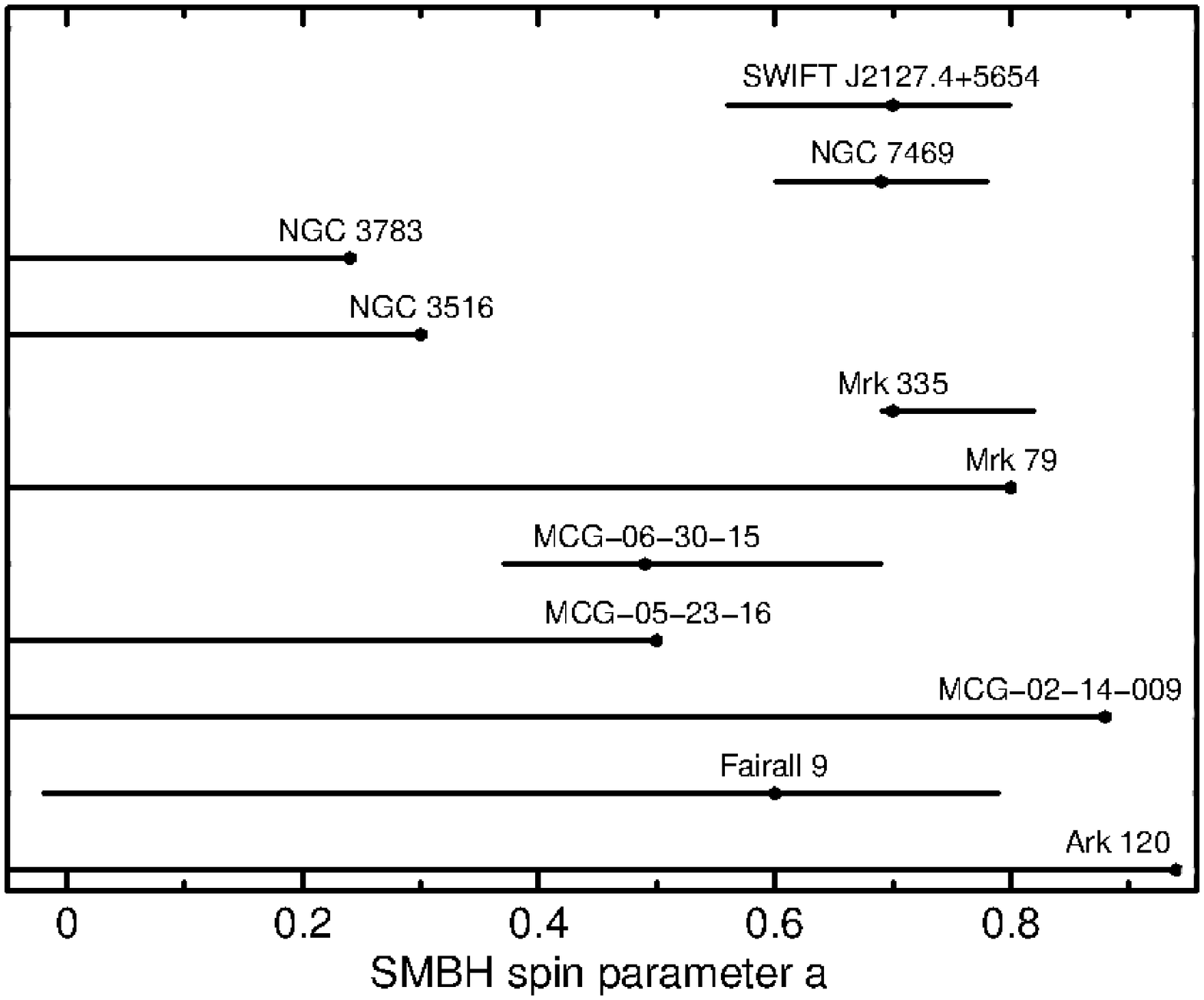}}
\rotatebox{-90}{\includegraphics[width=5cm]{Relline_EW_histo.ps}}
\rotatebox{-90}{\includegraphics[width=5cm]{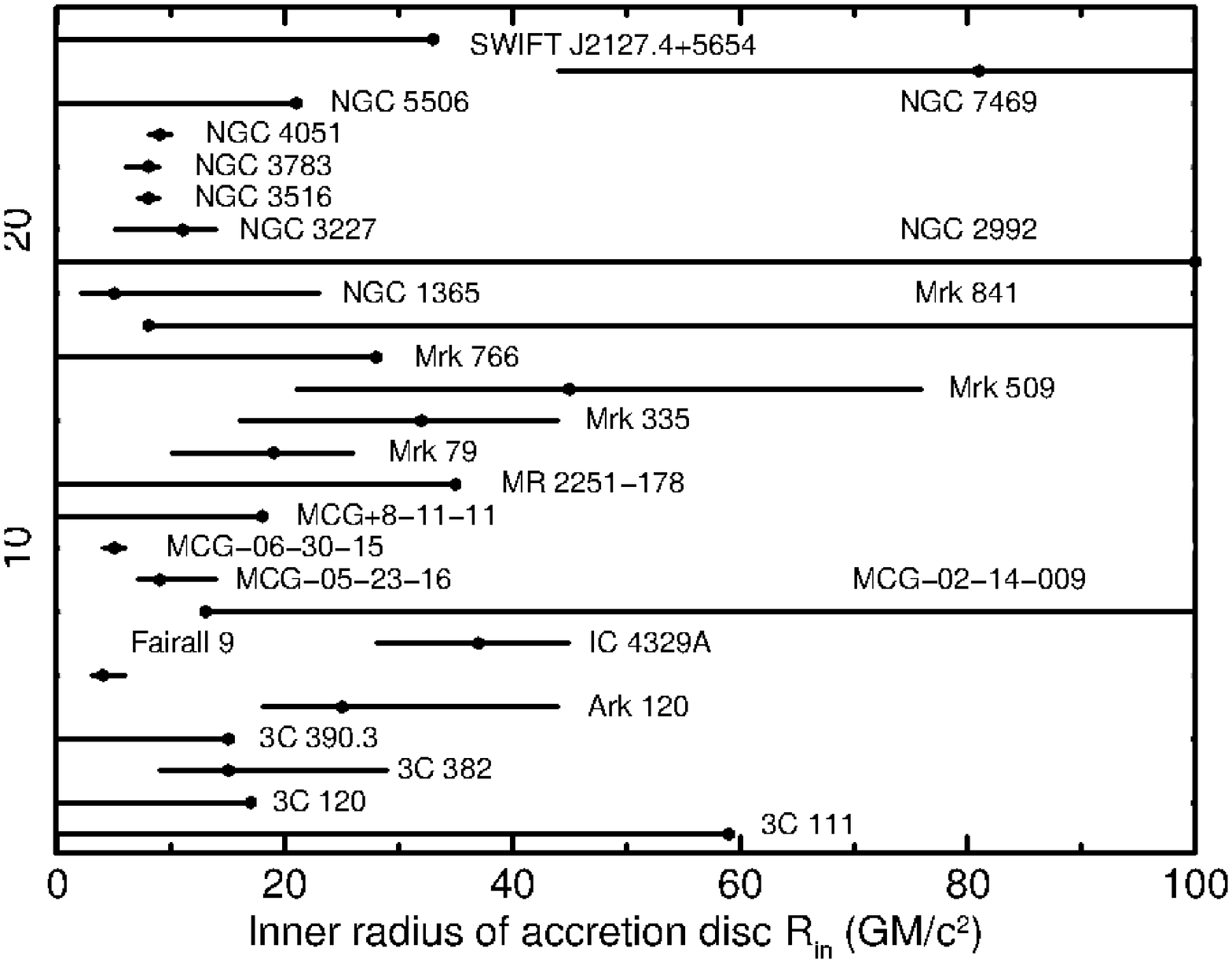}}
\caption{Figures of the distribution of emissivity index, \textsc{relline} equivalent width, SMBH spin and inner radius of emission. Note that the typical inner radius of the disc for the sample centres around tens of $R_{\rm g}$. }
\label{fig:relline_histos}
\end{figure*}

\subsection{Broad emission in the Fe\,K region}
After the formation of a complete baseline model, we introduce the physically motivated relativistic line emission model \textsc{relline} to account for any remaining broad residuals in the Fe\,K region. The criteria for the presence of a broad component is $\triangle\chi^{2}>7.8$ with the introduction of four additional free parameters ($q$, $a$, $i$ and normalisation), this is equivalent to the $90\%$ confidence level. Based upon this requirement, we find that 26/46 (i.e. $57\%$, Table \ref{tab:component_percentages}) of the AGN (by object, not observation) in this sample formally statistically require relativistic emission from the inner regions of the accretions disc. Furthermore, 23/46 ($50\%$, Table \ref{tab:relline}) require broadened line emission at $>99.5\%$ confidence ($\triangle\chi^{2}>15$). This fraction is somewhat lower (although not substantially so) than that obtained by Nandra et al. (2007) who find that 18/26 ($69\%$) of objects in their sample show evidence for broad line emission at $>99\%$ confidence. The slight disparity between the two surveys may be due to the energy range over which the analysis has been conducted: 0.6-100.0\,keV here, versus 2.5-10.0\,keV. This study also takes into account partial covering where required. However, the slight difference may be due to the smaller number statistics in the Nandra et al. (2007) sample. In a flux limited sample, de la Calle P\'{e}rez find 11/31 of objects (36\%) feature relativistic iron lines. 


\subsubsection{Average parameters of the disc}
Taking the parameters obtained with \textsc{relline} for the 26 objects featuring broadened line emission (Table \ref{tab:relline}) we can estimate the typical parameters of the accretion disc (Table \ref{tab:average_params}). The average inclination of the disc is $i=33^{\circ}\pm2^{\circ}$ for 21 objects, consistent with Nandra et al. (2007) who find $i=38^{\circ}\pm6^{\circ}$. The average emissivity index of the disc is measured at a low to moderate $q=2.4\pm0.1$ (for the 20 objects in which it can be constrained) and is consistent with an analysis of 6 `bare' Seyferts by Patrick et al. (2011a) and an {\sl XMM-Newton} survey by de la Calle P\'{e}rez et al. (2010) who find $q=2.4\pm0.4$ and $i=28^{\circ}\pm5^{\circ}$. This value is much lower compared to an often high emissivity ($q>5$) assumed by many light bending models (Miniutti \& Fabian 2004; Miniutti et al. 2009; Brenneman et al. 2011; Gallo et al. 2011; Nardini et al. 2011), suggesting that strong GR effects may not be present in the X-ray spectra of these Seyfert AGN. Figure \ref{fig:relline_histos} shows the distribution of the emissivity indices for the sample, note that there is a relatively small dispersion in $q$ with the majority of values centred around $q\sim2.4$. 

The average strength of the \textsc{relline} component is measured at $EW=96\pm10$\,eV for the total of 26 objects, consistent within errors with Patrick et al. (2011a) and Nandra et al. (2007). The typical equivalent widths of the line component are also largely consistent with Fukazawa et al. (2011) who parameterise all of the 6.4\,keV narrow and broad emission with a single Gaussian in a large {\sl Suzaku} sample of Seyfert 1 and 2 AGN. As can be seen from Figure \ref{fig:relline_histos}, there is a wide range in the measured equivalent widths of the \textsc{relline} component, revealing a possible bi-modal nature with a group centering around $EW\sim70$\,eV and a second stronger though less populated group at $EW\sim150$\,eV. Although there are insufficient statistics in this sample to draw any conclusions from this, indeed the distribution may be more continuous when taking error bars and different confidence levels into account. It is also interesting to note that in the cases where BH spin can be constrained, maximal spin is ruled out in all objects with the spin parameter $a$ preferring to take low to intermediate values, similarly the typical inner radius of emission is at tens of $R_{\rm g}$ with many clustered around $R_{\rm in}\sim20-30\,R_{\rm g}$ (see Figure \ref{fig:relline_histos}).

\begin{figure}
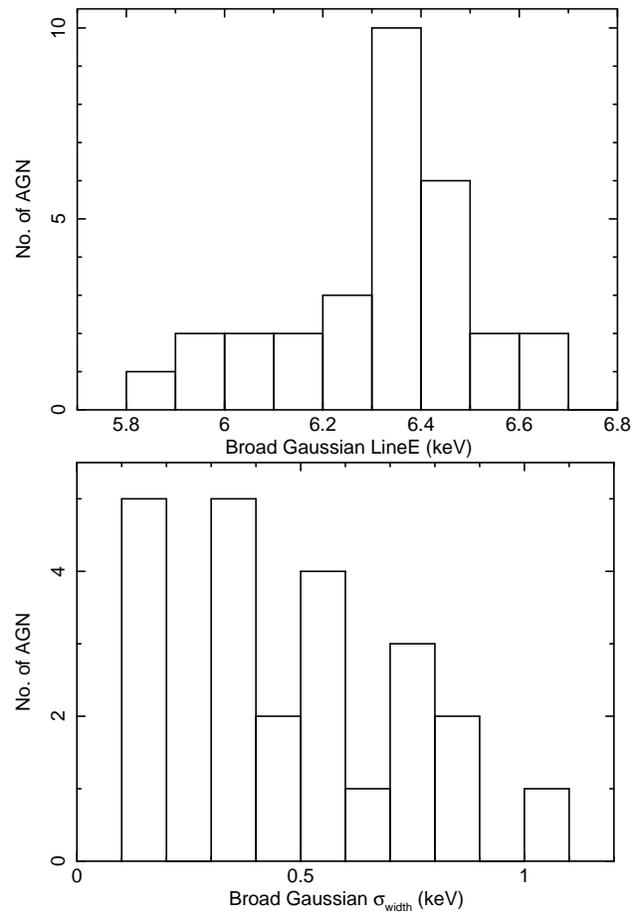

\rotatebox{-90}{\includegraphics[width=6cm]{bga_lineE_histo.ps}}
\rotatebox{-90}{\includegraphics[width=6cm]{bga_sigma_histo.ps}}
\caption{Histograms showing the distribution of line centroid energy and $\sigma_{\rm width}$ when broadened emission in the Fe\,K region is modelled with a Gaussian.}
\label{fig:bga_histos}
\end{figure}

When parameterised with a Gaussian, we find an average line centroid energy of ${\rm LineE}=6.32\pm0.04$\,keV (see Figure \ref{fig:bga_histos}), $\sigma_{\rm width}=0.47\pm0.05$ and equivalent width $EW=97\pm19$\,eV (Table \ref{tab:average_params}), both of which are also consistent with a similar parametrisation with a broad Gaussian by Nandra et al. (2007) who find an average ${\rm LineE}=6.27\pm0.07$\,keV, $EW=91\pm13$\,eV respectively. In an analysis of the 2-10\,keV spectrum of 149 radio-quiet Type 1 Seyfert AGN with {\sl XMM-Newton}, de la Calle P\'{e}rez (2010) estimate a mean broad line equivalent width of $EW=143\pm27$\,eV, which is higher than that obtained here, however, the authors note that this may be an overestimate due to an inability to detect weak lines and a bias towards higher values. 

Note that when obtaining measurements for the emissivity index of the disc and subsequently the spin of the SMBH, the way in which the soft excess is modelled can have a significant effect. For example, if the soft excess is not modelled according to a soft Comptonization origin (e.g. \textsc{compTT}, or to a lesser extent \textsc{diskbb}) and instead an atomic origin is preferred, the disc parameters are drastically altered, see Section 3.1.1 . 

\subsubsection{Shape of the line profile}
The emissivity index obtained here is based primarily upon any remaining residuals in the Fe\,K region i.e. \textsc{relline} is modelling a relativistic line profile which takes into account both red and blue-wings of a line profile rather than simply adding a broad `hump' to the data below 6.4\,keV (as, for example, a Gaussian might). It should be noted that for moderate to high inclinations ($i\gtrsim20^{\circ}$) the blue-wing of the line profile peaks at a higher flux than the red-wing peak, due to a boosting effect from material travelling at relativistic velocities in orbit `close' to the central black hole. This effect is accentuated with lower emissivities since with increasing $q$ (and hence more centrally concentrated emission) a greater proportion of the emitted photons are present in the red-wing of the line profile due to their proximity to the very inner regions and increased gravitational broadening. At particularly high emissivities, the blue-wing diminishes and the majority of the flux is contained within the red-wing of the line profile, forming a single peaked profile and an increasingly strong red-wing below 6.4\,keV (Fabian et al. 1989; Laor 1991; Dauser et al. 2010). Further still, at near maximal spin and high $q$ the entire line profile can produce an asymmetric featureless hump which simply creates additional curvature in the $\sim1-7$\,keV region.

Despite the narrow nature of the blue-wing at low $a$ and $q$, its strength relative to the red-wing (and continuum) is typically insufficient to provide an alternative explanation for narrow emission lines commonly attributed to distant Fe\,{\rm XXV} (at $\sim30^{\circ}$) or Fe\,{\rm XXVI} emission (at $\sim40^{\circ}$). However, there should be some caution to account for the presence of narrow ionized emission lines prior to applying a relativistic line model to ensure that the correct inclination of the disc is measured.

The driving factor behind the estimated emissivity and spin values is therefore primarily the strength and width of the red-wing. Figure \ref{fig:bga_histos} indicates that when modelled with a broad Gaussian, the typical line centroid energy of the broad line peaks at $6.3-6.4$\,keV which also suggests that any broad emission in the Fe\,K region is not strongly gravitationally redshifted. The low average emissivity index ($q=2.4\pm0.1$) of the disc found in this sample of 46 objects is as a result of a full and thorough modelling of the broad-band spectrum i.e. after including \textsc{xstar} grids for the required absorption zones which dominate the spectrum in many objects below 2.5\,keV, we note that the associated spectral curvature these zones add at higher energies (Miller et al. 2008; Zycki et al. 2010) leave little room for a significantly broadened relativistic line profile with high $q$ and maximal spin. The extremely centrally concentrated emission that emissivities of $q\gtrsim4$ suggest are simply no longer required and relativistic emission from the disc (if present) originates from regions more distant to the BH (typically tens of $R_{\rm g}$, see Table \ref{tab:relline} and Figure \ref{fig:relline_histos}).

The strength of the iron line profile has been noted to vary in a number of objects over long timescales and also over the course of a single observation (e.g. Matsumoto et al. 2003; Miyakawa et al. 2009; Risaliti et al. 2009). It is indeed possible that the analysis of the time-averaged spectrum may instead produce an artificial broad line profile which is simply an average line profile which could therefore result in inaccurately derived disc properties. It is likely, however, that the changes in the observed line profile are in fact a consequence of changes in the warm absorber or partial covering properties. For example, over the six {\sl Suzaku} observations of NGC 3227 the $EW$ of the broad iron line is seen to vary greatly when the absorber properties are fixed between each observation (i.e. by hundreds of eV), however, allowing the ionisation of the partial coverer to vary significantly reduces the requirement for a strong broad line, instead resulting in typical $EW\sim50$\,eV. Allowing for such absorber changes on medium to long term timescales to account for spectral variability can therefore drastically alter the measured Fe\,K line profile (Miyakawa et al. 2009; Risaliti et al. 2009).

\begin{table}
\caption{Table of main model components and the percentage of objects in the sample of 46 objects which feature those components based upon $>90\%$ confidence levels.}
\centering
\begin{tabular}{l c}
\hline
Component & Percentage \\
\hline
Soft excess & $67\pm6\%$ \\
Reflection & $85\pm6\%$ \\
Warm absorber & $59\pm5\%$ \\
Partial covering & $35\pm4\%$ \\
Fe\,K outflow & $30\pm4\%$ \\
Fe\,{\rm XXV} emission & $52\pm5\%$ \\
Fe\,{\rm XXVI} emission & $39\pm4\%$ \\
Broad Fe\,K$\alpha$ line & $57\pm5\%$ \\ 
\hline
\end{tabular}
\label{tab:component_percentages}
\end{table}

\subsubsection{SMBH spin}
With the use of broad-band data from observatories such as {\sl Suzaku} we can make tentative steps towards estimating the spin of the central supermassive black hole. The primary effect of the spin parameter $a$ upon the line profile is to set the degree to which the red-wing extends to the soft X-ray energies, i.e. a maximal Kerr black hole would have the smallest inner radius of the accretion disc, thereby subjecting emitted photons to a stronger red-shift and sharply extending the red-wing to well into the soft X-ray regime. A Schwarzschild black hole ($a=0$), however, truncates the red-wing of the line profile at $\sim4$\,keV due to a minimum inner radius of 6\,$R_{\rm g}$ and less pronounced red-shifting of emitted photons (indeed a maximally retrograde BH pushes the inner radius further out to $R_{\rm in}\sim9\,R_{\rm g}$). It is clear to see, therefore, that if the spectral curvature associated with various warm absorbing zones is not taken into account, a relativistic line emission model such as \textsc{relline} or \textsc{kerrdisk} may be forced to higher near maximal spin values to `take up the slack' of an improperly modelled continuum. 

Many of the prime candidates for measuring SMBH spin with {\sl Suzaku} have already been presented in previous papers, firstly the `bare' Seyferts (Patrick et al. 2011a) and a collection of deep {\sl Suzaku} observations (Patrick et al. 2011b), hence few new BH spin estimates are presented here. In Fairall 9, NGC 3516 and NGC 3783, however, we conduct for the first time an analysis of each of the multiple {\sl Suzaku} observations of each object, finding a baseline+\textsc{relline} model which can accurately describe the objects in different flux states with only subtle variations to the model over time e.g. changes in continuum flux. As stated in Section 3.3, all parameters of the \textsc{relline} component are tied between observations except the normalisation i.e. the flux of the line is allowed to vary appropriately with the flux of the other primary components of the baseline model. Finding a model which can adapt over time to fit each observation while retaining the same basic properties (e.g. emissivity, inclination, spin) arguably increases the robustness of the results. Encouragingly, the fits to Fairall 9 and NGC 3783 are consistent with the analysis of the individual observations by Schmoll et al. (2009), Emmanoulopoulos et al. (2011) and Patrick et al. (2011a, 2011b), finding $a=0.60^{+0.19}_{-0.62}$ in Fairall 9 ruling out maximal spin at $>95\%$ confidence with an emissivity index of $q=2.9^{+0.5}_{-0.4}$. 

In addition to the spin estimates from Patrick et al. (2011a, 2011b), the results of this paper suggest further low-to-moderate SMBH spin values on MCG--05-23-16 ($a<0.50$), Mrk 79 ($a<0.80$) and NGC 3516 ($a<0.30$). Out of the total of 46 objects we find that spin constraints can be placed on 11/46 AGN, while 5 of these allow us to place upper and lower bounds upon $a$ and hence constrain the spin to some degree. Albeit with this small number of SMBH spin estimates, we can take tentative steps towards starting to develop a basic picture of the spin distribution of these Seyfert 1 AGN (see Figure \ref{fig:relline_histos}). A maximal prograde SMBH is ruled out in all of these objects at a minimum of $90\%$ confidence and $>99.5\%$ confidence in 5/11 (MCG--6-30-15, NGC 3516, NGC 3783, NGC 7469 and SWIFT J2127.4+5654; Figure \ref{fig:relline_histos}). The dual reflector fits in Table \ref{tab:dual} also appear to suggest that 4 AGN in this sample may have retrograde SMBHs (i.e. $a<0$; Mrk 79, NGC 3227, NGC 3516 and NGC 3783), however, a more likely scenario is the possibility of a prograde rotating BH (or even Schwarzschild) with an accretion disc truncated short of the ISCO.

\subsubsection{NGC 3783 - high or low spin?}
In an analysis of the long 210\,ks 2009 {\sl Suzaku} observation of NGC 3783, 
Patrick et al. (2011b) (hereafter P2011b) 
concluded that the data appeared to rule out a 
maximal black hole in NGC 3783 (constraining spin to $a<0.31$ in a dual 
reflector model), while the fit 
could be achieved with approximate Solar abundances for the reflector. 
However, Brenneman et al. (2011) (hereafter Br2011) 
came to the opposite conclusion, appearing 
to require near maximal black hole spin from fits to the iron line data in 
NGC 3783 (with $a>0.98$ at $>90\%$ confidence) and obtaining a high iron 
abundance of $Z_{\rm Fe}=3.7^{+0.9}_{-0.9}$ and a steep inner emissivity law 
for the innermost disc. Subsequently Reynolds et al. 
(2012b) (hereafter Rey2012) also appeared to confirm the high spin, 
high abundance scenario in 
a re-analysis of the 2009 dataset, suggesting that the iron abundance of the 
reflector and the black hole spin may be degenerate upon each other and that a 
statistically preferred fit can be obtained with higher abundances and 
higher spin.

In this section we attempt to discuss the differences between these works, by 
considering the long 2009 {\sl Suzaku} observation of 
NGC 3783. Firstly the model of Br2011 was reconstructed, 
in order to understand the difference in the spectral modelling. 
The main difference between the Br2011 and P2011b models is the construction 
of the warm absorber; in Br2011 all 3 zones of the warm absorber only 
partially cover the AGN, where a fraction of $\sim17$\% of the direct 
continuum is unabsorbed by the warm absorber, or 
alternatively is scattered back into the line of sight.
On the other hand in P2011b and in this paper, the warm absorber fully covers 
the X-ray continuum emission. 
Indeed we note that previous analyses have not needed to invoke partial 
covering in order to model the warm absorber in NGC 3783 (Reeves et al. 2004; 
Yaqoob et al. 2005), including high resolution X-ray spectroscopy 
from {\sl XMM-Newton RGS} (Blustin et al. 2002) and during a 900\,ks 
{\sl Chandra HETG} observation (Kaspi et al. 2002).  

The other main difference in the model construction 
is that the \textsc{pexmon} neutral 
reflection model (Nandra et al. 2007) is used for the distant (narrow) 
reflector in Br2011. In P2011b and in this paper, 
the \textsc{reflionx} ionised reflection model (Ross \& Fabian 2005) 
is adopted, 
allowing the ionisation state to reach a low 
value of $\xi=1$\,erg\,cm\,s$^{-1}$ appropriate for low ionisation iron. 
For the disc (i.e. blurred) reflection, both analyses use the 
\textsc{reflionx} table, 
convolved with a relativistic blurring function such as 
\textsc{kerrconv} (Brenneman \& Reynolds 2006) 
or \textsc{relconv} (Dauser et al. 2010).  For simplicity in this section 
we use the 
\textsc{kerrconv} model, with spin allowed to vary between $a=0-0.998$ 
for a prograde black hole. The emissivity index is modelled 
as a broken powerlaw function (where $q_1$ 
is the inner emissivity and $q_2$ the outer emissivity and usually 
$q_{1}>q{2}$), breaking at a disk radius of $r_{\rm b}$ in units of 
$R_{\rm g}$. The inner disk radius is set equal to the ISCO, while the 
outer radius is set to $400R_{\rm g}$. We adopt here the 
Solar abundances of Anders \& Grevesse (1989) for the Galactic column, 
noting that there is little 
difference to the blurred reflector 
model parameters if the Wilms et al. (2000) ISM abundances are used instead.

In order to test the Br2011 model, we use the identical 
energy ranges for the XIS\,FI 
spectrum and HXD/PIN adopted by Br2011, from 0.7--45\,keV, ignoring the 
1.5--2.5\,keV band in the XIS. We also used the same warm absorber model 
and tables as per the Br2011 and Rey2012 papers.
The Br2011 model has a steep index for the inner emissivity law, a 
high black hole spin consistent with maximal and a high iron abundance of 
the inner reflector of $Z_{\rm Fe, inner}\sim3$, while the 
iron abundance of the distant reflector is initially fixed 
equal to Solar, i.e. $Z_{\rm Fe, outer}=1$.  
This model as detailed in Br2011 initially gives a poor fit 
($\chi^{2}_{\nu}=1501/1234$), however subsequently 
refitting the model parameters then gives an excellent fit to the data, 
with $\chi^{2}_{\nu}=1329/1234$. Nonetheless the model 
parameters are in good agreement with those obtained in Br2011, 
where we obtain a formal 90\% lower limit of $a>0.82$ to the black hole spin 
(for 1 interesting parameter), a disc inclination of $i=22\pm4$, 
while the emissivity indices have 
values of $q_1=4.4\pm1.2$ and $q_2=2.8\pm0.2$, with a break radius of 
$\sim6 R_{\rm g}$. 

However upon taking the same model and fixing the spin to $a=0$, 
then the fit statistic obtained is only slightly worse, where 
$\chi^{2}_{\nu}=1340/1237$. In this case only a single emissivity is 
required, where $q_{1}=q_{2}=3.0\pm0.5$, while the disc abundance is 
consistent 
with Solar. Given that the high black hole spin model only shows a marginal 
improvement in fit statistic ($\Delta\chi^2=11$ for 3 fewer 
degrees of freedom), then the claim of high black hole spin cannot 
be confirmed at a high confidence level, compared to the case where $a\sim0$. 
Furthermore if the Br2011 model is altered such that all 3 soft X-ray 
warm absorber zones fully cover the AGN, then this leads to 
a lower (and less constrained) value for the black hole spin, 
of $a=0.35^{+0.59}_{-0.10}$ and is then formally 
consistent with the upper-limit of $a<0.31$ from the dual reflector 
model in P2011b. 

The lower spin value may be due to the fact 
that the fully covering absorber adds greater spectral curvature to the 
model due to bound-free absorption, compared to the partial coverer. 
In the latter case, the absorption is diluted by an unabsorbed 
powerlaw, reducing the amount of spectral curvature, while the 
breadth of the iron line profile can increase to compensate. 
Finally we note that 
replacing the \textsc{pexmon} model with a 
\textsc{reflionx} table for the distant reflector 
made little difference to any of the blurred reflection 
parameters and this appears unaffected by the parametrisation of 
the distant reflection component.

We also attempted to recreate the maximal spin scenario subsequently 
presented by Rey2012, but applied to the model as constructed in P2011b and 
in this paper. Thus in this case we
assumed that all 3 zones of the warm 
absorber in NGC 3783 fully cover the AGN as per P2011b and 
use the same warm absorber grids as in this paper, although 
the choice of a particular absorption model grid appears to not effect the 
reflection models.  
For ease of comparison we also adopt the energy ranges 
used by Br2011 as above.   
We focus in particular on the iron abundances, 
which Rey2012 suggest is critical for determining the blurred 
reflection parameters. The values for the disc reflector parameters 
from Table 1 
in Rey2012 are used, namely the iron abundances, spin, emissivities and 
break radii, all with the 
inner and outer radii set to the ISCO and $400R_{\rm g}$ respectively. 
In summary we test three scenarios for the 2009 data: 
(i) $Z_{\rm Fe, inner}=\,Z_{\rm Fe, outer}=3.3$ with near maximal spin; 
(ii) $Z_{\rm Fe, inner}=\,Z_{\rm Fe, outer}=1$ with $a=0$;
(iii) $Z_{\rm Fe, inner}=4.2$, $Z_{\rm Fe, outer}=1$ with near maximal spin. 
These correspond closely to the models A, B and C respectively, 
as presented in Table\,1 of Rey2012. The \textsc{kerrconv} model is used 
to blur the disc reflection spectra, while an unblurred low ionisation 
\textsc{reflionx} grid is maintained to model the distant reflector. 
Aside from the blurred reflection parameters as per Rey2012, 
the fit parameters are allowed to 
vary.

All three scenarios give a good fit to the data, with reduced chi-squared 
values of $\chi^{2}_{\nu}=1300/1227, 1280/1227, 1284/1227$ for models (i), 
(ii) and (iii) respectively. Thus all the spin and abundance scenarios 
are statistically acceptable, while the $a=0$ and Solar abundance case 
formally gives the better fit. Allowing the disc parameters to vary for 
the Solar abundance case then gives a constraint on the spin of $a<0.45$ 
at 90\% confidence, while the disk emissivity is $q_{1}=q_{2}=3.0\pm0.5$ 
and no break radius is required. Indeed the same Solar abundance model when 
fitted over the energy ranges used in this paper also gives identical 
results and an equally good fit, with $\chi^{2}_{\nu}=1414/1374$, see 
Table~\ref{tab:dual}. 

\begin{figure}
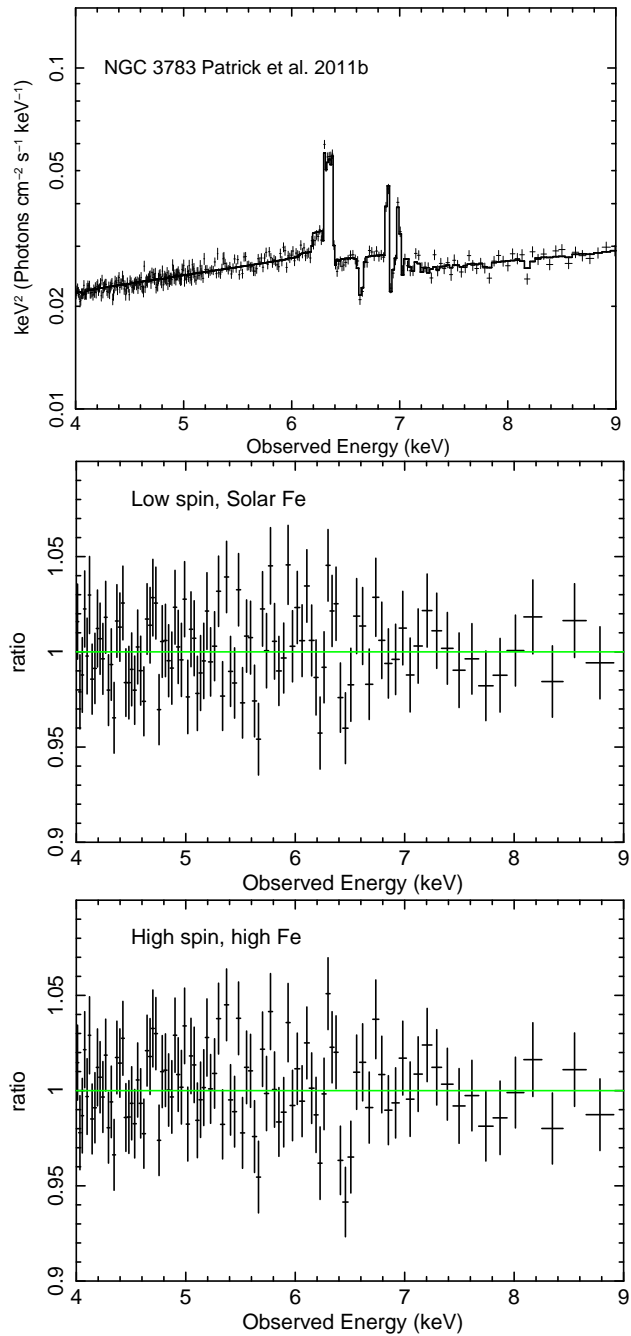

\rotatebox{-90}{\includegraphics[width=6cm]{Patrick_mydata_FeK_eeuf.ps}}
\rotatebox{-90}{\includegraphics[width=5.8cm]{NGC3783_lowspin.ps}}
\rotatebox{-90}{\includegraphics[width=5.8cm]{NGC3783_highspin.ps}}
\caption{{\sl Upper panel} $\nu\,F\nu$ $4-9$\,keV plot of the 
Patrick et al. (2011b) best-fit model (with $a=0$) 
to the 2009 NGC 3783 {\sl Suzaku} 
data, note that the Fe\,K region is well modelled in the fit as is 
the broad red-wing below 
6.4\,keV. {\sl Middle panel} Residuals to the best fit dual reflector 
model, as shown in the upper panel with $a=0$. 
{\sl Lower panel} Residuals from the model with
inner disc parameters as obtained by Reynolds 
et al. (2012b), i.e. with maximal spin, high emissivity and $Z_{\rm Fe}>3$. 
Both of the residual plots in the lower two panels show no noticeable 
difference 
in the iron K region, thus it appears difficult to discriminate between the 
two scenarios.}
\label{fig:NGC3783_2009_residuals}
\end{figure}

We also compare the fit focusing in on the iron 
K band in 
Figure~\ref{fig:NGC3783_2009_residuals}, where the top panel shows the 
best fit dual reflector model presented here (with $a=0$) 
and the lower two panels show the data/model ratios for the scenarios 
(i) and (ii) above, with high spin and low spin respectively. 
Clearly there is little difference in the residuals 
between either the high spin scenario with super Solar abundances and 
the low spin scenario with Solar abundance. The only marginal 
difference is the narrow core of the Fe K$\alpha$ line at 6.4 keV is slightly 
over-predicted in the high spin, high Fe abundance scenario.
This is in contrast to the 
plots shown in Figure 5 of Rey2012, where the $a=0$ scenario strongly 
over-predicts 
the red-wing of the line between 5-6 keV. This could be due to the fact that 
the blurred reflector has a higher normalisation in the high spin scenario, 
which when applied directly to the $a=0$ case could appear to over-predict the 
red-wing, unless the normalisation of the reflector is refitted 
accordingly.

Finally we note that if the \textsc{relline} model is used instead of 
a blurred reflection component, then a lower spin value is usually preferred. 
For example, the upper-limit on spin to NGC 3783 with this 
model is $a<0.24$ (also see Table 4). Indeed P2011b also noticed some 
tendency (although not statistically significant) for the line to favour 
a retrograde black hole spin. The lower spin value may be due to the fact 
that the model only fits the broad iron line and not the reflected continuum, 
where the latter can often be blurred to such an extent that it is hard to 
distinguish from the direct continuum.

Thus the overall conclusion would appear to be that it is very difficult to 
definitively determine the black hole spin in NGC 3783 with the present 
data, given the 
complexities of the models involved and the deep warm absorber present 
in this AGN. For instance as discussed above, 
the construction of overall the model and the warm absorber in particular 
can have an effect. In addition the spin value is indeed degenerate 
upon the iron abundance, as discussed in Rey2012 and thus the high spin 
cases always require 
high centrally concentrated iron abundances. Nonetheless a statistically 
good fit is obtained with a simple Solar abundance reflector and no black hole 
spin. The evidence for a more complex picture with high abundances, 
complex emissivity profiles and high spin, would appear not to formally 
be required at high confidence, although neither can the high spin
case be ruled out at present.

\subsubsection{Truncation of the disc}
As can be seen from Table \ref{tab:relline} and Figure \ref{fig:relline_histos}, the majority of AGN in this sample do not feature strong emission from the very inner regions of the accretion disc, both the emissivity index and spin of the central BH take typically low values. This suggests that the dominant regions producing the typical broad iron line profile are regions close to but more distant from the central BH than has been estimated in many analyses (Reynolds et al. 2005; Brenneman et al. 2006, 2011; Miniutti et al. 2007; Reynolds et al. 2012b), originating at typically tens of $R_{\rm g}$ and hence with a significant amount of flux present in the blue-wing of the line profile as opposed to the red-wing. The \textsc{relline} fits in Table \ref{tab:relline} are under the assumption that the inner radius of the accretion disc extends down the inner-most stable circular orbit (i.e. $R_{\rm ISCO}$), if the majority of the line profile originates from the outer regions of the disc, it is plausible that in some cases the accretion disc may be truncated at a few to tens of $R_{\rm g}$. For example, Lobban et al. (2010) suggest a truncated accretion disc in NGC 7213, stating a lack of a significant broad component  in the Fe\,K region and weak reflection. If the inner edge of the accretion disc does not extend down to the innermost stable circular orbit, it would make the determination of SMBH spin through spectral fitting very difficult. Essentially at $R_{\rm in}>9\,R_{\rm g}$ the spin could take any value $-0.998<a<0.998$ with the disc truncated at any $R_{\rm in}$ with little difference upon the relativistic line profile. 

\subsubsection{Alternative mechanisms}
Recent work by Tatum et al. (2012) suggests that an alternative mechanism to reflection from the inner regions of the accretion disc is emission from a Compton thick disc wind based upon a disc wind model by Sim et al. (2010). The authors take the cleanest sources possible i.e. the `bare' Seyfert sample from Patrick et al. (2011a) and assuming a relatively face-on viewing angle whereby only the scattered light from the wind is observed rather than any absorption features with a more edge-on viewing angle. The line profiles are therefore subject to velocity broadening due to Doppler velocity shear across the wind and Compton down-scattering of the Fe\,K$\alpha$ flux, these can combine to imitate a broad line profile from reflection off the inner regions of the accretion disc. Good fits are made to the broadband spectra for this model without the requirement for any further broadened emission from reflection off the inner regions of the disc. In this scenario, the full Fe\,K line profile is reproduced with Solar iron abundances at typically tens to hundreds of $R_{\rm g}$, however, in some objects an additional cold neutral reflector and neutral Fe\,K$\alpha$ line component is still required in order to model the Fe\,K edge and hard X-ray excess. 

While the disc wind model alone can start replicate many of the features in the broad-band spectrum, it is likely that the true scenario is a combination of absorption, reflection and scattered components off both the disc and a wind, a scenario which may not be resolved with current X-ray observatories without calorimeter resolution spectra in the Fe\,K band. It is, however, interesting to note that the tentative bi-modality of the broad line equivalent widths in this sample (Figure \ref{fig:relline_histos}) may represent such a scenario whereby a fraction of the weaker observed broad lines are as a result of disc wind geometries. Meanwhile objects which display stronger broad lines ($EW\gtrsim150$\,eV) may be the only objects in which we can see the inner tens of $R_{\rm g}$ of the disc. It must be stressed, however, that the apparent (albeit weak) bi-modal nature of the equivalent widths in this sample may simply due to insufficient statistics. 

\subsection{Distant reflection}
The use of the HXD instrument on {\sl Suzaku} and BAT onboard {\sl Swift} allows us to examine the hard X-ray spectra of the objects in this sample and to appropriately parameterise the distant reflection component. This may arise, for example, from the reflection of continuum X-rays off the cold ($T<10^{6}$\,K) outer regions of the accretion disc or via a parsec scale torus. 

The majority of AGN in this sample show evidence for a reflection component, only 7/46 (see Table \ref{tab:component_percentages}) do not require the addition of the unblurred \textsc{reflionx} model (3C 111, IRAS 13224--3809, MR 2251--178, Mrk 79, NGC 7314 and PG 1211+143). Neither are these objects best described with a high column density partially covering absorbing zone with $N_{\rm H}\sim10^{24}\,{\rm cm}^{-2}$ which can replicate a hard X-ray excess to some extent i.e. there is no apparent hard excess in these objects. Table \ref{tab:fluxes} also indicates that the large majority of AGN in this sample do feature excess emission in the 15-50\,keV region over the extrapolated 2-10\,keV intrinsic powerlaw. A number of the objects in the sample feature particularly strong reflection with 15-50\,keV reflector flux $F_{\rm reflector}\gtrsim3\times10^{-11}\,{\rm erg}\,{\rm cm}^{-2}\,{\rm s}^{-1}$, namely IC 4329A, MCG+8-11-11, NGC 3227, NGC 3783, NGC 4151 and NGC 5506. In addition to this, a large number of the AGN in this sample feature particularly hard X-ray spectra i.e. those with a high hardness ratio $F_{\rm 15-50\,keV}/F_{\rm 2-10\,keV}\gtrsim3$: 3C 445, Mrk 279, NGC 1365, NGC 3227 (Obs 2, 4 \& 6), NGC 3516 (Obs 1) and NGC 4151 (see Table \ref{tab:fluxes}). 

\begin{equation}
\centering
R_{15-50}=\frac{Reflected\,flux}{Full\,model\,flux-Reflected\,flux}
\label{eq:reflection_fraction}
\end{equation}

\begin{figure}
\rotatebox{-90}{\includegraphics[width=6cm]{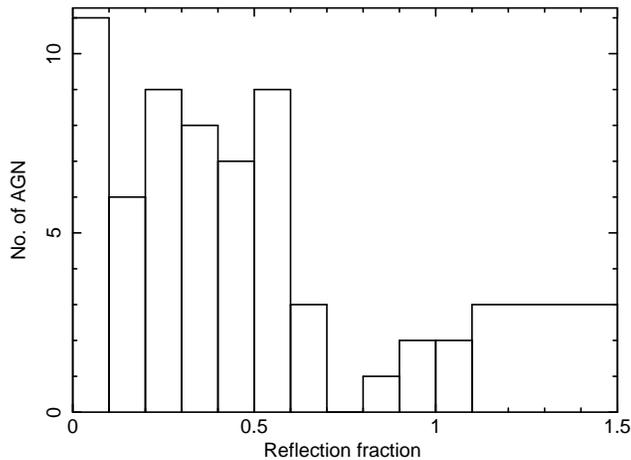}}
\caption{The distribution of reflection fractions for each observation in the sample with values obtained from Table \ref{tab:fluxes} whereby the fraction is the ratio of the 15-50\,keV reflected flux to the 15-50\,keV continuum flux (i.e. full model minus reflected flux). Note that the majority of AGN feature reflection fractions between 0 and 0.7. The final bin from 1.1 upwards contains only three observations of AGN with $R_{15-50}=1.6,\,3.5$ and $7.2$. $R_{15-50}=0.8$ is equivalent to a $R=1$ \textsc{pexrav} versus a simple unabsorbed \textsc{powerlaw}.}
\label{fig:reflection_fraction}
\end{figure}

Taken from values in Table \ref{tab:fluxes} we also calculate a reflection fraction over the 15-50\,keV range using Equation \ref{eq:reflection_fraction}, note that this $R_{15-50}$ is not the same as the reflection fraction $R$ included in models such as \textsc{pexrav} (Magdziarz \& Zdziarski 1995). For ease of comparison, we calculate the equivalent $R_{15-50}$ value for an $R=1$ \textsc{pexrav} versus a simple unabsorbed $\Gamma=2$ \textsc{powerlaw}, this yields $R_{15-50}\sim0.8$. The reflection fraction is particularly high in 8 observations in the sample and four objects (MCG--02-14-009, NGC 3227, NGC 3783 and NGC 4151). In these objects the reflector flux in the 15-50\,keV band contributes approximately half of the total 15-50\,keV flux, however, this appears to bear no correlation with properties such as covering fraction although each of these four objects do feature relatively high hardness ratios (ratio of 15-50\,keV to 2-10\,keV flux). The majority of AGN, however, feature reflection fractions fitting a smooth distribution between $R_{15-50}=0$ and $R_{15-50}=0.7$ (see Figure \ref{fig:reflection_fraction}). The measured strength of the reflection component can be somewhat degenerate with the method in which the partial coverer is modelled. However, the reflection continuum below 10\,keV can help to ensure that the correct amount of hard X-ray flux is produced by \textsc{reflionx} by measuring the strength of the narrow 6.4\,keV core, for an appropriate iron abundance, the self-consistent reflection strength is then modelled above 10\,keV. Any remaining residuals $>10$\,keV must then be due to a high column density partial coverer. That is, the strength of the narrow 6.4\,keV core can give a reasonable estimate of the contribution of the reflector above 10\,keV. 

A significant number of AGN in this sample are fit very well assuming a simple Solar iron abundance i.e. $Z_{\rm Fe}=1.0$ when allowed to vary as a free parameter within \textsc{reflionx}. Although the narrow Fe\,K$\alpha$ core is sufficiently strong in at least 9 objects that further neutral iron emission is required from material distant to the central SMBH, e.g. the broad line region in addition to reflection off the torus. Four objects strongly require sub-Solar iron abundance (IC 4329A, NGC 3227, NGC 5506 and SWIFT J2127.4+5654) while 4/39 statistically prefer a slight super-Solar iron abundance (typically $Z_{\rm Fe}\lesssim3$). This suggests that the hypothesis of Solar abundances throughout the disc and central engines of these AGN is a reasonable assumption in contrast to super-Solar $Z_{\rm Fe}$ based upon inner disc reflection (e.g. Brenneman et al. 2011; Fabian et al. 2012). The analysis of the 46 objects in this sample indicates that the reflection in an overwhelming majority of the Seyfert 1 type AGN in the {\sl Suzaku} archive is well described by reflection of a cold, distant, neutral material with Solar abundances throughout, also see Nandra \& Pounds (1994) and Rivers, Markowitz \& Rothschild (2011a) for similar findings regarding the recurrent presence of the Compton hump. We note that in higher 2-10\,keV luminosity AGN we find that the equivalent width of the narrow Fe\,K$\alpha$ emission line is low in comparison to other lower luminosity AGN, although the sample in this paper lacks the sufficient statistics to support the evidence for the X-ray Baldwin effect as suggested by Jiang et al. (2006) and Fukazawa et al. (2011).

\begin{table}
\caption{Table of average parameters of the sample, values quoted for a broad Gaussian and \textsc{relline} are for those objects in which those components were statistically significant at $>90\%$ confidence level.}
\centering
\begin{tabular}{l c}
\hline
Parameter & Average value \\
\hline
\multicolumn{2}{c}{\textsc{powerlaw}} \\
$\Gamma$ & $1.82\pm0.03$ \\
\multicolumn{2}{c}{Broad Gaussian} \\
Energy (keV) & $6.32\pm0.04$ \\
EW (eV) & $97\pm19$ \\
$\sigma_{\rm width}$ (keV) & $0.47\pm0.05$ \\
\multicolumn{2}{c}{\textsc{relline}} \\
EW (eV) & $96\pm10$ \\
$q$ & $2.4\pm0.1$ \\
$R_{\rm in}$ ($R_{\rm g}$) & $21\pm6$ \\
Inclination ($^{\circ}$) & $33\pm2$ \\
\hline
\end{tabular}
\label{tab:average_params}
\end{table}

\subsection{Warm absorbers and partial covering}
The way in which absorbing zones are modelled in X-ray spectra can have a significant affect upon the Fe\,K region and the estimation of properties relating to the central SMBH and the surrounding accretion disc. Within this {\sl Suzaku} sample, a large proportion of AGN feature complex absorbers, requiring one or more zones of an \textsc{xstar} grid. Only 12/46 ($26\%$) of objects in this sample can be considered `bare' for the purposes of this paper - featuring no statistically significant additional absorption zones in addition to the neutral Galactic column, these are 3C 390.3, Ark 120, Fairall 9, MCG--02-14-009, Mrk 110, Mrk 335, Mrk 359, NGC 3147, NGC 7213, NGC 7469 and RBS 1124. A few objects in the sample (6/46, $9\%$) feature an additional neutral zone of absorbing gas at the redshift of the object, while the remaining 27/46 ($59\%$) all require the application of complex warm (i.e. ionized) zones of gas to model the broad-band spectrum. 

\begin{figure}
\rotatebox{-90}{\includegraphics[width=6cm]{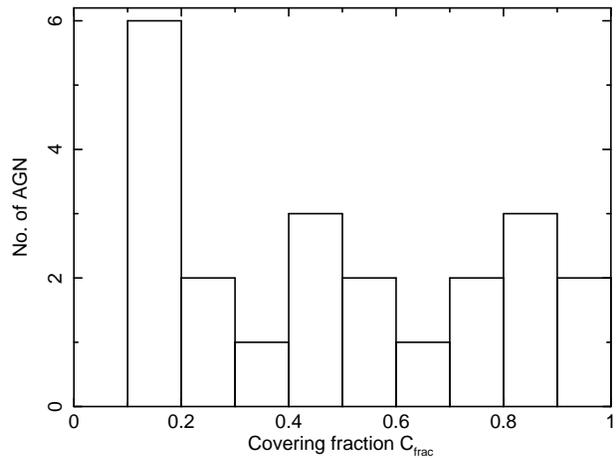}}
\caption{The distribution of partial covering covering fractions. Note that the full range is covered with a relatively even distribution.}
\label{fig:cfrac_histo}
\end{figure}

Just over a third of the sample (16 AGN) also statistically require a partially covering geometry (i.e. $35\%$ of the total sample, Table \ref{tab:component_percentages}). In this scenario one of the absorbing zones of gas (typically high column $N_{\rm H}\gtrsim5\times10^{23}\,{\rm cm}^{-2}$) partially obscures the nucleus in addition to one or more lower column density fully-covering absorbing zone. The majority of the partial covering models in this sample have a sufficiently high column density such that their predominant effect is to supplement and increase the hard X-ray flux. The distribution of covering fractions for the partial covering fits here cover the full range from low to high covering fractions (Figure \ref{fig:cfrac_histo}).

We note that only single layer partial covering geometries are used (where formally required) in this analysis since multiple layers can be somewhat ambiguous and care must be taken to ensure that layers are not added on an ad-hoc basis until all features are removed from the spectrum. None of the objects in this sample require more than a single layer of partial covering in order to obtain an acceptable fit to the broad-band data.

Partial covering scenarios in some previous analyses have been used to model spectral curvature in the Fe\,K region which may otherwise be attributed to relativistic line emission e.g. MCG--6-30-15 by Miller et al. (2009). However, careful examination of the effects of high column partial coverers such as those used here suggests that the measured strength of broad line emission in the Fe\,K region is simply reduced (i.e. lower $EW$) while the estimated accretion disc parameters and BH spin estimates remain relatively unchanged (for example MCG--06-30-15 in Patrick et al. 2011b, Section 3.3.1). During the analysis of this sample we have also noted that the application of high column density partial covering scenarios in fact has some bearing upon the iron abundance of the reflector derived from reflection models such as \textsc{reflionx} (or indeed \textsc{pexmon}). Since the partial coverer produces additional flux at hard X-ray energies, it is possible that the reflection continuum is subsequently under-predicted and the iron abundance is increased to compensate for the lack of narrow Fe\,K$\alpha$ core flux. However, the majority of the objects here in which we require additional narrow Fe\,K$\alpha$ emission through the use of an additional narrow Gaussian do not feature partial covering and therefore it seems likely that the additional narrow line flux in these particular objects originates from distant material, as opposed to being a consequence of improper modelling of the spectrum above 10\,keV.


\subsection{Inner reflection versus partial covering}
The baseline model constructed in the main models of this paper makes use of a high column partial coverer to account for any additional hard X-ray flux which may be present in the X-ray spectra of these AGN. An alternate scenario is to use a blurred reflection component from the inner regions of the disc in addition to an unblurred reflector from the outer regions of the disc can provide a means of supplementing the total hard X-ray flux at $>10$\,keV. As noted in Section 3.4, this scenario can provide a good description of the broad-band spectrum for MCG--06-30-15 and NGC 5506 whilst retaining relatively typical accretion disc parameters i.e. low to moderate emissivity index and inclination, similar to those obtained from an analysis of the broad Fe\,K$\alpha$ line profile (see Section 3.3 and Table \ref{tab:relline}). 

The fit to the remaining objects (Mrk 766, NGC 1365, NGC 3516, NGC 3783 and NGC 4051) is worse without  a partial covering geometry. This may be due to the inability of the inner reflector geometry to successfully describe the long term variability (e.g. variability in spectral curvature) of these AGN, all of which are noted to have varied between each of the observations analysed here. In each of these AGN the disc parameters are either poorly constrained or are forced to extreme values, for example, in order to describe the broad-band spectrum and long term variability of NGC 4051 the emissivity of the disc is increased to $q=6.1^{+0.3}_{-0.1}$ and spin to near maximal $a>0.99$, while producing an overall worse fit. This suggests that emission is required to be very centrally concentrated with the inner reflector proving the dominant component in the broad-band model. For example, the inner reflector for 2005 observation of NGC 4051 has 15-50\,keV flux $\sim8.8\times$ higher than the outer reflector ($F_{\rm 2005\,inner}=(1.49^{+0.02}_{-0.03})\times10^{-11}$erg\,{\rm cm}$^{-2}$\,s$^{-1}$ versus $F_{\rm 2005\,outer}=(0.17^{+0.01}_{-0.02})\times10^{-11}$erg\,{\rm cm}$^{-2}$\,s$^{-1}$). Comparing this to the alternative (i.e. with partial covering explaining the long term spectral variability, Table \ref{tab:dual}), the inner reflector 15-50\,keV flux is significantly reduced when partial covering is invoked: $F_{\rm 2005\,inner}=(0.29^{+0.03}_{-0.03})\times10^{-11}$erg\,{\rm cm}$^{-2}$\,s$^{-1}$ compared to $F_{\rm 2005\,outer}=(0.37^{+0.08}_{-0.05})\times10^{-11}$erg\,{\rm cm}$^{-2}$\,s$^{-1}$. 

We find here that in the majority of cases, the dual reflector scenario does not provide a reasonable alternative to a partial covering geometry particularly in objects which feature significant long term spectral variability (see Table \ref{tab:dual}). NGC 1365 is very poorly fit with $\chi^{2}_{\nu}\sim2.2$ above 2\,keV and $\chi^{2}_{\nu}\sim5.1$ over the 0.6-100\,keV range; NGC 3227 and NGC 3516 also featuring poor fits with $\chi^{2}_{\nu}\sim1.2$ and  $\chi^{2}_{\nu}\sim1.8$ respectively when fitted without partial covering. It should be noted that many of the iron line parameters in Tables \ref{tab:dual} are poorly constrained, this is due to the interplay between the inner reflector and outer reflector (and/ or partial covering) at hard X-ray energies and the \textsc{compTT} component at soft X-ray energies. 

\section{Conclusions}
Based upon an analysis of all publically available {\sl Suzaku} observations of Seyfert 1 AGN with observations longer than 50\,ks and greater than 30000 XIS counts, we conclude the following:
\begin{enumerate}
\item The majority ($59\%$) of AGN in this sample feature complex warm absorption which has a significant affect upon the Fe\,K region and any accretion disc parameters derived from it. The use of the full 0.6-100.0\,keV  broad-band data is therefore essential prior to any attempt to use relativistic line profile models as a diagnostic for the inner regions of the accretion disc. The mean photon index of the 46 objects on the sample is $\Gamma=1.82\pm0.03$. 
\item Absorption in the Fe\,K region due to highly ionized gas producing absorption features from Fe\,{\rm XXV} and Fe\,{\rm XXVI} are relatively common in AGN (30\%), most of which (71\%) are outflowing at high velocities. While a large fraction of the detected highly ionised winds are outflowing, there may be a larger number of low velocity winds which are not detected due to the possible presence of ionised Fe\,{\rm XXV} and Fe\,{\rm XXVI} emission lines. The additional curvature added to the region through modelling with an appropriate \textsc{xstar} grid, while subtle, has a notable effect upon the strength of any broad residuals which may remain below 6.4\,keV and be interpreted as strong relativistic emission.
\item A partial covering geometry is required in 35\% of all objects in the sample. These high column density zones primarily affect the hard X-ray spectrum above 7\,keV although reducing the strength of broad residuals in the Fe\,K region rather than removing them entirely. 
\item Narrow ionized emission in the Fe\,K region from Fe\,{\rm XXV} and Fe\,{\rm XXVI} are relatively common in these AGN, featuring in 24/46 and 18/46 of objects respectively. Of these AGN, 10/46 feature both Fe\,{\rm XXV} and Fe\,{\rm XXVI} emission. These lines are found to be much more common compared to an {\sl XMM-Newton} survey by Nandra et al. (2007) despite possible interplay with the blue-wing of more sophisticated relativistic line emission models which could reduce the number of narrow line detections. 
\item Examining the Fe\,K region after a complete modelling of the broad-band 0.6-100.0\,keV spectrum and all required absorption zones yields a range of weak to moderate strength broad residuals below 6.4\,keV. We find that 26/46 ($56\%$) of the objects in this sample require some degree of relativistic line emission in the Fe\,K region at $90\%$ confidence and 23/46 (50\%) at $>99.5\%$ confidence.
\item These broad residuals are well fit with the \textsc{relline} model and yield an average broad line strength of $EW=96\pm10$\,eV for the total of 26 objects. The line energy and $\sigma_{\rm width}$ of the broad residuals when modelled with a Gaussian are consistent with Nandra et al. (2007).
\item We estimate an average emissivity index of the accretion disc of $q=2.4\pm0.1$, suggesting that emission from the accretion disc responsible for relativistic lines is not extremely centrally concentrated when purely the line profile in the Fe\,K region is used as a diagnostic. The majority of the line flux therefore occurs from the blue-wing of the line profile with emission being insufficiently close to $R_{\rm ISCO}$ as to redshift a significant proportion of the X-ray flux into a strong red-wing. We also measure an average disc inclination of $i=33^{\circ}\pm2^{\circ}$ and inner radius of emission $R_{\rm in}=(21\pm6)\,R_{\rm g}$.
\item With the assumption that the inner radius of the accretion disc ($R_{\rm in}$) extends down to the innermost stable circular orbit ($R_{\rm ISCO}$), loose constraints upon the SMBH spin parameter $a$ can be made. The relativistic line emission profiles are sufficiently distinguished in 11/46 objects to place upper or lower bounds on the spin. After a broad-band analysis we make the following estimates: Ark 120, $a<0.94$; Fairall 9, $a=0.60^{+0.19}_{-0.63}$; MCG--02-14-009, $a<0.88$; MCG--05-23-16, $a<0.50$; MCG--06-30-15, $a=0.49^{+0.20}_{-0.12}$; Mrk 79, $a<0.80$; Mrk 335, $a=0.70^{+0.12}_{-0.01}$; NGC 3516, $a<0.30$; NGC 3783, $a<0.24$; NGC 7469, $a=0.69^{+0.09}_{-0.09}$ and SWIFT J2127.4+5654 with $a=0.70^{+0.10}_{-0.14}$. Under the assumption that $R_{\rm in}=R_{\rm ISCO}$, a maximally rotating SMBH is ruled out in each of these 11 objects. 
\end{enumerate}

\section*{Acknowledgements}
This research has made use of data obtained from the {\sl Suzaku} satellite, 
a collaborative mission between the space agencies of Japan (JAXA) and the 
USA (NASA). We acknowledge the use of public data from the {\sl Swift} data 
archive. We thank Laura Brenneman \& Chris Reynolds for discussions regarding 
the Suzaku spectrum of NGC 3783. 
We would also like to thank the referee for their helpful comments.

\appendix
\section{Selected individual objects}
\subsection{Fairall 9}
The Fe\,K region of the 'bare' Seyfert Fairall 9 has been well studied (Schmoll et al. 2009; Emmanoulopoulos et al. 2011; Patrick et al. 2011a, 2011b), here a simultaneous analysis of both the 2007 and 2010 observations provides a good fit to the data simply allowing the normalisation of the components to vary. In line with each of the previous analyses, we find an emissivity $q=2.9^{+0.5}_{-0.4}$ although a some what more poorly constrained spin parameter $a=0.60^{+0.19}_{-0.62}$ i.e. consistent with zero spin in this analysis of both Fairall 9 {\sl Suzaku} observations

\subsection{Mrk 205}
The best-fit model for Mrk 205 presented here features partial covering to account for additional flux at X-ray energies $>10$\,keV. The subsequent addition of a broad Gaussian or \textsc{relline} component has little affect upon the overall fit (an improvement of $\triangle\chi^{2}\sim4$) due to the spectral curvature introduced as a result of a partial coverer with column density $N_{\rm H}=(5.1^{+5.5}_{-2.5})\times10^{23}\,{\rm cm}^{-2}$, ionisation log$(\xi)=2.8^{+0.7}_{-0.5}$\,erg\,{\rm cm}\,s$^{-1}$ and covering fraction $C_{\rm frac}=13\%$. However, this AGN can be equally well described with a very broad iron line with equivalent width $EW=254^{+72}_{-59}$\,eV and an increase of $\triangle\chi^{2}\sim+23$ when the \textsc{relline} component is removed and the model refit without a partial coverer in place. This yields a relatively high emissivity index of $q=3.4^{+1.4}_{-0.5}$, an inclination of $i^{\circ}=30^{+10}_{-9}$ and a lower limit placed upon the spin parameter $a>0.1$.

\subsection{NGC 3227}
The six observations of NGC 3227 amount to nearly 500\,000 2-10\,keV counts and a relatively simple model is formed to describe each component and the variations between observations. The best-fit model for NGC 3227 consists of a high-column partial covering component of column $N_{\rm H}=(2.9^{+0.1}_{-0.1})\times10^{23}\,{\rm cm}^{-2}$ with a variable covering fraction of $C_{\rm frac}\sim32-89\%$ in addition to a single warm absorbing zone and both hard and soft excesses fitted with \textsc{reflionx} and \textsc{compTT} respectively. Examining the Fe\,K region there are clear absorption lines due to Fe\,{\rm XXV} and Fe\,{\rm XXVI} and small residuals indicative of weak to moderate emission from the inner regions of the accretion disc (with $EW$ generally higher in a lower continuum flux spectrum, also noted by Fukazawa et al. 2011). A good fit to each of the six observations is found simply by letting the strength of the \textsc{reflionx} and \textsc{compTT} to vary as well as the covering fraction of the partial coverer. All absorber properties such as ionization and column density are tied between observations other than the covering fraction and ionization of the partial coverer which is required to vary between log$\xi\sim0.50-1.53$ with each observation. Using the \textsc{relline} model to account for the broad residuals with parameters tied (but again allowing normalisation to vary) yields an improved fit to the data with $\chi^{2}_{\nu}=4465.6/4189$. The \textsc{relline} model yields a relatively typical emissivity of $q=2.7^{+0.5}_{-0.4}$ and an inclination of $i=33^{+2}_{-2}$\,$^{\circ}$ and $R_{\rm in}=11^{+3}_{-6}\,R_{\rm g}$ (Table \ref{tab:relline} and Figure \ref{fig:eeuf_var}). 

\subsection{NGC 3516}
The two observations of NGC 3516 differ in their shape significantly (Markowitz et al. 2008), the 2005 {\sl Suzaku} observation 2-10\,keV flux is a factor of 1.8 higher than in the 2009 observation. Here we have formed a model in which a good fit is found to both observations, allowing for changes in the absorber/ partial coverer properties and strength of reflection between data sets, although the basic geometry is maintained and a good overall fit is found. As noted in Patrick et al. (2011b) and Turner et al. (2011), the 2009 observation showed no strong indication of a broad red-wing, however, the Markowitz et al. (2008) analysis of the earlier 2005 data do suggest the presence of such a feature. The 2005 observation is more absorbed than the 2009 observation (i.e. $C_{\rm frac}=83\%$ compared to 18\% in the 2009 observation), yet it has a much higher continuum flux level throughout the spectrum above $\sim2$\,keV (Figure \ref{fig:eeuf_var}). This simultaneous analysis yields consistent results with previous work, retrieving the broad feature found by Markowitz et al. (2008) in the 2005 data. Fitting a \textsc{relline} component to both data sets with all parameters (other than normalisation) tied suggest a fairly typical and relatively weak broad line with $EW=58^{+9}_{-9}$\,eV and significantly weaker $EW=14^{+2}_{-2}$\,eV in the 2005 and 2009 observations respectively. An upper limit to the spin parameter is found $a<0.30$ and inclination $i<41^{\circ}$ at a disc emissivity of $q=3.1^{+0.4}_{-0.2}$.

\subsection{NGC 4051}
The three {\sl Suzaku} observations of NGC 4051 (as used in Lobban et al. 2011) included in this analysis vary significantly (see Figure \ref{fig:relline_eeuf} and \ref{fig:eeuf_var}). In particular, in the 2005 observation (OBSID: 700004010) the source dips into an extended period of low flux versus the two 2008 observations which show the object in a period of high flux ($F_{2-10\,keV 2008}\sim3.8\times\,F_{2-10\,keV 2005}$, see Figure \ref{fig:eeuf_var}, Table \ref{tab:fluxes}). Similarly to Lobban et al. (2011), we can successfully describe each observation and the long term spectral variability with a partial covering scenario whereby parameters such as the column density and ionization of the absorbing zones of gas remain approximately constant ($N_{\rm H}\sim9\times10^{22}$\,cm$^{-2}$, over these time scales at least). The differences in the broad-band spectrum of each observation can be accounted for simply by allowing the normalisation of the intrinsic powerlaw, soft excess and distant reflection component to vary, in addition to the covering fraction of the partially covering absorption zone. Here we find covering fractions of $67\%$, $11\%$ and $27\%$ for observations 1, 2 and 3 respectively albeit with a much flatter intrinsic powerlaw ($\Gamma\sim1.88$ here versus $\Gamma\sim2.49$ in Lobban et al. 2011), however, this is likely due to the treatment of the soft excess and reflection components. Here (for consistency with the analysis of other objects in the sample) we account for the soft excess with a \textsc{compTT} component which is akin to a second soft \textsc{powerlaw}, this could in part explain the discrepancies between the two different intrinsic powerlaw components. 

Similarly to Lobban et al. (2011), we find strong evidence for blue-shifted absorption in the Fe\,K region indicative of an outflowing highly ionized zone of gas. When described as such, we find an outflow velocity of $5800^{+1400}_{-1300}$\,${\rm km}\,{\rm s}^{-1}$ in each observation, consistent with Lobban et al. (2011). Both the Lobban et al. (2011) and the baseline models in this analysis describe the spectral variability with absorption dominated models. Replacing the partial coverer with a reflection component representing emission from the inner regions of the accretion disc as an alternate means of accounting for the hard excess and spectral curvature (i.e. a dual reflector) yields a fit worse by $\triangle\chi^{2}\sim+137$ for 4 fewer degrees of freedom. Parameters such as SMBH spin and the emissivity index of the disc are forced to high values i.e. $a\sim0.996$ and $q=6.1^{+0.3}_{-0.1}$. This is in order to smooth the reflection continuum to the extent to which it can successfully account for the long term spectral variability between observations i.e. accounting for both the hard excess and subtle continuum curvature changes. The partial covering scenario is therefore preferred both statistically and in terms of physical implications i.e. extreme parameters and relativistic blurring are not required which would otherwise significantly deviate from the sample norm. Note that there is still a evidence for an highly ionized outflow in the Fe\,K region, regardless of the application of a dual reflector or partial covering based model.

Based upon this, it is clear that the long term spectral variability of NGC 4051 cannot be produced purely by varying the reflected flux in a reflection dominated model; instead a partial covering scenario {\sl must} be invoked to some extent in order to accurately reproduce the differences between each of the three {\sl Suzaku} observations. Reintroducing a partial covering geometry (in addition to a dual reflector, Section 3.4) restores a fit similar to the baseline model. It is, therefore, perhaps more feasible that the broad-band spectrum and variability of NGC 4051 (in addition to other AGN) is primarily a result of variations in covering fraction in an absorption and not reflection dominated spectrum.

\section{}
\begin{table*}
\caption{Summary of observations for the  objects in the sample. $^1$ The observed {2--10\,keV} flux for XIS, 15--50\,keV flux for HXD and 20-100\,keV flux for BAT, in units 10$^{-11}$erg\,{\rm cm}$^{-2}$\,s$^{-1}$ from the baseline model. The XIS count rates listed are per XIS. We use BAT data from the 58 month BAT catalogue (Baumgartner et al. 2010).}
\begin{tabular}{l l l c c c c c c}
\hline
Object & Mission & Instrument & Date & Exposure (s) & Count rate & Flux$^{1}$ & Obs. ID & No. Counts \\
\hline
\multirow{5}{*}{1H 0419--577} & \multirow{2}{*}{\sl Suzaku} & XIS & \multirow{2}{*}{2007-07-25} & 205863 & $1.251\pm0.002$ & 1.75 & \multirow{2}{*}{702041010} & 519055 \\
& & HXD & & 142600 & $0.052\pm0.002$ & 2.93 & & 45303 \\
& \multirow{2}{*}{\sl Suzaku} & XIS & \multirow{2}{*}{2010-01-16} & 122835 & $0.866\pm0.02$ & 1.40 & \multirow{2}{*}{704064010} & 217440 \\
& & HXD & & 104900 & $0.028\pm0.001$ & 2.18 & & 22240 \\
& {\sl Swift} & BAT & -- & -- & $(2.4\pm0.3)\times10^{-4}$ & 1.71 & & 940 \\
\hline
\multirow{3}{*}{3C 111} & \multirow{2}{*}{\sl Suzaku} & XIS & \multirow{2}{*}{2008-08-22} & 122378 & $0.610\pm0.002$ & 1.95 & \multirow{2}{*}{703034010} & 152164 \\
& & HXD & & 101900 & $0.068\pm0.002$ & 3.29 & & 36150 \\
& {\sl Swift} & BAT & -- & -- & $(10.2\pm0.3)\times10^{-4}$ & 7.43 & \\
\hline
\multirow{9}{*}{3C 120} & \multirow{2}{*}{\sl Suzaku} & XIS & \multirow{2}{*}{2006-02-09} & 41932 & $3.082\pm0.005$ & 4.63 & \multirow{2}{*}{700001010} & 389949 \\
& & HXD & & 31870 & $0.099\pm0.004$ & 6.24 & & 13996 \\
& \multirow{2}{*}{\sl Suzaku} & XIS & \multirow{2}{*}{2006-02-16} & 42555 & $2.464\pm0.004$ & 4.01 & \multirow{2}{*}{700001020} & 309300\\
& & HXD & & 34540 & $0.144\pm0.005$ & 5.72 & & 21832 \\
& \multirow{2}{*}{\sl Suzaku} & XIS & \multirow{2}{*}{2006-02-23} & 40907 & $2.53\pm0.004$ & 4.04 & \multirow{2}{*}{700001030} & 312550 \\
& & HXD & & 36200 & $0.110\pm0.004$ & 6.07 & & 15360 \\
& \multirow{2}{*}{\sl Suzaku} & XIS & \multirow{2}{*}{2006-03-02} & 40905 & $2.351\pm0.004$ & 3.96 & \multirow{2}{*}{700001040} & 290680 \\
& & HXD & & 37870 & $0.093\pm0.003$ & 5.59 & & 14533 \\
& {\sl Swift} & BAT & -- & -- & $(8.5\pm0.2)\times10^{-4}$ & 5.82 \\
\hline
\multirow{3}{*}{3C 382} & \multirow{2}{*}{\sl Suzaku} & XIS & \multirow{2}{*}{2007-04-27} & 130580 & $2.596\pm0.003$ & 4.05 & \multirow{2}{*}{702125010} & 683288 \\
& & HXD & & 114300 & $0.097\pm0.002$ & 5.11 & & 47714 \\
& {\sl Swift} & BAT & -- & -- & $(7.6\pm0.2)\times10^{-4}$ & 5.11 \\
\hline
\multirow{3}{*}{3C 390.3} & \multirow{2}{*}{\sl Suzaku} & XIS & \multirow{2}{*}{2006-12-14} & 179800 & $1.831\pm0.003$ & 3.10 & \multirow{2}{*}{701060010} & 333573 \\
& & HXD & & 92060 & $0.096\pm0.003$ & 5.81 & & 37078 \\
& {\sl Swift} & BAT & -- & -- & $(9.09\pm0.18)\times10^{-4}$ & 6.82 & & \\
\hline
\multirow{3}{*}{3C 445} & \multirow{2}{*}{\sl Suzaku} & XIS & \multirow{2}{*}{2007-05-25} & 139769 & $0.159\pm0.001$ & 0.70 & \multirow{2}{*}{702056010} & 48098 \\
& & HXD & & 109500 & $0.049\pm0.002$ & 2.69 & & 37749 \\
& {\sl Swift} & BAT & -- & -- & $(3.7\pm0.2)\times10^{-4}$ & 2.35 \\
\hline
\multirow{3}{*}{4C 74.26} & \multirow{2}{*}{\sl Suzaku} & XIS & \multirow{2}{*}{2007-10-28} & 91583 & $1.493\pm0.003$ & 3.13 & \multirow{2}{*}{702057010} & 276179 \\
& & HXD & & 87340 & $0.084\pm0.002$ & 4.00 & & 34810 \\
& {\sl Swift} & BAT & -- & -- & $(4.6\pm0.2)\times10^{-4}$ & 3.37 & & \\
\hline
\multirow{3}{*}{Ark 120} & \multirow{2}{*}{\sl Suzaku} & XIS & \multirow{2}{*}{2007/04/01} & 100864 & $1.896\pm0.003$ & 3.05 & \multirow{2}{*}{702014010} & 384821 \\
& & HXD & & 89470 & $0.114\pm0.003$ & 3.46 & & 51795 \\
& {\sl Swift} & BAT & -- & 2453000 & $(6.9\pm0.4)\times10^{-4}$ & 4.89 & & 1690  \\
\hline
\multirow{2}{*}{Ark 546} & \multirow{2}{*}{\sl Suzaku} & XIS & \multirow{2}{*}{2007-06-26} & 99978 & $2.277\pm0.003$ & 1.84 & \multirow{2}{*}{702117010} & 458819 \\
& & HXD & & 81330 & $0.021\pm0.002$ & 1.54 & & 15343 \\
\hline
\multirow{5}{*}{Fairall 9} 
& \multirow{2}{*}{\sl Suzaku} & XIS & \multirow{2}{*}{2007/06/07} & 167814 & $1.718\pm0.002$ & 2.32 & \multirow{2}{*}{702043010} & 581331 \\
& & HXD & & 127310 & $0.089\pm0.002$ & 2.97 & & 46809 \\
& \multirow{2}{*}{\sl Suzaku} & XIS & \multirow{2}{*}{2010/08/26} & 229296 & $1.780\pm0.002$ & 2.17 & \multirow{2}{*}{705063010} & 742007 \\
& & HXD & & 162200 & $0.039\pm0.002$ & 3.51 & & 65230\\
& {\sl Swift} & BAT & -- & -- & $(4.7\pm0.2)\times10^{-4}$ & 3.34 & & 1619 \\
\hline
\end{tabular}
\label{tab:observations}
\end{table*}

\begin{table*}
\contcaption{Summary of observations for the  objects in the sample. $^1$ The observed {2--10\,keV} flux for XIS, 15--50\,keV flux for HXD and 20-100\,keV flux for BAT, in units 10$^{-11}$erg\,{\rm cm}$^{-2}$\,s$^{-1}$ from the baseline model. The XIS count rates listed are per XIS. We use BAT data from the 58 month BAT catalogue (Baumgartner et al. 2010).}
\begin{tabular}{l l l c c c c c c}
\hline
Object & Mission & Instrument & Date & Exposure (s) & Count rate & Flux$^{1}$ & Obs. ID & No. Counts \\
\hline
\multirow{11}{*}{IC 4329A} & \multirow{2}{*}{\sl Suzaku} & XIS & \multirow{2}{*}{2007-08-01} & 25453 & $5.212\pm0.010$ & 10.64 & \multirow{2}{*}{702113010} & 266796 \\
& & HXD & & 20050 & $0.289\pm0.006$ & 15.96 & & 11647 \\
& \multirow{2}{*}{\sl Suzaku} & XIS & \multirow{2}{*}{2007-08-06} & 30623 & $6.138\pm0.010$ & 12.58 & \multirow{2}{*}{702113020} & 377880 \\
& & HXD & & 24080 & $0.340\pm0.005$ & 17.97 & & 15349 \\
& \multirow{2}{*}{\sl Suzaku} & XIS & \multirow{2}{*}{2007-08-11} & 26896 & $5.554\pm0.010$ & 11.89 & \multirow{2}{*}{702113030} & 300624 \\
& & HXD & & 22110 & $0.345\pm0.006$ & 18.77 & & 15541 \\
& \multirow{2}{*}{\sl Suzaku} & XIS & \multirow{2}{*}{2007-08-16} & 24219 & $5.261\pm0.010$ & 10.88 & \multirow{2}{*}{702113040} & 256495 \\
& & HXD & & 18750 & $0.300\pm0.006$ & 16.47 & & 11923 \\
& \multirow{2}{*}{\sl Suzaku} & XIS & \multirow{2}{*}{2007-08-20} & 24026 & $3.221\pm0.008$ & 7.03 & \multirow{2}{*}{702113050} & 156141 \\
& & HXD & & 17520 & $0.242\pm0.006$ & 13.69 & & 9755\\
& {\sl Swift} & BAT & -- & -- & $(26.8\pm0.3)\times10^{-4}$ & 19.02 & & \\
\hline
\multirow{2}{*}{IRAS 13224--3809} & \multirow{2}{*}{\sl Suzaku} & XIS & \multirow{2}{*}{2007-01-26} & 197938 & $0.085\pm0.001$ & 0.06 & \multirow{2}{*}{701003010} & 39148 \\
& & HXD & & 158500 & $0.004\pm0.002$ & 0.015 & & 86214\\
\hline
\multirow{2}{*}{MCG--02-14-009} & \multirow{2}{*}{\sl Suzaku} & XIS & \multirow{2}{*}{2008/08/28} & 142152 & $0.216\pm0.001$ & 0.43 & \multirow{2}{*}{703060010} & 63436 \\
& & HXD & & 120028 & $0.017\pm0.002$ & 0.61 & & 40150 \\
\hline
\multirow{3}{*}{MCG--02-58-22} & \multirow{2}{*}{\sl Suzaku} & XIS & \multirow{2}{*}{2009-12-02} & 138969 & $2.921\pm0.003$ & 4.87 & \multirow{2}{*}{704032010} & 817420 \\
& & HXD & & 97980 & $0.163\pm0.002$ & 8.73 & & 44009 \\
& {\sl Swift} & BAT & -- & -- & $(9.8\pm0.2)\times10^{-4}$ & 7.48 & & \\
\hline
\multirow{3}{*}{MCG--05-23-16} & \multirow{2}{*}{\sl Suzaku} & XIS & \multirow{2}{*}{2005-12-07} & 95677 & $2.822\pm0.003$ & 8.93 & \multirow{2}{*}{700002010} & 815975 \\
& & HXD & & 79690 & $0.313\pm0.003$ & 14.52 & & 55913 \\
& {\sl Swift} & BAT & -- & -- & $(19.8\pm0.2)\times10^{-4}$ & 14.27 & & 4526\\
\hline
\multirow{7}{*}{MCG--06-30-15} & \multirow{2}{*}{\sl Suzaku} & XIS & \multirow{2}{*}{2006/01/09} & 143196 & $2.862\pm0.003$ & 4.38 & \multirow{2}{*}{700007010} & 1240090 \\
& & HXD & & 118900 & $0.094\pm0.002$ & 4.70 & & 53689 \\
& \multirow{2}{*}{\sl Suzaku} & XIS & \multirow{2}{*}{2006/01/23} & 98483 & $2.461\pm0.003$ & 3.81 & \multirow{2}{*}{700007020} & 734189\\
& & HXD & & 76800 & $0.103\pm0.003$ & 5.06 & & 36695 \\
& \multirow{2}{*}{\sl Suzaku} & XIS & \multirow{2}{*}{2006/01/27} & 96691 & $2.708\pm0.003$ & 4.16 & \multirow{2}{*}{700007030} & 792389 \\
& & HXD & & 83660 & $0.104\pm0.002$ & 4.98 & & 40408 \\
& {\sl Swift} & BAT & -- & -- & $(6.8\pm0.3)\times10^{-4}$ & 3.86 & \\
\hline
\multirow{3}{*}{MCG+8-11-11} & \multirow{2}{*}{\sl Suzaku} & XIS & \multirow{2}{*}{2007-09-17} & 98748 & $2.863\pm0.004$ & 6.48 & \multirow{2}{*}{702112010} & 568883 \\
& & HXD & & 82900 & $0.196\pm0.003$ & 10.44 & & 43023 \\
& {\sl Swift} & BAT & -- & -- & $(8.8\pm0.2)\times10^{-4}$ & 6.27 & & 2082 \\
\hline
\multirow{3}{*}{MR 2251--178} & \multirow{2}{*}{\sl Suzaku} & XIS & \multirow{2}{*}{2009-05-07} & 136924 & $2.089\pm0.003$ & 4.23 & \multirow{2}{*}{704055010} & 576608 \\
& & HXD & & 103800 & $0.101\pm0.002$ & 5.47 & & 40562 \\
& {\sl Swift} & BAT & -- & -- & $(9.1\pm0.2)\times10^{-4}$ & 6.02 & & \\
\hline
\multirow{3}{*}{Mrk 79} & \multirow{2}{*}{\sl Suzaku} & XIS & \multirow{2}{*}{2007-04-03} & 83704 & $0.768\pm0.002$ & 1.47 & \multirow{2}{*}{702044010} & 130639 \\
& & HXD & & 76920 & $0.027\pm0.002$ & 1.79 & & 19777 \\
& {\sl Swift} & BAT & -- & -- & $(4.1\pm0.2)\times10^{-4}$ & 2.87 & & \\
\hline
\multirow{3}{*}{Mrk 110} & \multirow{2}{*}{\sl Suzaku} & XIS & \multirow{2}{*}{2007-11-02} & 90871 & $1.220\pm0.003$ & 2.13 & \multirow{2}{*}{702124010} & 224098 \\
& & HXD & & 80370 & $0.053\pm0.002$ & 2.99 & & 26958 \\
& {\sl Swift} & BAT & -- & -- & $(5.4\pm0.3)\times10^{-4}$ & 3.95 & & 2128 \\
\hline
\multirow{3}{*}{Mrk 205} & \multirow{2}{*}{\sl Suzaku} & XIS & \multirow{2}{*}{2010-05-22} & 100961 & $0.501\pm0.002$ & 0.93 & \multirow{2}{*}{705062010} & 103464 \\
& & HXD & & 85280 & $0.019\pm0.001$ & 1.94 & & 13647 \\
& {\sl Swift} & BAT & -- & -- & $(1.4\pm0.2)\times10^{-4}$ & 1.05 & & \\
\hline
\multirow{3}{*}{Mrk 279} & \multirow{2}{*}{\sl Suzaku} & XIS & \multirow{2}{*}{2009-05-14} & 160351 & $0.263\pm0.001$ & 0.49 & \multirow{2}{*}{704031010} & 88122 \\
& & HXD & & 139800 & $0.024\pm0.001$ & 2.42 & & 29520 \\
& {\sl Swift} & BAT & -- & -- & $(5.2\pm0.3)\times10^{-4}$ & 3.93 & & 2617 \\
\hline
\multirow{3}{*}{Mrk 335} & \multirow{2}{*}{\sl Suzaku} & XIS & \multirow{2}{*}{2006/08/21} & 151296 & $1.324\pm0.002$ & 1.49 & \multirow{2}{*}{701031010} & 606927 \\
& & HXD & & 131744 & $0.012\pm0.001$ & 1.31 & 22385 \\
& {\sl Swift} & BAT & -- & -- & $(2.5\pm0.3)\times10^{-4}$ & 1.81 & & 830 \\
\hline
\multirow{3}{*}{Mrk 359} & \multirow{2}{*}{\sl Suzaku} & XIS & \multirow{2}{*}{2007-02-06} & 107507 & $0.329\pm0.001$ & 0.52 & \multirow{2}{*}{701082010} & 73004 \\
& & HXD & & 96130 & $0.007\pm0.001$ & 0.98 & & 13136 \\
& {\sl Swift} & BAT & -- & -- & $(1.2\pm0.1)\times10^{-4}$ & 0.92 & &  \\
\hline
\end{tabular}
\end{table*}

\begin{table*}
\contcaption{Summary of observations for the  objects in the sample. $^1$ The observed {2--10\,keV} flux for XIS, 15--50\,keV flux for HXD and 20-100\,keV flux for BAT, in units 10$^{-11}$erg\,{\rm cm}$^{-2}$\,s$^{-1}$ from the baseline model. The XIS count rates listed are per XIS. We use BAT data from the 58 month BAT catalogue (Baumgartner et al. 2010).}
\begin{tabular}{l l l c c c c c c}
\hline
Object & Mission & Instrument & Date & Exposure (s) & Count rate & Flux$^{1}$ & Obs. ID & No. Counts \\
\hline
\multirow{9}{*}{Mrk 509} & \multirow{2}{*}{\sl Suzaku} & XIS & \multirow{2}{*}{2006-04-25} & 24576 & $2.984\pm0.006$ & 4.68 & \multirow{2}{*}{701093010} & 221506 \\
& & HXD & & 14510 & $0.143\pm0.006$ & 7.50 & & 6647 \\
& \multirow{2}{*}{\sl Suzaku} & XIS & \multirow{2}{*}{2006-10-14} & 25930 & $3.802\pm0.007$ & 4.98 & \multirow{2}{*}{701093020} & 297900 \\
& & HXD & & 21880 & $0.128\pm0.005$ & 7.89 & & 9680 \\
& \multirow{2}{*}{\sl Suzaku} & XIS & \multirow{2}{*}{2006-11-15} & 24447 & $3.792\pm0.009$ & 4.74 & \multirow{2}{*}{701093030} & 186702 \\
& & HXD & & 17340 & $0.118\pm0.005$ & 7.57 & & 7260 \\
& \multirow{2}{*}{\sl Suzaku} & XIS & \multirow{2}{*}{2006-11-27} & 33094 & $3.278\pm0.007$ & 4.73 & \multirow{2}{*}{701093040} & 218248 \\
& & HXD & & 27560 & $0.110\pm0.004$ & 7.19 & & 11925 \\
& {\sl Swift} & BAT & -- & -- & $(8.4\pm0.2)\times10^{-4}$ & 6.53 & & \\
\hline
\multirow{5}{*}{Mrk 766} & \multirow{2}{*}{\sl Suzaku} & XIS & \multirow{2}{*}{2006-11-16} & 97869 & $0.927\pm0.002$ & 1.32 & \multirow{2}{*}{701035010} & 183736 \\
& & HXD & & 90500 & $0.025\pm0.002$ & 1.48 & & 26879 \\
& \multirow{2}{*}{\sl Suzaku} & XIS & \multirow{2}{*}{2007-11-17} & 59364 & $0.685\pm0.002$ & 1.36 & \multirow{2}{*}{701035020} & 82749 \\
& & HXD & & 47660 & $0.024\pm0.002$ & 1.67 & & 8974 \\
& {\sl Swift} & BAT & -- & -- & $(2.5\pm0.2)\times10^{-4}$ & 1.66 & &  \\
\hline
\multirow{5}{*}{Mrk 841} & \multirow{2}{*}{\sl Suzaku} & XIS & \multirow{2}{*}{2007-01-22} & 51753 & $0.819\pm0.003$ & 1.40 & \multirow{2}{*}{701084010} & 86781 \\
& & HXD & & 40710 & $0.039\pm0.003$ & 2.42 & & 11274 \\
& \multirow{2}{*}{\sl Suzaku} & XIS & \multirow{2}{*}{2007-07-23} & 50925 & $0.830\pm0.003$ & 1.44 & \multirow{2}{*}{701084020} & 86626 \\
& & HXD & & 44370 & $0.054\pm0.003$ & 2.82 & & 13282 \\
& {\sl Swift} & BAT & -- & -- & $(3.2\pm0.2)\times10^{-4}$ & 2.29 & & \\
\hline
\multirow{7}{*}{NGC 1365} & \multirow{2}{*}{\sl Suzaku} & XIS & \multirow{2}{*}{2008-01-21} & 160506 & $0.326\pm0.001$ & 1.28 & \multirow{2}{*}{702047010} & 107383 \\
& & HXD & & 136600 & $0.084\pm0.002$ & 4.72 & & 55436 \\
& \multirow{2}{*}{\sl Suzaku} & XIS & \multirow{2}{*}{2010-06-27} & 151613 & $0.159\pm0.001$ & 0.61 & \multirow{2}{*}{705031010} & 51909 \\
& & HXD & & 114300 & $0.070\pm0.002$ & 4.33 & & 38083 \\
& \multirow{2}{*}{\sl Suzaku} & XIS & \multirow{2}{*}{2010-07-15} & 302175 & $0.116\pm0.001$ & 0.39 & \multirow{2}{*}{705031020} & 77717 \\
& & HXD & & 231500 & $0.060\pm0.001$ & 3.49 & & 84887 \\
& {\sl Swift} & BAT & -- & -- & $(5.2\pm0.1)\times10^{-4}$ & 5.11 & & \\
\hline
\multirow{7}{*}{NGC 2992} & \multirow{2}{*}{\sl Suzaku} & XIS & \multirow{2}{*}{2005-11-06} & 37503 & $0.372\pm0.002$ & 1.04 & \multirow{2}{*}{700005010} & 43167 \\
& & HXD & & 29900 & $0.020\pm0.003$ &  1.71 & & 7137 \\
& \multirow{2}{*}{\sl Suzaku} & XIS & \multirow{2}{*}{2005-11-19} & 37494 & $0.486\pm0.002$ & 1.37 & \multirow{2}{*}{700005020} & 56041 \\
& & HXD & & 31890 & $0.042\pm0.003$ & 2.23 & & 10370 \\
& \multirow{2}{*}{\sl Suzaku} & XIS & \multirow{2}{*}{2005-12-13} & 46836 & $0.408\pm0.002$ & 1.15 & \multirow{2}{*}{700005030} & 59084 \\
& & HXD & & 41470 & $0.047\pm0.003$ & 2.44 & & 15070 \\
& {\sl Swift} & BAT & -- & -- & $(2.6\pm0.2)\times10^{-4}$ & 1.84 & & \\
\hline
\multirow{2}{*}{NGC 3147} & \multirow{2}{*}{\sl Suzaku} & XIS & \multirow{2}{*}{2010-06-03} & 150048 & $0.093\pm0.001$ & 0.17 & \multirow{2}{*}{705054010} & 31054 \\
& & HXD & & 122800 & $0.009\pm0.001$ & 0.29 & & 21785 \\
\hline
\multirow{13}{*}{NGC 3227} & \multirow{2}{*}{\sl Suzaku} & XIS & \multirow{2}{*}{2008-10-28} & 58917 & $1.475\pm0.004$ & 3.91 & \multirow{2}{*}{703022010} & 175651 \\
& & HXD & & 48070 & $0.145\pm0.003$ & 5.58 & & 21550\\
& \multirow{2}{*}{\sl Suzaku} & XIS & \multirow{2}{*}{2008-11-04} & 53700 & $0.496\pm0.002$ & 1.85 & \multirow{2}{*}{703022020} & 54717 \\
& & HXD & & 46740 & $0.123\pm0.003$ & 7.26 & & 18735 \\
& \multirow{2}{*}{\sl Suzaku} & XIS & \multirow{2}{*}{2008-11-12} & 56572 & $0.707\pm0.003$ & 2.58 & \multirow{2}{*}{703022030} & 81449 \\
& & HXD & & 46680 & $0.117\pm0.003$ & 6.78 & & 18919 \\
& \multirow{2}{*}{\sl Suzaku} & XIS & \multirow{2}{*}{2008-11-20} & 64568 & $0.278\pm0.002$ & 1.01 & \multirow{2}{*}{703022040} & 37518 \\
& & HXD & & 43430 & $0.083\pm0.003$ & 4.82 & & 17263 \\
& \multirow{2}{*}{\sl Suzaku} & XIS & \multirow{2}{*}{2008-11-27} & 79430 & $0.568\pm0.002$ & 2.16 & \multirow{2}{*}{703022050} & 92162\\
& & HXD & & 37420 & $0.104\pm0.003$ & 6.26 & & 14795 \\
& \multirow{2}{*}{\sl Suzaku} & XIS & \multirow{2}{*}{2008-12-02} & 51411 & $0.413\pm0.002$ & 1.61 & \multirow{2}{*}{703022060} & 43570\\
& & HXD & & 36910 & $0.082\pm0.003$ & 5.70 & & 13177 \\
& {\sl Swift} & BAT & -- & -- & $(10.3\pm0.2)\times10^{-4}$ & 7.68 & & \\
\hline
\multirow{5}{*}{NGC 3516} & \multirow{2}{*}{\sl Suzaku} & XIS & \multirow{2}{*}{2005/10/12} & 134469 & $0.681\pm0.001$ & 2.37 & \multirow{2}{*}{100031010} & 289936 \\
& & HXD & & 115400 & $0.125\pm0.002$ & 6.68 & & 49782 \\
& \multirow{2}{*}{\sl Suzaku} & XIS & \multirow{2}{*}{2009/10/28} & 251356 & $0.456\pm0.001$ & 1.35 & \multirow{2}{*}{704062010} & 235894\\
& & HXD & & 178200 & $0.059\pm0.001$ & 3.32 & & 58222 \\
& {\sl Swift} & BAT & -- & -- & $(10.5\pm0.2)\times10^{-4}$ & 7.76 & \\
\hline
\end{tabular}
\end{table*}

\begin{table*}
\contcaption{Summary of observations for the  objects in the sample. $^1$ The observed {2--10\,keV} flux for XIS, 15--50\,keV flux for HXD and 20-100\,keV flux for BAT, in units 10$^{-11}$erg\,{\rm cm}$^{-2}$\,s$^{-1}$ from the baseline model. The XIS count rates listed are per XIS. We use BAT data from the 58 month BAT catalogue (Baumgartner et al. 2010).}
\begin{tabular}{l l l c c c c c c}
\hline
Object & Mission & Instrument & Date & Exposure (s) & Count rate & Flux$^{1}$ & Obs. ID & No. Counts \\
\hline
\multirow{5}{*}{NGC 3783} & \multirow{2}{*}{\sl Suzaku} & XIS & 2006/06/24 & 75719 & $2.066\pm0.003$ & 4.59 & \multirow{2}{*}{701033010} & 416485 \\
& & HXD &  & 63930 & $0.167\pm0.003$ & 9.26 & & 33592 \\
& \multirow{2}{*}{\sl Suzaku} & XIS & \multirow{2}{*}{2009/07/10} & 209503 & $1.942\pm0.002$ & 5.92 & \multirow{2}{*}{704063010} & 805999 \\
& & HXD & & 162000 & $0.235\pm0.002$ & 12.04 & & 87319 \\
& {\sl Swift} & BAT & -- & -- & $(16.6\pm0.2)\times10^{-4}$ & 11.54 & \\
\hline
\multirow{7}{*}{NGC 4051} & \multirow{2}{*}{\sl Suzaku} & XIS & 2005/11/10 & 119578 & $0.406\pm0.001$ & 0.87 & \multirow{2}{*}{700004010} & 149284\\
& & HXD &  & 112600 & $0.037\pm0.002$ & 1.63 & & 29700 \\
& \multirow{2}{*}{\sl Suzaku} & XIS & 2008/11/06 & 274350 & $1.858\pm0.002$ & 2.46 & \multirow{2}{*}{703023020} & 1036060 \\
& & HXD &  & 204500 & $0.062\pm0.001$ & 3.05 & & 75619 \\
& \multirow{2}{*}{\sl Suzaku} & XIS & \multirow{2}{*}{2008/11/23} & 78385 & $1.276\pm0.003$ & 1.79 & \multirow{2}{*}{703023010} & 202942 \\
& & HXD & & 58530 & $0.048\pm0.002$ & 2.54 & & 21296 \\
& {\sl Swift} & BAT & -- & -- & $(4.0\pm0.2)\times10^{-4}$ & 2.80 & \\
\hline
\multirow{3}{*}{NGC 4151} & \multirow{2}{*}{\sl Suzaku} & XIS & \multirow{2}{*}{2006-12-18} & 124980 & $1.036\pm0.002$ & 4.35 & \multirow{2}{*}{701034010} & 262831 \\
& & HXD & & 123500 & $0.293\pm0.002$ & 15.90 & & 78564 \\
& {\sl Swift} & BAT & -- & -- & $(43.6\pm0.2)\times10^{-4}$ & 32.50 & &  \\
\hline
\multirow{3}{*}{NGC 4593} & \multirow{2}{*}{\sl Suzaku} & XIS & \multirow{2}{*}{2007-12-15} & 118842 & $0.524\pm0.002$ & 1.04 & \multirow{2}{*}{702040010} & 127503 \\
& & HXD & & 101600 & $0.033\pm0.002$ & 2.15 & & 30049\\
& {\sl Swift} & BAT & -- & -- & $(7.3\pm0.2)\times10^{-4}$ & 5.74 & & \\
\hline
\multirow{7}{*}{NGC 5506} & \multirow{2}{*}{\sl Suzaku} & XIS & \multirow{2}{*}{2006-08-08} & 47753 & $3.064\pm0.005$ & 10.05 & \multirow{2}{*}{701030010} & 441844 \\
& & HXD & & 38550 & $0.324\pm0.004$ & 17.30 & & 25416 \\
& \multirow{2}{*}{\sl Suzaku} & XIS & \multirow{2}{*}{2006-08-11} & 53296 & $3.230\pm0.005$ & 10.58 & \multirow{2}{*}{701030020} & 519889 \\
& & HXD & & 44830 & $0.343\pm0.004$ & 18.40 & & 30625 \\
& \multirow{2}{*}{\sl Suzaku} & XIS & \multirow{2}{*}{2007-01-31} & 57406 & $2.854\pm0.005$ & 9.90 & \multirow{2}{*}{701030030} & 330141 \\
& & HXD & & 44670 & $0.321\pm0.004$ & 17.11 & & 29399\\
& {\sl Swift} & BAT & -- & -- & $(23.2\pm0.2)\times10^{-4}$ & 15.95 & & \\
\hline
\multirow{15}{*}{NGC 5548} & \multirow{2}{*}{\sl Suzaku} & XIS & \multirow{2}{*}{2007-06-18} & 31119 & $0.369\pm0.003$ & 0.81 & \multirow{2}{*}{702042010} & 23857 \\
& & HXD & & 25590 & $0.046\pm0.004$ & 2.70 & & 9397 \\
& \multirow{2}{*}{\sl Suzaku} & XIS & \multirow{2}{*}{2007-06-24} & 35915 & $0.650\pm0.003$ & 1.34 & \multirow{2}{*}{702042020} & 47786 \\
& & HXD & & 31210 & $0.050\pm0.003$ & 3.12 & & 9880 \\
& \multirow{2}{*}{\sl Suzaku} & XIS & \multirow{2}{*}{2007-07-08} & 30700 & $1.349\pm0.005$ & 2.54 & \multirow{2}{*}{702042040} & 83863 \\
& & HXD & & 26960 & $0.064\pm0.004$ & 4.34 & & 9824 \\
& \multirow{2}{*}{\sl Suzaku} & XIS & \multirow{2}{*}{2007-07-15} & 30020 & $0.834\pm0.004$ & 1.74 & \multirow{2}{*}{702042050} & 51085 \\
& & HXD & & 24470 & $0.062\pm0.004$ & 4.16 & & 7612 \\
& \multirow{2}{*}{\sl Suzaku} & XIS & \multirow{2}{*}{2007-07-22} & 28918 & $1.656\pm0.005$ & 3.28 & \multirow{2}{*}{702042060} & 96936 \\
& & HXD & & 23060 & $0.095\pm0.004$ & 0.22 & & 8336 \\
& \multirow{2}{*}{\sl Suzaku} & XIS & \multirow{2}{*}{2007-07-29} & 31810 & $1.075\pm0.004$ & 2.19 & \multirow{2}{*}{702042070} & 69507 \\
& & HXD & & 27610 & $0.075\pm0.004$ & 4.71 & & 9377 \\
& \multirow{2}{*}{\sl Suzaku} & XIS & \multirow{2}{*}{2007-08-05} & 38776 & $0.558\pm0.003$ & 1.19 & \multirow{2}{*}{702042080} & 44486 \\
& & HXD & & 30380 & $0.054\pm0.003$ & 3.41 & & 9958 \\
& {\sl Swift} & BAT & -- & -- & $(6.1\pm0.2)\times10^{-4}$ & 4.56 & & \\
\hline
\multirow{3}{*}{NGC 7213} & \multirow{2}{*}{\sl Suzaku} & XIS & \multirow{2}{*}{2006-10-22} & 90736 & $1.528\pm0.002$ & 2.41 & \multirow{2}{*}{701029010} & 419604 \\
& & HXD & & 84290 & $0.064\pm0.002$ & 3.48 & & 33910 \\
& {\sl Swift} & BAT & -- & -- & $(3.6\pm0.2)\times10^{-4}$ & 5.13 & &  \\
\hline
\end{tabular}
\end{table*}

\begin{table*}
\contcaption{Summary of observations for the  objects in the sample. $^1$ The observed {2--10\,keV} flux for XIS, 15--50\,keV flux for HXD and 20-100\,keV flux for BAT, in units 10$^{-11}$erg\,{\rm cm}$^{-2}$\,s$^{-1}$ from the baseline model. The XIS count rates listed are per XIS. We use BAT data from the 58 month BAT catalogue (Baumgartner et al. 2010).}
\begin{tabular}{l l l c c c c c c}
\hline
Object & Mission & Instrument & Date & Exposure (s) & Count rate & Flux$^{1}$ & Obs. ID & No. Counts \\
\hline
\multirow{3}{*}{NGC 7314} & \multirow{2}{*}{\sl Suzaku} & XIS & \multirow{2}{*}{2007-04-25} & 109020 & $0.296\pm0.001$ & 0.88 & \multirow{2}{*}{702015010} & 67434 \\
& & HXD & & 85780 & $0.026\pm0.002$ & 1.43 & & 25675 \\
& {\sl Swift} & BAT & -- & -- & $(4.6\pm0.2)\times10^{-4}$ & 3.62 & & \\
\hline
\multirow{3}{*}{NGC 7469} & \multirow{2}{*}{\sl Suzaku} & XIS & \multirow{2}{*}{2008/06/24} & 112113 & $1.091\pm0.002$ & 2.10 & \multirow{2}{*}{703028010} & 248180 \\
& & HXD & & 85315 & $0.068\pm0.002$ & 3.22 & & 29831 \\
& {\sl Swift} & BAT & -- & 3286000 & $(6.6\pm0.3)\times10^{-4}$ & 4.87 & & 2179  \\
\hline
\multirow{2}{*}{PDS 456} & \multirow{2}{*}{Suzaku} & XIS & \multirow{2}{*}{2007-02-24} & 190600 & $0.237\pm0.001$ & 0.35 & \multirow{2}{*}{701056010} & 92499 \\
& & HXD & & 64150 & $0.012\pm0.003$ & 0.25 & & 20372 \\
\hline
\multirow{2}{*}{PG 1211+143} & \multirow{2}{*}{\sl Suzaku} & XIS & \multirow{2}{*}{2005-11-24} & 96324 & $0.268\pm0.001$ & 0.39 & \multirow{2}{*}{700009010} & 81584 \\
& & HXD & & 78800 & $0.004\pm0.002$ & 0.50 & & 29181 \\
\hline
\multirow{2}{*}{RBS 1124} & \multirow{2}{*}{\sl Suzaku} & XIS & \multirow{2}{*}{2007-04-14} & 86228 & $0.260\pm0.001$ & 0.49 & \multirow{2}{*}{702114010} & 46699 \\
& & HXD & & 82970 & $0.016\pm0.001$ & 0.91 & & 14206 \\
\hline
\multirow{3}{*}{SWIFT J2127.4+5654} & \multirow{2}{*}{\sl Suzaku} & XIS & \multirow{2}{*}{2007/12/09} & 91730 & $1.373\pm0.004$ & 3.35 & \multirow{2}{*}{702122010} & 127435 \\
& & HXD & & 83321 & $0.074\pm0.002$ & 3.33 & & 27970 \\
& {\sl Swift} & BAT & -- & -- & $(4.3\pm0.3)\times10^{-4}$ & 3.62 & & 1689 \\
\hline
\multirow{2}{*}{TON S180} & \multirow{2}{*}{\sl Suzaku} & XIS & \multirow{2}{*}{2006-12-09} & 120661 & $0.701\pm0.002$ & 0.56 & \multirow{2}{*}{701021010} & 172889 \\
& & HXD & & 102400 & $0.012\pm0.002$ & 0.77 & & 27241 \\
\hline
\end{tabular}
\end{table*}

\begin{landscape}
\begin{table}
\begin{minipage}{235mm}
\caption{Components for the baseline model (without broad or diskline emission) to the observations with {\sl Suzaku} XIS, HXD and BAT data from {\sl Swift}. $^{a}$ Unabsorbed \textsc{powerlaw} normalisation given in units $(10^{-2}\,{\rm ph\,keV^{-1}\,{\rm cm}^{-2}\,s^{-1}})$. $^{b}$ Flux for \textsc{compTT} quoted over the 0.6-10.0\,keV range and \textsc{reflionx} over the 2.0-100.0\,keV range in units 10$^{-11}$erg\,{\rm cm}$^{-2}$\,s$^{-1}$. $^{c}$ Ionization parameter given in units erg\,{\rm cm}\,s$^{-1}$. $^{d}$ Column density measured in units $10^{22}$\,${\rm cm}^{-2}$. $^{e}$ $v_{\rm turb}=1000$\,${\rm km}\,{\rm s}^{-1}$. $^{f}$ only the intrinsic powerlaw is absorbed. * indicates a frozen parameter. $\dagger$ indicates parameters are tied during the analysis of multiple observations.}
\begin{tabular}{p{2.5cm} c c c c c c c c c c c c c c}
\hline
Object & \multicolumn{2}{l}{Powerlaw} & \multicolumn{3}{l}{\textsc{compTT}} & \multicolumn{3}{l}{\textsc{reflionx}} & \multicolumn{4}{l}{Warm absorber} &  \\
& $\Gamma$ & Norm$^{a}$ & kT (keV) & $\tau$ & Flux$^{b}$ & $Z_{\rm Fe}$ & $\xi^{c}$ & Flux$^{b}$ & $N_{\rm H}^{d}$ & log$(\xi)^{c}$ & $C_{\rm frac}$ & BAT/XIS & $\chi^{2}_{\nu}$ \\
\hline
1H 0419--577 Obs 1,2& $1.70^{+0.01}_{-0.01}$ & $0.39^{+0.01}_{-0.01}$ & $7.5^{+0.4}_{-0.4}$ & $1.4^{+0.1}_{-0.1}$ & $0.59^{+0.01}_{-0.01}$ & 1.0* & $10.1^{+1.9}_{-3.0}$ & $0.36^{+0.21}_{-0.14}$ & $15.83^{+51.61}_{-15.68}$ & $>2.87$ & $100\%$ & $0.47^{+0.11}_{-0.11}$ & \multirow{4}{*}{1788.9/1750} \\
& & & & & & & & & $178.0^{+78.0}_{-54.0}$ & $2.05^{+0.43}_{-0.77}$ & $18\%$ & \\
1H 0419--577 & $1.70\dagger$ & $0.31^{+0.01}_{-0.01}$ & $7.5\dagger$ & $1.4\dagger$ &  $0.46^{+0.01}_{-0.01}$ & $1.0\dagger$ & $10.1\dagger$ & $0.45^{+0.28}_{-0.17}$ & $15.83\dagger$ & $>2.87\dagger$ & $100\%$ & $0.47\dagger$ & \\
& & & & & & & & & $178.0\dagger$ & $2.05\dagger$ & $15\%$ & \\
3C 111 & $1.58^{+0.01}_{-0.01}$ & $0.44^{+0.01}_{-0.01}$ & -- & -- & -- & -- & -- & -- & $0.61^{+0.01}_{-0.01}$ & Neutral & $100\%$ & $1.85^{+0.14}_{-0.14}$ & 1101.1/1098 \\
3C 120 Obs 1,2,3,4 & $1.65^{+0.01}_{-0.01}$ & $0.98^{+0.01}_{-0.01}$ & $6.9^{+0.3}_{-0.3}$ & $1.8^{+0.1}_{-0.1}$ & $1.82^{+0.03}_{-0.03}$ & $1.5^{+0.3}_{-0.2}$ & $<19.4$ & $1.92^{+0.30}_{-0.28}$ & $0.09^{+0.01}_{-0.01}$ & Neutral & $100\%$ & $0.83^{+0.04}_{-0.04}$ & \multirow{2}{*}{3630.5/3452} \\
3C 120 Obs 4 & $1.65\dagger$ & $0.86^{+0.01}_{-0.01}$ & $6.9\dagger$ & $1.8\dagger$ & $0.97^{+0.01}_{-0.01}$ & $1.5\dagger$ & $<19.4\dagger$ & $2.34^{+0.35}_{-0.30}$ & $0.09\dagger$ & Neutral & $100\%$ &$0.83\dagger$ \\
3C 382 & $1.78^{+0.01}_{-0.01}$ & $1.16^{+0.01}_{-0.01}$ & $<56.5$ & $1.0^{+1.4}_{-1.0}$ & $0.35^{+0.36}_{-0.31}$ & 1.0* & $<2.1$ & $1.67^{+0.14}_{-0.93}$ & $0.35^{+0.14}_{-0.09}$ & $2.74^{+0.09}_{-0.12}$ & $100\%$ & $0.91^{+0.06}_{-0.05}$ & 986.6/938 \\
3C 390.3 & $1.73^{+0.02}_{-0.02}$ & $0.75^{+0.01}_{-0.02}$ & $<55.9$ & $<25$ & $<0.429$ & $0.5^{+0.1}_{-0.1}$ & $2^{+1}_{-1}$ & $2.37^{+2.44}_{-0.65}$ & -- & -- & -- & $0.93^{+0.04}_{-0.04}$ & 1499.4/1486 \\
3C 445 & $1.43^{+0.01}_{-0.01}$ & $0.03^{+0.01}_{-0.01}$ & -- & -- & -- & 1.0* & $9.9^{+4.5}_{-7.8}$ & $2.79^{+10.14}_{-1.03}$ & $3.31^{+0.21\,e}_{-0.17}$ & $1.93^{+0.03}_{-0.03}$ & $100\%$ & $0.85^{+0.12}_{-0.11}$ & 458.1/422 \\
& & & & & & & & & $18.94^{+0.78}_{-0.75}$ & Neutral & $89\%$ & \\
4C 74.26 & $1.92^{+0.01}_{-0.01}$ & $1.06^{+0.01}_{-0.01}$ & -- & -- & -- & 1.0* & $<5.2$ & $1.23^{+1.36}_{-0.81}$ & $0.27^{+0.03}_{-0.01}$ & $1.00^{+0.10}_{-0.07}$ & $100\%$ & $0.72^{+0.05}_{-0.05}$ & 1344.8/1303 \\
Ark 120 & $1.90^{+0.01}_{-0.04}$ & $0.95^{+0.06}_{-0.06}$ & $<8.5$ & $<2.4$ & $1.05^{+0.05}_{-0.05}$ & $1.4^{+0.5}_{-0.3}$ & $<22$ & $0.93^{+0.44}_{-0.53}$ & -- & -- & -- & $1.07^{+0.12}_{-0.12}$ & 741.8/649 \\
Ark 564 & $2.34^{+0.01}_{-0.01}$ & $1.10^{+0.01}_{-0.01}$ & $4.7^{+27.3}_{-2.4}$ & $<2.0$ & $3.29^{+3.22}_{-2.53}$ & 1.0* & $<7.8$ & $0.12^{+0.19}_{-0.09}$ & $0.05^{+0.01}_{-0.01}$ & Neutral & $100\%$ & -- & 1148.9/1024 \\
& & & & & & & & & $<0.02^{e}$ & $1.77^{+0.18}_{-0.76}$ & $100\%$ & \\
& & & & & & & & & $<0.02^{e}$ & $0.63^{+0.11}_{-0.15}$ & $100\%$ & \\
& & & & & & & & & $>461.0$ & $2.34^{+0.67}_{-1.14}$ & $49\%$ & \\
Fairall 9 Obs 1 & $1.81^{+0.01}_{-0.01}$ & $0.61^{+0.01}_{-0.01}$ & $9.2^{+0.7}_{-0.6}$ & $1.1^{+0.1}_{-0.1}$ & $0.42^{+0.01}_{-0.01}$ & $2.2^{+1.3}_{-0.4}$ & $14.8^{+7.7}_{-9.0}$ & $7.94^{+1.83}_{-0.61}$ & -- & -- & -- & $0.83^{+0.10}_{-0.10}$ & \multirow{2}{*}{3604.6/3276} \\
Fairall 9 Obs 2 & $1.81\dagger$ & $0.56^{+0.01}_{-0.01}$ & $9.2\dagger$ & $1.1\dagger$ & $3.87^{+0.03}_{-0.03}$ & $2.2\dagger$ & $14.8\dagger$ & $7.13^{+5.47}_{-3.39}$ & -- & -- & -- & $0.83\dagger$ & \\
IC 4329A Obs 1,2, 3,4,5 & $1.91^{+0.01}_{-0.01}$ & $3.51^{+0.01}_{-0.01}$ & -- & -- & -- & $0.6^{+0.1}_{-0.1}$ & $<1.1$ & $9.15^{+0.18}_{-1.31}$ & $<0.02^{e}$ & $1.53^{+0.29}_{-0.48}$ & $100\%$ & $0.95^{+0.02}_{-0.02}$ & 2421.4/2200 \\
& & & & & & & & & $0.35^{+0.03}_{-0.04}$ & $1.84^{+0.05}_{-0.05}$ & $100\%$ & \\
& & & & & & & & & $0.60^{+0.02}_{-0.02}$ & $<0.01$ & $100\%$ & \\
IRAS 13224--3809 & $2.50^{+0.20}_{-0.16}$ & $<0.02$ & $<14.8$ & $<1.5$ & $0.33^{+0.01}_{-0.01}$ & -- & -- & -- & $<0.04^{e}$ & $>0.63$ & $100\%$ & -- & 484.86/447 \\
& & & & & & & & & $0.22^{+0.01}_{-0.02}$ & Neutral & $100\%$ & \\
& & & & & & & & & $5.00^{+2.17}_{-2.56}$ & $<0.52$ & $72\%$ & \\
MCG--02-14-009 & $1.86^{+0.01}_{-0.04}$ & $0.12^{+0.04}_{-0.03}$ & -- & -- & -- & $0.9^{+0.7}_{-0.3}$ & $<21.0$ & $0.41^{+0.09}_{-0.17}$ & -- & -- & -- & -- & 623.6/543 \\
MCG--02-58-22 & $1.64^{+0.01}_{-0.01}$ & $0.94^{+0.01}_{-0.01}$ & $10.5^{+148.4}_{-5.9}$ & $1.5^{+1.4}_{-1.5}$ & $1.61^{+1.89}_{-1.41}$ & 1.0* & $56.6^{+8.4}_{-5.2}$ & $2.74^{+0.47}_{-0.49}$ & $0.12^{+0.03}_{-0.02}$ & $2.22^{+0.09}_{-0.12}$ & $100\%$ & $0.63^{+0.03}_{-0.02}$ & 2023.7/1892 \\
MCG--05-23-16 & $1.84^{+0.01}_{-0.01}$ & $2.87^{+0.03}_{-0.03}$ & $5.3^{+0.3}_{-0.3}$ & $16.7^{+1.1}_{-1.0}$ & $0.45^{+0.02}_{-0.02}$ & 1.0* & $<1.0$ & $2.76^{+0.21}_{-0.27}$ & $1.46^{+0.01,f}_{-0.01}$ & Neutral & $100\%$ & $0.82^{+0.03}_{-0.03}$ & 1582.8/1466 \\
MCG--06-30-15 Obs 1,2,3 & $1.87^{+0.04}_{-0.04}$ & $1.62^{+0.05}_{-0.04}$ & $<8.8$ & $0.8^{+12.9}_{-0.1}$ & $0.72^{+0.18}_{-0.10}$ & 1.0* & $<12$ & $1.28^{+0.11}_{-0.35}$ & $0.22^{+0.06}_{-0.03}$ & $0.76^{+0.13}_{-0.09}$ & $100\%$ & $0.95^{+0.06}_{-0.07}$ & 2026.8/1823 \\
& & & & & & & & & $0.47^{+0.09}_{-0.03}$ & $1.76^{+0.05}_{-0.08}$ & $100\%$ & \\
& & & & & & & & & $341.0^{+24.0}_{-38.0}$ & $2.43^{+0.02}_{-0.05}$ & $50\%$ & \\
\hline
\end{tabular}
\label{tab:baseline}
\end{minipage}
\end{table}
\end{landscape}

\begin{landscape}
\begin{table}
\begin{minipage}{235mm}
\contcaption{Components for the baseline model (without broad or diskline emission) to the observations with {\sl Suzaku} XIS, HXD and BAT data from {\sl Swift}. $^{a}$ Unabsorbed \textsc{powerlaw} normalisation given in units $(10^{-2}\,{\rm ph\,keV^{-1}\,{\rm cm}^{-2}\,s^{-1}})$. $^{b}$ Flux for \textsc{compTT} quoted over the 0.6-10.0\,keV range and \textsc{reflionx} over the 2.0-100.0\,keV range in units 10$^{-11}$erg\,{\rm cm}$^{-2}$\,s$^{-1}$. $^{c}$ Ionization parameter given in units erg\,{\rm cm}\,s$^{-1}$. $^{d}$ Column density measured in units $10^{22}$\,${\rm cm}^{-2}$. $^{e}$ $v_{\rm turb}=1000$\,${\rm km}\,{\rm s}^{-1}$. $^{f}$ only the intrinsic powerlaw is absorbed. * indicates a frozen parameter. $\dagger$ indicates parameters are tied during the analysis of multiple observations.}
\begin{tabular}{p{2.5cm} c c c c c c c c c c c c c c}
\hline
Object & \multicolumn{2}{l}{Powerlaw} & \multicolumn{3}{l}{\textsc{compTT}} & \multicolumn{3}{l}{\textsc{reflionx}} & \multicolumn{4}{l}{Warm absorber} &  \\
& $\Gamma$ & Norm$^{a}$ & kT (keV) & $\tau$ & Flux$^{b}$ & $Z_{\rm Fe}$ & $\xi^{c}$ & Flux$^{b}$ & $N_{\rm H}^{d}$ & log$(\xi)^{c}$ & $C_{\rm frac}$ & BAT/XIS & $\chi^{2}_{\nu}$ \\
\hline
MCG+8-11-11 & $1.80^{+0.01}_{-0.01}$ & $1.86^{+0.01}_{-0.01}$ & -- & -- & -- & 1.0* & $8.0^{+3.5}_{-2.4}$ & $4.79^{+2.32}_{-1.57}$ & $0.79^{+0.08}_{-0.07}$ & $2.43^{+0.03}_{-0.03}$ & $100\%$ & $0.50^{+0.04}_{-0.04}$ & 1049.9/938 \\
& & & & & & & & & $0.06^{+0.01}_{-0.01}$ & Neutral & $100\%$ & \\
MR 2251--178 & $1.56^{+0.01}_{-0.01}$ & $0.90^{+0.01}_{-0.01}$ & $4.6^{+51.8}_{-2.3}$ & $2.2^{+1.4}_{-1.1}$ & $0.80^{+0.08}_{-0.08}$ & -- & -- & -- & $0.89^{+0.26}_{-0.20}$ & $2.88^{+0.07}_{-0.08}$ & $100\%$ & $1.04^{+0.06}_{-0.06}$ & 977.5/902 \\
& & & & & & & & & $0.54^{+0.03}_{-0.03}$ & $1.68^{+0.03}_{-0.03}$ & $100\%$ & \\
Mrk 79 & $1.55^{+0.02}_{-0.02}$ & $0.32^{+0.01}_{-0.01}$ & -- & -- & -- & -- & -- & -- & $0.06^{+0.04}_{-0.01}$ & $1.50^{+0.21}_{-0.14}$ & $100\%$ & $1.40^{+0.19}_{-0.17}$ & 601.7/545 \\
& & & & & & & & & $0.98^{+0.44}_{-0.28}$ & $2.66^{+0.13}_{-0.14}$ & $100\%$ & \\
Mrk 110 & $1.71^{+0.01}_{-0.01}$ & $0.52^{+0.01}_{-0.01}$ & $<48.7$ & $2.8^{+0.9}_{-0.1}$ & $0.31^{+0.17}_{-0.29}$ & 1.0* & $49.4^{+5.7}_{-31.5}$ & $0.42^{+0.88}_{-0.13}$ & -- & -- & -- & $1.05^{+0.11}_{-0.11}$ & 483.1/471 \\
Mrk 205 & $1.97^{+0.02}_{-0.02}$ & $0.31^{+0.01}_{-0.02}$ & $10.3^{+32.4}_{-8.0}$ & $<2.0$ & $0.08^{+0.20}_{-0.03}$ & 1.0* & $<1.5$ & $0.45^{+0.06}_{-0.17}$ & $50.94^{+54.76}_{-24.99}$ & $2.75^{+0.70}_{-0.52}$ & $13\%$ & $0.80^{+0.15}_{-0.15}$ & 167.7/165 \\
Mrk 279 & $1.76^{+0.01}_{-0.01}$ & $0.11^{+0.01}_{-0.01}$ & -- & -- & -- & 1.0* & $2.8^{+1.0}_{-0.5}$ & $1.14^{+0.29}_{-0.30}$ & $0.08^{+0.05}_{-0.03}$ & $1.39^{+0.30}_{-0.51}$ & $100\%$ & $1.15^{+0.11}_{-0.11}$ & 712.4/669 \\
& & & & & & & & & $>452.53$ & $1.24^{+0.96}_{-0.23}$ & $72\%$ & \\
Mrk 335 & $2.00^{+0.02}_{-0.02}$ & $0.51^{+0.01}_{-0.02}$ & $<8.3$ & $<2.2$ & $0.98^{+0.02}_{-0.02}$ & $2.2^{+1.1}_{-0.4}$ & $27.0^{+9.0}_{-4.0}$ & $0.46^{+0.14}_{-0.14}$ & -- & -- & -- & $1.03^{+0.23}_{-0.23}$ & 842.9/723 \\
Mrk 359 & $1.74^{+0.01}_{-0.01}$ & $0.12^{+0.01}_{-0.01}$ & $<151.5$ & $<4.3$ & $0.08^{+0.28}_{-0.07}$ & 1.0* & $21.8^{+5.8}_{-9.1}$ & $0.52^{+0.44}_{-0.17}$ & -- & -- & -- & $0.72^{+0.20}_{-0.20}$ & 610.1/562 \\
Mrk 509 Obs 1,2, 3,4 & $1.66^{+0.01}_{-0.01}$ & $0.89^{+0.01}_{-0.01}$ & $11.7^{+36.9}_{-5.3}$ & $1.2^{+1.0}_{-1.0}$ & $2.54^{+2.25}_{-1.86}$ & 1.0* & $21.4^{+30.9}_{-18.4}$ & $1.51^{+9.50}_{-0.94}$ & $0.24^{+0.03}_{-0.02}$ & $2.20^{+0.06}_{-0.07}$ & $100\%$ & $0.67^{+0.04}_{-0.03}$ & 1976.2/1870 \\
Mrk 766 Obs 1 & $1.99^{+0.01}_{-0.01}$ & $0.50^{+0.01}_{-0.01}$ & $7.1^{+3.0}_{-2.3}$ & $0.4^{+0.4}_{-0.2}$ & $0.13^{+0.02}_{-0.01}$ & 1.0* & $3.2^{+0.5}_{-0.4}$ & $0.50^{+0.11}_{-0.09}$ & $0.31^{+0.03}_{-0.02}$ & $1.20^{+0.06}_{-0.07}$ & $100\%$ & $0.98^{+0.12}_{-0.12}$ & \multirow{3}{*}{1057.1/1000} \\
Mrk 766 Obs 2 & $1.99\dagger$ & $<0.05$ & $7.1\dagger$ & $0.4\dagger$ & $0.15^{+0.02}_{-0.02}$ & $1.0\dagger$ & $3.2\dagger$ & $0.68^{+0.14}_{-0.12}$ & $2.41^{+0.10}_{-0.09}$ & $1.81^{+0.03}_{-0.03}$ & $100\%$ & $0.98\dagger$ & \\
& & & & & & & & & $5.88^{+1.40}_{-0.98}$ & $2.94^{+0.05}_{-0.05}$ & $>92\%$ & \\
Mrk 841 Obs 1,2 & $1.79^{+0.01}_{-0.01}$ & $0.36^{+0.01}_{-0.01}$ & $<116.9$ & $<3.9$ & $0.30^{+1.44}_{-0.23}$ & $0.8^{+0.2}_{-0.2}$ & $3.9^{+2.0}_{-1.4}$ & $1.43^{+0.86}_{-0.51}$ & $0.46^{+0.09}_{-0.10}$ & $2.17^{+0.10}_{-0.11}$ & $100\%$ & $0.71^{+0.09}_{-0.08}$ & 937.6/857 \\ 
NGC 1365 Obs 1 & $1.69^{+0.01}_{-0.01}$ & $0.03^{+0.01}_{-0.01}$ & -- & -- & -- & 1.0* & $24.3^{+1.0}_{-0.9}$ & $1.20^{+0.10}_{-0.09}$ & $2.71^{+0.69}_{-0.74}$ & Neutral & \multirow{2}{*}{$96\%$} & $0.76^{+0.04}_{-0.04}$ & \multirow{6}{*}{2172.5/1979} \\
& & & & & & & & & $29.57^{+3.35}_{-3.00}$ & $0.80^{+0.17}_{-0.16}$ & \\
NGC 1365 Obs 2 & $1.69\dagger$ & $0.02^{+0.01}_{-0.01}$ & -- & -- & -- & $1.0\dagger$ & $24.3\dagger$ & $1.36^{+0.11}_{-0.10}$ & $36.79^{+1.98}_{-1.92}$ & Neutral & \multirow{2}{*}{$96\%$} & $0.76\dagger$ \\
& & & & & & & & & $16.41^{+7.16}_{-6.72}$ & $2.16^{+0.18}_{-0.39}$ & \\
NGC 1365 Obs 3 & $1.69\dagger$ & $0.02^{+0.01}_{-0.01}$ & -- & -- & -- & $1.0\dagger$ & $24.3\dagger$ & $1.43^{+0.06}_{-0.08}$ & $65.38^{+6.18}_{-3.33}$ & Neutral & \multirow{2}{*}{$95\%$} & $0.76\dagger$ \\
& & & & & & & & & $17.99^{+6.29}_{-15.77}$ & $<1.90$ & \\
NGC 2992 Obs 1, 2,3 & $1.58^{+0.01}_{-0.01}$ & $0.26^{+0.01}_{-0.01}$ & $<16.6$ & $0.76^{+0.04}_{-0.10}$ & $1.45^{+0.31}_{-1.14}$ & 1.0* & $54.8^{+7.8}_{-8.3}$ & $1.59^{+0.29}_{-0.29}$ & $0.84^{+0.01}_{-0.01}$ & Neutral & $100\%$ & $0.79^{+0.13}_{-0.13}$ & 1088.2/1079 \\
NGC 3147 & $1.72^{+0.02}_{-0.03}$ & $0.04^{+0.01}_{-0.01}$ & -- & -- & -- & 1.0* & $49.3^{+23.4}_{-26.9}$ & $0.11^{+0.15}_{-0.06}$ & -- & -- & -- & -- & 280.3/266 \\
\hline
\end{tabular}
\end{minipage}
\end{table}
\end{landscape}

\begin{landscape}
\begin{table}
\begin{minipage}{235mm}
\contcaption{Components for the baseline model (without broad or diskline emission) to the observations with {\sl Suzaku} XIS, HXD and BAT data from {\sl Swift}. $^{a}$ Unabsorbed \textsc{powerlaw} normalisation given in units $(10^{-2}\,{\rm ph\,keV^{-1}\,{\rm cm}^{-2}\,s^{-1}})$. $^{b}$ Flux for \textsc{compTT} quoted over the 0.6-10.0\,keV range and \textsc{reflionx} over the 2.0-100.0\,keV range in units 10$^{-11}$erg\,{\rm cm}$^{-2}$\,s$^{-1}$. $^{c}$ Ionization parameter given in units erg\,{\rm cm}\,s$^{-1}$. $^{d}$ Column density measured in units $10^{22}$\,${\rm cm}^{-2}$. $^{e}$ $v_{\rm turb}=1000$\,${\rm km}\,{\rm s}^{-1}$. $^{f}$ only the intrinsic powerlaw is absorbed. * indicates a frozen parameter. $\dagger$ indicates parameters are tied during the analysis of multiple observations.}
\begin{tabular}{p{2.5cm} c c c c c c c c c c c c c c}
\hline
Object & \multicolumn{2}{l}{Powerlaw} & \multicolumn{3}{l}{\textsc{compTT}} & \multicolumn{3}{l}{\textsc{reflionx}} & \multicolumn{4}{l}{Warm absorber} &  \\
& $\Gamma$ & Norm$^{a}$ & kT (keV) & $\tau$ & Flux$^{b}$ & $Z_{\rm Fe}$ & $\xi^{c}$ & Flux$^{b}$ & $N_{\rm H}^{d}$ & log$(\xi)^{c}$ & $C_{\rm frac}$ & BAT/XIS & $\chi^{2}_{\nu}$ \\
\hline
NGC 3227 Obs 1 & $1.92^{+0.01}_{-0.01}$ & $1.11^{+0.02}_{-0.04}$ & $4.8^{+0.4}_{-0.4}$ & $1.2^{+0.1}_{-0.1}$ & $4.42^{+0.08}_{-0.08}$ & $0.2^{+0.1}_{-0.1}$ & $2.3^{+0.1}_{-0.1}$ & $15.77^{+0.88}_{-0.89}$ & $3.37^{+0.11}_{-0.07}$ & $2.05^{+0.02}_{-0.01}$ & $100\%$ & $0.82^{+0.03}_{-0.03}$ & \multirow{18}{*}{4585.7/4198} \\
& & & & & & & & & $0.43^{+0.01}_{-0.01}$ & Neutral & $100\%$ & \\
& & & & & & & & & $29.24^{+0.46}_{-0.46}$ & $0.50^{+0.18}_{-0.27}$ & $32\%$ & \\
NGC 3227 Obs 2 & $1.79\dagger$ & $0.13^{+0.01}_{-0.01}$ & $4.8\dagger$ & $1.2\dagger$ & $0.93^{+0.03}_{-0.03}$ & $0.2\dagger$ & $2.3\dagger$ & $13.274^{+0.37}_{-0.37}$ & $3.37\dagger$ & $2.05\dagger$ & $100\%$ & $0.82\dagger$ & \\
& & & & & & & & & $0.43\dagger$ & Neutral & $100\%$ & \\
& & & & & & & & & $29.24\dagger$ & $1.53^{+0.09}_{-0.10}$ & $81\%$ & \\
NGC 3227 Obs 3 & $1.79\dagger$ & $0.25^{+0.01}_{-0.01}$ & $4.8\dagger$ & $1.2\dagger$ & $0.76^{+0.03}_{-0.03}$ & $0.2\dagger$ & $2.3\dagger$ & $13.17^{+0.59}_{-0.59}$ & $3.37\dagger$ & $2.05\dagger$ & $100\%$ & $0.82\dagger$ & \\
& & & & & & & & & $0.43\dagger$ & Neutral & $100\%$ & \\
& & & & & & & & & $29.24\dagger$ & $0.50^{+0.03}_{-0.13}$ & $81\%$ & \\
NGC 3227 Obs 4 & $1.79\dagger$ & $0.07^{+0.01}_{-0.01}$ & $4.8\dagger$ & $1.2\dagger$ & $0.91^{+0.02}_{-0.02}$ & $0.2\dagger$ & $2.3\dagger$ & $8.13^{+0.22}_{-0.22}$ & $3.37\dagger$ & $2.05\dagger$ & $100\%$ & $0.82\dagger$ & \\
& & & & & & & & & $0.43\dagger$ &Neutral & $100\%$ & \\
& & & & & & & & & $29.24\dagger$ & $1.50^{+0.12}_{-0.13}$ & $75\%$ & \\
NGC 3227 Obs 5 & $1.79\dagger$ & $0.13^{+0.01}_{-0.01}$ & $4.8\dagger$ & $1.2\dagger$ & $0.79^{+0.02}_{-0.02}$ & $0.2\dagger$ & $2.3\dagger$ & $11.79^{+0.48}_{-0.48}$ & $3.37\dagger$ & $2.05\dagger$ & $100\%$ & $0.82\dagger$ & \\
& & & & & & & & & $0.43\dagger$ & Neutral & $100\%$ & \\
& & & & & & & & & $29.24\dagger$ & $0.70^{+0.05}_{-0.05}$ & $89\%$ & \\
NGC 3227 Obs 6 & $1.79\dagger$ & $0.10^{+0.01}_{-0.01}$ & $4.8\dagger$ & $1.2\dagger$ & $0.64^{+0.02}_{-0.02}$ & $0.2\dagger$ & $2.3\dagger$ & $9.46^{+0.47}_{-0.46}$ & $3.37\dagger$ & $2.05\dagger$ & $100\%$ & $0.82\dagger$ & \\
& & & & & & & & & $0.43\dagger$ & Neutral & $100\%$ & \\
& & & & & & & & & $29.24\dagger$ & $1.41^{+0.12}_{-0.23}$ & $89\%$ & \\
NGC 3516 Obs 1 & $1.68^{+0.01}_{-0.01}$ & $0.15^{+0.01}_{-0.01}$ & -- & -- & -- & 1.0* & $6.3^{+0.3}_{-0.3}$ & $3.50^{+0.22}_{-0.20}$ & $13.40^{+9.06}_{-6.40}$ & $3.55^{+0.22}_{-0.10}$ & $100\%$ & $0.82^{+0.02}_{-0.02}$ & \multirow{8}{*}{1263.2/1126} \\
& & & & & & & & & $2.76^{+0.30}_{-0.17}$ & $1.93^{+0.09}_{-0.06}$ & $100\%$ & \\
& & & & & & & & & $0.12^{+0.02}_{-0.01}$ & Neutral & $100\%$ & \\
& & & & & & & & & $24.83^{+1.74}_{-1.69}$ & $0.60^{+0.08}_{-0.09}$ & $83\%$ & \\
NGC 3516 Obs 2 & $1.68\dagger$ & $0.29^{+0.01}_{-0.01}$ & -- & -- & -- & $1.0\dagger$ & $6.3\dagger$ & $1.96^{+0.12}_{-0.12}$ & $0.35^{+0.09}_{-0.06}$ & $1.02^{+0.12}_{-0.13}$ & $100\%$ & $0.82\dagger$ & \\
& & & & & & & & & $1.54^{+0.10}_{-0.26}$ & $2.20^{+0.05}_{-0.06}$ & $100\%$ & \\
& & & & & & & & & $0.08^{+0.02}_{-0.02}$ & Neutral & $100\%$ & \\
& & & & & & & & & $69.18^{+25.66}_{-24.57}$ & $2.80^{+0.10}_{-0.10}$ & $18\%$ & \\
NGC 3783 Obs 1 & $1.79^{+0.01}_{-0.01}$ & $1.30^{+0.01}_{-0.01}$ & $7.4^{+0.5}_{-0.4}$ & $1.1^{+0.1}_{-0.1}$ & $0.78^{+0.02}_{-0.02}$ & $0.9^{+0.1}_{-0.1}$ & $4.4^{+0.1}_{-0.1}$ & $7.99^{+0.26}_{-0.26}$ & $0.16^{+0.02}_{-0.02}$ & $0.27^{+0.14}_{-0.09}$ & $100\%$ & $0.99^{+0.03}_{-0.03}$ & \multirow{4}{*}{2554.2/2305} \\
& & & & & & & & & $3.02^{+0.07}_{-0.08}$ & $2.05^{+0.01}_{-0.01}$ & $100\%$ & \\
NGC 3783 Obs 2 & $1.79\dagger$ & $1.73^{+0.01}_{-0.01}$ & $7.4\dagger$ & $1.1\dagger$ & $2.58^{+0.08}_{-0.08}$ & $0.9\dagger$ & $4.4\dagger$ & $8.26^{+0.27}_{-0.26}$ & $0.27^{+0.03}_{-0.03}$ & $0.83^{+0.05}_{-0.06}$ & $100\%$ & $0.99\dagger$ \\
& & & & & & & & & $3.81^{+0.08}_{-0.08}$ & $2.09^{+0.01}_{-0.01}$ & $100\%$ & \\
NGC 4051 Obs 1 & $1.88^{+0.01}_{-0.01}$ & $0.10^{+0.01}_{-0.01}$ & $6.9^{+0.2}_{-0.2}$ & $1.5^{+0.1}_{-0.1}$ & $0.13^{+0.01}_{-0.01}$ & $0.9^{+0.1}_{-0.1}$ & $9.6^{+0.4}_{-0.5}$ & $1.07^{+0.07}_{-0.06}$ & $0.29^{+0.06}_{-0.05}$ & $2.97^{+0.05}_{-0.04}$ & $100\%$ & $0.70^{+0.05}_{-0.05}$ & \multirow{9}{*}{3204.7/2944} \\
& & & & & & & & & $0.31^{+0.01}_{-0.01}$ & $1.85^{+0.01}_{-0.01}$ & $100\%$ & \\
& & & & & & & & & $8.96^{+0.47}_{-0.26}$ & $1.95^{+0.03}_{-0.03}$ & $67\%$ & \\
NGC 4051 Obs 2 & $1.88\dagger$ & $0.61^{+0.01}_{-0.01}$ & $6.9\dagger$ & $1.5\dagger$ & $1.56^{+0.02}_{-0.02}$ & $0.9\dagger$ & $9.6\dagger$ & $1.54^{+0.10}_{-0.08}$ & $0.29\dagger$ & $2.97\dagger$ & $100\%$ & $0.70\dagger$ & \\
& & & & & & & & & $0.31\dagger$ & $1.85\dagger$ & $100\%$ & \\
& & & & & & & & & $8.96\dagger$ & $1.95\dagger$ & $11\%$ & \\
NGC 4051 Obs 3 & $1.88\dagger$ & $0.39^{+0.01}_{-0.01}$ & $6.9\dagger$ & $1.5\dagger$ & $1.06^{+0.01}_{-0.01}$ & $0.9\dagger$ & $9.6\dagger$ & $1.40^{+0.12}_{-0.13}$ & $0.29\dagger$ & $2.97\dagger$ & $100\%$ & $0.70\dagger$ & \\
& & & & & & & & & $0.31\dagger$ & $1.85\dagger$ & $100\%$ & \\
& & & & & & & & & $8.96\dagger$ & $1.95\dagger$ & $27\%$ & \\
\hline
\end{tabular}
\end{minipage}
\end{table}
\end{landscape}

\begin{landscape}
\begin{table}
\begin{minipage}{235mm}
\contcaption{Components for the baseline model (without broad or diskline emission) to the observations with {\sl Suzaku} XIS, HXD and BAT data from {\sl Swift}. $^{a}$ Unabsorbed \textsc{powerlaw} normalisation given in units $(10^{-2}\,{\rm ph\,keV^{-1}\,{\rm cm}^{-2}\,s^{-1}})$. $^{b}$ Flux for \textsc{compTT} quoted over the 0.6-10.0\,keV range and \textsc{reflionx} over the 2.0-100.0\,keV range in units 10$^{-11}$erg\,{\rm cm}$^{-2}$\,s$^{-1}$. $^{c}$ Ionization parameter given in units erg\,{\rm cm}\,s$^{-1}$. $^{d}$ Column density measured in units $10^{22}$\,${\rm cm}^{-2}$. $^{e}$ $v_{\rm turb}=1000$\,${\rm km}\,{\rm s}^{-1}$. $^{f}$ only the intrinsic powerlaw is absorbed. * indicates a frozen parameter. $\dagger$ indicates parameters are tied during the analysis of multiple observations.}
\begin{tabular}{p{2.5cm} c c c c c c c c c c c c c c}
\hline
Object & \multicolumn{2}{l}{Powerlaw} & \multicolumn{3}{l}{\textsc{compTT}} & \multicolumn{3}{l}{\textsc{reflionx}} & \multicolumn{4}{l}{Warm absorber} &  \\
& $\Gamma$ & Norm$^{a}$ & kT (keV) & $\tau$ & Flux$^{b}$ & $Z_{\rm Fe}$ & $\xi^{c}$ & Flux$^{b}$ & $N_{\rm H}^{d}$ & log$(\xi)^{c}$ & $C_{\rm frac}$ & BAT/XIS & $\chi^{2}_{\nu}$ \\
\hline
NGC 4151 & $1.56^{+0.01}_{-0.01}$ & $0.27^{+0.01}_{-0.01}$ & $3.4^{+0.1}_{-0.1}$ & $10.6^{+0.2}_{-0.2}$ & $3.63^{+0.07}_{-0.07}$ & 1.0* & $43.6^{+7.1}_{-9.9}$ & $12.63^{+3.01}_{-1.48}$ & $0.09^{+0.02}_{-0.01}$ & $<0.12$ & $100\%$ & $1.57^{+0.02}_{-0.02}$ & 647.1/583 \\
& & & & & & & & & $14.45^{+0.19}_{-0.18}$ & $2.16^{+0.01}_{-0.01}$ & $100\%$ & \\
& & & & & & & & & $45.82^{+3.58}_{-7.78}$ & $0.52^{+0.26}_{-0.16}$ & $58\%$ & \\
NGC 4593 & $1.60^{+0.01}_{-0.01}$ & $0.21^{+0.01}_{-0.01}$ & -- & -- & -- & 1.0* & $5.4^{+14.7}_{-5.4}$ & $0.61^{+3.24}_{-0.46}$ & $0.53^{+0.09}_{-0.10}$ & $2.19^{+0.11}_{-0.11}$ & $100\%$ & $1.87^{+0.12}_{-0.11}$ & 293.62/289 \\
NGC 5506 Obs 1,2 & $2.08^{+0.01}_{-0.01}$ & $5.58^{+0.02}_{-0.02}$ & -- & -- & -- & $0.5^{+0.1}_{-0.1}$ & $10.5^{+0.2}_{-0.2}$ & $9.69^{+0.35}_{-0.34}$ & $2.23^{+0.55}_{-0.45}$ & $<0.05$ & $100\%$ & $0.79^{+0.01}_{-0.01}$ & \multirow{8}{*}{3189.6/2872} \\
& & & & & & & & & $3.26^{+0.31}_{-0.33}$ & $0.59^{+0.10}_{-0.06}$ & $100\%$ & \\
& & & & & & & & & $1.51^{+0.02\,e}_{-0.02}$ & $1.30^{+0.02}_{-0.02}$ & $100\%$ & \\
& & & & & & & & & $449.0^{+40.0}_{-36.0}$ & $<1.20$ & $19\%$ & \\
NGC 5506 Obs 3 & $2.08\dagger$ & $5.36^{+0.02}_{-0.02}$ & -- & -- & -- & $0.5\dagger$ & $10.5\dagger$ & $8.48^{+0.31}_{-0.31}$ & $2.23\dagger$ & $0.05\dagger$ & $100\%$ & $0.79\dagger$ & \\
& & & & & & & & & $3.26\dagger$ & $0.59\dagger$ & $100\%$ & \\
& & & & & & & & & $1.51^{e}\dagger$ & $1.30\dagger$ & $100\%$ & \\
& & & & & & & & & $448.76\dagger$ & $1.20\dagger$ & $23\%$ & \\
NGC 5548 Obs 1,2, 3,4,5,6,7 & $1.70^{+0.01}_{-0.01}$ & $0.44^{+0.01}_{-0.01}$ & -- & -- & -- & $1.0^{+0.1}_{-0.1}$ & $<2.2$ & $2.12^{+1.27}_{-0.75}$ & $0.59^{+0.09}_{-0.13}$ & $2.28^{+0.07}_{-0.08}$ & $100\%$ & $0.86^{+0.04}_{-0.04}$ & 1772.0/1734 \\
& & & & & & & & & $0.26^{+0.06}_{-0.03}$ & $0.98^{+0.14}_{-0.08}$ & $100\%$ & \\
NGC 7213 & $1.74^{+0.01}_{-0.01}$ & $0.61^{+0.01}_{-0.01}$ & $<61.8$ & $2.1^{+0.5}_{-1.6}$ & $0.03^{+0.08}_{-0.02}$ & 1.0* & $29.6^{+26.9}_{-16.9}$ & $0.33^{+0.63}_{-0.20}$ & -- & -- & -- & $0.61^{+0.06}_{-0.06}$ & 703.5/707 \\
NGC 7314 & $1.68^{+0.01}_{-0.01}$ & $0.22^{+0.01}_{-0.01}$ & $<15.8$ & $<0.8$ & $0.34^{+1.03}_{-0.26}$ & -- & -- & -- & $0.74^{+0.02}_{-0.02}$ & Neutral & $100\%$ & $1.84^{+0.16}_{-0.16}$ & 592.3/545 \\
NGC 7469 & $1.78^{+0.07}_{-0.10}$ & $0.57^{+0.04}_{-0.06}$ & $<8.4$ & $0.8^{+0.2}_{-0.5}$ & $0.22^{+0.03}_{-0.03}$ & $1.6^{+0.3}_{-0.3}$ & $<11.0$ & $1.69^{+0.14}_{-0.29}$ & -- & -- & -- & $1.03^{+0.23}_{-0.23}$ & 840.2/812 \\
PDS 456 & $2.41^{+0.03}_{-0.03}$ & $0.23^{+0.01}_{-0.01}$ & $<33.4$ & $1.4^{+0.6}_{-1.1}$ & $<0.204$ & -- & -- & -- & $421.40$ & $3.16^{+0.19}_{-0.08}$ & $50\%$ & -- & 146.9/132 \\
& & & & & & & & & $3.82^{+1.90}_{-1.09}$ & Neutral & $50\%\dagger$ & \\
PG 1211+143 & $1.82^{+0.01}_{-0.01}$ & $0.12^{0.01}_{-0.01}$ & $<18.3$ & $0.8^{+0.3}_{-0.7}$ & $<0.11$ & -- & -- & -- & $3.05^{+0.95}_{-0.67}$ & $2.85^{+0.06}_{-0.08}$ & $100\%$ & -- & 813.62/702 \\
RBS 1124 & $1.71^{+0.01}_{-0.01}$ & $0.12^{+0.01}_{-0.01}$ & $<52.9$ & $0.8^{+11.3}_{-0.4}$ & $<0.19$ & $0.7^{+0.5}_{-0.3}$ & $<15.1$ & $0.32^{+4.34}_{-0.21}$ & -- & -- & -- & -- & 386.8/364 \\
SWIFT J2127.4+5654 & $2.11^{+0.03}_{-0.02}$ & $1.52^{+0.04}_{-0.05}$ & -- & -- & -- & $0.5^{+0.1}_{-0.1}$ & $<13.0$ & $2.28^{+0.37}_{-1.11}$ & $0.08^{+0.02}_{-0.02}$ & Neutral & $100\%$ & $0.70^{+0.10}_{-0.10}$ & 871.6/867 \\
TON S180 & $2.14^{+0.02}_{-0.01}$ & $0.23^{+0.01}_{-0.01}$ & $6.3^{+54.9}_{-4.1}$ & $1.2^{+1.8}_{-1.2}$ & $<1.14$ & $2.4^{+0.8}_{-0.5}$ & $269.6^{+65.0}_{-30.5}$ & $0.10^{+0.03}_{-0.03}$ & $0.02^{+0.01}_{-0.01}$ & $1.44^{+0.34}_{-1.06}$ & $100\%$ & -- & 750.0/692 \\
& & & & & & & & & $>415.20$ & $1.99^{+0.54}_{-0.74}$ & $60\%$ \\
\hline
\end{tabular}
\end{minipage}
\end{table}
\end{landscape}

\begin{table*}
\caption{Fe\,K region properties - distant emission lines and ionized absorption zones in the baseline model. We typically model the absorption zones with an \textsc{xstar} grid with $v_{\rm turb}=1000$\,${\rm km}\,{\rm s}^{-1}$. $^{a}$ $v_{\rm turb}=3000$\,${\rm km}\,{\rm s}^{-1}$. Line flux given in units 10$^{-5}$erg\,{\rm cm}$^{-2}$\,s$^{-1}$. Column density given in units $10^{22}$\,${\rm cm}^{-2}$. $\dagger$ indicates that parameters are tied in multiple observations. * denotes a frozen parameter. }
\begin{tabular}{l c c c c c c c c}
\hline
Object & \multicolumn{4}{c}{Emission lines} & High $\xi$ zone \\
& LineE & $EW$ (eV) & Flux & $\triangle\chi^{2}$ & $N_{\rm H}$ & log$\xi$ & $v_{\rm out}$\,(${\rm km}\,{\rm s}^{-1}$) & $\triangle\chi^{2}$ \\
\hline
3C 111 & $6.50^{+0.06}_{-0.05}$ & $19^{+23}_{-11}$ & $0.53^{+0.62}_{-0.30}$ & -9 & $>3.60$ & $4.40^{+1.11}_{-0.37}$ & $84600^{+42600}_{-42400}$ & -28 \\
& $6.78^{+0.11}_{-0.08}$ & $11^{+8}_{-8}$ & $0.27^{+0.21}_{-0.21}$ & -5 & \\
3C 120 & $6.76^{+0.07}_{-0.07}$ & $8^{+4}_{-4}$ & $0.42^{+0.21}_{-0.20}$ & -12 & -- & -- & -- \\
3C 445 & $6.68^{+0.07}_{-0.10}$ & $17^{+402}_{-9}$ & $0.29^{+6.74}_{-0.15}$ & -6 & $>22.88$ & $4.95^{+0.05}_{-0.60}$ & $4000^{+600}_{-500}$ & -8 \\
4C 74.26 & $6.68^{+0.09}_{-0.08}$ & $10^{+7}_{-7}$ & $0.43^{+0.31}_{-0.31}$ & -5 & -- & -- & -- \\
Ark 120 & $6.66^{+0.04}_{-0.05}$ & $20^{+7}_{-7}$ & $0.72^{+0.25}_{-0.25}$ & -20 & -- & -- & -- \\
& $6.95^{+0.03}_{-0.03}$ & $31^{+9}_{-9}$ & $0.92^{+0.27}_{-0.27}$ & -31 & \\
Ark 564 & 6.63* & $20^{+16}_{-13}$ & $0.31^{+0.25}_{-0.20}$ & -7 & -- & -- & -- \\ 
Fairall 9 & $6.71^{+0.03}_{-0.02}$ & $20^{+5}_{-6}$ & $0.58^{+0.15}_{-0.17}$ & -32 & -- & -- & -- \\
& $6.96^{+0.03}_{-0.03}$ & $20^{+6}_{-6}$ & $0.48^{+0.13}_{-0.15}$ & -25 & \\
IC 4329A & $6.95^{+0.05}_{-0.06}$ & $9^{+5}_{-5}$ & $0.93^{+0.53}_{-0.51}$ & -9 & -- & -- & -- \\
IRAS 13224--3809 & $6.72^{+0.06}_{-0.06}$ & $101^{+67}_{-67}$ & $0.06^{+0.04}_{-0.04}$ & -6 & -- & -- & -- \\
MCG--02-14-009 & $6.94^{+0.05}_{-0.11}$ & $39^{+24}_{-23}$ & $0.16^{+0.10}_{-0.10}$ & -7 & -- & -- & -- \\
MCG--5-23-16 & $6.40^{+0.01}_{-0.01}$ & $54^{+7}_{-6}$ & $6.02^{+0.73}_{-0.68}$ & -82 & -- & -- & -- \\
MCG--6-30-15 & 6.97* & $10^{+4}_{-4}$ & $0.41^{+0.18}_{-0.18}$ & -7 & $3.99^{+3.65}_{-1.28}$ & $3.94^{+0.08}_{-0.25}$ & $3200^{+400}_{-500}$ & -298 \\
MCG+8-11-11 & $6.93^{+0.03}_{-0.03}$ & $21^{+6}_{-6}$ & $1.39^{+0.40}_{-0.40}$ & -17 & -- & -- & -- \\
MR 2251--178 & $6.48^{+0.06}_{-0.06}$ & $12^{+9}_{-9}$ & $0.66^{+0.47}_{-0.50}$ & -4 & $0.29^{+0.31}_{-0.09}$ & $3.12^{+0.23}_{-0.08}$ & $33600^{+7000}_{-7700}$ & -39 \\
Mrk 79 & $6.39^{+0.01}_{-0.01}$ & $110^{+15}_{-15}$ & $2.05^{+0.28}_{-0.28}$ & -45 & -- & -- & -- \\
& $6.60^{+0.05}_{-0.06}$ & $26^{+14}_{-13}$ & $0.51^{+0.27}_{-0.25}$ & -11 & -- & -- & -- \\
& $6.98^{+0.06}_{-0.07}$ & $20^{+17}_{-17}$ & $0.30^{+0.25}_{-0.25}$ & -5 & --& -- & -- \\
Mrk 110 & $6.66^{+0.12}_{-0.11}$ & $11^{+10}_{-10}$ & $0.26^{+0.25}_{-0.24}$ & -4 & -- & -- & -- \\
Mrk 335 & $6.68^{+0.03}_{-0.05}$ & $40^{+8}_{-8}$ & $0.64^{+0.13}_{-0.13}$ & -67 & -- & -- & -- \\
& $6.96^{+0.06}_{-0.16}$ & $21^{+9}_{-9}$ & $0.28^{+0.12}_{-0.12}$ & -14 \\
Mrk 359 & $6.74^{+0.05}_{-0.04}$ & $32^{+18}_{-19}$ & $0.19^{+0.11}_{-0.11}$ & -6 & -- & -- & -- \\
Mrk 509 & $6.43^{+0.03}_{-0.02}$ & $23^{+17}_{-11}$ & $1.29^{+0.98}_{-0.51}$ & -9 & -- & -- & -- \\
& $6.67^{+0.08}_{-0.10}$ & $8^{+12}_{-6}$ & $0.45^{+0.71}_{-0.37}$ & -4 & \\
Mrk 766 & $6.64^{+0.04}_{-0.05}$ & $27^{+10}_{-10}$ & $0.37^{+0.14}_{-0.14}$ & -28 & $>3.90$ & $5.44^{+0.41}_{-1.16}$ & $5200^{+2900}_{-2900}$ & -108 \\
Mrk 766 & $6.64\dagger$ & $27\dagger$ & $0.37\dagger$ & -- & $7.83^{+3.54}_{-2.61}$ & $3.68^{+0.08}_{-0.11}$ & $5200\dagger$ \\
Mrk 841 & $6.69^{+0.06}_{-0.06}$ & $19^{+12}_{-13}$ & $0.32^{+0.20}_{-0.22}$ & -5 & -- & -- & -- \\
NGC 1365 & $6.64^{+0.01}_{-0.01}$ & $26^{+5}_{-5}$ & $0.71^{+0.14}_{-0.15}$ & -11 & $11.07^{+5.93\,a}_{-2.19}$ & $3.74^{+0.07}_{-0.06}$ & $<300$ & -2052\\
& $6.89^{+0.01}_{-0.01}$ & $28^{+6}_{-5}$ & $0.41^{+0.09}_{-0.08}$ & -32 & $52.84^{+7.43\,a}_{-13.95}$ & $3.96^{+0.02}_{-0.04}$ & $4900^{+400}_{-500}$ \\
NGC 1365 & $6.64\dagger$ & $24^{+5}_{-5}$ & $0.71\dagger$ & -- & $41.53^{+17.64\,a}_{-11.58}$ & $4.00^{+0.21}_{-0.03}$ & $<300\dagger$ \\
& $6.89\dagger$ & $41^{+9}_{-8}$ & $0.41\dagger$ & -- & $<6.72^{a}$ & $3.46^{+0.02}_{-0.02}$ & $<500$ \\
NGC 1365 & $6.64\dagger$ & $37^{+7}_{-8}$ & $0.71\dagger$ & -- & $<190.40^{a}$ & $>4.11$ & $<300\dagger$ \\
& $6.89\dagger$ & $42^{+9}_{-8}$ & $0.41\dagger$ & -- & $5.98^{+0.49\,a}_{-0.48}$ & $3.41^{+0.02}_{-0.02}$ & $<200$ \\
NGC 2992 & $6.40^{+0.01}_{-0.01}$ & $154^{+18}_{-18}$ & $2.60^{+0.30}_{-0.30}$ & -23 & -- & -- & -- \\
NGC 3147 & $6.97^{+0.05}_{-0.06}$ & $84^{+43}_{-54}$ & $0.13^{+0.07}_{-0.08}$ & -6 & -- & -- & -- \\
NGC 3227 & $6.40^{+0.01}_{-0.01}$ & $59^{+3}_{-4}$ & $2.98^{+0.17}_{-0.19}$ & -80 & $20.88^{+2.48}_{-3.04}$ & $4.35^{+0.06}_{-0.08}$ & $<2100$ & -67\\
& $6.83^{+0.01}_{-0.01}$ & $28^{+3}_{-3}$ & $1.20^{+0.13}_{-0.13}$ & -25 & \\
NGC 3227 & $6.40\dagger$ & $107^{+6}_{-7}$ & $2.98\dagger$ & -- & $20.88\dagger$ & $4.35\dagger$ & $<2100\dagger$ \\
& $6.83\dagger$ & $56^{+6}_{-6}$ & $1.20\dagger$ & -- & \\
NGC 3227 & $6.40\dagger$ & $75^{+4}_{-5}$ & $2.98\dagger$ & -- & $20.88\dagger$ & $4.35\dagger$ & $<2100\dagger$ \\
& $6.83\dagger$ & $37^{+4}_{-4}$ & $1.20\dagger$ & -- & \\
NGC 3227 & $6.40\dagger$ & $168^{+9}_{-10}$ & $2.98\dagger$ & -- & $20.88\dagger$ & $4.35\dagger$ & $<2100\dagger$ \\
& $6.83\dagger$ & $95^{+11}_{-10}$ & $1.20\dagger$ & -- & \\
NGC 3227 & $6.40\dagger$ & $85^{+5}_{-5}$ & $2.98\dagger$ & -- & $20.88\dagger$ & $4.35\dagger$ & $<2100\dagger$ \\
& $6.83\dagger$ & $42^{+5}_{-5}$ & $1.20\dagger$ & -- & \\
NGC 3227 & $6.40\dagger$ & $109^{+6}_{-7}$ & $2.98\dagger$ & -- & $20.88\dagger$ & $4.35\dagger$ & $<2100\dagger$ \\
& $6.83\dagger$ & $56^{+6}_{-6}$ & $1.20\dagger$ & -- & \\
NGC 3516 & $6.42^{+0.01}_{-0.01}$ & $51^{+4}_{-4}$ & $2.19^{+0.18}_{-0.18}$ & -15 & $2.12^{+1.24}_{-0.95}$ & $3.87^{+0.18}_{-0.14}$ & $<4500$ & -23 \\
& $6.69^{+0.03}_{-0.04}$ & $8^{+3}_{-3}$ & $0.39^{+0.14}_{-0.13}$ & -8 \\
NGC 3516 & $6.42\dagger$ & $108^{+9}_{-9}$ & $2.19\dagger$ & -- & $2.33^{+2.79}_{-0.82}$ & $3.80^{+0.25}_{-0.12}$ & $<9600$ \\
& $6.69\dagger$ & $19^{+7}_{-6}$ & $0.9\dagger$ & -- \\
NGC 3783 & $6.96^{+0.01}_{-0.02}$ & $30^{+7}_{-8}$ & $2.37^{+0.58}_{-0.64}$ & -21 & $5.46^{+21.31}_{-2.02}$ & $4.22^{+0.66}_{-0.19}$ & $<400$ & -50 \\
NGC 3783 & $6.96\dagger$ & $26^{+7}_{-7}$ & $2.37\dagger$ & -- & $2.40^{+0.75}_{-0.58}$ & $3.87^{+0.11}_{-0.11}$ & $<400\dagger$ \\
\hline
\end{tabular}
\label{tab:FeK}
\end{table*}
\begin{table*}
\contcaption{Fe\,K region properties - distant emission lines and ionized absorption zones in the baseline model. We typically model the absorption zones with an \textsc{xstar} grid with $v_{\rm turb}=1000$\,${\rm km}\,{\rm s}^{-1}$. $^{a}$ $v_{\rm turb}=3000$\,${\rm km}\,{\rm s}^{-1}$. Line flux given in units 10$^{-5}$erg\,{\rm cm}$^{-2}$\,s$^{-1}$. Column density given in units $10^{22}$\,${\rm cm}^{-2}$. $\dagger$ indicates that parameters are tied in multiple observations. * denotes a frozen parameter.}
\begin{tabular}{l c c c c c c c c}
\hline
Object & \multicolumn{4}{c}{Emission lines} & High $\xi$ zone \\
& LineE & $EW$ (eV) & Flux & $\triangle\chi^{2}$ & $N_{\rm H}$ & log$\xi$ & $v_{\rm out}$\,(${\rm km}\,{\rm s}^{-1}$) & $\triangle\chi^{2}$ \\
\hline
NGC 4051 & $6.43^{+0.03}_{-0.02}$ & $42^{+2}_{-2}$ & $0.53^{+0.03}_{-0.03}$ & -5 & $1.17^{+0.27}_{-0.26}$ & $3.68^{+0.05}_{-0.05}$ & $5800^{+1400}_{-1300}$ & -51 \\
& $6.62^{+0.03}_{-0.04}$ & $21^{+9}_{-9}$ & $0.26^{+0.11}_{-0.11}$ & -3 \\
NGC 4051 & $6.43\dagger$ & $19^{+1}_{-1}$ & $0.53\dagger$ & -5 & $1.17\dagger$ & $3.68\dagger$ & $5800^\dagger$ \\
& $6.62\dagger$ & $10^{+4}_{-4}$ & $0.26\dagger$ & -3 \\
NGC 4051 & $6.43\dagger$ & $25^{+1}_{-1}$ & $0.53\dagger$ & -5 & $1.17\dagger$ & $3.68\dagger$ & $5800^\dagger$ \\
& $6.62\dagger$ & $12^{+5}_{-5}$ & $0.26\dagger$ & -3 \\
NGC 4151 & $6.38^{+0.01}_{-0.01}$ & $91^{+6}_{-6}$ & $8.79^{+0.56}_{-0.56}$ & -193 & $2.58^{+41.80}_{-2.53}$ & $>3.57$ & $12800^{+1800}_{-4800}$ & -164 \\
& & & & & $145.27^{+66.56}_{-107.01}$ & $>4.41$ & $12800\dagger$ \\
NGC 4593 & $6.42^{+0.01}_{-0.01}$ & $173^{+21}_{-21}$ & $2.23^{+0.27}_{-0.28}$ & -24 & -- & -- & -- \\
& $6.71^{+0.13}_{-0.09}$ & $20^{+10}_{-10}$ & $0.29^{+0.14}_{-0.14}$ & -8 \\
NGC 5506 & 6.63* & $<25$ & $<1.10$ & -15 & -- & -- & -- \\
& $6.98^{+0.06}_{-0.10}$ & $10^{+11}_{-8}$ & $0.88^{+0.97}_{-0.68}$ & -36 \\
NGC 5506 & 6.63* & $<25\dagger$ & $<1.10\dagger$ & -- & -- & -- & -- \\
& $6.98\dagger$ & $10\dagger$ & $0.88\dagger$ & -- \\
NGC 5548 & -- & -- & -- & -- & $0.88^{+0.74}_{-0.54}$ & $3.73^{+0.23}_{-0.29}$ & $<1700$ & -12 \\
NGC 7213 & $6.39^{+0.01}_{-0.01}$ & $67^{+10}_{-11}$ & $1.78^{+0.27}_{-0.28}$ & -45 & -- & -- & -- \\
& $6.61^{+0.04}_{-0.35}$ & $<27$ & $<0.75$ & -31 & \\
& $6.96^{+0.13}_{-0.02}$ & $31^{+15}_{-20}$ & $0.71^{+0.34}_{-0.45}$ & -31 & \\
NGC 7314 & $6.38^{+0.02}_{-0.02}$ & $116^{+16}_{-16}$ & $1.15^{+0.16}_{-0.16}$ & -131 & -- & -- & -- \\
& $6.98^{+0.05}_{-0.06}$ & $38^{+17}_{-17}$ & $0.33^{+0.15}_{-0.15}$ & -12 & \\
PDS 456 & $6.75^{+0.06}_{-0.07}$ & $25^{+17}_{-19}$ & $0.12^{+0.08}_{-0.09}$ & -5 & $>6.31$ & $4.98^{+0.35}_{-0.90}$ & $82200^{+5500}_{-5500}$ & -27 \\
& $7.00^{+0.48}_{-0.13}$ & $19^{+19}_{-17 }$ & $0.09^{+0.09}_{-0.08}$ & -4 & $0.22^{+0.19}_{-0.14}$ & $3.03^{+0.14}_{-0.33}$ & $<163200$  \\
PG 1211+143 & $6.40^{+0.10}_{-0.09}$ & $22^{+16}_{-16}$ & $0.14^{+0.10}_{-0.10}$ & -4 & $7.13^{+14.22}_{-3.04}$ & $3.79^{+0.29}_{-0.09}$ & $18800^{+4800}_{-5300}$ & -30 \\
& $6.75^{+0.07}_{-0.07}$ & $44^{+34}_{-21}$ & $0.24^{+0.18}_{-0.11}$ & -3 & \\
SWIFT J2127.4+5654 & $6.66^{+0.07}_{-0.05}$ & $25^{+11}_{-11}$ & $0.94^{+0.41}_{-0.41}$ & -16 & -- & -- & -- \\
& $6.98^{+0.09}_{-0.26}$ & $16^{+13}_{-13}$ & $0.49^{+0.40}_{-0.40}$ & -4 & \\
\hline
\end{tabular}
\end{table*}

\begin{table*}
\caption{Components for the dual reflector fit to the observations with {\sl Suzaku} XIS, HXD and BAT data from {\sl Swift}. \textsc{comptt} and warm absorber parameters are consistent with those in the baseline model. Here we quote the inner blurred reflector properties, some of which are tied to the out unblurred \textsc{reflionx}. Fits include partial covering geometries where required as per the baseline model. Where a reasonable fit can also be obtained without the use of a partial coverer, both scenarios are tabulated, for those without partial covering we quote the change in $\chi^{2}$ with respect to the dual reflector plus partial covering fit. Note that in some objects accretion disc parameters cannot be constrained, denoted by --. $^{a}$ Ionization parameter given in units erg\,{\rm cm}\,s$^{-1}$.* Denotes a frozen parameter. $\dagger$ $\triangle\chi^{2}$ in relation to the dual reflector fit without partial covering (which for these objects has not been tabulated) compared to the fit with partial covering tabulated here. $^{b}$ Best-fit parameters to the 2009 NGC 3783 {\sl Suzaku} data only for the Solar abundance model presented in Section 4.3.4.}
\begin{tabular}{l c c c c c c c c}
\hline
Object & $\Gamma$ & $Z_{\rm Fe}$ & $\xi^{a}$ & $q$ & $a$ & $i^{\circ}$ & $\triangle\chi^{2}$ & $\chi^{2}_{\nu}$ \\
\hline
3C 111 & $1.60^{+0.02}_{-0.03}$ & $0.8^{+2.4}_{-0.5}$ & $<102$ & $<2.69$ & -- & -- & & 1087.8/1092 \\
3C 120 & $1.62^{+0.01}_{-0.01}$ & 1.0* & $19^{+1}_{-1}$ & $1.7^{+0.3}_{-0.4}$ & -- & $18^{+2}_{-1}$ & & 3564.8/3448 \\
3C 382 & $1.79^{+0.03}_{-0.02}$ & 1.0* & $<1.4$ & $<3$ & -- & $30^{+31}_{-6}$ & & 972.3/933 \\
3C 390.3 & $1.72^{+0.01}_{-0.01}$ & $1.4$ & $<2$ & $1.8^{+1.0}_{-0.5}$ & -- & $39^{+20}_{-7}$ & & 1478.7/1482 \\
Ark 120 & $1.98^{+0.04}_{-0.04}$ & $2.2^{+0.1}_{-0.6}$ & $<21$ & $2.2^{+0.1}_{-0.6}$ & -- & $45^{+6}_{-5}$ & & 715.5/644 \\
Fairall 9 & $1.91^{+0.01}_{-0.01}$ & $1.7^{+0.1}_{-0.1}$ & $3^{+1}_{-1}$ & $2.7^{+0.6}_{-0.5}$ & $<0.95$ & $42^{+3}_{-2}$ & & 3558.3/3271 \\
IC 4329A & $1.93^{+0.02}_{-0.01}$ & $0.8^{+0.1}_{-0.1}$ & $7^{+2}_{-1}$ & 2.4* & $<0.73$ & 36* & & 2342.8/2199 \\
MCG--02-14-009 & $1.90^{+0.02}_{-0.02}$ & $0.7^{+0.5}_{-0.3}$ & $<13$ & $<2.0$ & -- & $>29$ & & 601.5/539 \\
MCG--05-23-16 & $1.84^{+0.01}_{-0.02}$ & 1.0* & $<1.0$ & $1.6^{+0.5}_{-0.7}$ & -- & 24* & & 1989.5/1890 \\
MCG--06-30-15 (p/c) & $2.05^{+0.01}_{-0.01}$ & 1.0* & $<11$ & $2.3^{+0.2}_{-0.1}$ & $0.61^{+0.15}_{-0.17}$ & $35^{+2}_{-2}$ & & 2061.4/1823 \\
MCG--06-30-15 (no p/c) & $2.09^{+0.01}_{-0.01}$ & 1.0* & $<10$ & $2.9^{+0.2}_{-0.1}$ & -- & $36^{+1}_{-1}$ & --41& 2020.2/1823 \\
MCG+8-11-11 & $1.82^{+0.01}_{-0.02}$ & 1.0* &  $7^{+3}_{-2}$ & 1.9* & -- & -- & & 1034.0/933 \\
MR 2251--178 & $1.54^{+0.06}_{-0.10}$ & 1.0* & -- & 3.0* & -- & -- & & 969.0/897 \\
Mrk 79 & $1.61^{+0.02}_{-0.02}$ & 1.0* & $984^{+319}_{-396}$ & 3.0* & $<-0.25$ & $<25$ & & 573.2/542 \\
Mrk 335 & $2.04^{+0.02}_{-0.03}$ & $2.5^{+1.1}_{-0.6}$ & $25^{+7}
_{-3}$ & $2.0^{+0.2}_{-0.3}$ & -- & $50^{+9}_{-13}$ & & 820.0/721 \\
Mrk 509 & $1.69^{+0.03}_{-0.05}$ & 1.0* & $23^{+2}_{-4}$ & $1.5^{+0.8}_{-1.3}$ & -- & 35* & & 1959.4/1867 \\
Mrk 766 (p/c) & $1.99^{+0.01}_{-0.01}$ & 1.0* & $3^{+1}_{-1}$ & -- & -- & -- & & 1055.1/995 \\
Mrk 766 (no p/c) & $2.01^{+0.01}_{-0.01}$ & 1.0* & $4^{+1}_{-1}$ & 3.0* & -- & $80^{+2}_{-3}$ & +54 & 1109.6/1002 \\
Mrk 841 & $1.96^{+0.05}_{-0.04}$ & $1.0^{+0.3}_{-0.2}$ & $<2$ & $2.7^{+0.3}_{-0.2}$ & $>-0.40$ & -- & & 910.8/853 \\
NGC 1365 & $1.69^{+0.01}_{-0.01}$ & 1.0* & $25^{+1}_{-1}$ & 3.0* & $>0.70$ & $83^{+7}_{-4}$ & +7995$\dagger$ & 2157.0/1974 \\
NGC 2992 & $1.60^{+0.04}_{-0.05}$ & 1.0* & $59^{+16}_{-14}$ & $<1.8$ & -- & -- & & 1082.3/1076 \\
NGC 3227 & $1.80^{+0.01}_{-0.01}$ & $0.4^{+0.1}_{-0.1}$ & $6^{+1}_{-1}$ & $2.5^{+0.3}_{-0.3}$ & $<-0.35$ & $47^{+3}_{-2}$ & +433$\dagger$ & 4373.9/4075 \\
NGC 3516 & $1.70^{+0.01}_{-0.01}$ & 1.0* & $6^{+1}_{-1}$ & $2.6^{+0.1}_{-0.3}$ & $<-0.50$ & -- & +814$\dagger$ & 1228.6/1121 \\
NGC 3783 & $1.79^{+0.01}_{-0.01}$ & $1.1^{+0.1}_{-0.1}$ & $4^{+1}_{-1}$ & $3.9^{+1.4}_{-0.6}$ & $<-0.35$ & $19^{+4}_{-7}$ & & 2502.6/2300 \\
NGC 3783$^{b}$ & $1.84^{+0.05}_{-0.01}$ & $1.0\pm0.2$ & $<11$ & $3.0^{+0.5}_{-0.5}$ & $<0.45$ & $<13$ & & 1413.7/1374 \\
NGC 4051 (p/c) & $1.89^{+0.01}_{-0.01}$ & $1.1^{+0.2}_{-0.1}$ & $18^{+2}_{-3}$ & 3.0* & -- & -- & & 3177.1/2939 \\
NGC 4051 (no p/c) & $1.93^{+0.01}_{-0.01}$ & $0.8^{+0.1}_{-0.1}$ & $12^{+1}_{-1}$ & $6.1^{+0.3}_{-0.1}$ & $>0.99$ & $13^{+2}_{-2}$ & +137 & 3321.9/2943 \\
NGC 5506 (p/c) & $2.08^{+0.02}_{-0.03}$ & $0.7^{+0.1}_{-0.1}$ & $10^{+1}_{-1}$ & 3.0* & -- & $26^{+5}_{-4}$ & & 3191.2/2868 \\
NGC 5506 (no p/c) & $2.07^{+0.01}_{-0.01}$ & $0.8^{+0.1}_{-0.1}$ & $9^{+1}_{-1}$ & $1.7^{+0.3}_{-0.8}$ & -- & $48^{+19}_{-4}$ & -12 & 3179.5/2871 \\
NGC 7469 & $1.80^{+0.01}_{-0.01}$ & $0.9^{+0.5}_{-0.2}$ & $<13$ & $1.7^{+0.1}_{-0.8}$ & $0.72^{+0.18}_{-0.17}$ & $80^{+8}_{-5}$ & & 780.9/805 \\
SWIFT J2127.4+5654 & $2.19^{+0.05}_{-0.03}$ & $0.9^{+0.3}_{-0.3}$ & $<18$ & $2.2^{+0.4}_{-0.9}$ & -- & $42^{+16}_{-7}$ & & 830.8/866 \\
\hline
\end{tabular}
\label{tab:dual}
\end{table*}


\begin{table*}
\caption{List of 15-50\,keV component fluxes for each object and the full model 2-10\,keV flux. The full model flux includes all model components, reflector flux is the the 15-50\,keV flux of the \textsc{reflionx} component in the HXD whereas the continuum flux is the model flux minus any contribution from reflection or partial covering. The hardness ratio is the ratio of the full model 15-50\,keV to 2-10\,keV flux, a powerlaw with slope $\Gamma=2.0$ gives a hardness ratio of 0.75. The reflection fraction, $R_{15-50}$, is the ratio of the 15-50\,keV reflector flux to the 15-50\,keV continuum flux i.e. to the full model minus reflector flux, Equation \ref{eq:reflection_fraction}. Flux given in units 10$^{-11}$erg\,{\rm cm}$^{-2}$\,s$^{-1}$.}
\begin{tabular}{l c c c c c c}
\hline
Object &  Full model & Reflector & Continuum & Full model 2-10\,keV flux & Hardness ratio & $R_{15-50}$ \\
\hline
1H 0419--577 Obs 1 & $2.848\pm0.110$ & $0.237^{+0.138}_{-0.092}$ & $2.156\pm0.055$ & 1.753 & $1.625\pm0.063$ & $0.091^{+0.053}_{-0.036}$ \\
1H 0419--577 Obs 2 & $2.315\pm0.083$ & $0.301^{+0.187}_{-0.114}$ & $1.719\pm0.055$ & 1.374 & $1.685\pm0.060$ & $0.149^{+0.094}_{-0.058}$ \\
3C 111 & $3.292\pm0.097$ & 0 & $3.292\pm0.097$ & 1.947 & $1.691\pm0.050$ & -- \\
3C 120 Obs 1 & $6.236\pm0.239$ & $0.889^{+0.139}_{-0.130}$ & $5.347\pm0.055$ & 4.629 & $1.347\pm0.052$ & $0.166^{+0.027}_{-0.026}$ \\
3C 120 Obs 2 & $5.788\pm0.187$ & $1.082^{+0.162}_{-0.139}$ & $4.706\pm0.055$ & 3.963 & $1.461\pm0.047$ & $0.230^{+0.036}_{-0.032}$ \\
3C 382 & $5.108\pm0.105$ & $1.093^{+0.092}_{-0.609}$ & $4.015\pm0.035$ & 4.045 & $1.263\pm0.026$ & $0.272^{+0.025}_{-0.157}$ \\
3C 390.3 & $5.812\pm0.182$ & $1.444^{+1.487}_{-0.396}$ & $4.368^{+0.058}_{-0.116}$ & 3.101 & $1.874\pm0.059$ & $0.331^{+0.359}_{-0.096}$ \\
3C 445 & $2.692\pm0.110$ & $0.902^{+3.278}_{-0.333}$ & $1.790\pm0.597$ & 0.699 & $3.851\pm0.157$ & $0.504^{+2.051}_{-0.211}$ \\
4C 74.26 & $3.996\pm0.095$ & $0.821^{+0.908}_{-0.541}$ & $3.175\pm0.030$ & 3.127 & $1.278\pm0.030$ & $0.259^{+0.295}_{-0.176}$ \\
Ark 120 & $3.858\pm0.102$ & $1.188^{+0.562}_{-0.677}$ & $2.670\pm0.169$ & 3.051 & $1.263\pm0.033$ & $0.445^{+0.231}_{-0.278}$ \\
Ark 564 & $1.536\pm0.146$ & $0.078^{+0.124}_{-0.059}$ & $0.822\pm0.007$ & 1.837 & $0.836\pm0.079$ & $0.053^{+0.085}_{-0.041}$ \\
Fairall 9 Obs 1 & $3.370\pm0.076$ & $1.269^{+0.091}_{-0.084}$ & $2.101\pm0.031$ & 2.317 & $1.454\pm0.033$ & $0.604^{+0.055}_{-0.052}$ \\
Fairall 9 Obs 2 & $3.511\pm0.180$ & $1.043^{+0.070}_{-0.070}$ & $2.468\pm0.034$ & 2.165 & $1.622\pm0.083$ & $0.423^{+0.044}_{-0.044}$ \\
IC 4329A & $16.778\pm0.139$ & $5.309^{+0.104}_{-0.760}$ & $11.469\pm0.033$ & 10.711 & $1.566\pm0.013$ & $0.463^{+0.011}_{-0.073}$ \\
IRAS 13224--3809 & $0.021\pm0.011$ & 0 & $<0.009$ & 0.055 & $0.382\pm0.200$ & -- \\
MCG--02-14-009 & $0.789\pm0.093$ & $0.413^{+0.091}_{-0.171}$ & $0.376^{+0.125}_{-0.094}$ & 0.430 & $1.835\pm0.216$ & $1.098^{+0.451}_{-0.728}$ \\
MCG--02-58-22 & $8.726\pm0.107$ & $1.752^{+0.301}_{-0.313}$ & $6.974\pm0.074$ & 4.872 & $1.791\pm0.022$ & $0.251^{+0.045}_{-0.046}$ \\
MCG--05-23-16 & $14.518\pm0.139$ & $1.797^{+0.137}_{-0.321}$ & $12.721\pm0.133$ & 8.929 & $1.626\pm0.016$ & $0.141^{+0.011}_{-0.026}$ \\ 
MCG--06-30-15 & $5.013\pm0.067$ & $0.584^{+0.050}_{-0.160}$ & $3.093^{+0.095}_{-0.076}$ & 4.160 & $1.205\pm0.016$ & $0.132^{+0.012}_{0.036}$ \\
MCG+8-11-11 & $10.440\pm0.160$ & $3.111^{+1.507}_{-1.020}$ & $7.329\pm0.039$ & 6.480 & $1.611\pm0.025$ & $0.424^{+0.224}_{-0.151}$ \\
MR 2251--178 & $5.465\pm0.108$ & 0 & $5.365\pm0.108$ & 4.230 & $1.292\pm0.026$ & -- \\
Mrk 79 & $1.875\pm0.139$ & 0 & $1.875\pm0.139$ & 1.466 & $1.279\pm0.095$ & -- \\
Mrk 110 & $2.990\pm0.113$ & $0.267^{+0.559}_{-0.083}$ & $2.723\pm0.052$ & 2.126 & $1.406\pm0.053$ & $0.098^{+0.206}_{-0.031}$ \\
Mrk 205 & $1.184\pm0.062$ & $0.299^{+0.040}_{-0.113}$ & $0.711^{+0.023}_{-0.046}$ & 0.934 & $1.268\pm0.066$ & $0.338^{+0.053}_{-0.137}$ \\
Mrk 279 & $2.322\pm0.101$ & $0.736^{+0.187}_{-0.194}$ & $0.554\pm0.050$ & 0.489 & $4.748\pm0.207$ & $0.464^{+0.133}_{-0.138}$ \\
Mrk 335 & $1.502\pm0.125$ & $0.288^{+0.088}_{-0.088}$ & $1.214^{+0.024}_{-0.048}$ & 1.490 & $1.008\pm0.084$ & $0.237^{+0.078}_{-0.078}$ \\
Mrk 359 & $0.984\pm0.141$ & $0.337^{+0.285}_{-0.110}$ & $0.647\pm0.054$ & 0.515 & $1.912\pm0.274$ & $0.521^{+0.509}_{-0.223}$ \\
Mrk 509 & $7.186\pm0.162$ & $0.966^{+6.077}_{-0.601}$ & $6.220\pm0.070$ & 4.723 & $1.521\pm0.034$ & $0.161^{+0.989}_{-0.098}$ \\
Mrk 766 Obs 1 & $1.476\pm0.118$ & $0.329^{+0.072}_{-0.059}$ & $1.140\pm0.023$ & 1.325 & $1.181\pm0.094$ & $0.287^{+0.072}_{-0.061}$ \\
Mrk 766 Obs 2 & $1.846\pm0.154$ & $0.414^{+0.091}_{-0.078}$ & 0 & 1.363 & $1.354\pm0.113$ & $0.289^{+0.073}_{-0.065}$ \\
Mrk 841 & $2.618\pm0.105$ & $0.948^{+0.570}_{-0.338}$ & $1.670\pm0.046$ & 1.416 & $1.849\pm0.074$ & $0.568^{+0.394}_{-0.235}$ \\
NGC 1365 Obs 1& $4.880\pm0.116$ & $0.752^{+0.263}_{-0.056}$ & $0.162\pm0.054$ & 1.284 & $6.489\pm0.154$ & $0.182^{+0.065}_{-0.015}$ \\
NGC 1365 Obs 2 & $4.428\pm0.127$ & $0.860^{+0.070}_{-0.063}$ & $0.143\pm0.073$ & 0.610 & $7.259\pm0.208$ & $0.241^{+0.022}_{-0.020}$ \\
NGC 1365 Obs 3 & $3.615\pm0.060$ & $0.909^{+0.038}_{-0.051}$ & $0.146\pm0.075$ & 0.385 & $9.390\pm0.116$ & $0.336^{+0.017}_{-0.021}$ \\
NGC 2992 & $2.106\pm0.101$ & $0.620^{+0.113}_{-0.113}$ & $1.486\pm0.057$ & 1.182 & $1.782\pm0.085$ & $0.417^{+0.087}_{-0.087}$ \\
NGC 3147 & $0.294\pm0.033$ & $0.072^{+0.098}_{-0.098}$ & $0.223\pm0.056$ & 0.165 & $1.782\pm0.200$ & $0.324^{+0.467}_{-0.049}$ \\
NGC 3227 Obs 1 & $7.564\pm0.156$ & $3.010^{+0.167}_{-0.169}$ & $3.209^{+0.043}_{-0.012}$ & 3.956 & $2.357\pm0.049$ & $0.661^{+0.049}_{-0.030}$ \\
NGC 3227 Obs 2 & $6.705\pm0.164$ & $5.218^{+0.147}_{-0.144}$ & $0.365^{+0.011}_{-0.014}$ & 1.853 & $3.618\pm0.089$ & $3.509^{+0.529}_{-0.524}$ \\
NGC 3227 Obs 3 & $6.858\pm0.170$ & $3.419^{+0.153}_{-0.152}$ & $0.682\pm0.016$ & 2.551 & $2.688\pm0.067$ & $0.994^{+0.079}_{-0.079}$ \\
NGC 3227 Obs 4 & $4.704\pm0.170$ & $4.127^{+0.113}_{-0.111}$ & $0.228\pm0.013$ & 0.996 & $4.723\pm0.171$ & $7.153^{+2.538}_{-2.524}$ \\
NGC 3227 Obs 5 & $6.194\pm0.179$ & $3.128^{+0.129}_{-0.127}$ & $0.360\pm0.014$ & 2.138 & $2.897\pm0.084$ & $1.020^{+0.085}_{-0.084}$ \\
NGC 3227 Obs 6 & $5.283\pm0.193$ & $3.250^{+0.160}_{-0.158}$ & $0.221^{+0.011}_{-0.009}$ & 1.571 & $3.363\pm0.123$ & $1.599^{+0.212}_{-0.211}$ \\
\hline
\end{tabular}
\label{tab:fluxes}
\end{table*}
\begin{table*}
\contcaption{List of 15-50\,keV component fluxes for each object and the full model 2-10\,keV flux. The full model flux includes all model components, reflector flux is the the 15-50\,keV flux of the \textsc{reflionx} component in the HXD whereas the continuum flux is the model flux minus any contribution from reflection or partial covering. The hardness ratio is the ratio of the full model 15-50\,keV to 2-10\,keV flux, a powerlaw with slope $\Gamma=2.0$ gives a hardness ratio of 0.75. The reflection fraction, $R_{15-50}$, is the ratio of the 15-50\,keV reflector flux to the 15-50\,keV continuum flux i.e. to the full model minus reflector flux. Flux given in units 10$^{-11}$erg\,{\rm cm}$^{-2}$\,s$^{-1}$.}
\begin{tabular}{l c c c c c c}
\hline
Object &  Full model & Reflector & Continuum & Full model 2-10\,keV flux & Hardness ratio & $R_{15-50}$ \\
\hline
NGC 3516 Obs 1 & $7.321\pm0.117$ & $2.093^{+0.132}_{-0.120}$ & $0.920\pm0.061$ & 2.370 & $3.089\pm0.049$ & $0.400^{+0.029}_{-0.026}$ \\
NGC 3516 Obs 2 & $3.340\pm0.057$ & $1.198^{+0.073}_{-0.073}$ & $0.920\pm0.061$ & 2.370 & $1.409\pm0.024$ & $0.559^{+0.042}_{-0.042}$ \\
NGC 3783 Obs 1 & $9.917\pm0.178$ & $4.465^{+0.145}_{-0.145}$ & $5.452\pm0.042$ & 4.586 & $2.162\pm0.039$ & $0.819^{+0.044}_{-0.026}$ \\
NGC 3783 Obs 2 & $12.003\pm0.100$ & $4.650^{+0.152}_{-0.146}$ & $7.353\pm0.043$ & 5.921 & $2.027\pm0.017$ & $0.632^{+0.026}_{-0.025}$ \\
NGC 4051 Obs 1 & $1.648\pm0.089$ & $0.592^{+0.039}_{-0.033}$ & $0.339\pm0.034$ & 0.873 & $1.888\pm0.102$ & $0.561^{+0.063}_{-0.059}$ \\
NGC 4051 Obs 2 & $3.223\pm0.052$ & $0.887^{+0.058}_{-0.046}$ & $2.116\pm0.035$ & 2.464 & $1.308\pm0.021$ & $0.380^{+0.028}_{-0.023}$ \\
NGC 4051 Obs 3 & $2.591\pm0.110$ & $0.823^{+0.071}_{-0.076}$ & $1.322\pm0.034$ & 1.794 & $1.444\pm0.061$ & $0.465^{+0.033}_{-0.036}$ \\
NGC 4151 & $15.901\pm0.109$ & $7.825^{+1.865}_{-0.917}$ & $4.512\pm0.167$ & 4.352 & $3.654\pm0.025$ & $0.969^{+0.322}_{-0.159}$ \\
NGC 4593 & $2.152\pm0.130$ & $0.388^{+2.061}_{-0.293}$ & $1.764\pm0.084$ & 1.041 & $2.067\pm0.124$ & $0.220^{+1.196}_{-0.171}$ \\
NGC 5506 Obs 1 & $17.800\pm0.153$ & $6.279^{+0.227}_{-0.220}$ & $9.755\pm0.036$ & 10.379 & $1.715\pm0.015$ & $0.545^{+0.024}_{-0.023}$ \\
NGC 5506 Obs 2 & $17.114\pm0.210$ & $5.952^{+0.218}_{-0.218}$ & $9.375\pm0.035$ & 9.890 & $1.730\pm0.021$ & $0.533^{+0.024}_{-0.024}$ \\
NGC 5548 & $3.998\pm0.087$ & $1.364^{+0.817}_{-0.483}$ & $2.624\pm0.060$ & 1.841 & $2.172\pm0.047$ & $0.518^{+0.350}_{-0.207}$ \\
NGC 7213 & $3.477\pm0.109$ & $0.208^{+0.397}_{-0.126}$ & $3.269\pm0.054$ & 2.410 & $1.443\pm0.045$ & $0.064^{+0.222}_{-0.039}$ \\
NGC 7314 & $1.427\pm0.110$ & 0 & $1.427\pm0.110$ & 0.879 & $1.623\pm0.125$ & -- \\
NGC 7469 & $3.478\pm0.102$ & $0.923^{+0.076}_{-0.158}$ & $2.555^{+0.179}_{-0.269}$ & 2.102 & $1.655\pm0.049$ & $0.361^{+0.035}_{-0.007}$ \\
PDS 456 & $0.249\pm0.062$ & 0 & $0.130\pm0.006$ & 0.353 & $0.795\pm0.176$ & -- \\
PG 1211+143 & $0.498\pm0.249$ & 0 & $0.498\pm0.249$ & 0.391 & $1.274\pm0.637$ & -- \\
RBS 1124 & $0.905\pm0.057$ & $0.211^{+2.862}_{-0.138}$ & $0.694\pm0.058$ & 0.494 & $1.832\pm0.115$ & $0.304^{+4.310}_{-0.209}$ \\
SWIFT J2127.4+5654 & $3.533\pm0.095$ & $1.276^{+0.207}_{-0.621}$ & $2.257^{+0.059}_{-0.074}$ & 3.348 & $1.055\pm0.028$ & $0.565^{+0.108}_{-0.317}$ \\
TON S180 & $0.773\pm0.129$ & $0.052^{+0.016}_{-0.147}$ & $0.328\pm0.014$ & 0.564 & $1.371\pm0.229$ & $0.072^{+0.026}_{-0.072}$ \\
\hline
\end{tabular}
\end{table*}


\begin{thebibliography}{100}
\bibitem{Anders Grevesse 1989}
Anders E., Grevesse N., Geochimica et Cosmochimica Acta, 1989, 53, 197

\bibitem{Arnaud 1996}
Arnaud K.A., 1996, {\sl Astronomical Data Analysis Software and Systems V}, eds. Jacoby, G., Barnes, J., pg17, ASP Conf. Series Volume 101

\bibitem{Bania 1991}
Bania T., et al, 1991, ApJ, 101, 2147

\bibitem{Baumgartner 2010} 
Baumgartner W.H. et al., 2010, ApJS submitted

\bibitem{Berti Volonteri 2008}
Berti E., Volonteri M., 2008, ApJ, 684, 822

\bibitem{Bianchi et al 2004}
Bianchi S., Matt G., Balestra I., Guainazzi M., Perola G.C., 2004, A\&A, 422, 65

\bibitem{Bianchi et al 2009}
Bianchi S., Guainazzi M., Matt G., Fonseca Bonilla N., Ponit G., 2009, A\&A, 495, 421

\bibitem{Blandford Znajek 1977}
Blandford R.D., Znajek R.L., 1977, MNRAS, 179, 433

\bibitem{Blustin 2002}
Blustin A.J., Branduardi-Raymont G., Behar E., Kaastra J.S., Kahn S.M., Page M.J., Sako M., Steenbrugge K.C., 2002, A\&A, 392, 453

\bibitem{Braito 2007}
Braito V., et al., 2007, ApJ, 670, 978

\bibitem{Brenneman Reynolds 2006}
Brenneman L.W., Reynolds C.S., 2006, ApJ, 652, 1028

\bibitem{Brenneman 2011}
Brenneman L.W., et al., 2011, ApJ, 736, 103

\bibitem{Dauser 2010}
Dauser T., Wilms J., Reynolds C.S., Brenneman L.W., 2010, MNRAS, 409, 1534

\bibitem{de la Calle Perez}
de la Calle P\'{e}rez I., et al., 2010, A\&A, 524, A50

\bibitem{Dovciak 2004}
Dov\v{c}iak M., Karas V., Yaqoob T., 2004, ApJS, 153, 205 

\bibitem{Emmanoulopoulos 2011}
Emmanoulopoulos D., Papadakis I.E., McHardy I.M., Nicastro F., Bianchi S., Ar\'{e}valo P., 2011, MNRAS, 415, 1895

\bibitem{Fabian et al 1989}
Fabian A.C., Rees M.J., Stella L., White N.E., 1989, MNRAS, 238, 729

\bibitem{Fabian et al 2012}
Fabian A.C., et al., 2012, MNRAS, 419, 116

\bibitem{Fukazawa et al 2009}
Fukazawa Y., et al., 2009, PASJ, 61, 17

\bibitem{Fukazawa et al 2011}
Fukazawa Y., et al., 2011, ApJ, 727, 19

\bibitem{Gallo 2011}
Gallo L.C., Miniutti G., Miller J.M., Brenneman L.W., Fabian A.C., Guainazzi M., Reynolds C.S., 2011, MNRAS, 411, 607


\bibitem{George Fabian 1991}
George I.M., Fabian A.C., 1991, MNRAS, 249, 352

\bibitem{Grevesse 1998}
Grevesse N., Sauval, A.J., 1998, Space Science Reviews, 85, 161

\bibitem{Gruber 1999}
Gruber D. E., Matteson J. L., Peterson L. E., Jung G. V., 1999, ApJ, 520, 124

\bibitem{Gierlinski Done 2004}
Gierli\'{n}ski M., Done C., 2004, MNRAS, 349, L7

\bibitem{Gofford 2012}
Gofford J.A., et al., 2012, in prep

\bibitem{Hughes Blandford 2003}
Hughes S.A., Blandford R.D., 2003, ApJ, 585, 101

\bibitem{Jiang et al 2006}
Jiang P., Wang J.X., Wang T.G., 2006, ApJ, 644, 725

\bibitem{Kalberla 2005}
Kalberla P.M.W., et al., 2005, A\&A, 440, 775

\bibitem{Kallman 2004}
Kallman T.R., Palmeri P., Bautista M. A., Mendoza C., Krolik J. H., 2004, ApJS, 155, 675

\bibitem{Kaspi 2002}
Kaspi S., et al., 2002, ApJ, 574, 643

\bibitem{King 2005}
King A.E., Lubow A.H., Ogilvie G.I., Pringle J.E., 2005, MNRAS, 363, 49

\bibitem{King Pringle 2007}
King A.E., Pringle J.E., 2007, MNRAS, 377, 25

\bibitem{Koyama et al 2007}
Koyama K. 
et al., 2007, PASJ, 59S, 23

\bibitem{Krolik and Kallman 1987}
Krolik J.H., Kallman T.R., 1987, ApJ, 320, 5

\bibitem{Laor 1991}
Laor A., 1991, ApJ, 376, 90

\bibitem{Lobban 2010}
Lobban A.P., Reeves J.N., Porquet D., Braito V., Markowitz A.G., Miller L., Turner T.J., 2010, MNRAS, 508, 551

\bibitem{Lobban 2011}
Lobban A.P., Reeves J.N, Miller L., Turner T.J., Braito V., Kraemer S.B., Crenshaw D.M., 2011, MNRAS, 414, 1965

\bibitem{Magdziarz  Zdziarski 1995} 
Magdziarz P., Zdziarski A.A., 1995, MNRAS, 273, 837

\bibitem{Malkan Sargent 1982}
Malkan M.A., Sargent W.L.W., 1982, ApJ, 254, 22

\bibitem{Markowitz 2008}
Markowitz A.G. et al., 2008, PASJ, 60S, 277

\bibitem{Matsumoto et al 2003}
Matsumoto C., Inoue H., Fabian A.C., Iwasawa K., 2003, PASJ, 55, 615

\bibitem{Mehdipour 2011}
Mehdipour M., et al., 2011, A\&A, 534, 39

\bibitem{Miller 2008}
Miller L., Turner T.J., Reeves J.N., 2008, A\&A, 483, 437

\bibitem{Miller 2009}
Miller L., Turner T.J., Reeves J.N., 2009, MNRAS, 399, 69

\bibitem{Miniutti Fabian 2004}
Miniutti G., Fabian A.C., 2004, MNRAS, 349, 1435

\bibitem{Miniutti 2007}
Miniutti G., et al., 2007, PASJ, 59S, 315

\bibitem{Miniutti 2009}
Miniutti G., Panessa F., De Rosa A., Fabian A.C., Malizia A., Molina M., Miller J.M., Vaughan S., 2009, MNRAS, 398, 255

\bibitem{Miniutti 2010}
Miniutti G., Piconcelli E., Bianchi S., Vignali C., Bozzo E., 2010, MNRAS, 401, 1315

\bibitem{Miyakawa et al 2009}
Miyakawa T., Ebisawa K., Terashima Y., Tsuchihashi F., Inoue H., Zycki P., 2009, PASJ, 61, 1355

\bibitem{Morrison McCammon}
Morrison R., McCammon D., 1983, ApJ, 270, 119

\bibitem{Nandra Pounds 1994}
Nandra K., Pounds K.A., 1994, MNRAS, 268, 405

\bibitem{Nandra 2006}
Nandra K., 2006, MNRAS, 368, 62

\bibitem{Nandra et al 2007}
Nandra K., O'Neill P.M., George I.M., Reeves J.N., 2007, MNRAS, 382, 194

\bibitem{Nardini et al 2010}
Nardini E., Fabian A.C., Reis R.C., Walton D.J., 2011, MNRAS, 410, 1251

\bibitem{Patrick 2011a}
Patrick A.R., Reeves J.N., Porquet D., Markowitz A.G., Lobban A.P., Tershima Y., 2011a, MNRAS, 411, 2353

\bibitem{Patrick 2011b}
Patrick A.R., Reeves J.N., Lobban A.P., Porquet D., Markowitz A.G., 2011b, MNRAS, 416, 2725

\bibitem{Porquet et al 2004}
Porquet D., Reeves J.N., O'Brien P., Brinkmann W., 2004, A\&A, 422, 85

\bibitem{Reeves 2004}
Reeves J.N., Nandra K., George I.M., Pounds K.A., Turner T.J., Yaqoob T., 2004, ApJ, 602, 648

\bibitem{Reynolds et al 2005}
Reynolds C.S., Brenneman L.W., Garofalo D., 2005, Ap\&SS, 300, 71

\bibitem{Reynolds 2012a}
Reynolds C. S., Brenneman, L. W., Lohfink A. M., Trippe M. L., Miller J. M., Reis R. C., Nowak M. A., Fabian A. C., 2012a, AIPC, 1427, 157

\bibitem{Reynolds 2012b}
Reynolds C. S., Brenneman, L. W., Lohfink A. M., Trippe M. L., Miller J. M., Fabian A. C., Nowak M. A., 2012b, ApJ, 755, 88

\bibitem{Rezzolla 2008}
Rezzolla L., Barausse E., Dorband E.N., Pollney D., Reisswig C., Seiler J., Husa S., 2008, Phys. Rev. D, 78, 044002

\bibitem{Risaliti et al 2009}
Risaliti G., et al., 2009, ApJ, 696, 160

\bibitem{Rivers Markowitz Rothschild 2011a}
Rivers E., Markowitz A.G., Rothschild R., 2011a, ApJS, 193, 3

\bibitem{Rivers Markowitz Rothschild 2011b}
Rivers E., Markowitz A.G., Rothschild R., 2011b, ApJ, 742, 29

\bibitem{Ross Fabian Ballantyne 2002}
Ross R.R., Fabian A.C., Ballantyne D.R., 2002, MNRAS, 336, 315

\bibitem{Ross Fabian 2005}
Ross R.R., Fabian A.C., 2005, MNRAS, 358, 211

\bibitem{Schmoll 2009}
Schmoll S. 
et al., 2009, ApJ, 703, 2171

\bibitem{Scott 2011}
Scott A. E., Stewart G. C., Mateos S., Alexander D. M., Hutton S., 
Ward M. J., 2011, MNRAS, 417, 992

\bibitem{Sim 2010}
Sim S.A., Miller L., Long K.S., Turner T.J., Reeves J.N., 2010, MNRAS, 404, 1369

\bibitem{Takahashi et al 2007}
Takahashi T. 
et al., 2007, PASJ, 59S, 35

\bibitem{Tatum 2012}
Tatum M.M., Turner T.J., Sim S.A., Miller L., Reeves J.N., Patrick A.R., Long K.S., 2012, ApJ, 752, 94

\bibitem{Titarchuk 1994}
Titarchuk L., 1994, ApJ, 434, 313

\bibitem{Tombesi 2010a}
Tombesi F., Sambruna R.M, Reeves J.N., Braito V., Ballo L., Gofford J., Cappi M., Mushotzky R.F., 2010a, ApJ, 719, 700

\bibitem{Tombesi 2010b}
Tombesi F., Cappi M., Reeves J.N., Palumbo G.G.C., Yaqoob T., Braito V., Dadina M., 2010b, A\&A, 521, 57

\bibitem{Tombesi 2011}
Tombesi, F., Cappi, M., Reeves, J. N., Palumbo, G. G. C., Braito, V., 
Dadina, M., 2011, ApJ, 742, 44

\bibitem{Turner 2005}
Turner T.J., Kraemer S.B., George I.M., Reeves J.N., Bottorff M.C., 2005, ApJ, 618, 155

\bibitem{Turner 2009}
Turner T.J., Miller L., Kraemer S.B., Reeves J.N., Pounds K.A., 2009, ApJ, 698, 99

\bibitem{Turner et al. 2011}
Turner T.J., Miller L., Kraemer S.B., Reeves J.N., 2011, ApJ, 733, 48

\bibitem{Volonteri 2005}
Volonteri M. et al., 2005, ApJ, 620, 69

\bibitem{Wilms 2000}
Wilms J., Allen A., McCray R., 2000, ApJ, 542, 91

\bibitem{Yaqoob 2005}
Yaqoob T., Reeves J. N., Markowitz A., Serlemitsos P. J., Padmanabhan U., 2005, ApJ, 627, 156

\bibitem{Zycki 2010}
Zycki P.T., Ebisawa K., Niedzwiecki A., Miyakawa T., 2010, PASJ, 62, 1185

\end{thebibliography}
\end{document}